\pdfoutput=1

\documentclass[12pt]{antonio}

\usepackage{graphicx}

\def\theequation{\thesection.\arabic{equation}}

\newcommand{\bea} {\begin{eqnarray*}}
\newcommand{\beq} {\begin{equation}}
\newcommand{\bey} {\begin{eqnarray}}
\newcommand{\eea} {\end{eqnarray*}}
\newcommand{\eeq} {\end{equation}}
\newcommand{\eey} {\end{eqnarray}}

\newcommand{\dif} {{\rm d}}
\newcommand{\dps} {\displaystyle}

\newcommand{\lge} {\langle}

\newcommand{\meio} {{1 \over 2}}

\newcommand{\nd} {\noindent}
\newcommand{\ovl} {\overline}
\newcommand{\ptl} {\partial}

\newcommand{\rge} {\rangle}

\newcommand{\vv} {\vspace{4mm}}
\newcommand{\vpq} {\vspace{2mm}}

\newcommand{\af} {\alpha}
\newcommand{\bt} {\beta}
\newcommand{\dt} {\delta}
\newcommand{\Dt} {\Delta}
\newcommand{\eps} {\epsilon}

\newcommand{\lb} {\lambda}

\newcommand{\Rho} {\varrho}

\newcommand{\ta} {\theta}

\newcommand{\vf} {\varphi}
\newcommand{\vps} {\varepsilon}
\newcommand{\zt} {\zeta}

\newcommand{\cala} {{\cal A}}
\newcommand{\calc} {{\cal C}}
\newcommand{\cald} {{\cal D}}
\newcommand{\cale} {{\cal E}}
\newcommand{\calh} {{\cal H}}
\newcommand{\caln} {{\cal N}}
\newcommand{\calt} {{\cal T}}

\begin{document}

\pagestyle{empty}

\vspace*{2cm}

\begin{center}

{\LARGE\bf

SOLUTION TO THE GRAD-SHAFRANOV \\ BOUNDARY VALUE PROBLEM \\FOR A
THERMONUCLEAR PLASMA CONTAINED IN A TOROIDAL VASE WITH A
CONDUCTING WALL \\UNDER THE ASSUMPTION OF CONSTANT SOURCES TO THE
EQUILIBRIUM IN FLUX SPACE BY THE METHOD OF EXPANSION OF THE
POLOIDAL FLUX FUNCTION IN SERIES OF MULTIPOLE SOLUTIONS

}

\vspace{1.5cm}

{\Large\it by}

\vspace{1.5cm}

{\LARGE\bf A. FERREIRA}

\vfill


\end{center}

\newpage


\begin{center}

{\baselineskip=10mm

{\Large\bf Solution to the Grad-Shafranov Boundary Value Problem for
a Thermonuclear Plasma Contained in a Toroidal Vase with a
Conducting Wall under the Assumption of Constant Sources to the
Equilibrium in Flux Space by the Method of Expansion of the Poloidal
Flux Function in Series of Multipole Solutions }

}

\vspace{1cm}

{\Large\bf A. Ferreira}

\vspace{5mm}

\textit{R. Goi\'as, 1021, Jardim Santa Cruz}

\textit{18700-140 \ Avar\'e, S\~ao Paulo, Brazil}

\end{center}

\vspace{5mm}\vv

\centerline{{\large\bf Abstract}}

\vspace{5mm}

\nd \textit{A method is proposed to solve the Grad-Shafranov partial
differential equation for the poloidal flux function associated with
the equilibrium of a plasma magnetically confined in an axisymmetric
torus under the assumption that the sources to the equilibrium (the
gradient of the plasma pressure and the gradient of half the squared
toroidal field function in flux space) are constant, and subjected
to the condition of constant value of the poloidal flux at the
toroidal boundary. The solution to the equation is written as the
sum of the particular solution and a combination of $N$ multipole
solutions of the lowest orders in the variables of the
toroidal-polar coordinate system having the pole at the centre of
the torus cross section. The term of the multipole solution of the
zeroth order does not bring about any magnetic field and its
inclusion in the combination fulfills the purpose of providing it
with a constant term whose value can be adjusted so as to make equal
to zero the value taken by the poloidal flux at the boundary; the
term of the multipole solution of the first order is associated with
a (constant) vertical field, such as required for the attainment of
the balance of forces in a confinement system of toroidal geometry,
and has the effect of introducing a shift (the Shafranov shift) to
the line of null magnetic poloidal field inside the plasma with
regard to the centre line of the torus; the terms containing the
multipole solutions of higher orders tailor the contours of the flux
surfaces, modifying the shape they would take from the sole presence
of the particular solution. Imposition of the boundary condition on
the assumed solution leads to a system of linear equations for the
constants multiplying the multipole solutions, in which the driving
terms are derived from the expression for the particular solution at
the boundary. The number of equations exceeds the number of unknowns
in this system, and, in principle, it admits of no solution.
Relaxation of the strict terms of the condition of nullity and
application of the criterium that the magnitude of the poloidal flux
at the boundary be the least possible decides which of the equations
are to be abandoned in order that a solution can be extracted from
the set of the remaining ones. To make the Shafranov shift appear as
a parameter in the final expression of the flux function and to have
it included in the solving process since the first stages of its
development, an equation, obtained by setting equal to zero the
gradient of the poloidal flux function on the equator line of the
torus cross section, in which the radial coordinate is identified
with that of the magnetic axis (and thus with the measure of the
Shafranov shift), is then added to the set of equations. The
information concerning the equilibrium sources is made present in
the system so constituted only through a parameter measuring the
ratio these sources keep between themselves, which one is formally
treated as an unknown while the Shafranov shift itself is considered
to be a given quantity. The coefficients of the $N$ multipole
solutions in the proposed solution to the Grad-Shafranov boundary
value problem follow from the solution to this truncated and
enlarged system of algebraic equations for which the matrix of
coefficients is a known function of the inverse aspect ratio and of
the Shafranov shift. The Shafranov shift ultimately appears as the
solution of an isolated algebraic equation whose coefficients are
functions of the squared inverse aspect ratio and whose independent
term is the ratio of the equilibrium sources. This equation can be
seen, conversely, as an explicit definition of the ratio between the
equilibrium sources in terms of the values ascribed to the inverse
aspect ratio and to the Shafranov shift. The poloidal flux function
that results from the application of the method is expressed in
terms of elementary functions only and satisfies exactly the
Grad-Shafranov equation while satisfying the boundary condition in
an approximate way. For the purpose of illustration solutions are
derived for three configurations as defined by the signals and
strengths of the equilibrium sources, respectively paramagnetic,
diamagnetic and magnetically neutral. It is concluded that the error
carried by the expression found for the poloidal flux function in
general decreases as the number $N$ of multipole solutions entering
its composition is increased. Other findings are: 1) there is a
non-null value of the Shafranov shift even for a paramagnetic
configuration of vanishingly small plasma pressure; 2) the magnetic
axis is always placed between the centre line of the cross section
and the outer edge of the torus, irrespective of the magnetic
character of the configuration; 3) although the outermost magnetic
surface shows to become a separatrix for a sufficiently high value
of the Shafranov shift, all indications are that, by increasing the
number $N$ of multipole solutions contained in the representation of
the poloidal flux function, it reassumes the shape of a surface that
is topologically equivalent to that of a torus.}

\vfill


\newpage

\pagestyle{plain} \setcounter{page}{1} \baselineskip=8.5mm

\centerline{{\bf I. INTRODUCTION}}

\setcounter{section}{1}

\vv

The mathematical tool of commonest use to investigate the plasma
equilibrium in the toroidal pinch (the tokamak) is the equation that
came to be known as the Grad-Shafranov equation, a second order
partial differential equation of the elliptic type for the flux
associated with the poloidal magnetic field, which can be solved
under proper boundary conditions when the plasma pressure function
and the toroidal field function (the sources to the poloidal flux)
are specified in flux space.

In vector notation, the Grad-Shafranov writes as \cite{um}: \beq R^2
\nabla.\left(\frac{1}{R^2}\nabla\Psi\right) = -\mu_0R^2\frac{\dif
p}{\dif\Psi} - I\frac{\dif I}{\dif \Psi} \label{eq1.1} \eeq where
$\Psi$ is the poloidal field magnetic flux divided by $2\pi$; $p =
p(\Psi)$ is the plasma pressure; $I = I(\Psi)$ is the toroidal field
function, also called poloidal current function, defined as: \beq I
\equiv RB_\phi, \label{eq1.2} \eeq $B_\phi$ being the toroidal
magnetic field; $R$ is the distance from the axis of rotational
symmetry of the physical system to the point where the poloidal flux
function is $\Psi$; $\mu_0$ is the absolute permeability of vacuum.

An exact analytic solution in closed form to the Grad-Shafranov
equation under the double assumption of a constant pressure gradient
and a constant squared toroidal field function gradient in flux
space, which has the effect of suppressing any dependence of the
terms in the equation coming from the poloidal field sources on the
poloidal flux function, has been presented in Ref. \cite{dois}. This
solution, however, does not fit the geometry of any boundary we may
expect to find in laboratory devices. In Ref. \cite{tres} the form
assumed for the source functions in flux space is such that the
terms they contribute to the Grad-Shafranov equation are
proportional to the flux function. A perturbative solution is
obtained by developing an expansion of the poloidal flux function in
powers of the reciprocal of the aspect ratio ($\eps$) of the torus
about the corresponding cylindrical configuration (that is, the one
with vanishing $\eps$), which satisfies the requirement that the
normal component of the magnetic field vanishes at the wall of the
plasma container, supposed to be made of a perfectly conducting
material. These are examples of solutions to a Grad-Shafranov
differential equation which, because of the postulations regarding
the form of dependence of the source functions on the flux function,
is linear. We have no knowing of references in the literature on
plasma equilibrium that report the use of Fourier's method to solve
the boundary value problem constituted by any of the linear versions
that can be given to the Grad-Shafranov equation and a set of local
conditions on the solution, possibly because the equation has been
found to be separable in none of the coordinate systems of common
knowledge in which one family of coordinate surfaces comprises a
surface that fits the toroidal boundary of the physical system.
Regarding the nonlinear problem, the usual procedure consists in
resorting first to the expansion of the source functions in Taylor
series in $\Psi$ about the magnetic axis where the flux function is
assigned an arbitrary value $\Psi = \Psi_M$; the solution to the
equation at a point is then obtained in the form of a power series
of the distance from the considered point to the magnetic axis with
coefficients that are functions of the poloidal angle. The domain of
validity of such a solution is restricted to the neighborhood of the
magnetic axis and this makes it unsuitable to supporting the
imposition of boundary conditions.

In the present paper we consider the equilibrium of a plasma
enclosed in a toroidal chamber as a boundary value problem. The
equilibrium is supposed to be governed by the Grad-Shafranov
equation; the plasma pressure gradient and the gradient of half the
squared toroidal field function are assumed to obey a flat profile
in flux space; the wall of the containing vase, with which the
plasma is in contact, to be made of a perfectly conducting material.
We have already observed that for a Grad-Shafranov equation shaped
by identical hypotheses concerning the sources of equilibrium the
solution presented in Ref. \cite{dois} is not able to meet the
boundary conditions that appertain to any of the physical situations
currently found in practice, and, in particular, to the one we
propound to consider here.

The method we shall develop in the following pages takes as
``particular'' a solution to the equation whose set of level curves
on a plane $\phi =$ constant, analogously to that representing the
Solovev's, does not include any that fits the contour of the torus
cross section; to get a solution to the equation that also satisfies
the boundary condition, we add to it a ``complementary'' solution,
which we build as a linear combination of functions belonging to the
infinite set of those that solve the sourceless Grad-Shafranov
equation. These functions, usually referred to in the literature of
Plasma Physics by the name of multipole solutions, are available to
us thanks to the methods expounded in Ref. \cite{quatro}.

This procedure of building the ``complete'' solution is permissible,
of course, because the problem is linear; also because the multipole
equation and the homogeneous equation associated with the version of
the Grad-Shafranov's for the case, under present consideration, of
constant sources to the equilibrium in flux space, are the same.

The equilibrium of a plasma contained in a vase with a metallic
lining on the internal side of its wall leads to just one boundary
condition, and this is that the poloidal flux function remains
constant when its coordinate variables are made to describe the
surface of the wall. By applying this sole requirement to the
complete solution we expect to be able to determine the whole set of
constants that multiply the functions of reference participating in
the constitution of the complementary solution.

In fact it is impossible to have it satisfied in the strictest
terms. As imposed on the complete solution, the boundary condition
translates by a set of constraints on the constants, and, since the
number of constraints exceeds that of the constants, there is no
possible choice for the constants that would make them capable to
satisfy simultaneously all the constraints in the set -- the form
proposed for the solution, which does indeed solve the equation,
seems to suffer, after all, from a prohibitive inconsistency with
the boundary condition. In an effort to save that form for the
advantages that it however presents, we accede to satisfy, not all
of the constraints, but the largest possible number of them -- and
thus to give only an approximate solution to the boundary value
problem. The criterium to decide which constraints are to be kept
and which are to be abandoned in fixing the values of the constants
is that the abandoned be the ones within the whole set that, in
remaining unsatisfied, give margin to the arising of only the least
possible error at the boundary (that is, the smallest departure of
the flux function from a constant value thereat).

A thorough understanding of the last statement passes through the
recognition that the need of using only a finite number of multipole
solutions in the construction of the flux function is an inherent
feature of the method, and this not solely because of the lack of a
general expression that would give a unified representation to the
infinite set of multipole solutions, but, more significantly than
that, because of the very \textit{modus operandi} of the
mathematical machinery of the method, which involves the resolution
of algebraic equations of high degrees and the evaluation of
functional determinants of high orders. We have found nonetheless
that the error at the boundary (and by consequence at all internal
points) is consistently diminished as the number of multipole
solutions that are included in the complementary solution is
increased, suggesting that in the limit of infinitely many ones we
should have generated a convergent series in which each term would
satisfy the differential equation separately and whose sum at the
boundary would reproduce the exact value the flux function is to
take there. A theorem regarding the existence and uniqueness of the
solution to a Grad-Shafranov boundary value problem of the same kind
as that which is considered in the present paper \cite{cinco} makes
us then be sure that the solution we have found is the only possible
one.

This paper is organized as follows. Section II is devoted to
introducing a convenient normalization of the physical and
geometrical quantities that enter the formulation of the problem,
and to state it in precise mathematical terms; the coordinate system
of election is the toroidal-polar one. In Section III the particular
solution to the Grad-Shafranov equation with sources to the
equilibrium that are constant in flux space is derived by the method
of expansion of the flux function in series of Chebyshev polynomials
of the cosine of the polar angle.  Section IV is the core of the
paper, the one in which the method of solution to the Grad-Shafranov
boundary value problem for constant equilibrium sources by series of
multipole solutions is developed. Section V states certain
conditions on the values that can be assumed by the Shafranov shift
(or, rather, by a related quantity) and by the poloidal flux
function at the magnetic axis in order that the solutions provided
by the method describe true equilibrium configurations (in the
physical sense) and not merely field lines structurings which might
disregard some obvious requirements for plasma confinement; it also
discusses the complex question of the appearance of branching points
on the plasma bounding surface in connection with the limits of
validity of the method, and concludes by defining the quantities
appropriate to evaluate the accurateness of the approximate
solutions that can be obtained through the use of it. The method is
then applied in Section VI to solve the Grad-Shafranov boundary
value problem for a number of equilibrium situations by using a
combination of the least possible number of multipole solutions,
which is three. It is solved again in Section VII under the same
physical terms of the previous Section by considering a combination
of four multipole solutions, for the uses of an odd number and of an
even number of multipole solutions imply a difference in the
structures of an auxiliary function (called equilibrium function)
that participates in the solving process, and there is need to show
that this difference is not a source of incongruities between
solutions to the same problem as obtained with distinct compositions
of the complementary solution to the equation. Section VIII is
devoted to expounding some general conclusions regarding the
mathematical structures of the flux function and of some auxiliary
functions that are required for the derivation of the former, and to
discussing the accuracy that can be gained with the increase of the
number of multipole solutions that are made to enter the combination
aimed at representing the flux function. In Section IX a few
equilibrium configurations characterized by having representative
values of the equilibrium sources are described by using the
solutions for the flux function obtained with the help of the
method. Finally, Section X recapitulates the main results and
findings of the paper.

\vv\vv

\begin{center}
{\bf II. STATEMENT OF THE PROBLEM}
\end{center}

\setcounter{section}{2} \setcounter{equation}{0}

\vv

The solutions of the Grad-Shafranov equation in which we shall be
interested in this paper are those that give the flux surfaces in
the plasma region as nested tori, that is to say, a surface inside
the other, the innermost one degenerating into a line, the magnetic
axis, where the poloidal field vanishes. We place the pole of the
toroidal-polar coordinate system on a meridian plane at the centre
of the torus cross section, which, designated by the letter $C$ in
Fig. 2.1, is located at the distance $R_C$ from the axis of
rotational symmetry of the physical system (in Fig. 1 in Ref.
\cite{quatro} the pole of the toroidal-polar coordinate system on a
meridian plane is a point arbitrarily chosen, which we have
designated by the letter $A$ and located at a distance $R_A$ from
the axis of rotational symmetry. In borrowing the mathematical
expressions, as they become required by the needs of the present
paper, from the collections appearing in Ref. \cite{quatro}, we
shall, therefore, be implicitly taking $A$ as $C$ and identifying
$R_A$ with $R_C$.). In Fig. 2.1 we have also represented the
intersection of the magnetic axis with the meridian plane as a point
that we have denoted by $M$ and placed at a distance $R_M$ from the
axis of rotational symmetry. The distance that separates the
magnetic axis from the centre line of the torus is $\Delta$. The
coordinates of a generic point $P$ in this coordinate system are
$r$, $\ta$ and $\phi$, as shown in Fig. 2.1 of the present paper and
also in Fig. 1 in Ref. \cite{quatro}, and then its distance to the
axis of rotational symmetry is given by: \beq R = R_C + r\cos\ta\ .
\label{eq2.1} \eeq

\vv\vv

\begin{center}

\vspace{2mm} {\unitlength=1mm
\begin{picture}(100,70)
\put(0,0) {\includegraphics[width=12cm]{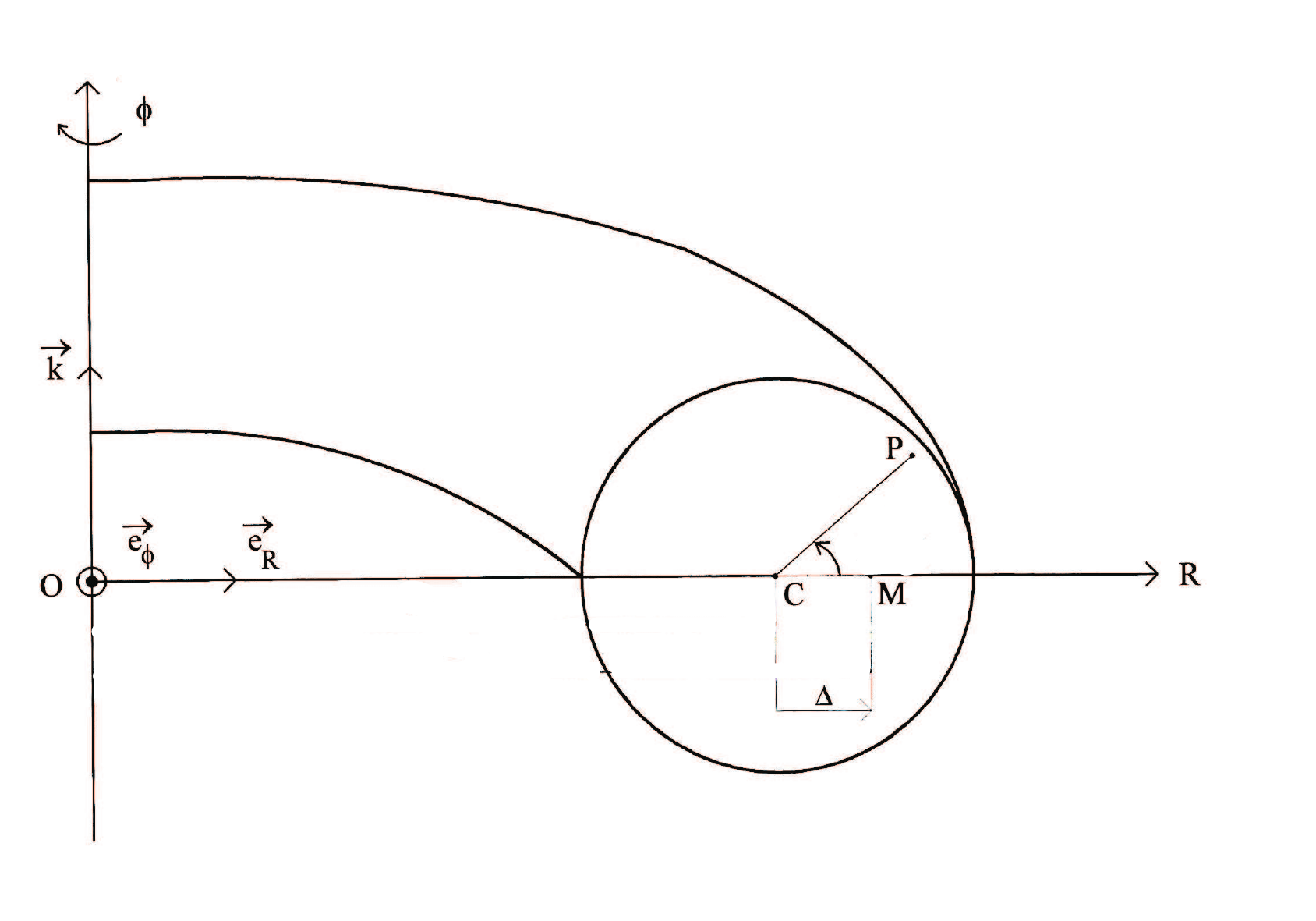}}
\put(37,28.5){{\footnotesize R}} \put(39,27.5){{\tiny C}}
\put(41,24){{\footnotesize R}} \put(43,23){{\tiny M}}
\put(75,38){{\footnotesize r}} \put(77,33.4){{\footnotesize
$\theta$}} \put(8.5,42.5){\vector(1,0){73}}
\put(8.5,27.1){\vector(1,0){62}}\put(8.5,22.6){\vector(1,0){71}}
\put(45,44){{\footnotesize R}} \put(10,76){{\footnotesize z}}
\end{picture}}

\begin{tabular}{lp{12cm}}
{\bf FIG. 2.1} & The toroidal-polar coordinate system $(r, \ta,
\Phi)$ for which the pole $C$ is located at the intersection of the
torus centre line with the meridian plane coinciding with the plane
of the figure. The intersection of the magnetic axis with the
meridian plane is the point denoted by $M$. The distance from the
axis of rotational symmetry $Oz$ to $C$ is $R_C$ and that to $M$ is
$R_M$. The Shafranov shift for the configuration represented in the
figure is $\Dt = R_M - R_C$.
\end{tabular}

\end{center}

\nd We also introduce here the notation $B_{\phi_C}$ for the
strength of the toroidal magnetic field at the torus centre line.

To write the plasma equilibrium equation in dimensionless form, we
introduce the normalized radial coordinate, the normalized flux, the
normalized plasma pressure and the normalized toroidal field
function according to the definitions respectively: \bey x &\equiv&
\frac{r}{R_C},
\label{eq2.2} \\
\psi &\equiv& \frac{\Psi}{B_{\phi_C}R_C^2} , \label{eq2.3} \\
\hat p &\equiv& \frac{p}{B_{\phi_C}^2/2\mu_0}, \ \ \hbox{ and} \label{eq2.4} \\
\hat I &\equiv& \frac{I}{R_CB_{\phi_C}}\ . \label{eq2.5} \eey

Taking $\mu \equiv \cos\ta$ as variable in place of the polar angle
$\ta$, Eq. (\ref{eq1.1}) then writes in the toroidal-polar
coordinate system as: \bey &&x^2(1 + x\mu)\frac{\ptl^2\psi}{\ptl
x^2} + x\frac{\ptl\psi}{\ptl x} + (1 +
x\mu)(1-\mu^2)\frac{\ptl^2\psi}{\ptl\mu^2} - (x +
\mu)\frac{\ptl\psi}{\ptl\mu} \nonumber \\
&&= x^2(1 + x\mu)\left[(1+ x\mu)^2s_P + s_I\right]\ ,
\label{eq2.6} \eey where \beq s_P \equiv
-\frac{\dif}{\dif\psi}\left(\frac{\hat p}{2}\right) \label{eq2.7}
\eeq and \beq s_I \equiv -\frac{\dif}{\dif\psi}\left(\frac{\hat
I^2}{2}\right) \label{eq2.8} \eeq are the sources to the
equilibrium, which ones we shall consider in this paper to be
constant.

We shall assume that the plasma is in contact with the wall of the
containing vessel, which is made of a perfectly conducting
material\footnote{It may seem physically artificial to assume that
the plasma is in contact with the wall of the enclosing shell. As an
alternative situation of more realistic appeal, which still
preserves the terms in which the problem is formulated, it can be
imagined that the plasma is surrounded by vacuum and that a system
of coils applies an external magnetic field with a configuration
such designed that it exactly cancels the field the currents
circulating in the plasma bulk generate in the outside vacuum
region, while \textit{ipso facto} constraining the plasma surface to
conform to the shape assumed in the boundary condition stated in Eq.
(\ref{eq2.9}). See Reference \cite{seis}.}. Thus there is no
magnetic field perpendicular to the internal surface of the wall,
the field lines run tangent to it, and the plasma-wall interface
coincides with a plasma flux surface: \beq \psi(\hbox{plasma
boundary}) = \hbox{ constant}. \label{eq2.9} \eeq Since the flux
function is undetermined by a constant, we are free to choose the
constant in Eq. (\ref{eq2.9}) as we please, and, following the
common practice, which will prove in the course of future
developments to be indeed the most convenient for our purposes, we
take it to be zero. We shall see in later Sections that with such a
choice the solutions for the flux function can assume only one sign
in the plasma region, either positive or negative, being this no
more than a matter of working preference, but cannot change sign.

In which regards the interior of the conducting wall and the outer
side of the containing vessel, the magnetic field is assumed to be
zero. The supposition that a perfectly conducting wall is placed at
the plasma boundary has to be followed physically by the recognition
that a surface current is induced at the plasma-wall interface by
the fields arising from the currents flowing in the bulk of the
plasma, and that this surface current in its turn creates a magnetic
field which cancels the former ones in the interior of the
conducting wall and in the outside of the container.

With this we may state the problem as represented by Eq.
(\ref{eq2.6}) subjected to the boundary condition of Eq.
(\ref{eq2.9}); the region of physical interest is restricted to the
domain of integration of the partial differential equation. In terms
of the variables of the toroidal-polar coordinate system, this
region is defined by \beq 0 \le x \le \eps \ , \ \ -\pi \le \ta \le
\pi\ , \label{eq2.10} \eeq where \beq \eps \equiv \frac{a}{R_C}
\label{eq2.11} \eeq is the inverse aspect ratio of a torus of minor
radius $a$ and major radius $R_C$.

\vv\vv

\begin{center}
{\bf III. DERIVATION OF THE PARTICULAR SOLUTION TO THE
GRAD-SHAFRANOV EQUATION WITH CONSTANT EQUILIBRIUM SOURCES}
\end{center}

\setcounter{section}{3}  \setcounter{equation}{0}

\vv

We shall derive the particular solution to Eq. (\ref{eq2.6}) by the
method of expansion of the solution in series of Chebyshev
polynomials. As a preliminary step to accomplish this end we shall
take advantage of the formula that gives the representation of an
integral power of a variable $\mu$ in terms of the Chebyshev
polynomials of the first kind $T_j(\mu)$ in that variable, to know
\cite{sete}:
 \beq \mu^k = \frac{1}{2^{k-1}}\left[T_k(\mu) +
{k\choose 1}T_{k-2}(\mu) + {k\choose 2}T_{k-4}(\mu) + \cdots\right]\
, \label{eq3.1} \eeq where the series in brackets terminates with
${k\choose m}T_1(\mu)$ for $k = 2m+1$ or with $\meio{k\choose
m}T_0(\mu)$ for $k = 2m$, and reexpress the right hand side of Eq.
(\ref{eq2.6}) as: \bey RHS &=& (s_P + s_I)T_0(\mu)x^2 + (3s_P + s_I)
T_1(\mu)x^3 + 3s_P\left[\meio
T_2(\mu) + \meio T_0(\mu)\right]x^4 \nonumber \\
&&+ s_P\left[\frac{1}{4}T_3(\mu) + \frac{3}{4}T_1(\mu)\right]x^5 \ .
\label{eq3.2} \eey

In which regards its dependence on the radial variable, we may
assume an expression of the form: \beq \psi_p(x,\mu) =
\psi_2(\mu)x^2 + \psi_3(\mu)x^3 + \psi_4(\mu)x^4 \label{eq3.3} \eeq
for the flux function, since the inclusion of powers of $x$ lower
than the second and higher than the fourth would give rise to terms
on the left hand side of Eq. (\ref{eq2.6}) that would not be
balanced by terms of equal powers of $x$ on the right hand side. It
should be noted that the process of derivation of the solution based
on an expression like the one of Eq. (\ref{eq3.3}) will generate not
only the particular solution, but also the multipole solution of the
second order (see Eq. (\ref{a8}) in Appendix A and Eq. (B.2) in
Appendix B in Ref. \cite{quatro}), in which we have no present
interest, and, to have it suppressed from the analytical
developments, we must choose the arbitrary constant by which it
would otherwise appear multiplied along these to be zero at the step
we first recognize its emergence in the calculations.

We start by inserting Eq. (\ref{eq3.3}) in Eq. (\ref{eq2.6}). By
recalling the definitions of the operators $\calt_l$ and $\calh_l$,
given by Eqs. (3.19) and (3.10) respectively in Ref. \cite{quatro},
we can write the left hand side of Eq. (\ref{eq2.6}) as: \beq LHS =
\sum_{k=2}^4 x^k\calt_k\{\psi_k(\mu)\} + \sum_{k=3}^5
x^k\calh_{k-1}\{\psi_{k-1}(\mu)\}\ . \label{eq3.4} \eeq

We now expand the angular functions $\psi_2(\mu)$, $\psi_3(\mu)$ and
$\psi_4(\mu)$ into series of Chebyshev polynomials $T_n(\mu)$ as:
\beq \psi_k(\mu) = \sum_{n=0}^\infty C_{k,n}T_n(\mu)\ , \ \ \ (k =
2, 3, 4)\label{eq3.5} \eeq where the coefficients $C_{k,n}$ are
constants. By having recourse to the property of the operator
$\calt_l$ stated in Eq. (3.20) and to those of the operator
$\calh_l$ stated in Eqs. (3.14) and (3.15) in Ref. \cite{quatro}, we
may easily prove that \beq \calt_k\{\psi_k(\mu)\} =
\sum_{n=0}^\infty (k^2 - n^2)C_{k,n}T_n(\mu) \label{eq3.6} \eeq and
that \bey &&\hspace{-5mm}\calh_k\{\psi_k(\mu)\} \!=\!
\meio(k\!+\!1)(k\!-\!2)C_{k,1}T_0(\mu) \!+\! \left[k(k\!-\!1)C_{k,0}
\!+\! \meio(k\!+\!2)(k\!-\!3)C_{k,2}\right]T_1(\mu)
\nonumber \\
&&\hspace{-5mm}-\meio\sum_{n=2}^\infty\left[(n\!+\!k\!+\!1)(n\!-\!k\!+\!2)C_{k,n+1}
\!+\! (n\!-\!k\!-\!1)(n\!+\!k\!-\!2)C_{k,n-1}\right]T_n(\mu)\ .
\label{eq3.7} \eey

\vpq

We next equate the terms of equal powers of $x$ on the left and
right hand sides of Eq. (\ref{eq2.6}). From those in $x^2$ we
obtain: \bey \calt_2\{\psi_2(\mu)\} &=& \sum_{n=0}^\infty (4-n^2)
C_{2,n}T_n(\mu)
\nonumber \\
&=& (s_P + s_I)T_0(\mu)\ , \label{eq3.8} \eey from which we conclude
that \beq C_{2,1} = C_{2,3} = C_{2,4} = \cdots = 0 \label{eq3.9}
\eeq and \beq 4C_{2,0} = s_P + s_I \ , \label{eq3.10} \eeq while
$C_{2,2}$ remains indeterminate. It is seen, however, that this
constant, which multiplies $x^2$ and the Chebyshev polynomial
$T_2(\mu)$ in the expression proposed for the particular solution,
is precisely the one that, kept in the subsequent analytical
developments, would appear at the end, still indetermined,
multiplying the multipole solution of the second order. We thus take
$C_{2,2}$ to be zero, as we have anticipated in the sequel of Eq.
(\ref{eq3.3}).

From the terms in $x^3$ in Eq. (\ref{eq2.6}) we get: \beq
\calt_3\{\psi_3(\mu)\} + \calh_2\{\psi_2(\mu)\} = (3s_P +
s_I)T_1(\mu)\ . \label{eq3.11} \eeq

Since all the coefficients $C_{2,n}$ vanish with the exception of
$C_{2,0}$, Eq. (\ref{eq3.7}), with $k$ taken iqual to 2, shows that
the second term on the left hand side of Eq. (\ref{eq3.11}) consists
of a multiple of a single Chebyshev polynomial, $T_1(\mu)$. Using
Eq. (\ref{eq3.6}) with $k = 3$ \ for the first term, Eq.
(\ref{eq3.11}) then becomes: \beq
\sum_{n=0}^\infty(9-n^2)C_{3,n}T_n(\mu) + 2C_{2,0}T_1(\mu) = (3s_P +
s_I)T_1(\mu) \ , \label{eq3.12} \eeq from which we obtain: \bey
&&8C_{3,1} + 2C_{2,0} = 3s_P + s_I, \label{eq3.13} \\
&&C_{3,0} = C_{3,2} = C_{3,4} = C_{3,5} = \cdots = 0\ ,
\label{eq3.14} \eey and no relation involving $C_{3,3}$.

The next equation to be considered is the one coming from the
equality of the terms in the fourth power of $x$ on both sides of
Eq. (\ref{eq2.6}). Using the results already known for the
coefficients $C_{3,k}$ to reduce the expression for
$\calh_3\{\psi_3(\mu)\}$, this equation writes as: \beq
\sum_{n=0}^\infty (16 - n^2)C_{4,n}T_n(\mu) \!+\! 2C_{3,1}T_0(\mu) -
(3C_{3,3} - 3C_{3,1})T_2(\mu) = \frac{3s_P}{2}T_0(\mu) \!+\!
\frac{3s_P}{2}T_2(\mu), \label{eq3.15} \eeq and implies the
conditions: \bey &&16C_{4,0} + 2C_{3,1} = \frac{3s_P}{2},
\label{eq3.16} \\
&&12C_{4,2} - 3C_{3,3} + 3C_{3,1} = \frac{3s_P}{2}\ ,
\label{eq3.17} \eey and \beq C_{4,1} = C_{4,3} = C_{4,5} = C_{4,6}
= \cdots = 0\ . \label{eq3.18} \eeq

We finally arrive at the balance of the terms in $x^5$, the
highest power of $x$ to appear in Eq. (\ref{eq2.6}). Taking into
account the results for the coefficients $C_{4,k}$ expressed in
Eq. (\ref{eq3.18}), the left hand side of the balancing equation
simplifies to: \bey LHS &=& \calh_4\{\psi_\mu(\mu)\} \nonumber \\
&=& (12C_{4,0} + 3C_{4,2})T_1(\mu) - (4C_{4,4} - 5C_{4,2})T_3(\mu)
\ , \label{eq3.19} \eey while the right hand side is given by:
\beq RHS = \frac{3s_P}{4}T_1(\mu) + \frac{s_P}{4}T_3(\mu) \ .
\label{eq3.20} \eeq

We have therefore: \bey &&12C_{4,0} + 3C_{4,2} = \frac{3s_P}{4},
\label{eq3.21} \\
&&4C_{4,4} - 5C_{4,2} = -\frac{s_P}{4}\ , \label{eq3.22} \eey and at
this last step, in which the operator $\calt_l$ is not called to a
participation, no coefficient $C_{i,j}$ remains indeterminate, and
the cycle of equations for them is closed.

Having established which coefficients $C_{k,n}$ of the series in Eq.
(\ref{eq3.5}) must vanish, we can write the angular functions
$\psi_k(\mu)$ ($k = 2, 3, 4$) as finite combinations of Chebyshev
polynomials, to know: \bey
\psi_2(\mu) &=& C_{2,0}T_0(\mu), \label{eq3.23} \\
\psi_3(\mu) &=& C_{3,1}T_1(\mu) + C_{3,3}T_3(\mu), \ \ \
\hbox{and} \label{eq3.24}
\\
\psi_4(\mu) &=& C_{4,0}T_0(\mu) + C_{4,2}T_2(\mu) + C_{4,4}T_4(\mu)\
. \label{eq3.25} \eey The six constants $C_{i,j}$ that appear in the
above equations can be determined by solving the linear system
constituted by Eqs. (\ref{eq3.10}), (\ref{eq3.13}), (\ref{eq3.16}),
(\ref{eq3.17}), (\ref{eq3.21}) and (\ref{eq3.22}).

This having been accomplished, we return to Eq. (\ref{eq3.3}).
Recalling the expressions for the Chebyshev polynomials $T_1(\mu)$,
$T_2(\mu)$, $T_3(\mu)$ and $T_4(\mu)$ as given by Eqs. (2.9),
(2.21), (4.49) and (4.50) respectively in Ref. \cite{quatro}, we can
finally state the particular solution to the Grad-Shafranov equation
with constant sources as: \bey \psi_p(x,\mu) &=& \frac{1}{4}(s_P +
s_I)x^2 + \left[\meio(s_P -
s_I)\mu + \frac{1}{4}(-s_P + 3s_I)\mu^3\right]x^3 \nonumber \\
&&+ \left[\frac{1}{4}(s_P - s_I)\mu^2 + \frac{1}{16}(-3s_P +
5s_I)\mu^4\right]x^4\ . \label{eq3.28} \eey An expression
alternative to this that takes the form of a linear combination of
trigonometric terms can be written down immediately by replacing the
Chebyshev polynomials in the angular dependent coefficient functions
of the particular solution as they appear defined in Eqs.
(\ref{eq3.23}) -- (\ref{eq3.25}) by their trigonometric
representations, to know, $T_n(\cos\ta) = \cos n\ta$ ($n = 0, 1, 2,
3, 4$).

Because of the terms in which the problem has been posed, the
expression obtained for the flux function as Eq. (\ref{eq3.28}) has
the form of a linear combination of two constants ($s_P$ and $s_I$)
with no independent term. This means that the functional dependence
of the particular solution on the source terms, assumed to be
constant in flux space, is actually given shape singly by the ratio
between them, with one of the constants assuming the role of a mere
overall multiplicative factor. In fact this assertion applies not
only to the particular but also to the complete solution, and,
trivial as it seems to be, it warrants the problem to be treated as
one of a single input. We thus divide the expressions we have for
the particular solution by one of the constants, which we choose to
be $s_P$, and pass to consider a newly normalized flux function,
which, in the case of the representation by a finite trigonometric
series, assumes the form: \bey \ovl\psi_p(x,\ta) &=&
\frac{1}{4}(1\!+\!\lb)x^2 \!+\! \frac{1}{128}(7\!-\!\lb)x^4 \!+\!
\frac{1}{16}(5\!+\!\lb)x^3\cos\ta \!+\!
\frac{1}{32}(1\!+\!\lb)x^4\cos 2\ta \nonumber \\
&&+ \frac{1}{16}(-1+3\lb)x^3\cos 3\ta +
\frac{1}{128}(-3+5\lb)x^4\cos 4\ta\ , \label{eq3.29nova} \eey where
\beq \ovl\psi_p(x,\ta) \equiv \frac{\psi_p(x,\ta)}{s_P} \ ,
\label{eq3.29} \eeq and $\lb$ is the ratio between constants: \beq
\lb \equiv \frac{s_I}{s_P}\ , \label{eq3.30} \eeq which we shall
refer to in future discussions as the \textit{equilibrium
parameter}.

In future developments we shall find it convenient to express the
normalized particular solution as: \beq \ovl\psi_p(x,\ta) = S_0(x) +
S_1(x)\cos\ta + S_2(x)\cos 2\ta + S_3(x)\cos 3\ta + S_4(x)\cos 4\ta\
, \label{eq3.32nova} \eeq and the $x$-dependent functions that
appear as coefficients in the above representation as: \beq S_i(x) =
P_i(x) + \lb Q_i(x) \ \ \ (i = 0, 1, 2, 3, 4)\ . \label{eq3.33nova}
\eeq The explicit forms of the $P_i(x)$'s and $Q_i(x)$'s are
collected in Appendix A.

We shall also face the need to have an expression for the particular
solution in the cylindrical coordinate system $(R, \phi, z)$
represented in Fig. 1 in Ref. \cite{quatro}. The conversion of that
in terms of the coordinates $r,\ta$ and $\phi$ of the toroidal-polar
system we have just derived is most easily accomplished by resorting
to the representation it finds as a trigonometrical polynomial in
the variable $\mu = \cos\ta$. In this way, by introducing the
transformation relations of Eq. (5.21) in Ref. \cite{quatro} in Eq.
(\ref{eq3.28}), we obtain:
 \beq \ovl\psi_p(\rho, Z) = \frac{1}{16}(1+\lb)(\rho^2 - 1)^2 +
\frac{1}{4}Z^2[(1-\lb)\rho^2 + 2\lb]\ , \label{eq3.31} \eeq where
$\rho \equiv R/R_C$ \ and \ $Z \equiv z/R_C$.$\,$\footnote{The
Solovev solution, in a form equivalent to that in which it is quoted
in Ref. \cite{dois}, is obtained by combining the particular
solution, as given by Eq. (\ref{eq3.31}), with the multipole
solutions of the three lowest orders: $\ovl\psi_{Sol}(\rho, Z) =
\ovl \psi_p + K_0\vf^{(0)}(\rho, Z) + K_1\vf^{(1)}(\rho, Z) +
K_2\vf^{(2)}(\rho, Z)$, and by taking the value of the constant
multiplying the multipole solution of order $n = 2$ as $K_2 = (\lb -
1)/4$. The values of the constants $K_0$ and $K_1$ remain free, so
that they can be chosen as to fix the flux surface where the
pressure is zero (the plasma border) and the position of the
magnetic axis.} Note that the expressions for both the flux function
and its gradient -- and thus for the magnetic field associated with
$\ovl{\psi}_p(x,\ta)$ -- in the toroidal-polar coordinate system
vanish at the point $x = 0$. In cylindrical coordinates the point of
concurrent vanishing of these quantities translates as $R = R_C$, $z
= 0$.

\vv\vv

\begin{center}
{\bf IV. THE PRINCIPLES OF THE METHOD OF SOLUTION TO THE
GRAD-SHAFRANOV BOUNDARY VALUE PROBLEM BY SERIES OF MULTIPOLE
SOLUTIONS}
\end{center}

\setcounter{section}{4} \setcounter{equation}{0}

\vv

This method applies only to the linear version of the problem in
which the sources to the equilibrium are constant.

The magnetic poloidal fields that are derived from the flux
functions corresponding to the multipole solutions vanish all at a
single point which is located by the coordinate $x = 0$, exception
being taken to the multipole solution $\varphi^{(0)}(x,\ta)$, the
constant one, which gives a null field everywhere in space, and to
that of the first order, $\varphi^{(1)}(x,\ta)$, which gives a
uniform vertical field. As we have seen in Section III, the poloidal
field that is associated with the particular solution to the
Grad-Shafranov equation for constant equilibrium sources also
vanishes at that point where the multipole fields vanish. Thus the
sum of the particular solution $\psi_p(x,\ta)$ and a combination of
a number of multipole solutions $\vf^{(n)}(x,\ta)$ of the orders $n
= 0, 2, 3, \ldots$ is a solution to the Grad-Shafranov equation and
describes a magnetic configuration that admits of a null poloidal
magnetic field at the pole of the toroidal-polar coordinate system,
which we have placed at the centre of the torus cross section. The
occurrence of a line inside the torus lying on its equatorial plane
and extending all the way around the axis of rotational symmetry
where the poloidal field vanishes is an essential feature of the
equilibrium solutions in which we have a dominant interest, these
solutions being the ones that appear as a system of level surfaces
encircling a magnetic axis in the usual pictorial representation of
a magnetic configuration. It is well known, however, that the
toroidal equilibrium of the physical arrangement we are
considerering is accomplished in general by effect of a shift of
this ``centre'' of the family of nested flux surfaces (\textit{id
est}, the magnetic axis) from the geometrical centre line towards
the outer side of the torus \cite{oito}, usually referred to in the
literature of Plasma Physics as the Shafranov shift. In a tokamak
experiment this requirement for equilibrium is achieved by applying
a vertical field to the plasma, this field being generated either by
a set of external coils or by the image current induced by the
currents circulating in the plasma bulk on the surface of a
perfectly conducting shell placed around the plasma \cite{oitolin}.
In the mathematical treatment of the equilibrium problem, taking
after the experimental procedure, we shall include the multipole
solution of order $n = 1$ in the combination of multipole solutions,
such that, by giving origin to a uniform vertical field, it will
produce the desired effect of shifting the locus of null poloidal
field from the centre line of the torus at $R = R_C$ to some other
circle with centre at the major axis, having the radius larger than
$R_C$, and also lying on the midplane of the torus. The
representation we obtain in this way for the flux function is: \beq
\ovl\psi^{(N)}(x,\ta) = \ovl\psi_p(x,\ta) + K_0 +
\sum_{i=1}^{N-1}K_i\vf^{(i)}(x,\ta)\ , \label{eqnova41} \eeq where
the $K_i$'s ($i = 0, 1, 2, \ldots, N-1$) are constants, and,
consistent with the normalization adopted for the particular
solution, we assume the complete solution to be equally normalized
by the constant $s_P$: \beq \ovl\psi(x,\ta) \equiv
\frac{\psi(x,\ta)}{s_P}\ . \label{eq4.2nova} \eeq We shall refer to
the representation of Eq. (\ref{eqnova41}), which comprises the
multipole solutions of the $N$ lowest orders in combination, as the
\textit{approximation of order $N$ to the normalized flux function}
or the \textit{normalized partial flux function of order} $N$.

In general, by the reasons we have discussed in Section I, there is
no choice of the $N$ constants $K_0, K_1, \ldots, K_{N-1}$ capable
to make the solution stated as Eq. (\ref{eqnova41}) satisfy the
requirement that it vanishes at the boundary for any finite $N$. We
shall show, however, that a criterium can be set up to determine the
$K_i$ constants ($i = 0, 1, 2, \ldots, N-1$) such that the boundary
condition $\ovl\psi^{(N)} = 0$, at least for some equilibria among
which are included those of the greatest theoretical and practical
importance, can be approximately satisfied, and that the deviation
of the partial flux function from the null value at the boundary can
be reduced to any prescribed magnitude by increasing the number $N$
of multipole solutions that enter the combination of Eq.
(\ref{eqnova41}) to a sufficiently high value.

We write the multipole solutions in Eq. (\ref{eqnova41}) in the form
they are stated in Eq. (A.1) in Ref. \cite{quatro}, here reproduced:
\beq \vf^{(i)}(x,\ta) = \sum_{j=0}^{2i}M_{ji}(x)\cos j\ta\ ,
\label{eqnova42} \eeq and then, by interchanging the order of
summations over the indices $i$ and $j$, the double sum into which
the last term on the right hand side is thereby converted can be
brought to assume the following form: \beq
\sum_{i=1}^{N-1}K_i\vf^{(i)}(x,\ta) =
\sum_{n=1}^{2N-2}\left[\sum_{i=1}^{N-1}K_iM_{ni}(x)\right]\cos n\ta\
. \label{eqnova43} \eeq

Writing further the particular solution as in Eq. (\ref{eq3.32nova})
we are conducted to reexpress Eq. (\ref{eqnova41}) as: \beq
\ovl\psi^{(N)}(x,\ta) = G_0^{(N)}(x) + \sum_{n=1}^{2N-2}
G_n^{(N)}(x)\cos n\ta\ . \label{eqnova44} \eeq

In this representation of the normalized partial flux of order $N$
the angle independent term is given by: \beq G_0^{(N)}(x) = S_0(x) +
K_0 + \sum_{i=1}^{N-1}M_{0i}(x)K_i\ , \label{eqnova45} \eeq where,
as in Eq. (\ref{eq3.32nova}), we have written $S_0(x)$ for $S_0(\lb;
x)$; the coefficients of the harmonics of the poloidal angle can be
conveniently represented by a single formula: \beq G_n^{(N)}(x) =
S_n(x) + \sum_{i=1}^{N-1} M_{ni}(x)K_i\ \ (n = 1, 2, \ldots, 2N-2)\
, \label{eqnova46} \eeq whose generality of scope in terms of the
index $n$ presupposes that the following definitions apply to the
two sets of functions $S_n(x) \equiv S_n(\lb;x)$ and $M_{ni}(x)$
entering its composition: \beq S_n(x) = 0 \ \hbox{ for } \ n \ge 5
\label{eqnova47} \eeq and \beq M_{ni}(x) = 0 \ \hbox{ for } n > 2i\
, \label{eqnova48} \eeq respectively, in compliance, the first with
the expression for the particular solution in Eq.
(\ref{eq3.32nova}), and the second with that of the multipole
solutions in Eq. (\ref{eqnova42}).

Now, at the boundary, which is defined in terms of the normalized
radial coordinate by $x = \eps$, the normalized partial flux must
vanish: \beq \ovl\psi^{(N)}(\eps,\ta) = 0\ , \label{eqnova49} \eeq a
requirement that can be satisfied only if \beq G_0^{(N)}(\eps) =
G_1^{(N)}(\eps) = \cdots = G_{2N-2}^{(N)}(\eps) = 0 \ .
\label{eqnova410} \eeq

Considering the expressions for $G_0(x)$ and for $G_n^{(N)}(x)$ ($n
= 1, 2, \ldots, 2N-2$) as given by Eq. (\ref{eqnova45}) and by Eq.
(\ref{eqnova46}) respectively, this last condition translates by:
\bey &&\hspace{-12mm}K_0 \nonumber \\
&&\hspace{-12mm}+M_{01}(\eps)K_1 + M_{02}(\eps) K_2 + \cdots +
M_{0,N-1}(\eps)K_{N-1} = -S_0(\eps) \label{eqnova411} \\
&& \nonumber \\
&&\hspace{-12mm}\left.\begin{array}{l} M_{11}(\eps)K_1 +
M_{12}(\eps)K_2 + \cdots
+ M_{1,N-1}(\eps)K_{N-1} = -S_1(\eps) \\
M_{21}(\eps)K_1 + M_{22}(\eps)K_2 + \cdots
+ M_{2,N-1}(\eps)K_{N-1} = -S_2(\eps) \\
\ \ \vdots \hspace{2.5cm} \vdots \hspace{3.5cm} \vdots  \hspace{3cm} \vdots \\
M_{N-1,1}(\eps)K_1 + M_{N-1,2}(\eps)K_2 + \cdots
+ M_{N-1,N-1}(\eps)K_{N-1} = -S_{N-1}(\eps) \\
\\
\rule{1mm}{0.1mm}\ \rule{1mm}{0.1mm}\ \rule{1mm}{0.1mm}\
\rule{1mm}{0.1mm}\ \rule{1mm}{0.1mm}\ \rule{1mm}{0.1mm}\
\rule{1mm}{0.1mm}\ \rule{1mm}{0.1mm}\ \rule{1mm}{0.1mm}\
\rule{1mm}{0.1mm}\ \rule{1mm}{0.1mm}\ \rule{1mm}{0.1mm} \
\rule{1mm}{0.1mm}\ \rule{1mm}{0.1mm}\ \rule{1mm}{0.1mm}\
\rule{1mm}{0.1mm}\ \rule{1mm}{0.1mm}\ \rule{1mm}{0.1mm}\
\rule{1mm}{0.1mm}\ \rule{1mm}{0.1mm}\ \rule{1mm}{0.1mm}\
\rule{1mm}{0.1mm}\ \rule{1mm}{0.1mm}\ \rule{1mm}{0.1mm}\
\rule{1mm}{0.1mm}\ \rule{1mm}{0.1mm}\ \rule{1mm}{0.1mm}\
\rule{1mm}{0.1mm}\ \rule{1mm}{0.1mm}\ \rule{1mm}{0.1mm}\
\rule{1mm}{0.1mm}\ \rule{1mm}{0.1mm}\ \rule{1mm}{0.1mm}\
\rule{1mm}{0.1mm}\ \rule{1mm}{0.1mm}\ \rule{1mm}{0.1mm}\
\rule{1mm}{0.1mm}\ \rule{1mm}{0.1mm}\ \rule{1mm}{0.1mm}\
\rule{1mm}{0.1mm}\ \rule{1mm}{0.1mm}\ \rule{1mm}{0.1mm}\
\rule{1mm}{0.1mm}\ \rule{1mm}{0.1mm}\ \rule{1mm}{0.1mm}\
\rule{1mm}{0.1mm}\ \rule{1mm}{0.1mm}\ \rule{1mm}{0.1mm}\
\rule{1mm}{0.1mm}\ \rule{1mm}{0.1mm}\ \rule{1mm}{0.1mm}  \\
\\
M_{N,1}(\eps)K_1 + M_{N,2}(\eps)K_2 + \cdots
+ M_{N,N-1}(\eps)K_{N-1} = -S_{N}(\eps) \\
\ \ \vdots \hspace{2.5cm} \vdots \hspace{3.5cm} \vdots  \hspace{3cm} \vdots \\
M_{2N-2,1}(\eps)K_1 \!+\! M_{2N-2,2}(\eps)K_2 \!+\! \cdots \!+\!
M_{2N-2, N-1}(\eps)K_{N-1} \!=\! -S_{2N-2}(\eps),  \end{array}
\!\!\right\} \label{eqnova412} \eey

\vspace{5mm}

\nd a system of linear algebraic equations for the constants $K_0$,
$K_1$, $K_2, \ldots, K_{N-1}$. The constant $K_0$ may be considered
separately from the set of the remaining ones since it appears
solely in the equation generated by the condition imposed on
$G_0^{(N)}(\eps)$, and its value is accordingly to be determined
from Eq. (\ref{eqnova411}) in isolation once the constants $K_1,
K_2, \ldots, K_{N-1}$ have been found by solving the system of
equations associated with Eq. (\ref{eqnova412}). This decoupling of
the equation for $K_0$ from the equations for the other constants is
of course an aspect of the arbitrariness of the value ascribable to
the flux function at the boundary.

Since the $K$'s \ in Eq. (\ref{eqnova412}) count as $N-1$ and the
equations as $2N-2$, the system we have in hand comprehends $N-1$
equations in excess to the number of unknowns, and we may expect it
to be incompatible, that is to say, not admitting of a solution. Of
course, this is just a manifestation of the inability of the
coordinate system we have adopted in fulfilling the purpose of
serving as supporting frame to the generation of a reference set of
solutions to the Grad-Shafranov equation that be capable to
accommodate themselves to the shape of the given boundary.

Now, from the general expressions for the coefficients of the
harmonics that combine to make up the multipole solutions, given by
Eqs. (A.2) and (A.4) in Ref. \cite{quatro}, it is apparent that, for
a fixed first suffix $i$, the dominant coefficient is that for which
the second suffix is $j = i$, and such that \beq M_{ii}(\eps) \sim
\eps^i\ . \label{eqnova413} \eeq Thus, considered two any equations
in the set defined by Eq. (\ref{eqnova412}), for the same departures
of the values respective to the constants $K_1, K_2, \ldots,
K_{N-1}$ from those by which they can be both simultaneously
satisfied, the smallest between the errors that will result for the
values of the terms on the right hand side of one and the other
equations will be associated with the equation of the highest order
(that is to say, the equation having the largest suffix $i$ for the
coefficients $M_{ij}$, $j = 1, 2, \ldots, N-1$). In this trend of
the error to diminish with the increasing of the order of the
equation we find the possibility of extracting a meaningful
information concerning the boundary value problem for the flux
function from the system represented by Eq. (\ref{eqnova412}).
Indeed, if we shall abandon the equations of the highest orders in
this system, we can arrive at a reduced system in which the number
of equations will equal the number of unknowns, which will be
solvable, and whose solution for the coefficients $K_i$ when
substituted in Eq. (\ref{eqnova41}) will cause the annihilation at
the boundary of those trigonometric components of the flux function
having the largest amplitudes. In this way we shall have an exact
solution to the partial differential equation that will satisfy
approximately the condition at the boundary; the error it will carry
along the frontier of the domain of integration will be given by the
sum of the trigonometrical components of the partial flux function
whose amplitudes will not have contributed to the generation of
equations in the system for the constants $K_i$, the leading term of
which will be dominated by a factor having the form of the inverse
aspect ratio raised to a power equal to the lowest of the orders of
the equations that will have been neglected in the system.

We thus disregard the equations of the orders $i = N$ to $i = 2N-2$,
which are below the dashed line in Eq. (\ref{eqnova412}), and
consider those of the orders $i = 1$ to $i = N-1$ as the ones that
contain the profitable information concerning the constants $K_1,
K_2, \ldots, K_{N-1}$. For the system of $N-1$ equations and $N-1$
unknowns to which Eq. (\ref{eqnova412}) is in this way reduced, the
determinant is: \beq D_A^{(N)}(\eps) = \left|\begin{array}{cccc}
M_{11}(\eps) &
M_{12}(\eps) & \cdots & M_{1,N-1}(\eps) \\
M_{21}(\eps) & M_{22}(\eps) & \cdots & M_{2,N-1}(\eps) \\
\vdots & \vdots & & \vdots \\
M_{N-1,1}(\eps) & M_{N-1,2}(\eps) & \cdots & M_{N-1,N-1}(\eps)
\end{array}\right| \ , \label{eqnova414}
\eeq

\nd and the solution, given by Cramer's rule, is:

\beq K_j^{(N)}\!=\! \frac{\left|\begin{array}{ccc|c|ccc} \cline{4-4}
M_{11}(\eps) & \cdots & M_{1,j\!-\!1}(\eps) & -S_1(\lb;\eps) &
M_{1,j\!+\!1}(\eps)
& \cdots & M_{1,N\!-\!1}(\eps) \\
M_{21}(\eps) & \cdots & M_{2,j\!-\!1}(\eps) & -S_2(\lb;\eps)&
M_{2,j\!+\!1}(\eps) & \cdots & M_{2,N\!-\!1}(\eps) \\
\vdots & & \vdots & \vdots & \vdots & & \vdots \\
M_{N\!-\!1,1}(\eps) & \cdots & M_{N\!-\!1,j\!-\!1}(\eps) &
-\!S_{N\!-\!1}(\lb;\eps)& M_{N\!-\!1,j\!+\!1}(\eps) & \cdots &
M_{N\!-\!1,N\!-\!1}(\eps) \\ \cline{4-4}
\end{array}\right|}{D_A^{(N)}(\eps)} \label{eqnova415}
\eeq

\vspace{2mm}

\nd ($j = 1, 2, \ldots, N-1$), where the superscripts attached to
symbols (here $D_A^{(N)}(\eps)$ and $K_j^{(N)}$) in general denote
the number of multipole solutions that enter the evaluation of the
quantities the symbols represent through the expression adopted for
the partial flux function. The integer corresponding to the order of
the column constituted by the driving terms to the equations being
solved simultaneously (put within a frame) in the determinant
appearing as the numerator of the expression for $K_j^{(N)}$
according to Cramer's equals the suffix $j$ of this constant. Note
that the constants $K_j^{(N)}$, as given by Eq. (\ref{eqnova415}),
depend linearly on $\lb$ through the dependence the functions
$S_n(\lb; \eps)$ keep on this last quantity.

Besides being calculable by means of Eq. (\ref{eqnova411}) it is
possible to give $K_0$ an expression that does not presume the
knowledge of the constants $K_1, K_2, \ldots, K_{N-1}$. Indeed, if
we shall add Eq. (\ref{eqnova411}) as first equation to the ensemble
made up by the $N-1$ equations of the lowest orders in Eq.
(\ref{eqnova412}), we shall have a composite system for which the
determinant will still be given by Eq. (\ref{eqnova414}), since its
first column will be constituted by unity for the top element and by
zero for all the ones below it; the solution for $K_0$ will be
written as:

\beq K_0^{(N)} = \frac{\left|\begin{array}{cccc}
-S_0(\lb;\eps) & M_{01}(\eps) & \cdots & M_{0,N-1}(\eps) \\
-S_1(\lb;\eps) & M_{11}(\eps) & \cdots & M_{1,N-1}(\eps) \\
\vdots & \vdots & & \vdots \\
-S_{N-1}(\lb;\eps) & M_{N-1,1}(\eps) & \cdots & M_{N-1,N-1}(\eps)
\end{array}\right|}{D_A^{(N)}(\eps)}\ . \label{eqnova416}
\eeq

\vv

When the expressions for $\ovl\psi_p(x,\ta)$, for $K_0$ and for
$K_i$ ($i = 1, 2, \ldots, N-1$), as given by Eqs.
(\ref{eq3.29nova}), (\ref{eqnova416}) and (\ref{eqnova415})
respectively, are used in Eq. (\ref{eqnova41}), the expression that
will result for the flux function will display an explicit
dependence on $\lb$, which is the input parameter of the problem
(besides the one having a purely geometrical meaning, $\eps$). From
a physical standpoint, however, a parameter having a greater
interest than $\lb$ (and one that can always be associated with a
configuration irrespectively of the shapes of the profiles that can
be assumed for the sources to the equilibrium in flux space) is the
Shafranov shift $\Dt$. This appears as one of the equilibrium
quantities that can be determined once the expression for the flux
function is known. We shall show next, however, that it is also
possible to conduct the solving process to the boundary value
problem in a way that the Shafranov shift is made present since the
very first stages of it, and such that the expression it generates
for the flux function contains an explicit reference to $\Dt$ (or
rather to a normalized version of $\Dt$) instead of to $\lb$.

In the magnetic configurations in which we have interest the
innermost of the flux surfaces degenerates into a line, the magnetic
axis, which, because of the symmetry up-down of the physical system
with regard to the equatorial plane of the torus, is always located
on that plane. The field lines on any flux surface intercept the
equatorial plane at right angles, and, correspondingly, the radial
component of the magnetic poloidal field, as evaluated with the help
of the expression for the flux function, must vanish at all points
lying on that plane between the inner and the outer edge of the
torus. This condition, which translates by  \beq
\frac{\ptl\psi}{\ptl\ta}(x, \ta = 0 \hbox{ or } \pi) = 0 \ ,
\label{eqnova417} \eeq is automatically satisfied by the flux
function by virtue of the properties of symmetry, consequent to the
geometrical assumptions on the boundary, with which this function is
endowed. The location of the magnetic axis hence follows solely from
the proposition that the polar component of the magnetic poloidal
field must vanish at the points along the circumference of a circle
 lying on the equatorial plane. This last condition writes as:
 \beq \frac{\ptl\psi}{\ptl x}(x = \dt, \ta = 0) = 0 \
, \label{eqnova418} \eeq where \beq \dt \equiv \frac{\Dt}{R_C}
\label{eqnova419} \eeq is the Shafranov shift normalized to the
torus major radius. Note that in Eq. (\ref{eqnova418}) we have
specified the angular coordinate as $\ta = 0$, anticipating a
property of the equilibrium pinch to be evinced in later Sections,
to know, that the shifting of the magnetic axis always proceeds from
the centre line towards the outer edge of the torus.

From Eq. (\ref{eqnova44}), the normalized partial flux linked with
the annular region $1 \le x \le 1 + \eps$ distributes itself along
the radial coordinate according to: \beq \ovl\psi^{(N)}(x,\ta = 0) =
G_0^{(N)}(x) + \sum_{n=1}^{2N-2}G_n^{(N)}(x)\ . \label{eqnova420}
\eeq

Introducing $G_0^{(N)}(x)$ and $G_n^{(N)}(x)$ as given by Eq.
(\ref{eqnova45}) and (\ref{eqnova46}) respectively in Eq.
(\ref{eqnova420}), the resulting expression for $\ovl\psi^{(N)}(x,
\ta = 0)$ can be put in the form: \beq \ovl\psi^{(N)}(x,\ta=0) =
\sum_{n=0}^{2N-2}S_n(x) + K_0 +
\sum_{n=0}^{2N-2}\left[\sum_{j=1}^{N-1} M_{nj}(x)K_j\right]\ .
\label{eqnova421} \eeq

Recalling Eq. (\ref{eq3.33nova}) for $S_n(x)$ we can write the first
summation term on the right hand side of Eq. (\ref{eqnova421}) as:
\beq \sum_{n=0}^{2N-2} S_n(x) = P(x) + \lb Q(x)\ , \label{eqnova422}
\eeq where we have made use of the following definitions: \bey P(x)
&=& \sum_{n=0}^{2N-2}P_n(x)\ , \label{eqnova423} \\
Q(x) &=& \sum_{n=0}^{2N-2}Q_n(x)\ , \label{eqnova424} \eey in which,
consistent with the values specified for $S_n(x)$ in Eq.
(\ref{eqnova47}), there are implied the assumptions respectively:
\beq
\left.\begin{array}{c} P_n(x) = 0 \\ \\
\hbox{and} \\ \\Q_n(x) = 0 \end{array}\right\} \ \ \hbox{ for } n
\ge 5\ . \label{eqnova425} \eeq

Note that, since the least number of multipole solutions to enter a
combination aimed to approximate the flux function is $N = 3$, the
sums defining $P(x)$ and $Q(x)$ in Eqs. (\ref{eqnova423}) and
(\ref{eqnova424}), for any $N$, include the totality of the
functions $P_n(x)$ and $Q_n(x)$ that are non-null according to Eq.
(\ref{eqnova425}), and there is no need to attach a superscript $N$
to $P(x)$ and $Q(x)$ in order to have the order of the partial flux
from which they proceed indicated. In Appendix A expressions for
both these functions show that $P(x) = Q(x)$.

With regard to the term of double summation that appears on the
right hand side of Eq. (\ref{eqnova421}), by interchanging the order
in which sums over the indices $j$ and $n$ are performed, it
becomes: \beq
\sum_{n=0}^{2N-2}\left[\sum_{j=1}^{N-1}M_{nj}(x)K_j\right] =
\sum_{j=1}^{N-1} V_j^{(N)}(x)K_j\ , \label{eqnova426} \eeq where we
have introduced the functions: \beq V_j^{(N)}(x) =
\sum_{n=0}^{2N-2}M_{nj}(x)\ \ \ (j = 1, 2, \ldots, N-1).
\label{eqnova427} \eeq Substituting the two summation terms on the
right hand side of Eq. (\ref{eqnova421}) by the representations of
Eq. (\ref{eqnova422}) and Eq. (\ref{eqnova426}) respectively, we
reach: \beq \ovl\psi^{(N)}(x, \ta = 0) = P(x) + \lb Q(x) + K_0 +
\sum_{j=1}^{N-1}K_jV_j^{(N)}(x)\ . \label{eqnova428} \eeq
Application of the condition stated as Eq. (\ref{eqnova418}) to Eq.
(\ref{eqnova428}) leads us to an algebraic relation connecting the
normalized radial coordinate $x = \dt$ of the magnetic axis, the
equilibrium parameter $\lb$ and the $N-1$ constants $K_1, K_2,
\ldots, K_{N-1}$. Written in the form that most conveniently adapts
itself to our future purposes, this relation is: \beq
\sum_{j=1}^{N-1}V_j^{(N)'}(\dt)K_j + Q'(\dt)\lb = -P'(\dt)\ ,
\label{eqnova429} \eeq where primes denote derivatives with respect
to the argument. Note that, since the highest power of $x$ to appear
in the multipole solution $\vf^{(i)}(x,\ta)$ is $2i$ according to
Eq. (2.32) in Ref. \cite{quatro}, the partial flux of order $N$ in
Eq. (\ref{eqnova428}) is a complete polynomial of the degree $2N-2$
in $x$, and Eq. (\ref{eqnova429}) is an algebraic relation of the
$(2N-3)$th degree in $\dt$ whose coefficients exhibit linear
dependences on the parameter $\lb$ and on the constants $K_i$ ($i =
1, 2, \ldots, N-1$). Concrete forms of the last mentioned equation
for any particular $N$ show in general that its terms on both of its
sides contain $1 + \dt$ as a common factor\footnote{This is a
consequence of the fact that the dependences of the functions
$V^{(N)}(\dt)$, $Q(\dt)$ and $P(\dt)$ on $\dt$ are mediated by the
function of $\dt$ that defines the variable $\chi$ in Eq.
(\ref{eqnova437}) ahead.}.

Equation (\ref{eqnova429}) can be used to find a representation of
$\lb$ in terms of $\dt$, and then, by substituting it in the
expression for the particular solution $\ovl\psi_p(x,\ta)$ and in
those for the constants $K_j$'s ($j = 0, 1, 2, \ldots, N-1$) in the
forms they are originally generated by application of Eqs.
(\ref{eqnova416}) and (\ref{eqnova415}), the flux function can be
made to appear as dependent on $\dt$ instead of on $\lb$. A more
direct approach than this to accomplish the same end is however
possible and in point of fact preferable.

We start by viewing the parameter $\lb$ momentarily as an unknown of
the problem, on the same footing as the constants $K_j$ ($j = 0, 1,
2, \ldots, N-1$), and take $\dt$ as a given quantity. Writing the
driving terms in the equations placed above the dashed line in Eq.
(\ref{eqnova412}) in accordance with the pattern of representation
of the functions $S_i(x)$ in  Eq. (\ref{eq3.33nova}), these same
equations can be concisely restated as: \beq
\sum_{j=1}^{N-1}M_{ij}(\eps)K_j + Q_i(\eps)\lb = -P_i(\eps)\ \ \ (i
= 1, 2, \ldots, N-1)\ , \label{eqnova430} \eeq which is a form
appropriate to make them appear as a system of $N-1$ equations for
the $N$ unknowns $K_1, K_2, \ldots, K_{N-1}$ and $\lb$. To close the
set, we add Eq. (\ref{eqnova429}) to it as last equation. The
determinant of the linear system of $N$ unknowns and $N$ equations
in this way composed is: \beq D_B^{(N)}(\eps, \dt) =
\left|\begin{array}{ccccc} M_{11}(\eps) &
M_{12}(\eps) & \cdots & M_{1,N-1}(\eps) & Q_1(\eps) \\
M_{21}(\eps) &
M_{22}(\eps) & \cdots & M_{2,N-1}(\eps) & Q_2(\eps) \\
\vdots & \vdots & & \vdots & \vdots \\
M_{N-1,1}(\eps) &
M_{N-1,2}(\eps) & \cdots & M_{N-1,N-1}(\eps) & Q_{N-1}(\eps) \\
&&&& \\
V_1^{(N)'}(\dt) & V_2^{(N)'}(\dt) & \cdots & V_{N-1}^{(N)'}(\dt) &
Q'(\dt) \end{array}\right| \ . \label{eqnova431} \eeq

We write the solution for $\lb$ in the form: \beq F^{(N)}(\eps, \dt)
= \lb\ , \label{eqnova432} \eeq where \beq F^{(N)}(\eps,\dt) =
\frac{\left|\begin{array}{ccccc} M_{11}(\eps) & M_{12}(\eps) &
\cdots & M_{1,N-1}(\eps) & -P_1(\eps) \\
M_{12}(\eps) & M_{22}(\eps) &
\cdots & M_{2,N-1}(\eps) & -P_2(\eps) \\
\vdots & \vdots & & \vdots & \vdots \\
M_{N-1,1}(\eps) & M_{N-1,2}(\eps) &
\cdots & M_{N-1,N-1}(\eps) & -P_{N-1}(\eps) \\
&&&& \\
V_1^{(N)'}(\dt) & V_2^{(N)'}(\dt) & \cdots & V_{N-1}^{(N)'}(\dt) &
-P'(\dt) \end{array}\right|}{D_B^{(N)}(\eps, \dt)}\ .
\label{eqnova433} \eeq

\vv

Equation (\ref{eqnova432}) of course gives us the solution for $\lb$
in the same form that it would be provided by Eq. (\ref{eqnova429})
if in this latter equation the constants $K_j$ would first be
expressed in the form that stems from their evaluation according to
the formula in Eq. (\ref{eqnova415}). Since the terms on the one and
the other side of Eq. (\ref{eqnova429}), as we have seen, include $1
+ \dt$ as a common factor, which would be stricken out in a process
of obtaining the solution for $\lb$, we conclude that
$F^{(N)}(\eps,\dt)$, as given by Eq. (\ref{eqnova433}), is a
rational function of $\dt$ whose numerator and denominator are
polynomials of the degree $2N-4$ in $\dt$.

To express the particular solution in terms of the relative
Shafranov shift instead of in terms of the equilibrium parameter, we
substitute $\lb$ by the function $F^{(N)}(\eps,\dt)$ in Eq.
(\ref{eq3.29nova}). This substitution procedure can also be applied
to achieve analogous end with respect to the constants $K_1, K_2,
\ldots, K_{N-1}$ as evaluated according to the rule of Eq.
(\ref{eqnova415}), but a more direct approach than such is here made
possible by appeal to that same linear system that provides us with
the connection between $\lb$ and $\dt$. Indeed, the solution for the
constants $K_j$ ($j = 1, 2, \ldots, N-1$), as extracted from the
system constituted by Eq. (\ref{eqnova430}) and Eq.
(\ref{eqnova429}), writes as:

{\arraycolsep=1.1mm \bey K_j^{(N)} &\!=\!&
\frac{\left|\begin{array}{cccccccc} M_{11}(\eps) & \cdots &
M_{1,j-1}(\eps) & -P_1(\eps) &
M_{1,j+1}(\eps) & \cdots & M_{1,N-1}(\eps) & Q_1(\eps) \\
M_{21}(\eps) & \cdots & M_{2,j-1}(\eps) & -P_2(\eps) &
M_{2,j+1}(\eps) & \cdots & M_{2,N-1}(\eps) & Q_2(\eps) \\
\vdots & & \vdots & \vdots & \vdots & & \vdots & \vdots \\
M_{N\!-\!1,1}\!(\eps) & \cdots & M_{N\!-\!1,j\!-\!1}\!(\eps) &
-\!P_{N\!-\!1}\!(\eps) & M_{N\!-\!1,j\!+\!1}\!(\eps) & \cdots &
M_{N\!-\!1,N\!-\!1}\!(\eps) & Q_{N\!-\!1}\!(\eps) \\
&&&&&&& \\
V_1^{(N)'}(\dt) & \cdots & V_{j-1}^{(N)'}(\dt) & -P'(\dt) &
V_{j+1}^{(N)'}(\dt) & \cdots & V_{N-1}^{(N)'}(\dt) & Q'(\dt)
\end{array}\right|}{D_B^{(N)}(\eps,\dt)} \nonumber \\
&& (j = 1, 2, \ldots, N-1), \label{eqnova434} \eey }  and contains
reference solely to $\dt$ in place of to $\lb$. In what regards
$K_0$, the kind of representation that is now desired can be most
simply obtained by resorting to a procedure parallel to that adopted
to derive its representation in terms of $\lb$. We consider a system
of $N + 1$ equations consisting of those expressed by Eqs.
(\ref{eqnova430}) and (\ref{eqnova429}) plus the following one,
taken as the first of the set: \beq K_0 +
\sum_{j=1}^{N-1}M_{0j}(\eps)K_j + Q_0(\eps)\lb = -P_0(\eps)\ ,
\label{eqnova435} \eeq which is not another one but Eq.
(\ref{eqnova411}) with $S_0(\eps,\lb)$ replaced by its definition
according to Eq. (\ref{eq3.33nova}), and written in a manner that
brings $\lb$ to appear as an unknown the same as the $N$ constants
$K_0, K_1, \ldots, K_{N-1}$. Since Eq. (\ref{eqnova435}) is the only
equation within the whole set to contain $K_0$, the determinant of
the system reduces to that given by Eq. (\ref{eqnova431}). The
solution for $K_0$ is then:

\beq K_0^{(N)} = \frac{\left|\begin{array}{cccccc} -P_0(\eps) &
M_{01}(\eps) &
M_{02}(\eps) & \cdots & M_{0,N-1}(\eps) & Q_0(\eps) \\
-P_1(\eps) & M_{11}(\eps) & M_{12}(\eps) &
\cdots & M_{1,N-1}(\eps) & Q_1(\eps) \\
-P_2(\eps) & M_{21}(\eps) & M_{22}(\eps) &
\cdots & M_{2,N-1}(\eps) & Q_2(\eps) \\
\vdots & \vdots & \vdots & & \vdots & \vdots \\
-P_{N-1}(\eps) & M_{N-1,1}(\eps) & M_{N-1,2}(\eps) &
\cdots & M_{N-1,N-1}(\eps) & Q_{N-1}(\eps) \\
&&&& \\
-P'(\dt) & V_1^{(N)'}(\dt) & V_2^{(N)'}(\dt) & \cdots &
V_{N-1}^{(N)'}(\dt) & Q'(\dt) \end{array}\right|}{D_B^{(N)}(\eps,
\dt)} \ . \label{eqnova436} \eeq

With the constants $K_0, K_1, \ldots, K_{N-1}$ and the equilibrium
parameter in the particular solution expressed in terms of the
relative Shafranov shift, all elements are given to make possible a
representation of the flux function in which $\dt$ appears as the
one physical parameter to characterize globally the equilibrium,
assuming a role that belonged to $\lb$ in the original formulation
of the problem. An appreciable simplification in the mathematical
development of the solution and in the form of the flux function can
however be achieved if, in place of $\dt$, another quantity will be
taken for reference, which is: \beq \chi \equiv (1 + \dt)^2 - 1 \ .
\label{eqnova437} \eeq

Because of the central role it shows to play in the theory of
representation of equilibrium configurations by series of multipole
solutions, we shall reserve the name of \textit{displacement
variable} to $\chi$. The relative Shafranov shift can be obtained
from $\chi$ through the relation: \beq \dt = \sqrt{1 + \chi} - 1 \ .
\label{eqnova438} \eeq Note that, being $\dt$ positive, as it indeed
is, $\chi$ is also a positive quantity, and that, in the limit of
small values, the two quantities obey a relation of proportionality
between themselves: $\chi \simeq 2\dt$, vanishing one when the other
vanishes.

A simple geometrical interpretation can be given to $\chi$ as
follows. The defining relation for the displacement variable can be
stated in the form: \beq \chi = \dt \left(1 +
\frac{R_{M}}{R_C}\right)\ . \label{eqnova439} \eeq Also, if we
introduce a quantity $Z_{MC}$ with the dimension of length by means
of the relation: \beq \chi \equiv \frac{Z_{MC}^2}{R_C^2}\ ,
\label{eqnova440} \eeq from the above definition of $\chi$ we find
that \beq Z_{MC}^2 = R_M^2 - R_C^2\ , \label{eqnova441} \eeq which
shows that $Z_{MC}$ is the measure of half the segment determined on
a vertical line passing through $C$ by an arc of circumference of
radius $R_M$ drawn with centre at $O$ in Fig. 2.1.

To accomplish the transformation of $F^{(N)}(\eps, \dt)$,
$K_j^{(N)}$ ($j = 1, 2, \ldots, N-1$) and $K_0^{(N)}$ from functions
of $\dt$ to functions of $\chi$ we have first to proceed to the
transformations: \vspace{1mm} \beq \left.\begin{array}{l}
V_j^{(N)}(\dt) \to V_j^{(N)}(\chi) \ \ \ (j =
1, 2, \ldots, N-1) \\ \\
P(\dt) \to P(\chi) \\ \\
Q(\dt) \to Q(\chi) \end{array}\right\} \label{eqnova442} \eeq
\vspace{2mm}

\nd by using Eq. (\ref{eqnova438}) for $\dt$ in the expressions for
the functions listed on the left hand side of Eq. (\ref{eqnova442});
next, taking the derivatives of the functions listed on the right
hand side with respect to $\chi$, we obtain the new elements that
are to substitute those forming the bottom rows of the determinants
in Eqs. (\ref{eqnova431}), (\ref{eqnova433}) (\ref{eqnova434}) and
(\ref{eqnova436}). Note that in this scheme of substitution we have
omitted the ``inverse scale factor'' of the transformation of
derivatives with respect to $\dt$ to derivatives with respect to
$\chi$, given by \bey \frac{\dif \chi}{\dif \dt} &=& 2(1 + \dt) \nonumber \\
&=& 2\sqrt{1 + \chi}\ , \label{eqnova443} \eey which is indeed
unnecessary to be written out, for it would appear in both the
numerator and the denominator of the expression for each of the
quantities we are seeking to have transformed from one
representation to the other, and would be cancelled. Note also that
in Eq. (\ref{eqnova442}) we have used, in conformity with the common
practice, the same symbol for a function of $\dt$ and for its
transformed version as a function of $\chi$, although the
dependences they keep each on the respective variables are
different. In Appendix A the reader will find the expressions for
$P(\chi)$, $Q(\chi)$, $P'(\chi)$ and $Q'(\chi)$, and in Appendix C
those for $V_j^{(N)'}(\dt)$ and $V_j^{(N)'}(\chi)$ ($j = 1, 2,
\ldots, 9$).

One of the advantages offered by the displacement variable with
respect to the relative Shafranov shift as working quantity is that
the former is able to reduce the degrees of the polynomials entering
the expression of $F^{(N)}(\eps,\dt)$ to the half. That is to say,
if in the function defined by Eq. (\ref{eqnova433}) we replace $\dt$
by $\chi$ according to Eq. (\ref{eqnova438}), the resulting
function, which we shall denote by $F^{(N)}(\eps,\chi)$, will be
given by the quotient of two polynomials, each of the degree $N-2$
in $\chi$. In substitution to Eq. (\ref{eqnova432}), the solution
for $\lb$ will be written as: \beq F^{(N)}(\eps, \chi) = \lb\ .
\label{eqnova444} \eeq

We shall call the function $F^{(N)}(\eps,\chi)$ the
\textit{equilibrium function of order} $N$ and Eq. (\ref{eqnova444})
the \textit{equilibrium equation in the approximation of $N$
multipoles}. Reverting now to the view of $\lb$ as a parameter given
and considering that $\eps$ is also a datum of the problem, then Eq.
(\ref{eqnova444}) is an algebraic equation of the $(N-2)$-th degree
for $\chi$. In a graphical interpretation of this equation, the
equilibrium value (or possibly the equilibrium values) of the
displacement variable $\chi$ for a fixed aspect ratio $\eps$ of the
toroidal chamber appears as determined by the intersection of the
curve $F^{(N)}(\eps, \chi)$ \textit{versus} $\chi$ with the straight
line drawn parallel to the axis of the abscissas at the ordinate
$\lb$. Once a value of the displacement variable $\chi$ is known,
the corresponding Shafranov shift can be evaluated by means of Eq.
(\ref{eqnova438}), and the location of a magnetic axis be set.

The representation of the flux function that takes the equilibrium
parameter as the reference quantity to characterize an equilibrium
configuration depends linearly on $\lb$, whereas the representation
that expresses itself in terms of the displacement variable exhibits
a more complicated pattern of dependence on the quantity it takes
for reference than the linear one, being as it is a rational
function of $\chi$. The latter representation, however, results in
general simpler in aspect than the former, or at least more compact.
In applying in later Sections the method of solution to the
Grad-Shafranov boundary value problem by expansion of the poloidal
flux function in series of multipole solutions to obtain the
description of a particular equilibrium configuration we shall have
occasion to illustrate the use of one and the other of these two
possible representations of the flux function.

It is opportune to note here that, since the particular solution
$\ovl\psi_p(x,\ta)$ and the multipole solutions $\vf^{(n)}(x,\ta)$
($n = 1, 2, \ldots$) vanish all at the coordinate $x = 0$, Eq.
(\ref{eqnova41}) shows that $K_0^{(N)}$ physically means the
approximation of order $N$ to the normalized magnetic poloidal flux
at the centre of the torus cross section: \beq K_0^{(N)} =
\ovl{\psi_C^{(N)}} \equiv \ovl{\psi^{(N)}}(x = 0, \ta)\ .
\label{eqnova445} \eeq

We have noted, in the sequel of Eq. (\ref{eq3.31}), that the two
components of the gradient of the particular solution, and thus of
the magnetic poloidal field generated by it, vanish at the
coordinate $x = 0$. Also noticed, in Ref. \cite{quatro}, that the
poloidal fields derived from the multipole solutions vanish all at
the coordinate $x = 0$, exception being taken to the field
proceeding from the solution of order $n = 1$.  The multipole
solution of this last mentioned order, which is expressed in
cylindrical coordinates by: \beq \vf^{(1)}(R,z) =
\frac{1}{4}\left(\frac{R^2}{R_C^2} - 1 \right)\ , \label{eqnova446}
\eeq gives origin to a poloidal magnetic field whose radial
component is null and whose axial component is constant everywhere
in space. We thus conclude, by Eq. (\ref{eqnova41}), that the
poloidal magnetic field associated with the approximation of order
$N$ to the poloidal flux function at the centre of the torus cross
section is given by (see Eq. (\ref{eq10.6lin}) farther on): \beq
\vec B_v^{(N)} = \frac{1}{2}s_PK_1^{(N)}B_{\phi_C}\vec k\ ,
\label{eqnova447} \eeq an expression for which, to have it written
in a way  such as it has been here, we have to have remembered that
$\ovl\psi^{(N)}(x,\ta)$ embodies a double normalization, that by
$B_{\phi_C}R_C^2$ (introduced by Eq. (\ref{eq2.3})) and that by
$s_P$ (introduced by Eq. (\ref{eq4.2nova})). The field of Eq.
(\ref{eqnova447}) is obviously the vertical magnetic field that has
to be applied to the toroidal pinch by external means to ajust the
position of the plasma column inside the containing chamber.

It is also of interest to have an expression for the magnetic
poloidal flux at the magnetic axis. From the expression for the flux
function that identifies itself with the particular solution to the
Grad-Shafranov equation, given in cylindrical coordinates by Eq.
(\ref{eq3.31}),  and from the expressions that are identical with
those for the multipole solutions, given in cylindrical coordinates
in Appendix C in Ref. \cite{quatro}, we see that, for $z = 0$, $R =
R_M$ and $R_A \equiv R_C$, they reduce all to a term proportional to
some power of $(R_M^2 - R_C^2)/R_C^2$. By referring to Eqs.
(\ref{eqnova440}) and (\ref{eqnova441}) we learn that this quantity
equals $\chi$. We can thus write the expression for the normalized
magnetic poloidal flux at the magnetic axis within the approximation
of order $N$ to the poloidal flux function as: \beq
\ovl{\psi_M^{(N)}} = \frac{1}{16}\left[1 + F^{(N)}(\eps,
\chi)\right]\chi^2 + K_0^{(N)} + \sum_{i=1}^{N-1}K_i^{(N)}
(-1)^{i-1}(N.F.)_i\chi^i\ , \label{eqnova449} \eeq where we have
replaced $\lb$ by $F^{(N)}(\eps, \chi)$ according to Eq.
(\ref{eqnova444}), and $(N.F.)_i$ is the numerical factor
multiplying the first term on the right hand side of the expression
for the multipole solution $\vf^{(i)}(\xi,\nu)$ in Appendix C in
Ref. \cite{quatro}: $(N.F.)_1 = 1/4$, $(N.F.)_2 = 1/4$, $(N.F.)_3 =
1/8$, $(N.F.)_4 = 1/20$, etc.

\vv\vv

\begin{center}
{\bf V. THE CONSTRAINTS ON THE VALUES OF THE DISPLACEMENT VARIABLE
AND THE LIMITATIONS OF THE METHOD OF SERIES OF MULTIPOLE SOLUTIONS}
\end{center}

\setcounter{section}{5} \setcounter{equation}{0}

\vv

Although the physical parameters that appear as input to the
Grad-Shafranov equation are the pressure gradient and the gradient
of half the squared toroidal field function, which, if uniform in
flux space as we are considering them to be in this paper, can be
expressed under proper normalizations as the constants $s_P$ and
$s_I$, and although for the resolution by the method of series of
multipole solutions  the relevant physical parameter shows to be the
ratio of $s_I$ to $s_P$, which we have called the equilibrium
parameter and denoted by $\lb$, the analysis and reasonings on the
relations between the conceptual tools and operational devices of
the method on one side and the physical properties of the
equilibrium on the other are better conducted in terms of the
parameter that we have denominated the displacement variable and
have denoted by $\chi$. The value of $\chi$ for a given order $N$ of
approximation to the flux function is fixed by the value of $\lb$
through Eq. (\ref{eqnova444}), which, however, given the
mathematical complexity of the equilibrium function, is not a
connection such as to admit of a prompt interpretation. Under this
perspective, the relative Shafranov shift $\dt$, which relates to
$\chi$ in a more transparent way than $\lb$ does, becomes the most
significant single physical parameter to spring out from the method.

In the remaining Sections of this paper we shall be pursuing the
objective of illustrating the method of series of multipole
solutions by applying it to obtain approximate solutions to the
Grad-Shafranov boundary value problem that satisfy certain
preconditions, the effect of which is to place them in the class of
those that partake of the greatest physical interest. Such
preconditions, which can in a way be considered as complementary to
the boundary conditions and can be stated in simple mathematical
terms, prevent from the onset of the solving process the appearance
of solutions corresponding to magnetic configurations that in some
manner are unable to restrain the plasma from drifting to the walls
of the containing vase and thus do not contrive to achieve its
confinement. It should be remembered, however, that the equilibrium
quasi-states we shall be deriving and describing still may or may
not be dynamically stable to small perturbations, a concern that
falls outside the scope of the present paper.

The first precondition the solutions we are interested in must
satisfy is that they admit a magnetic axis that falls inside the
chamber. Since the position of the null poloidal field with respect
to the centre of the torus cross section is directly specified by
the displacement variable, this precondition signifies that there is
a maximal allowed value $\chi_{\max}$ for $\chi$, at which the
magnetic axis, on the equatorial plane, would be placed at the outer
edge of the torus: \beq \chi < \chi_{\max} \equiv (1 + \eps)^2 - 1\
. \label{eq6.1} \eeq

The second precondition stems from the recognition that the
incorporation of the pressure gradient parameter into the quantity
the problem is solved for by way of the relation stated in Eq.
(\ref{eq4.2nova}) has the effect of uniquely defining the sign that
the solution must take in the domain enclosed by the boundary.
Indeed, Eq. (\ref{eq2.7}), which introduces the constant $s_P$ as
the rate of decay of the pressure with the flux coordinate $\psi$,
transforms to \beq \frac{\dif}{\dif\ovl\psi}\left(\frac{\hat
p}{2}\right) = -s_P^2  \label{eq3.35} \eeq when the coordinate
variable is changed from $\psi$ to $\ovl\psi$. Equation
(\ref{eq3.35}) above defines the quantity that we have placed on its
left hand side as negative, independently of the sign of $s_P$.
Thus, if we agree that the pressure is greater in the interior than
at the edge of the plasma, we must have $\ovl\psi$ negative within
the frontiers of the domain of the solution, and, in particular, at
the point associated with the location of the magnetic axis. If the
sign of $s_P$ is supposed to be positive, with this definition for
the sign of $\ovl\psi$, both $\psi$ and $\Psi$ are negative
quantities (except at the boundary, where they are null), while a
negative sign of $s_P$ would give them positive signs. The sign of
$s_P$ is ultimately irrelevant as it has no physical implication
whatever. We shall see in the Sections to come that the requirement
of a negative sign for $\ovl\psi$ in the plasma region has the
double consequence that the magnetic axis is always shifted from the
geometrical centre line of the torus outwardly, and that even for a
vanishingly small plasma pressure a nonvanishing shift is needed for
equilibrium. In other words, this means that, considered all
possible physical conditions for equilibrium, there is a minimal
positive value $\chi_{\min}$ that can be taken by the displacement
variable, \beq \chi \ge \chi_{\min}
> 0\ , \label{eq53nova} \eeq which is determined by the geometry of
the confining field (that is, by the inverse aspect ratio of the
magnetic configuration).

Finally, the third precondition is that the surface found for $\psi
= 0$ does not exhibit singular points, to borrow the denomination by
which they are referred to in Ref. \cite{nove}, at which the
gradient of the flux function vanishes: \beq \nabla\psi = 0\ ,
\label{eq6.3} \eeq and where a field line, as projected on the plane
of the torus cross section, branches into two field lines, since
this would lead to loss of the ability of the magnetic configuration
to confine the plasma.

The meaning of this last precondition is subtler than those of the
two first ones. For a sequence of equilibrium quasi-states generated
by increasing gradually the displacement variable since small values
while the inverse aspect ratio is kept constant we have found that
branching points break out on the boundary for some critical shift
of the magnetic axis relative to the centre of the torus cross
section as the culmination of a process of distortion of the surface
$\psi = 0$, which then becomes a separatrix of the magnetic
configuration, and ceases to be topologically equivalent to a
toroidal surface. There are reasons, however, to believe that this
evolvement  of the boundary surface into a separatrix we have
observed might not be inherent to the solutions that comply with the
first two preconditions, but merely a product of an unduly low
number of multipole solutions that have been employed to approximate
the flux function for such a high value of $\chi$, and, in this
sense, an artificial effect of the method. Indeed, as the order of
approximation is increased, we observe that the critical value of
$\chi$ is also increased, the distortion of the flux surfaces in the
vicinity of the boundary, which precedes the emergence of the
branching points, becomes less and less pronounced, and when they
finally break out the loop described by the segments of field lines
comprised between them shows a tendency to become more and more
shortened and tighter, suggesting that they would fuse into a single
point that at last would disappear if the number $N$ of multipole
solutions entering the composition of the flux function would become
infinite. In brief, all indications are that the occurrence of
points on the surface $\psi = 0$ whose coordinates verify the
singularity condition stated in Eq. (\ref{eq6.3}) can be suppressed
by simply including an adequate number of multipole solutions in the
approximation to the flux function. Thus the true content of the
third precondition is the setting of an upper limit to the allowed
range of variation of the displacement variable for a given order
$N$ of the partial flux as the value of $\chi$ that triggers off
structural alterations in the geometry of the boundary as they
appear portrayed by the method itself. We shall see in later
Sections that the limits we encountered according to this criterium
are much more stringent than the one fixed by Eq. (\ref{eq6.1}) for
all values of $N$ that have been considered.

A question that naturally arises concerns the properties of
convergence of the series of multipoles, to know, if by increasing
the number $N$ of multipole solutions the error associated with the
partial flux always decreases and the approximate solution such flux
represents then tends to a well defined limit (in which case the
series is convergent), or if the error diminishes with the increase
of $N$ up to an optimal value of the number of multipole solutions
and starts to grow without limit thereafter by further increase of
$N$ (which typifies an asymptotic series), or, if regardless of the
value of the approximate solution that is obtained for the minimal
value of $N$ (which is 3), any new term that is added to the series
makes the error amplify (in which case the series is divergent).

The reply we may give to this question as supported by examples
worked out in the present paper that apply the method to a variety
of equilibrium situations with a succession of values of $N$ to
approximate the flux function is that the third alternative (the
divergence of the series) should be excluded and that of the two
remaining ones the results consistently favour that of convergence.

The error carried by the partial flux of order $N$ is of course to
be evaluated at the boundary, where the values that would be assumed
by an exact solution are known. Two are the quantities we shall use
to measure the accuracy of the solutions obtained for the flux
function. The first, which we shall call simply the error, is given
by the value of the flux itself at the normalized radial coordinate
$x = \eps$, since were $\ovl\psi(x,\ta; \chi, \eps)$ an exact
solution to the boundary value problem, this would be zero. As we
intend to compare the accurateness reached by solutions to different
equilibria among themselves and also compare mutually those for the
same equilibria as obtained with distinct orders of approximation to
the flux function, we find it convenient to map the range of
variation of $\ovl\psi$, which is changeable according to the value
of $\lb$ (or $\chi$) and $N$, onto a fixed one, and introduce a new
normalization to the flux function, namely: \beq
\widehat{\ovl\psi}(x, \ta; \chi, \eps) \equiv \frac{\ovl\psi(x,\ta;
\chi, \eps)}{(-\ovl\psi_M(\chi, \eps))}\ , \label{eq6.4} \eeq where
$\ovl\psi_M(\chi, \eps)$, we recall, is the (normalized) flux at the
magnetic axis and must be a negative quantity according to the
second precondition. In this way the interval of variation of
$\widehat{\ovl\psi}(x,\ta; \chi,\eps)$ is comprised between $-1$ (at
the point where it takes its minimal value, the magnetic axis) and
zero (at the boundary, where it takes its maximal value). The value
of $\widehat{\ovl\psi}(x,\ta; \chi, \eps)$ at $x = \eps$, which
turns out to be zero only for an exact solution, is a function of
the angle $\ta$ for the approximate ones and gives us the
\textit{normalized or relative error}, which we shall denote by
$\cale$, defined over a common scale of reference: \beq \cale \equiv
\cale (\ta; \chi, \eps) \equiv \widehat{\ovl\psi}(x = \eps, \ta;
\chi, \eps)\ . \label{eq6.5} \eeq

The second quantity we shall use to characterize the degree of
precision of a solution to the Grad-Shafranov boundary value
problem, which we shall refer to by the name of \textit{relative
deviation}, is the ratio between the departure of the contour of the
surface $\widehat{\ovl\psi} = 0$, as it is described by the actual
approximate solution, from the one of circular shape, such as it
would be obtained at the boundary if $\widehat{\ovl\psi}$ were the
exact solution, and the minor radius of the torus cross section. In
terms of the normalized radial coordinate this writes as: \beq \cald
\equiv \frac{x(\widehat{\ovl\psi} = 0) - \eps}{\eps}\ .
\label{eq6.6} \eeq

Note that the relative deviation, as much as the relative error, is
a function of the poloidal angle $\ta$.

To illustrate the use of the method of solution to the
Grad-Shafranov boundary value problem by series of multipole
solutions we have elected a tokamak of inverse aspect ratio $\eps =
2/5$, which, as it can be seen from Fig. 5.1, makes neither a thin
nor a compact torus, but one of an intermediate thickness such as
that which can be found in typical laboratory devices. For this
tokamak we have chosen three equilibrium situations to study,
characterized by having $\lb$ equal to 0, 1 and $-1/5$ respectively.
The first value of these for the equilibrium parameter corresponds
to an equilibrium that has been widely studied \cite{um} and can
serve therefore as a ground reference to estimate the merits and
limitations of the method. The second value ($\lb = 1$), positive,
provides us with the example of an equilibrium in which the plasma
as a whole displays the behaviour of a paramagnetic body, while by
the third $(\lb = -1/5)$, negative, we are led to contemplate the
case of an equilibrium that is reached with the plasma column
assuming the characteristics of a diamagnetic body. All of them will
be studied in the pages to follow with expressions for the partial
flux that include combinations of a variable number, ranking from $N
= 3$ to $N = 10$, of multipole solutions of the orders $n = 0, 1, 2,
\ldots, 9$, which are the ones made available to us by Ref.
\cite{quatro}. We anticipate that, of the three equilibria
considered, the largest relative error and the largest relative
deviation are found for the one with $\lb = -1/5$, and that, for a
combination of $N = 10$ multipole solutions in the making up of the
partial flux, at the angular positions where they attain their
respective maximum values, these two measures of departure from
exactness are: \beq \cale \cong 0.66 \times 10^{-2}, \ \ \ \cald
\cong 0.21 \times 10^{-2}\ . \label{eq6.7} \eeq

Besides for these three we have determined the solutions for one or
two more equilibria at each of the orders of approximation
considered, as these were found to afford the possibility of
obtaining expressions for the flux functions in which the terms
carrying the dominant sources of error could be eliminated by a
judicious choice of $\lb$, and appeared then as equilibria that
could be described with exceptional accuracy within the
approximation of a restricted number of multipole solutions.

All calculations, analytical and numerical, were performed with the
help of a computer and the use of the program Maple (TM) \cite{dez}.

\begin{center}

\includegraphics[width=5cm,angle=-90]{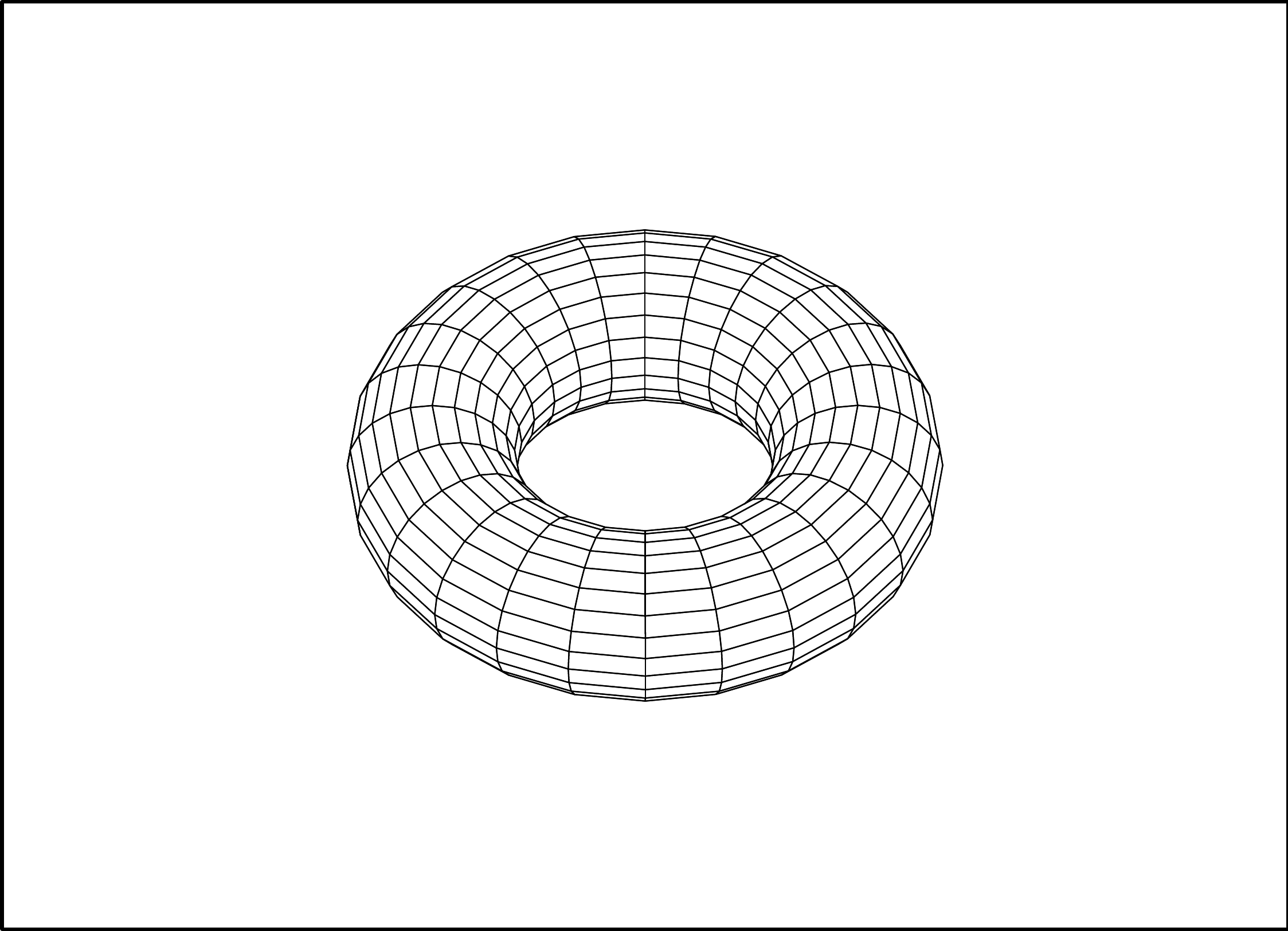}

\vv

{\bf FIG. 5.1} \ The torus of inverse aspect ratio $2/5$.

\end{center}

\vv\vv

\begin{center}
{\bf VI. SOLUTION TO THE GRAD-SHAFRANOV BOUNDARY VALUE PROBLEM IN
THE APPROXIMATION OF $N = 3$ MULTIPOLE SOLUTIONS}
\end{center}

\setcounter{section}{6} \setcounter{equation}{0}

\vv

In this and in the next Section we shall apply the method of series
of multipole solutions to obtain solutions to the Grad-Shafranov
boundary value problem in the approximations of order $N = 3$ and $N
= 4$ respectively, a task we shall carry out in some detail to take
full advantage of the relatively simple mathematical terms in which
it presents itself, and to illustrate techniques to obtain results
of interest that are in general also employable when the order of
the approximation is higher than those of the two lowest ones.

According to the pattern of development proposed by Eq.
(\ref{eqnova41}), the poloidal flux function in the order $N = 3$ of
approximation consists of the sum of the particular solution and a
combination of the multipole solutions of the three lowest orders,
to know: \beq \ovl\psi^{(3)}(x,\ta) = \ovl\psi_p(x,\ta) +
K_0\vf^{(0)}(x,\ta) + K_1\vf^{(1)}(x,\ta) + K_2\vf^{(2)}(x,\ta)\ .
\label{eq61} \eeq

Note that this is the least number of multipole solutions to be
included in a combination with the particular solution in order to
obtain an approximate representation of the flux function that be
minimally meaningful, since the term involving the zeroth order
multipole solution, being a constant, gives zero magnetic field, and
the term comprising the first order multipole solution merely
generates a uniform vertical field, without modifying the shape of
the field lines associated with the field brought about by the
particular solution. Note also that the second is the lowest of the
orders of the multipole solutions for which one of these shows to
contain a power of $x$ equal to the highest power, and a harmonic of
the poloidal angle of order equal to the highest order, of $x$ and
of the harmonics of the poloidal angle, respectively, to be present
in the particular solution. In fact, the particular solution and the
multipole solution of the second order are bivariate polynomials
with terms respectively of the same degrees both in $x$ and $\mu$.

We shall start by searching an expression for the poloidal flux
function that takes $\lb$ as the quantity defining the input datum
to an equilibrium problem. In this case the formulae we resort to as
the appropriate ones for the evaluation of the constants
$K_0^{(3)}$, $K_1^{(3)}$ and $K_2^{(3)}$ are those expressed by Eqs.
(\ref{eqnova416}) and (\ref{eqnova415}). We have in this way: \bey
K_0^{(3)} &=& \frac{\left|\begin{array}{ccc} -S_0(\lb,\eps) &
M_{01}(\eps) &
M_{02}(\eps) \\
-S_1(\lb,\eps) & M_{11}(\eps) &
M_{12}(\eps) \\
-S_2(\lb,\eps) & M_{21}(\eps) &
M_{22}(\eps)\end{array}\right|}{D_A^{(3)}(\eps)}\ , \label{eq62} \\
&& \nonumber \\
K_1^{(3)} &=& \frac{\left|\begin{array}{cc} -S_1(\lb,\eps) &
M_{12}(\eps) \\
-S_2(\lb, \eps) & M_{22}(\eps) \end{array} \right|}{D_A^{(3)}(\eps)} \ , \label{eq63} \\
&& \nonumber \\
K_2^{(3)} &=& \frac{\left|\begin{array}{cc} M_{11}(\eps) &
-S_1(\lb,\eps) \\
M_{21}(\eps) & -S_2(\lb,\eps)
\end{array}\right|}{D_A^{(3)}(\eps)}\ , \label{eq64} \eey the
denominator $D_A^{(3)}(\eps)$ common to these three formulae, as we
learn from Eq. (\ref{eqnova414}), being given by: \beq
D_A^{(3)}(\eps) = \left|\begin{array}{cc} M_{11}(\eps) &
M_{12}(\eps) \\ M_{21}(\eps) & M_{22}(\eps)
\end{array}\right|\ . \label{eq65}
\eeq

The expressions for the functions $S_0(\lb, \eps)$, $S_1(\lb, \eps)$
and $S_2(\lb, \eps)$ and the expressions for the several
$M_{ij}(\eps)$'s that enter Eqs. (\ref{eq62}) -- (\ref{eq65}) as
elements of the determinants on their right hand sides can be
written down by referring to those that constitute the matter of
Appendices A and B respectively. With this the determinant
$D_A^{(3)}(\eps)$, for example, assumes the definite form: \beq
D_A^{(3)}(\eps) =
\left|\begin{array}{ccc} \dps\frac{\eps}{2} & \dps-\frac{\eps^3}{4} \\
& \\
\dps \frac{\eps^2}{8} & \dps -\frac{1}{8}\eps^4 - \eps^2
\end{array}\right|\ , \label{eq66} \eeq from which we extract:
\beq D_A^{(3)}(\eps) = -\frac{\eps^3}{32}(\eps^2 + 16)\ .
\label{eq67} \eeq

We merely quote the results that are obtained for the three
constants: \bey K_0^{(3)} &=& -\frac{\eps^2}{32(\eps^2 +
16)}\left[(\lb + 1)(128 - 4\eps^2) - 3\eps^4\right]\ , \label{eq68}
\\
K_1^{(3)} &=& -\frac{\eps^2}{\eps^2 + 16}(2\lb + \eps^2 + 10)\ ,
\label{eq69} \\
K_2^{(3)} &=& \frac{\eps^2}{4(\eps^2 + 16)}(\lb - 3)\ .
\label{eq610} \eey

Inserting the expression for $\ovl\psi_p(x,\ta)$, as given by Eq.
(\ref{eq3.29nova}), those for $K_0^{(3)}$, $K_1^{(3)}$ and
$K_2^{(3)}$, as given by Eqs. (\ref{eq68}), (\ref{eq69}) and
(\ref{eq610}), and the formulae for the multipole solutions
$\vf^{(0)}(x,\ta)$, $\vf^{(1)}(x,\ta)$ and $\vf^{(2)}(x,\ta)$, found
in Appendix A in Ref. \cite{quatro}, in Eq. (6.1), we obtain the
expression of the flux function normalized to $s_P$ having $\lb$ as
the characteristic equilibrium parameter in the approximation of
three multipole solutions: \bey \ovl\psi^{(3)}(x,\ta) &=&
\frac{\eps^2 - x^2}{\eps^2 + 16}\left\{\frac{1}{32} \left[(4\lb -
\eps^2 - 28)x^2 - (\lb + 1)(128
- 4\eps^2) + 3\eps^4\right]\right. \nonumber \\
&&\left. - \frac{1}{2}(2\lb + \eps^2 + 10)x\cos\ta -
\frac{1}{8}(4\lb + \eps^2 + 4)x^2\cos 2\ta\right\} \nonumber \\
&&+ \frac{1}{2}\frac{(6\lb + \eps^2 - 2)}{\eps^2 + 16}x^3\cos 3\ta +
\frac{1}{32}\frac{(20\lb + 3\eps^2 - 12)}{\eps^2 + 16}x^4\cos 4\ta\
. \label{eq611} \eey

Notice the factor $\eps^2 - x^2$ making the terms independent of the
poloidal angle and those dependent on $\cos\ta$ and $\cos 2\ta$
respectively, which are grouped inside the keys, vanish at the
boundary, where $x = \eps$.

To find the position of the magnetic axis we have first to determine
the equilibrium function. Choosing to represent it in terms of the
relative Shafranov shift, we refer to Eq. (\ref{eqnova433}), which,
when the order of approximation to the flux function is the third,
takes the form: \beq F^{(3)}(\eps,\dt) =
\frac{\left|\begin{array}{ccc}
M_{11}(\eps) & M_{12}(\eps) & -P_1(\eps) \\
M_{21}(\eps) & M_{22}(\eps) & -P_2(\eps) \\
V_1'(\dt) & V_2'(\dt) & -P'(\dt)
\end{array}\right|}{D_B^{(3)}(\eps,\dt)}\ , \label{eq612}
\eeq

\vv

\nd where the denominator, following Eq. (\ref{eqnova431}), is: \beq
D_B^{(3)}(\eps,\dt) = \left|\begin{array}{ccc} M_{11}(\eps) &
M_{12}(\eps)
& Q_1(\eps) \\
M_{21}(\eps) & M_{22}(\eps) & Q_2(\eps) \\
V_1'(\dt) & V_2'(\dt) & Q'(\dt) \end{array}\right|\ . \label{eq613}
\eeq

\vv

Using for $P_1(x = \eps)$, $P_2(x = \eps)$, $Q_1(x = \eps)$, $Q_2(x
= \eps)$, $P'(\dt)$ and $Q'(\dt)$ the expressions given in Appendix
A, for the several $M_{ij}(x = \eps)$'s those given in Appendix B
and for $V_1'(\dt)$ and $V_2'(\dt)$ those in Appendix C, we are able
to evaluate the equilibrium function as: \beq F^{(3)}(\eps, \dt) =
-\frac{1}{2}\frac{2(4+\eps^2)\dt(\dt + 2) - 10\eps^2 -
\eps^4}{4\dt(\dt + 2) - \eps^2}\ . \label{eq614} \eeq

By replacing $\dt(\dt + 2)$ by $\chi$ in conformity with the
definition introduced in Eq. (\ref{eqnova437}), we obtain the
equilibrium function in terms of the displacement variable, which we
can write in a convenient form as: \beq F^{(3)}(\eps, \chi) = -(1 +
\chi_P^{(3)}) \frac{\chi - \chi_Z^{(3)}}{\chi - \chi_P^{(3)}} \ ,
\label{eq615} \eeq where \beq \chi_Z^{(3)} =
\frac{5}{4}\eps^2\left(\frac{1+\dps\frac{\eps^2}{10}}{1 +
\dps\frac{\eps^2}{4}}\right) \label{eq616} \eeq and \beq
\chi_P^{(3)} = \frac{\eps^2}{4}\ . \label{eq617} \eeq

It is useful to consider also the representation of $F^{(3)}(\eps,
\chi)$ in partial fractions: \beq F^{(3)}(\eps, \chi) = -\left(1 +
\frac{\eps^2}{4}\right) + \frac{\eps^2\left(1 +
\dps\frac{\eps^2}{16}\right)}{\chi - \dps\frac{\eps^2}{4}}\ ,
\label{eq618} \eeq which shows that the equilibrium function of the
third order is graphically described by an equilateral hyperbola
having a vertical asymptote passing through the point that locates
the pole $\chi_P = \eps^2/4$ on the horizontal axis, and a
horizontal asymptote that intercepts the vertical axis at the
ordinate $-(1 + \eps^2/4)$.

Before exploring the information contained in the equilibrium
function, as a necessary preliminary step, we determine the
expression of the poloidal flux at the magnetic axis. Resorting to
Eq. (\ref{eqnova449}), we write for the third order of approximation
to the flux function: \beq \ovl{\psi_M^{(3)}}(\eps,\chi) =
\frac{1}{16}\left[1 + F^{(3)}(\eps, \chi)\right]\chi^2 + K_0^{(3)} +
\frac{1}{4}K_1^{(3)}\chi - \frac{1}{4}K_2^{(3)}\chi^2 \ ,
\label{eq619} \eeq from which, by using Eq. (\ref{eq618}) for
$F^{(3)}(\eps,\chi)$ and Eqs. (\ref{eq68}), (\ref{eq69}) and
(\ref{eq610}) for $K_0^{(3)}$, $K_1^{(3)}$ and $K_2^{(3)}$
respectively with $\lb$ replaced by $F^{(3)}(\eps, \chi)$, we
obtain: \beq \ovl{\psi_M^{(3)}}(\eps, \chi) = -\frac{\eps^2}{32}
\frac{8\chi(\chi - \eps^2) + 32\eps^2 + \eps^4}{4\chi - \eps^2}\ .
\label{eq620} \eeq

For $\eps = 2/5$, the above expression becomes: \beq
\ovl{\psi_M^{(3)}}(\eps = \frac{2}{5}, \chi) = -\frac{402 - 100\chi
+ 625\chi^2}{2500(25\chi - 1)}\ . \label{eq621} \eeq

\begin{figure}
\begin{center}

{\unitlength=1mm

\begin{picture}(80,65)
\put(0,0){\includegraphics[width=8cm]{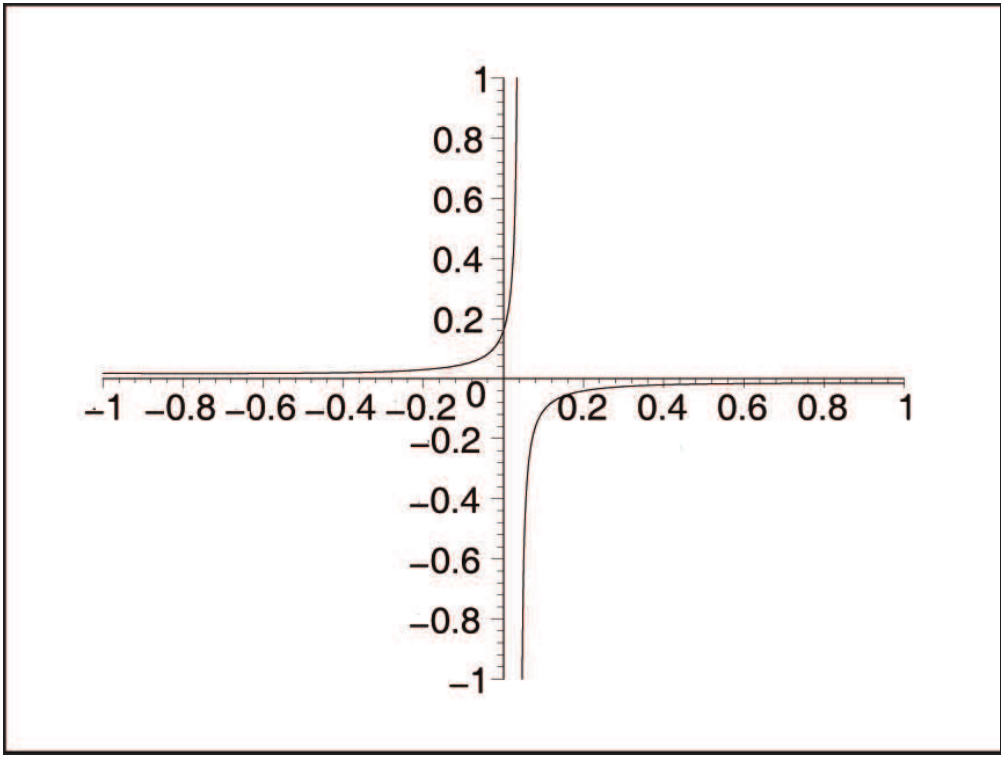}}
\put(8,55){{\footnotesize $\ovl{\psi_M^{(3)}}(\eps = 2/5, \chi)$}}
\put(72,33){{\footnotesize $\chi$}}
\end{picture}

}

\vv

\begin{tabular}{lp{12cm}}
{\bf FIG. 6.1} &The poloidal flux at the magnetic axis as a function
of the displacement variable for a torus of inverse aspect ratio
equal to 2/5, according to the approximation of order $N=3$ to the
flux function.
\end{tabular}

\end{center}
\end{figure}

From the pictorial representation of the function on the right hand
side of Eq. (\ref{eq621}), how it appears in Fig. 6.1, it becomes
evident that negative values of the flux at the magnetic axis are
reached only in the domain of the displacement variable extending
beyond the point of abscissa $\chi_P^{(3)}$ that locates  its single
pole (and also that of the equilibrium function $F^{(3)}(\eps,
\chi)$) on the $\chi$-axis. Thus, combining the condition,
formulated in mathematical terms through Eq. (\ref{eq6.1}), that the
magnetic axis must fall inside the plasma containing chamber with
the one that the flux must be negative at the magnetic axis, we may
state that, in so far these two criteria are concerned, the range of
variation of the displacement variable has to be restricted to: \beq
\chi_P < \chi < (1+\eps)^2 - 1\ , \label{eq622} \eeq in order that a
solution to the flux function be considered physically acceptable.
We have abstained from writing the superscript that designates the
order of approximation employed in the evaluation of $\chi_P$,
since, as we are going to see in the Sections to come, the lower
limit of the interval expressed by Eq. (\ref{eq622}) is fixed in
general by the value of the abscissa of the one real pole of the
equilibrium function that is, among those lying on the positive side
of the $\chi$-axis, placed in the tightest closeness to the origin.
For a tokamak of inverse aspect ratio equal to 2/5, as the one we
are taking as the recipient of applications of our theoretical
results, and a description of the equilibrium based on a solution to
the flux function of the third order of approximation, Eq.
(\ref{eq622}) translates to numerical terms as: \beq 0.04 < \chi <
0.96\ . \label{eq623} \eeq

In general, corresponding to a vanishingly small pressure gradient,
there is an equilibrium ``ground state'', defined, in terms of the
equilibrium parameter, by the limit  $\lb \to +\infty$. For such a
state, which can be shown to be paramagnetic in character, the
magnitude of the displacement variable under any order $N$ of
approximation to the flux function can be estimated as: \beq
\chi_{\min} \simeq \frac{\eps^2}{4}\ , \label{eq624} \eeq provided
that $\eps$ is not too close to unity. This means that it is not
possible to have an equilibrium configuration for which the relative
Shafranov shift assumes a value smaller than that expressed
generically by: \beq \dt_{\min} \simeq \sqrt{1 + \frac{\eps^2}{4}} -
1\ . \label{eq625} \eeq

For $\eps = 2/5$ we find the minimal value of the displacement
variable and the minimal relative Shafranov shift to be: \beq \left.
\begin{array}{l}
\chi_{\min} \simeq 0.04 \\
\\
\dt_{\min} \simeq 0.0198 \end{array}\right\}\ . \label{eq626} \eeq

We now resume the considerations on the equilibrium function. To fix
ideas, we refer to the particular form it assumes for a numerically
defined aspect ratio of the torus, which we again choose to be 2/5.
From Eqs. (\ref{eq615}), (\ref{eq616}) and (\ref{eq617}) we have:
\beq F^{(3)}(\eps = \frac{2}{5}, \chi) = -\frac{1}{25}\frac{650\chi
- 127}{25\chi - 1}\ . \label{eq627} \eeq

\begin{figure}
\begin{center}

{\unitlength=1mm

\begin{picture}(80,65)
\put(0,0){\includegraphics[width=8cm]{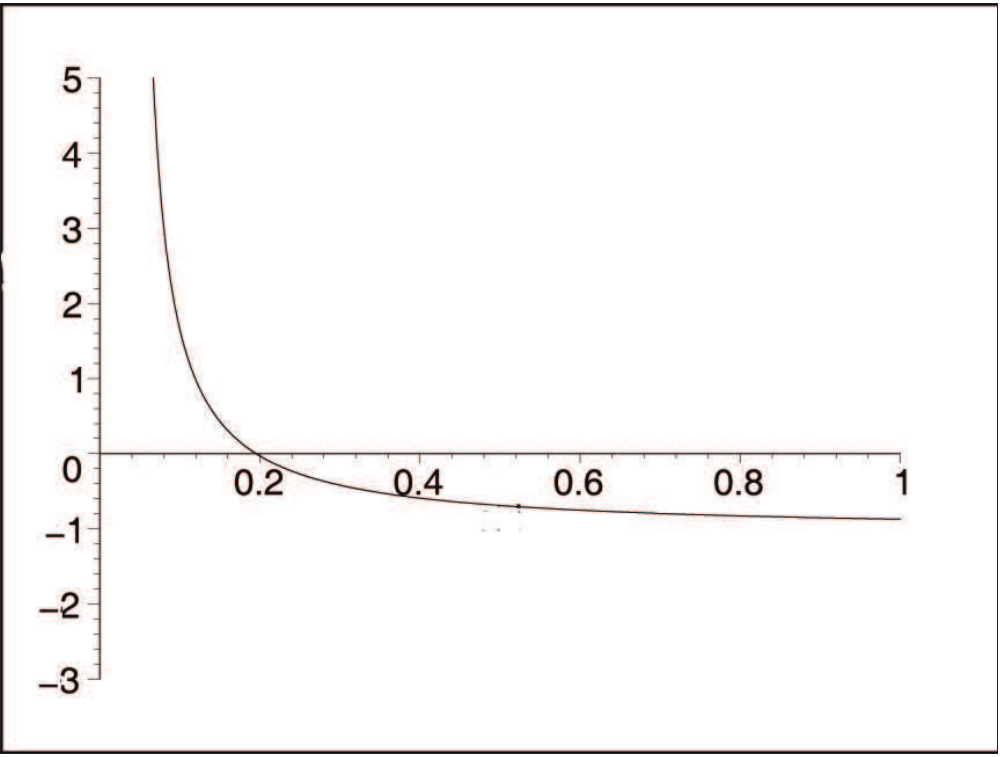}}
\put(13,57){{\footnotesize $F^{(3)}(\eps = 2/5, \chi)$}}
\put(73,27){{\footnotesize $\chi$}}
\end{picture}

}

\vv

\begin{tabular}{lp{12cm}}
{\bf FIG. 6.2} &The equilibrium function of the third order
$F^{(3)}(\eps,\chi)$ as a function of the displacement variable for
the inverse aspect ratio $\eps$ equal to 2/5. Only the portion of
the curve corresponding to the physically significant range of
values of $\chi$ ($0 < \chi_P < \chi < \infty$) is represented.
\end{tabular}

\end{center}
\end{figure}

A graph of the above function is displayed in Fig. 6.2. It is seen
that, in the domain of $\chi$ extending from $\chi = \chi_P = 0.04$
to plus infinity, $F^{(3)}(\eps, \chi)$ decreases monotonically from
plus infinity to its minimal value, which is worth \beq \left.
\begin{array}{ll}
F^{(3)}(\eps = \frac{2}{5}, \chi = +\infty) &= \dps -\frac{650}{625} \\
& \\
&= -1.04 \end{array}\right\}\ . \label{eq628} \eeq No local extremum
occurs between these two points to allow for the existence of two
different values of $\chi$ for a given value of $F^{(3)}(\eps,
\chi)$. Since according to Eq. (\ref{eqnova444}) the equilibrium
equation in the third order of approximation to the flux function
writes as \beq \lb = F^{(3)}(\eps, \chi)\ , \label{eq629} \eeq we
conclude that, in general, the allowed range of variation of the
equilibrium parameter is: \beq \lb_\infty < \lb < \infty
\label{eq630} \eeq where $\lb_\infty$, as obtained by carrying out
the limit $\chi\to\infty$ in Eq. (\ref{eq618}), is: \beq \lb_\infty
= -\left(1 + \frac{\eps^2}{4}\right)\ . \label{eq631} \eeq If the
restriction on the maximal value of $\chi$, as it is imposed by Eq.
(\ref{eq622}), is taken into account, then the minimal allowed value
of $\lb$ shifts from $\lb_\infty$ to this one: \beq \lb_{\min} =
-\frac{16 - 2\eps + 4\eps^2 + \eps^3}{2(8 + 3\eps)}\ . \label{eq632}
\eeq The numerical value of this quantity for $\eps = 2/5$ is: \beq
\left.\begin{array}{ll} \lb_{\min}(\eps = 2/5) &= \dps
-\frac{497}{575} \\
& \\
&\simeq -0.864 \end{array}\right\} \ , \label{eq633} \eeq about 17\%
smaller in absolute value than that predicted by Eq. (\ref{eq628}).

We now establish a direct connection between the displacement
variable and the equilibrium parameter through the equilibrium
equation, Eq. (\ref{eq629}). Equating the expression for the
$F^{(3)}(\eps, \chi)$, as given by Eq. (\ref{eq618}), to $\lb$, and
then solving the resulting equation for $\chi$, we obtain: \beq \chi
= \frac{\eps^2}{2} \frac{2\lb + \eps^2 + 10}{4\lb + \eps^2 + 4}\ .
\label{eq634} \eeq

Table 6.1 exhibits the values of the displacement variable, as
furnished by this relation, for the three values of the equilibrium
parameter associated with the configurations we are taking as of
reference; also the corresponding values of the relative Shafranov
shift, as evaluated according to Eq. (\ref{eqnova438}).


\begin{center}

{\renewcommand{\arraystretch}{2.5}
\begin{tabular}{|c||c|c|} \hline
$\lb$ & $\chi^{(3)}$ & $\dt^{(3)}$ \\ \hline\hline 1 & $\dps
\frac{152}{1275} \simeq 0.1192$ & $\dps \frac{\sqrt{72777}}{255} - 1
\simeq 0.05793$
\\ \hline
0 & $\dps \frac{127}{650} \simeq 0.1954$ & $\dps
\frac{\sqrt{20202}}{130} - 1 \simeq 0.09334$ \\ \hline $\dps
-\frac{1}{5}$ & $\dps \frac{122}{525} \simeq 0.2324$ & $\dps
\frac{\sqrt{13587}}{105} - 1 \simeq 0.1101$ \\ \hline
\end{tabular}
}

\vv

\begin{tabular}{lp{12cm}} \textbf{Table 6.1} & The values of the
displacement variable ($\chi$) and of the relative Shafranov shift
($\dt$) according to the third order of approximation to the flux
function for $\eps = 2/5$ and the three reference values of the
equilibrium parameter ($\lb$).
\end{tabular}

\end{center}


For the purpose of drawing flux maps the expression for the flux
function of Eq. (\ref{eq611}) is perfectly adequate, being enough to
substitute $\eps$ and $\lb$ by the numerical values relative to the
configurations being considered. The position of the magnetic axis
is always a datum of relevance and must be supplied independently by
evaluating the relative Shafranov shift with the help of the
equilibrium equation. In this paper, however, for the reasons we
have expounded in Section V, we shall favour in general the use of
the flux function normalized to the absolute value of the magnetic
flux at the magnetic axis, and this requires that the expression of
Eq. (\ref{eq611}) be divided by minus that of Eq. (\ref{eq620}). We
should then have a statement for the doubly normalized flux function
having reference to two physical parameters, $\lb$ and $\chi$,
which, however, are not independent between themselves but
interrelated through the equilibrium equation. It would be clearly
preferable to this situation to dispose of an expression for the
normalized flux function depending on only one characteristic
parameter, as it is the equilibrium problem itself, either $\lb$ or
$\chi$. In the case of the approximation of the third order to the
flux function, which we are presently considering, the connection
between these two quantities can be translated into an algebraic
equation that is linear both in $\lb$ and $\chi$, from which
explicit definitions of either parameter in terms of the other can
be extracted, but in approximations of higher orders the equilibrium
function is a quotient of polynomials in the displacement variable
of degrees higher than the first, and this makes it impossible in
general to have $\chi$ stated explicitly as a function of $\lb$. The
inverse alternative (to express $\lb$ in terms of $\chi$) is however
always possible, being as it is the statement of the equilibrium
equation itself. This is one of the reasons why the convenient
parameter to which to refer in defining an equilibrium problem be
the displacement variable rather than the equilibrium parameter,
although it is the second of these two quantities the one to
participate ostensively in the mathematical formulation of it.

We thus substitute $\lb$ by $F^{(3)}(\eps, \chi)$, as given by Eq.
(\ref{eq618}), in Eq. (\ref{eq611}), divide the result by minus the
expression of the poloidal flux at the magnetic axis, as given by
Eq. (\ref{eq620}), and obtain the normalized flux function as: \bey
\widehat{\ovl\psi}^{(3)}(x, \ta; \chi,\eps) &=&
\frac{\eps^2-x^2}{d^{(3)}} \left[(3\eps^2 \!-\! 8\chi)x^2 \!+\!
8\chi\eps^2 \!-\! 32\eps^2 \!-\! \eps^4 \!-\! 32\chi x\cos\ta \!-\!
4\eps^2x^2\cos2\ta\right]
\nonumber \\
&&+ \frac{1}{d^{(3)}}\left[32(\eps^2 - \chi)x^3\cos 3\ta + (7\eps^2
- 8\chi)x^4 \cos 4\ta\right]\ , \label{eq635} \eey where \bey
d^{(3)} &\equiv& d^{(3)}(\eps, \chi) \nonumber \\
&=& \eps^2(8\chi^2 - 8\chi\eps^2 + 32\eps^2 + \eps^4)\ ,
\label{eq636} \eey and we have used the notation introduced in Eq.
(\ref{eq6.4}) for the left hand side.

We now replace $\eps$ by $2/5$ and $\chi$ by the three numerical
values appearing in Table 6.1 in Eq. (\ref{eq635}) to obtain the
expressions for the flux function corresponding to the three values
we are considering for the equilibrium parameter. For the purposes
of the present paper it is sufficient to have reproduced here just
the expression we get relative to the equilibrium with $\lb = 0$\,:
\bey && \widehat{\ovl\psi}^{(3)}\left(x,\ta; \lb =
0,\eps=\frac{2}{5}\right) = -\frac{16250}{274673}\left[\left(-x +
\frac{2}{5}\right)\left(x + \frac{2}{5}\right)\left(\frac{2486}{25}
+ 22x^2 \right.\right.
\nonumber \\
&& \qquad \qquad\qquad+ 127x\cos \ta + 13x^2\cos 2\ta\biggr) +
23x^3\cos 3\ta + 9x^4\cos 4\ta\biggr]\ . \label{eq637} \eey

\begin{figure}

\begin{center}

\begin{tabular}{ll}
(a) $\lb = 1$ & (b) $\lb = 0$ \\
& \\
\includegraphics[width=7.08cm]{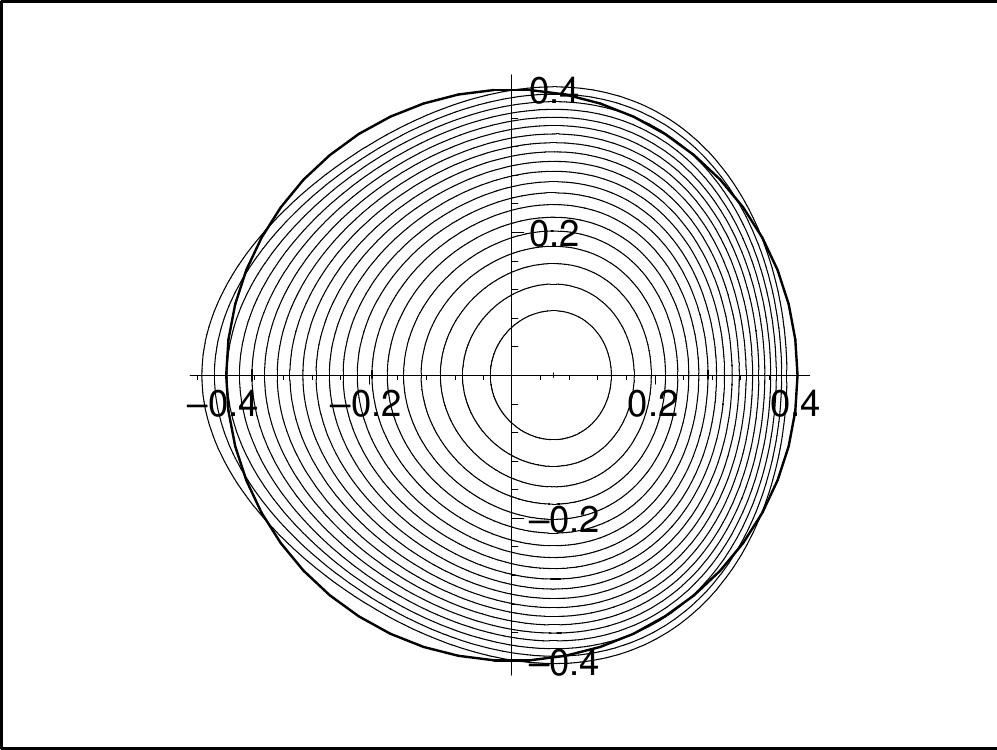} &
\includegraphics[width=7.08cm]{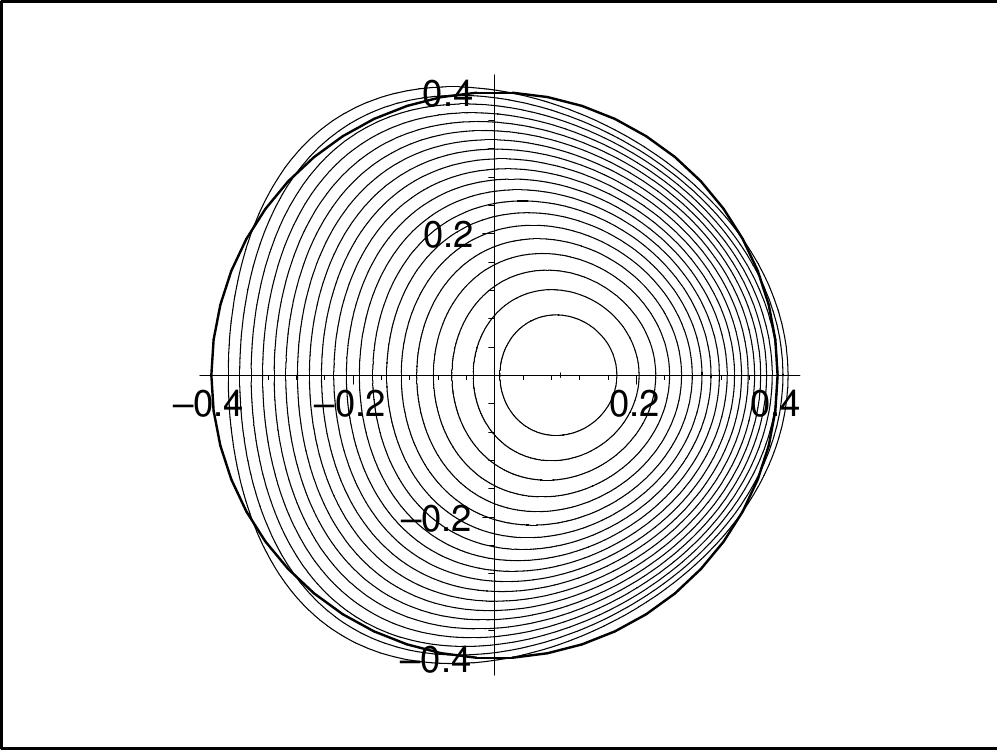} \\
& \\
& \\
\multicolumn{2}{c} {(c) $\lb = -1/5$} \\
& \\
\multicolumn{2}{c} {\includegraphics[width=7.08cm]{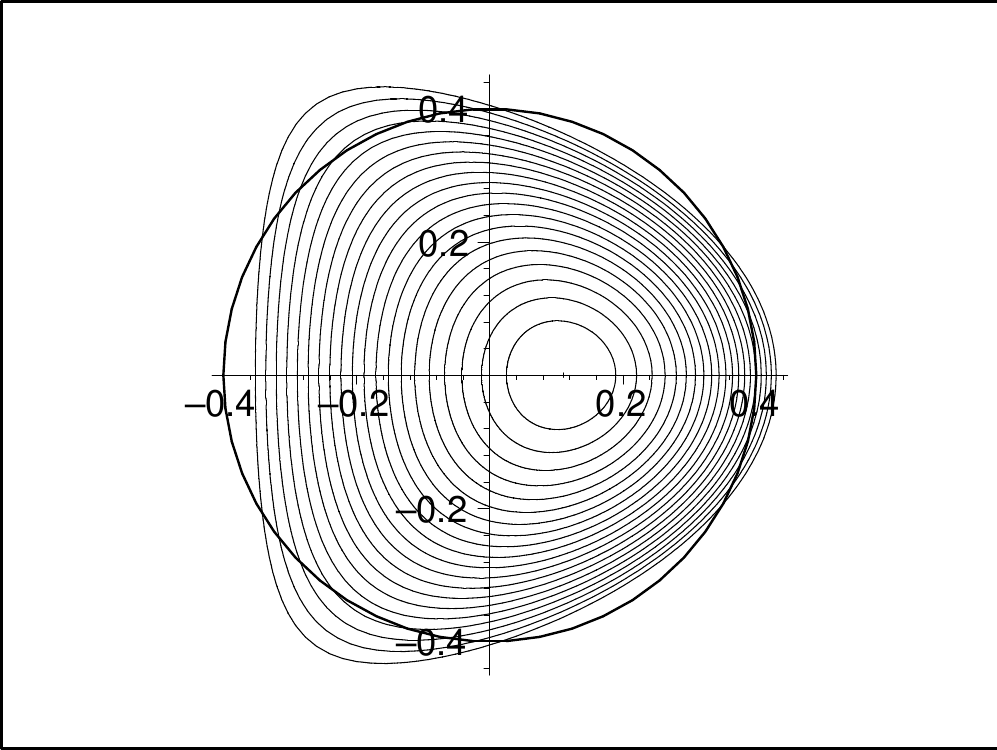}}
\end{tabular}

\vv

\begin{tabular}{lp{12cm}}
{\bf FIG. 6.3} & Flux surfaces in the approximation of order $N=3$
to the flux function for the equilibria in a torus of inverse aspect
ratio 2/5 and equilibrium parameter $\lb$ equal to (a) 1; (b) 0; (c)
$-1/5$. In terms of the normalized flux function, the spacing
between two neighboring flux surfaces in flux space is $\Dt
\widehat{\ovl\psi}^{(3)}(x,\ta)=1/20$. The outermost of the flux
surfaces, as seen in the figure, corresponds to the flux
$\widehat{\ovl\psi}^{(3)}(x,\ta)= 0$ and the innermost one, which
defines the magnetic axis, corresponds to the flux
$\widehat{\ovl\psi}^{(3)}(x,\ta)=-1$. The circle drawn in thick line
represents the contour of the cross section of a torus of inverse
aspect ratio equal to 2/5 and centre line located at $x = 0$.
\end{tabular}

\end{center}

\end{figure}

Plots of the level curves portraying the flux function in the
approximation of the third order in the domain comprised between the
surface identified by $\widehat{\ovl\psi}^{(3)}(x,\ta) = -1$ (which
reduces to the magnetic axis) and that identified by
$\widehat{\ovl\psi}^{(3)}(x,\ta) = 0$ for $\eps = 2/5$ \ and $\lb =
1$, 0 and $-1/5$ are given in Figs. 6.3(a), (b) and (c)
respectively. Superimposed on each flux map, a circle allows a
visual comparison between the contour of the surface
$\widehat{\ovl\psi}^{(3)}(x,\ta)= 0$ and that of the boundary of the
cross section of a torus of inverse aspect ratio equal to $2/5$ and
centre line placed at $x = 0$. The most salient discrepancy between
both contours is seen to belong to the map for $\lb = -1/5$, and can
be attributed to the high value assumed by the displacement variable
in this case. The error carried by the flux function in the
approximation of the third order, as by that of any order of
approximation in general, is not, however, a monotonic function of
the value assumed by $\chi$.

The function that expresses the normalized error associated with the
flux function $\widehat{\ovl\psi}^{(3)}(x,\ta; \chi,\eps)$,
according to the definition introduced by Eq. (\ref{eq6.5}), is:
\beq \cale^{(3)}(\ta; \chi,\eps) = -\frac{\eps}{8\chi^2 -
8\chi\eps^2 + 32\eps^2+ \eps^4} \left[32(\chi- \eps^2)\cos 3\ta +
\eps(8\chi - 7\eps^2)\cos 4\ta\right]\ . \label{eq638} \eeq

A graph of $\cale^{(3)}(\ta; \chi,\eps)$ for $\eps = 2/5$ and the
value of $\chi$ equal to that corresponding to $\lb = 0$ \ is
displayed in Fig. 6.4. Note that the maximum error in absolute value
occurs at the angle $\ta = 0$. Also for the equilibrium
configurations labelled by $\lb = 1$ and $\lb = -1/5$ (with $\eps =
2/5$) the errors of maximum magnitudes occur at the outer edge of
the torus. Table 6.2 displays the values reached by the function
$\cale^{(3)} (\ta; \chi, \eps=2/5)$ at the angular position $\ta =
0$ for the three values we are taking as of reference for the
equilibrium parameter. As it should be expected from the inspection
of the flux maps, the largest error is that which accompanies the
configuration with $\lb = -1/5$. Table 6.2 also shows that, when
$\lb$ decreases from unity to zero, the maximum error in absolute
value decreases, although the relative Shafranov shift be then
increased.

Of course there is nothing of necessity in that the error attains
its maximum absolute value at $\ta = 0$ as it happens to attain for
the three cases just examined. It is possible, for example, to make
$\cale ^{(3)}(\ta; \chi,\eps)$ vanish at this same angular position
by choosing the displacement variable to be:
\setcounter{equation}{0}
\def\theequation{\thesection.39\alph{equation}}
\beq \left.
\begin{array}{ll} \chi&=
{\dps \frac{\eps^2(32 +7\eps)}{32 + 8\eps}} \\
& \\
&= {\dps \frac{87}{550}} \simeq 0.1582, \end{array} \right\}
\label{eq639a} \eeq the numerical values applying to the value $\eps
= 2/5$ for the inverse aspect ratio. For such an equilibrium the
values of $\dt$ and $\lb$ are:
\bey \dt &=& \frac{7\sqrt{286}}{110} -1 \simeq 0.07619\ , \label{eq639b} \\
\lb &=& \frac{532}{1625} \simeq 0.3274\ . \label{eq639c} \eey

\begin{center}

{\unitlength=1mm
\begin{picture}(80,65)
\put(0,0){\includegraphics[width=8cm]{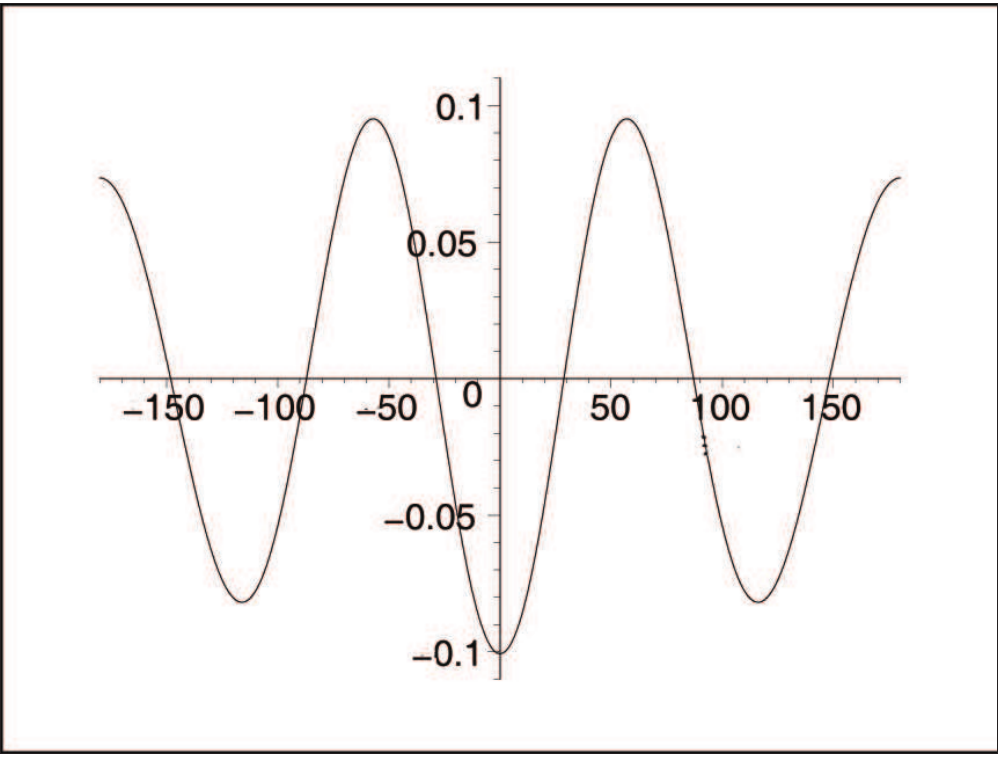}}
\put(25,57){{\footnotesize $\cale(\ta; \eps = 2/5, \lb=0)$}}
\put(70,32){{\footnotesize $\ta$ (degrees)}}
\end{picture}

}

\vspace{2mm}

\begin{tabular}{lp{12cm}}
{\bf FIG. 6.4} &The poloidal angle $(\ta)$ dependence of the
function $\cale^{(3)}(\ta; \chi,\eps)$ describing the normalized
error relative to the $N = 3$ order of approximation to the flux
function along the circular contour $x=\eps$ of the torus cross
section for the equilibrium characterized by having $\eps=2/5$ and
$\lb=0$.
\end{tabular}

\end{center}

\vspace{3mm}

\begin{center}
\begin{tabular}{|c||c|c|c|} \hline
$\lb$ & 1 & 0 & $-1/5$ \\
\hline\hline $\cale^{(3)}(\ta = 0; \chi,\eps)$ & 0.1074 & $-0.1007$
& $-0.1979$  \\ \hline
\end{tabular}

\vspace{2mm}

\begin{tabular}{lp{12cm}}
\textbf{Table 6.2} & The values of the function describing the
normalized errors associated with the third order of approximation
to the flux function ($\cale^{(3)}$) at the outer border of the
torus ($\ta = 0$) for $\eps=2/5$ and the three reference values of
the equilibrium parameter ($\lb$).
\end{tabular}
\end{center}

\setcounter{equation}{39}
\def\theequation{\thesection.\arabic{equation}}

A graph depicting the behaviour of the error as a function of the
poloidal angle in this case is given in Fig. 6.5. The maximum
absolute value is reached at $\ta  = 180^{\rm o}$ and it is worth
only 0.9050\% of the flux at the magnetic axis. The flux function
writes as:

\begin{center}

{\unitlength=1mm
\begin{picture}(80,63)
\put(0,0){\includegraphics[width=8cm]{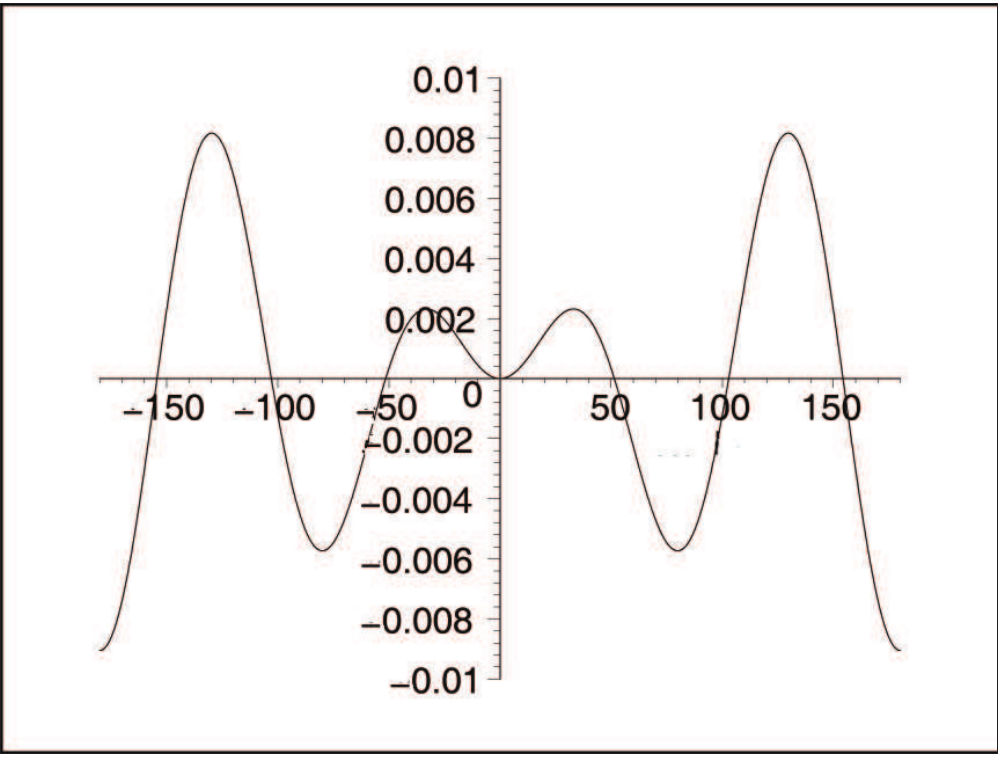}}
\put(22,57){{\footnotesize $\cale(\ta; \eps = 2/5, \chi=87/550)$}}
\put(70,32){{\footnotesize $\ta$ (degrees)}}
\end{picture}

}

\begin{tabular}{lp{12cm}}
{\bf FIG. 6.5} &The normalized error $\cale^{(3)}(\ta; \eps = 2/5,
\chi=87/550)$ as a function of the poloidal angle.
\end{tabular}

\end{center}

\bey &&\widehat{\ovl\psi}^{(3)}\left(x, \ta; \chi=\frac{87}{550},
\eps=\frac{2}{5}\right) = \hspace{-1mm}-\frac{275}{7203}
\left(\frac{4}{25} - x^2\right)\left(\frac{472}{3} + 25x^2 +
\frac{1450}{9}x\cos \ta \right. \nonumber \\
&&\qquad\qquad \qquad\qquad\left. \hspace{-1mm}+\frac{550}{27}x^2
\!\cos 2\ta\!\right)\!+\! \frac{6875}{194481}\!\left(2x^3\!\cos 3\ta
\!-\! 5x^4\!\cos 4\ta\!\right)\!. \label{eq640} \eey

A map of the flux surfaces that is implied by this expression is
given in Fig. 6.6; on account of the smallness of the error it is
barely possible to distinguish visually the contour of the surface
$\widehat{\ovl\psi}^{(3)} = 0$ from the one of circular shape this
surface would assume if the boundary condition were exactly
satisfied.

Other situations of interest are those in which one of the two terms
of the normalized error, by appropriate choice of $\chi$, is made to
vanish. Thus, if we take $\chi = \eps^2$, we obtain from Eq.
(\ref{eq638}): \beq \cale^{(3)}(\ta; \chi = \eps^2) =
-\frac{\eps^2}{32 + \eps^2}\cos 4\ta \ , \label{eq641} \eeq the term
of the third harmonic in the poloidal angle of the flux function
having been suppressed at the boundary.

\vv

\begin{center}

\includegraphics[width=8cm]{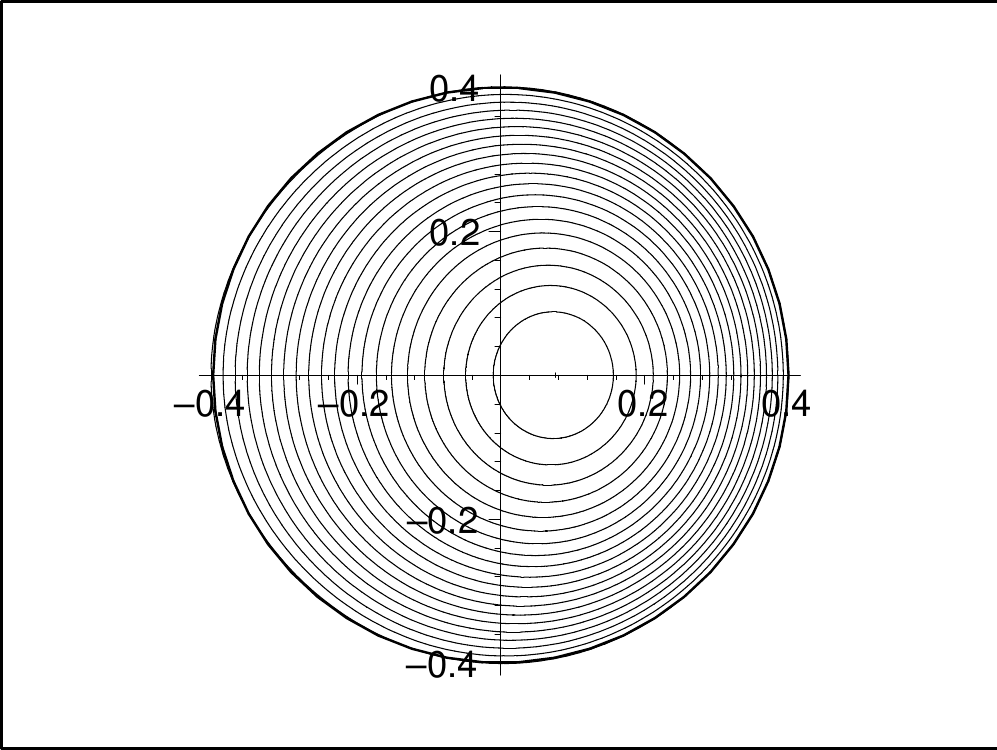}

\vv

\begin{tabular}{lp{12cm}}
{\bf FIG. 6.6} &Flux surfaces as described by the flux function in
the approximation of order $N=3$ for the equilibrium characterized
by having $\eps=2/5$ and $\chi = 87/550$.
\end{tabular}

\end{center}

\vv

Similarly, by choosing $\dps\chi = \frac{7\eps^2}{8}$, the term of
the fourth harmonic of the poloidal angle in the expression for the
flux function at $x = \eps$ is made to vanish, and Eq. (\ref{eq638})
for the error becomes: \beq \cale^{(3)}\left(\ta;\eps, \chi =
\frac{7\eps^2}{8}\right) = \frac{32\eps}{256 + \eps^2}\cos 3\ta \ .
\label{eq642} \eeq

Table 6.3 exhibits the numerical values of the main parameters
characterizing these two equilibria whose flux functions contain
just one harmonic of the poloidal angle at the boundary. As it could
be expected, the maximum error drops sensibly when the term of the
third harmonic is absent from the function $\cale^{(3)}(\ta;
\chi,\eps)$ in comparison with that which results when the harmonic
made to be absent is the fourth.

\begin{figure}

\begin{center}
\begin{tabular}{|p{4cm}|c|c|c|c|} \hline
Number of the harmonic of the poloidal angle present in the function
$\cale^{(3)}(\ta; \chi,\eps)$ & \raisebox{-9mm}{$\lb$} &
\raisebox{-9mm}{$\chi$}
& \raisebox{-9mm}{$\dt$} & \raisebox{-9mm}{$|\cale^{(3)}(\ta, \chi,\eps)|_{\max}$} \\
\hline\hline &&&& \\
\multicolumn{1}{|c|}{3} & ${\dps \frac{72}{125}} \simeq 0.5760$ &
0.1400 & 0.06771 & 0.04997 \\ &&&& \\
&&&& \\
\multicolumn{1}{|c|}{4} & ${\dps
\frac{23}{75}} \simeq 0.3067$ & 0.1600 & 0.07703 & 0.004975 \\
&&&& \\ \hline
\end{tabular}

\vv

\begin{tabular}{lp{12cm}}
\textbf{Table 6.3} & Data concerning the equilibrium configurations
and the maximum values of the relative errors associated with the
approximation of the order $N=3$ to the flux function when the
function $\cale^{(3)}(\ta; \chi,\eps)$ contains either one of the
third and the fourth harmonics of the poloidal angle.
\end{tabular}
\end{center}

\end{figure}

It may be of some interest to determine the local extrema of the
function $\cale^{(3)}(\ta; \chi,\eps)$, their magnitudes, and the
angles where they occur. Because of the symmetry of the flux
function with respect to the equatorial plane of the torus, two of
such positions are defined by the angles $\ta = 0$ and $\ta = \pi$.
The remaining ones can be obtained by deriving Eq. (\ref{eq638})
with respect to $\ta$ and equating the result to zero; after
dividing by $\sin\ta$ and expressing the angular functions of the
arcs multiple of $\ta$ in terms of $\cos\ta$, we are led to the
following algebraic equation for $\cos\ta \equiv \mu$: \beq
2\eps\af_0\mu^3 + 24\af_1\mu^2 - \eps\af_0\mu - 6\af_1 = 0\ ,
\label{eq647} \eeq where \beq \af_0 \equiv 8\chi - 7\eps^2
\label{eq648} \eeq and \beq \af_1 \equiv \chi - \eps^2\ .
\label{eq649} \eeq

As an example of results obtained from the use of this equation,
Table 6.4 exhibits the data regarding the local extrema of the
function $\cale^{(3)}(\ta; \chi,\eps)$ when the value attributed to
the equilibrium parameter is $\lb = 0$.

Another measure of the degree of accuracy reached by an approximate
solution, and one which bears a closer correspondence with its
visual representation as provided by the flux map than does the
relative error, is the relative deviation, already introduced in
Section V by Eq. (\ref{eq6.6}).

\vv

\begin{center}
\begin{tabular}{|c||c|c|c|c|} \hline
$\ta$ (degrees) & 0 & 57.02 & 116.11 & 180 \\ \hline
$\cale^{(3)}(\ta; \chi,\eps)$ & $-0.1007$ & 0.09513 & $-0.08189$ & 0.07346 \\
\hline
\end{tabular}

\vv

\begin{tabular}{lp{12cm}}
\textbf{Table 6.4} & Measures of the poloidal angle and values
assumed by $\cale^{(3)}(\ta; \chi,\eps)$ for $\eps = 2/5$ and $\lb =
0$ at the angular positions where this function reaches a local
extremum.
\end{tabular}

\end{center}

\vv

Figure 6.7 depicts the dependence of the normalized radial
coordinate $x$ on the po\-loi\-dal angle $\ta$ at the flux surface
specified by $\widehat{\ovl\psi}^{(3)}(x, \ta; \chi,\eps) = 0$ for
the equilibrium having $\eps = 2/5$, $\lb = 0$\,; it is just the
plot of $x$, taken as a function implicitly defined by the flux
function of Eq. (\ref{eq637}) under the constraint that the value of
this last-mentioned function be kept at its boundary value,
\textit{versus} the angle $\ta$, but while in drawing the flux
contour the variables of the couple $(x, \ta)$ are interpreted as
polar coordinates, here the cartesian ones are employed. Of course,
if $\widehat{\ovl\psi}(x,\ta)$ in general were the exact solution to
the boundary value problem, a graph like this would reduce to a
straight line parallel to the horizontal axis passing through the
ordinate $x = \eps$; with the fluctuations, as those seen in Fig.
6.7, it shows the extent to which the contour of the boundary that
is implied by the solution actually found departs from the shape of
the one it intends to be an approximation.

\begin{figure}
\begin{center}

{\unitlength=1mm
\begin{picture}(80,65)
\put(0,0){\includegraphics[width=8cm]{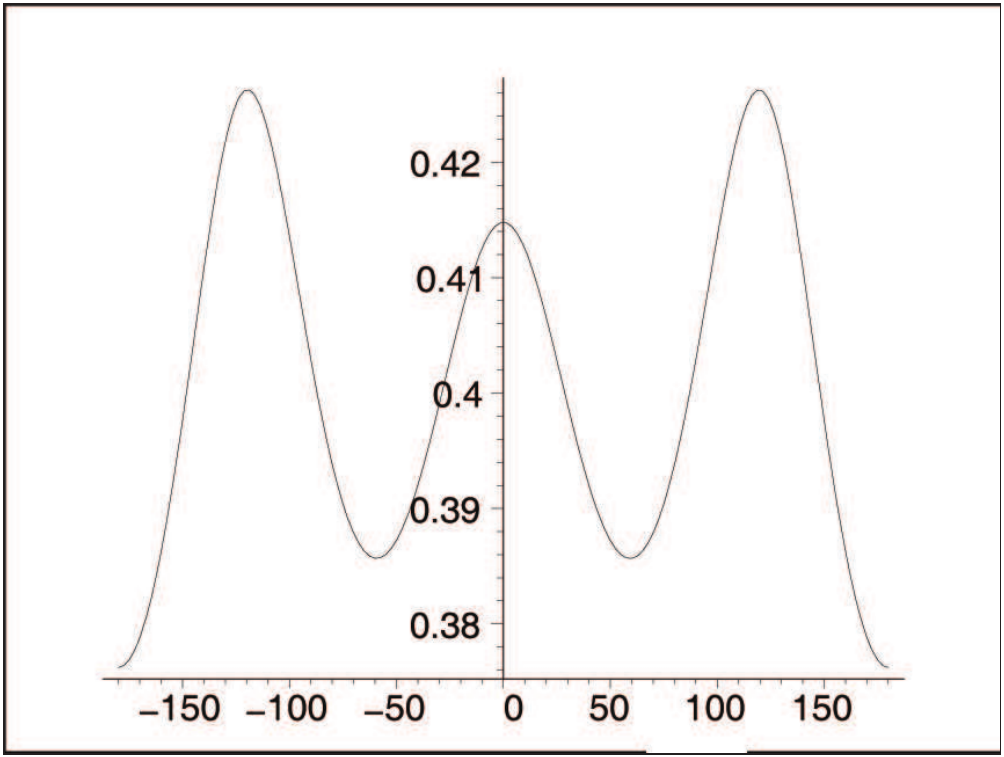}}
\put(26,57){{\footnotesize $x(\ta; \eps = 2/5, \lb = 0)$}}
\put(61,9){{\footnotesize $\ta$ (degrees)}}
\end{picture}

}

\vv

\begin{tabular}{lp{12cm}}
{\bf FIG. 6.7} &Variation of the radial coordinate $x$ with the
poloidal angle $\ta$ at the flux surface
$\widehat{\ovl\psi}^{(3)}(x,\ta; \chi,\eps)=0$ for the equilibrium
specified by $\eps = 2/5$, $\lb = 0$.
\end{tabular}

\end{center}
\end{figure}

The same as for the relative error, it is of interest to determine
the local maxima and minima of the relative deviation. Again by
reasons of symmetry the outer and the inner edges of the torus are
two positions at which $x$ and $\cald$ go through extrema of values.
To find which are these values we solve the equations: \beq 8x^4 \pm
32x^3 + 16(2 - \chi)x^2 \mp 32\chi x - [8(4-\chi) + \eps^2]\eps^2 =
0\ , \label{eq650} \eeq which follow from Eq. (\ref{eq635}) with
$\widehat{\ovl\psi}^{(3)}(x,\ta; \chi,\eps)$ taken to be zero, the
one with the upper signs by putting $\ta = 0$ and the one with the
lower signs by putting $\ta = \pi$. The solutions with physical
meaning for the one and the other equations are respectively: \beq x
= \mp 1 \pm \sqrt{1 + \chi \pm \sqrt{\chi^2 - \chi\eps^2 + 4\eps^2 +
\frac{\eps^4}{8}}}\ . \label{eq651} \eeq

Not always, however, the dominant extremum of $x$ is located at $\ta
= 0$ or $\ta = \pi$. To determine the local extrema of $x$ inside
the interval comprised between these two angular positions we note
that, upon infinitesimal variations $\dif x$ and $\dif\ta$ of $x$
and $\ta$, a flux function of general order of approximation
$\widehat{\ovl\psi}^{(N)}(x, \ta; \chi,\eps)$ undergoes an
infinitesimal change $\dif\widehat{\ovl\psi}^{(N)}$ given by: \beq
\dif\widehat{\ovl\psi}^{(N)} =
\frac{\ptl\widehat{\ovl\psi}^{(N)}}{\ptl x}\Bigl|_\ta \dif x +
\frac{\ptl\widehat{\ovl\psi}^{(N)}}{\ptl\ta}\Bigl|_x\dif\ta\ .
\label{eq652} \eeq  Now, the flux surface we are referring to is
defined by: \beq \widehat{\ovl\psi}^{(N)}(x, \ta; \chi,\eps) =
\hbox{ constant } = 0\ , \label{eq653} \eeq so that the differential
$\dif\widehat{\ovl\psi}^{(N)}$ on the left hand side of Eq.
(\ref{eq652}) is zero. If $x$ is to pass through an extreme value as
$\ta$ is increased by $\dif\ta$, then $\dif x$ in Eq. (\ref{eq652})
also vanishes. We are thus left with: \beq
\frac{\ptl\widehat{\ovl\psi}^{(N)}}{\ptl\ta}\Bigl|_x = 0\ .
\label{eq654} \eeq

Equations (\ref{eq653}) and (\ref{eq654}) form a system of two
equations for $x$ and $\ta$ (or trigonometric functions of $\ta$)
which specify both the angular positions where $x$ reaches local
extrema and the values of these on the fixed surface
$\widehat{\ovl\psi}^{(N)} = 0$.

We now consider specifically the case $N = 3$ using Eq.
(\ref{eq635}) for the normalized flux function. Expressing the
trigonometric functions whose arguments are multiples of the
poloidal angle in terms of $\cos\ta \equiv \mu$, taken as the free
variable, the equation that translates the condition that
$\widehat{\ovl\psi}^{(N)}(x, \ta;\chi, \eps)$ be the outermost
magnetic surface can be put in the form: \bey &&8\af_0x^4\mu^4 +
128\af_1x^3\mu^3 -8(2\af_2x^2 -
\eps^4)x^2\mu^2 - 32(\af_2x^2 - \chi\eps^2)x\mu \nonumber \\
&&+8\eps^2(2\chi - 4 - \eps^2)x^2 - (8\chi- 32 - \eps^2)\eps^4 = 0\
, \label{eq655} \eey where \beq \begin{array}{lll}\af_2
&\equiv& 4\chi - 3\eps^3 \\
&=& \af_0 - 4\af_1 \end{array} \label{eq656} \eeq and $\af_0$ and
$\af_1$ are as defined in Eqs. (\ref{eq648}) and (\ref{eq649}).

Under similar manipulation the equation that expresses the
requirement that $x$ be an extremum, after being divided by
$\sin\ta$, can be conducted to assume the shape: \beq 2\af_0x^3\mu^3
+ 24\af_1x^2\mu^2 - (2\af_2x^2 - \eps^4)x\mu - 2\af_2x^2 +
2\chi\eps^2 = 0\ . \label{eq657} \eeq

It is possible to obtain the solutions for $x$ and $\mu$ of the
system constituted by Eqs. (\ref{eq655}) and (\ref{eq657}) by
solving just one equation. We first introduce a transformation of
the unknown $\mu$ according to: \beq \mu = \frac{y}{x}\ ,
\label{eq658} \eeq obtaining in place of Eqs. (\ref{eq655}) and
(\ref{eq657}) two equations for $y$ and $x$ in which $x$ appears
only to the second power. Isolating $x^2$ from the transformed
version of Eq. (\ref{eq657}) we have: \beq x^2 = \frac{2\af_0y^3 +
24\af_1y^2 + \eps^4y + 2\chi\eps^2}{2\af_2(y + 1)}\ . \label{eq659}
\eeq

Upon substitution of $x^2$ as given by this expression in the
transformed version of Eq. (\ref{eq655}), we are led to the
following equation for the quantity $y$: \beq c_0y^5 +c_1y^4 +
c_2y^3 + c_3y^2 + c_4y + c_5 = 0 \ , \label{eq660} \eeq where \beq
\left. \begin{array}{l} c_0 = 8\af_0\af_2 \ , \\
c_1 = 8\af_2(32\chi  - 29\eps^2) \ , \\
c_2 = 128(8-\eps^2)\chi ^2 - 16\eps^2(96-11\eps^2)\chi  +
8\eps^4(68-7\eps^2) \ , \\
c_3 = -8\eps^2[32\chi ^2 - 2(24 + 23\eps^2)\chi  + 48\eps^2 + 15\eps^4] \ , \\
c_4 = \eps^4(4\chi  - 16 - \eps^2)\af_0  \ , \\
c_5 = \eps^4[16\chi ^2 - 4(24 + 5\eps^2)\chi  + 3\eps^2(32 +
\eps^2)]\ .
\end{array}\right\}  \label{eq661}
\eeq

\vv

Once Eq. (\ref{eq660}) has been solved, Eq. (\ref{eq659}) can be
used to determine the values of $x$ associated with those found for
$y$, and then, from Eq. (\ref{eq658}), the values of $\mu$ (and
$\ta$) corresponding to the pairs $(y,x)$. Of course not all of the
five solutions obtained for $y$ will give rise to physically
meaningful solutions for the pair $(x,\mu)$.

Numerical programs like the ones built in Maple 7 are able to handle
the system constituted by Eqs. (\ref{eq655}) and (\ref{eq657})
keeping its original form of two coupled equations for $x$ and
$\mu$, but as a rule they require that the intervals where the roots
are located be specified. For this a graphical solution to the
system can be helpful. For example, for $\eps = 2/5$ and $\lb = 0$,
the two equations corresponding to Eq. (\ref{eq655}) and Eq.
(\ref{eq657}) can be written respectively as: \beq x^4\mu^4 \!+\!
\frac{23}{18}x^3\mu^3 \!+\! \left(\frac{13}{225} -
\frac{49}{36}x^2\right)x^2\mu^2 \!+\! \left(\frac{127}{450} -
\frac{49}{18}x^2\right)x\mu \!+\! \frac{1243}{5625} -
\frac{49}{36}x^2 \!=\! 0 \label{eq662} \eeq and \beq x^3\mu^3 +
\frac{23}{24}x^2\mu^2 + \left(\frac{13}{450} -
\frac{49}{72}x^2\right)x\mu + \frac{127}{1800} - \frac{49}{72}x^2 =
0\ . \label{eq663} \eeq

Figure 6.8 displays the two curves for $x$, as they are each
implicitly defined by one and the other of these two equations as a
function of $\mu$, on the same graph. By inspection we are able to
find that one of the roots of $x$ of interest belongs to the
interval $(0.42, 0.44)$ and the other to the interval $(0.38,
0.40)$. These informations are sufficient for the program to solve
the system.

\begin{center}

{\unitlength=1mm
\begin{picture}(80,65)
\put(10,0){\includegraphics[width=8cm]{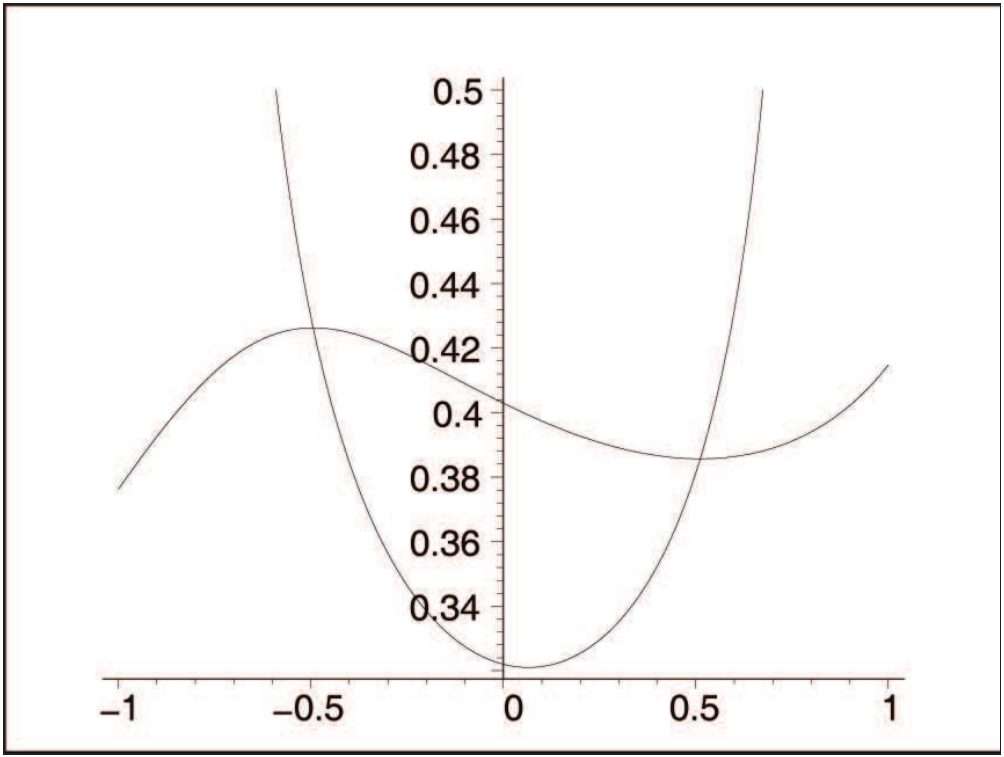}}
\put(47,57){{\footnotesize $x(\mu)$}} \put(85,9){{\footnotesize
$\mu$}}
\end{picture}

}

\vv

\begin{tabular}{lp{12cm}}
{\bf FIG. 6.8} &Graphical solution to the system of equations
constituted by $\widehat{\ovl\psi}^{(3)}(x,\mu)=0$ and
$\ptl\widehat{\ovl\psi}^{(3)}(x,\mu)/\ptl\ta =0$ for the pair
$(x,\mu)$, where $\mu=\cos\ta$, in the case $\eps=2/5$ and $\lb =
0$.
\end{tabular}

\end{center}

\vv

As an example of the results obtained by use of these equations,
Table 6.5 exhibits the angular positions and values of the extrema
of the relative deviation associated with the third order of
approximation to the flux function for the equilibrium characterized
by having $\eps = 2/5$ and $\lb = 0$. Table 6.6 is a collection of
the data regarding the maximum absolute values reached by the
relative deviations pertaining to this same order of approximation
to the flux function for the equilibria labelled by $\eps = 2/5$ and
the several values of $\lb$ we have been considering in this
Section.

For those equilibria for which $|\cald|_{\max} \ll 1$ it is possible
to find an explicit representation of the coordinate $x$ in terms of
the poloidal angle $\ta$ that approximates the description of the
surface defined by $\widehat{\ovl\psi}(x, \ta; \chi,\eps) = 0$ and
thus an approximate expression for the deviation as a function of
$\ta$. Indeed, if the quantity $x - \eps$ remains small at all
points along the contour $\widehat{\ovl\psi}(x, \ta; \chi,\eps) =$
constant, then it suffices to retain the terms up to the linear one
in the Taylor expansion of $\widehat{\ovl\psi}(x, \ta; \chi,\eps)$
about the toroidal surface on which $x = \eps$; that is to say, we
may approximate: \beq \widehat{\ovl\psi}(x, \ta; \chi,\eps) \simeq
\widehat{\ovl\psi}(x = \eps, \ta; \chi,\eps) + (x - \eps)
\frac{\ptl\widehat{\ovl\psi}(x, \ta; \chi,\eps)}{\ptl x}\Bigl|_{x =
\eps}\ . \label{eq664} \eeq

\vv

\begin{center}

\begin{tabular}{|c||c|c|c|c|} \hline
$\ta$ (degrees) & 0 & 59.2466 & 119.6028 & 180 \\ \hline
$\cald^{(3)}$ & 0.03702 & $-0.03580$ & 0.06573 & $-0.05942$ \\
\hline
\end{tabular}

\vv

\begin{tabular}{lp{12cm}}
\textbf{Table 6.5} & Angular positions ($\ta$) and values of the
extrema of the relative deviation $\cald$ associated with the flux
function $\psi^{(3)}(x, \ta; \chi,\eps)$ for the equilibrium $\eps =
2/5$, $\lb = 0$.
\end{tabular}

\end{center}

\vv

\begin{center}

\begin{tabular}{|c||c|c|} \hline
$\lb$ & $\ta$ (degrees) & $|\cald^{(3)}|_{\max}$ \\ \hline\hline && \\
1 & 180 &
0.08522 \\ && \\
0 & 119.6028 & 0.06573 \\ && \\ ${\dps -\frac{1}{5}}$ & 121.8675
& 0.2078 \\  && \\
${\dps \frac{23}{75}}$ & 180 & 0.004167 \\
&& \\ ${\dps \frac{72}{125}}$ & 180 & 0.04280 \\ && \\ ${\dps
\frac{532}{1625}}$ & 180 & 0.007595 \\ && \\ \hline
\end{tabular}

\vv

\begin{tabular}{lp{12cm}}
\textbf{Table 6.6} & Angular positions ($\ta$) and magnitudes of the
maxima of the extrema reached by the relative deviations
($|\cald|_{\max}$) associated with the flux functions $\psi^{(3)}(x,
\ta; \chi,\eps)$ in the angular interval $[0^{\rm o}, 180^{\rm o}]$
for the equilibria having $\eps = 2/5$ and the values of $\lb$
displayed in the left column.
\end{tabular}

\end{center}

\vv

Now, if the flux function on the left hand side is made to coincide
with the function specifying the surface $\widehat{\ovl\psi}(x, \ta;
\chi,\eps)= 0$, then an approximate value of the coordinate $x$
along this surface is obtained from the above expression as: \beq
\left. x \simeq \eps -\frac{\widehat{\ovl\psi}(x, \ta;
\chi,\eps)}{{\dps\frac{\ptl\widehat{\ovl\psi}(x, \ta;
\chi,\eps)}{\ptl x}}}\right|_{x = \eps}\ , \label{eq665} \eeq and
the relative deviation is uniformly approximated by: \beq \left.
\cald \equiv \cald(\ta; \chi,\eps) \simeq -\frac{1}{x{\dps
\frac{\ptl\ln\widehat{\ovl\psi}(x,\ta; \chi,\eps)}{\ptl
x}}}\right|_{x = \eps}\ . \label{eq666} \eeq

This expression holds valid whatever be the order of approximation
to be used for the flux function. If that is the third one, taking
for $\widehat{\ovl\psi}(x, \ta; \chi,\eps)$ the expression given by
Eq. (\ref{eq635}), the function $\cald(\ta; \chi,\eps)$ can be
evaluated to be: \bey &&\hspace{-15mm}\cald^{(3)}(\ta; \chi,\eps)
\simeq
\nonumber \\
&&\hspace{-15mm}\frac{\eps}{4} \frac{32(\chi - \eps^2)\cos 3\ta +
\eps(8\chi - 7\eps^2)\cos 4\ta}{\eps(16 \!-\! \eps^2) \!+\!
16\chi\cos\ta \!+\! 2\eps^3\cos 2\ta \!-\! 24(\chi \!-\! \eps^2)\cos
3\ta \!-\! 8\eps(8\chi \!-\! 7\eps^2)\cos 4\ta}\,\, . \label{eq667}
\eey

\vv

The above expression can be used to determine the extrema of the
deviation in an approximate way without the need of solving a pair
of coupled equations, as it is required by the exact calculation.
The content of Table 6.7 is an example of the results obtained by
such an application of the explicit representation of the deviation
given by Eq. (\ref{eq667}) in which the equilibrium parameters were
taken to be $\eps = 2/5$ and $\lb = 0$, and should be compared with
that of Table 6.5, which shows the results of the exact calculation.

\begin{figure}

\begin{center}

\begin{tabular}{|c||c|c|c|c|} \hline
$\ta$ (degrees) & 0 & 59.3234 & 119.3787 & 180 \\ \hline
$\cald^{(3)}$ & 0.03823 & $-0.03481$ & 0.06613 & $-0.05955$ \\
\hline
\end{tabular}

\vv

\begin{tabular}{lp{12cm}}
\textbf{Table 6.7} & Angular positions ($\ta$) and values of the
extrema of the relative deviation $\cald$ in the interval $[0^{\rm
o}, 180^{\rm o}]$ for the equilibrium $\eps = 2/5$, $\lb = 0$ as
evaluated with the help of the approximate formula for $\cald$ given
by Eq. (\ref{eq667}).
\end{tabular}

\end{center}

\end{figure}

To conclude this Section we determine the condition of breakdown of
the boundary condition for the flux function as expressed according
to its third order of approximation, that is to say, the critical
value of $\chi$ at which the surface $\psi^{(3)}(x, \ta; \eps, \chi)
= 0$ ceases to be topologically equivalent to that of a torus, and
becomes a separatrix within the set of all magnetic surfaces implied
by the flux function. For this purpose we find it convenient to have
the flux function represented in terms of the normalized cylindrical
coordinate variables $\rho$ and $Z$ rather than in terms of the
coordinate variables of the toroidal-polar system we have been using
throughout. By using the transformation relations stated in Eq.
(5.21) in Ref. \cite{quatro}, Eq. (\ref{eq635}) assumes the form:
\bey \widehat{\ovl\psi}^{(3)}\!(\rho, Z; \chi,\eps) &\!\!\!=\!\!\!&
\frac{1}{\eps^2[8\chi(\chi \!-\! \eps^2) \!+\! 32\eps^2 \!+\!
\eps^4]}\Biggl\{8\eps^2\rho^4 \!+\!
16[(4\chi-3\eps^2)Z ^2 - (\chi \!+\! 1)\eps^2]\rho^2 \nonumber \\
&&-\!8[2\chi(4\!+\!\eps^2) \!-\! 10\eps^2 \!-\! \eps^4]Z ^2 \!+\!
\eps^2[8(2 \!+\! \eps^2)\chi \!+\! 8 \!-\! 32\eps^2 \!-\!
\eps^4]\Biggr\}\, . \label{eq669} \eey

For the kind of magnetic configuration we are considering, a
magnetic surface can be identified as a separatrix if it includes
points (called singular points) where the poloidal field (and thus
the gradient of the flux function) vanishes \cite{nove}. The
critical value $\chi_c$ of the displacement variable at which the
plasma outermost surface becomes a separatrix of the magnetic
configuration and the coordinates of the singular points are
therefore determined by: \bey
&&\frac{\ptl\widehat{\ovl\psi}^{(3)}(\rho, Z ; \chi,\eps)}{\ptl
\rho} =
0\ , \label{eq670} \\
&&\frac{\ptl\widehat{\ovl\psi}^{(3)}(\rho, Z ; \chi,\eps)}{\ptl Z }
= 0 \label{eq671} \eey and \beq \widehat{\ovl\psi}^{(3)}(\rho, Z ;
\chi,\eps) = 0\ . \label{eq672} \eeq

Using Eq. (\ref{eq669}) for $\widehat{\ovl\psi}^{(3)}(\rho, Z ;
\chi,\eps)$, Eqs. (\ref{eq670}) and (\ref{eq671}) become
respectively: \beq (4\chi - 3\eps^2)Z ^2 + \eps^2\rho^2 - \eps^2(1 +
\chi) = 0 \label{eq673} \eeq and \beq 2(4\chi - 3\eps^2)\rho^2 - 2(4
+ \eps^2)\chi + 10\eps^2 + \eps^4 = 0\ , \label{eq674} \eeq while
the equation corresponding to the condition stated in Eq.
(\ref{eq672}) is obtained by putting the expression between keys on
the right hand side of Eq. (\ref{eq669}) equal to zero.

In general, for other orders of approximation to the flux function,
the algebraic equations that translate the conditions analogous to
those given by Eq. (\ref{eq670}), Eq. (\ref{eq671}) and Eq.
(\ref{eq672}) have to be solved numerically in the form they are
originally stated of three simultaneous equations for the unknowns
$\chi$, $\rho$ and $Z $. For the present case of the third order of
approximation, however, it is possible to reduce the problem to the
solution of a single equation. Indeed, from Eq. (\ref{eq674}), in
which no term containing reference to the variable $Z$ appears, we
can isolate $\rho^2$ as: \beq \rho^2 = \frac{(8 + 2\eps^2)\chi -
10\eps^2 - \eps^4}{2(4\chi - 3\eps^2)}\ . \label{eq675} \eeq

Substituting this expression for $\rho^2$ in Eq. (\ref{eq673}) and
then solving it for $Z ^2$ we obtain: \beq Z ^2 =
\frac{\eps^2(8\chi^2 - 8\eps^2\chi + 4\eps^2 + \eps^4)}{2(4\chi -
3\eps^2)}\ . \label{eq676} \eeq

Finally, the expression in terms of $\chi$ solely that we have been
able to determine for $\rho^2$ is inserted in the equation that
expresses the vanishing of $\widehat{\ovl\psi}^{(3)}(\rho, Z ;
\chi,\eps)$. The terms depending on $Z$ in this equation are
cancelled out and we are led to the following equation for $\chi$:
\beq \chi^3 - 3\left(2 + \frac{5}{8}\eps^2\right)\chi^2 + \eps^2(10
+ \eps^2)\chi + \frac{\eps^2}{64}(32 - 272\eps^2 - 7\eps^4) = 0\ .
\label{eq677} \eeq

An approximate solution to the above equation for the root with
physical significance in the form of a series of powers of the
inverse aspect ratio can be obtained as: \beq \chi = \frac{\sqrt
3}{6}\eps + \frac{121}{144}\eps^2 - \frac{103\sqrt 3}{20736}\eps^3 -
\frac{49}{15552}\eps^4 - \frac{30715\sqrt 3}{47775744}\eps^5 +
\cdots \label{eq678} \eeq For $\eps = 2/5$ \ Eq. (\ref{eq677})
becomes: \beq \chi^3 - \frac{63}{10}\chi^2 + \frac{1016}{625}\chi -
\frac{457}{15625} = 0\ . \label{eq679} \eeq

The root of interest of Eq. (\ref{eq679}) to four decimal places,
obtained either by solving it numerically or by using the expansion
about $\eps = 0$ given by Eq. (\ref{eq678}), is: \beq \chi \equiv
\chi_c^{(3)} \simeq 0.2493\ . \label{eq680} \eeq

Substituting this value for $\chi$ and \ $2/5$ for $\eps$ in Eqs.
(\ref{eq675}) and (\ref{eq676}) we determine the normalized
cylindrical coordinates of the singular point as: \beq \left.
\begin{array}{l} \rho_c \simeq
0.6584 \\
Z _c \simeq  0.5024\ . \end{array}\right\} \label{eq681} \eeq

Comparing the upper limit of $\chi$ as dictated by the requirement
that the magnetic axis be internal to the plasma containing vase
with the critical value $\chi_c^{(3)}$ just determined we conclude
that the allowed domain of variation of the displacement variable
previously stated in Eq. (\ref{eq623}) has to be replaced by this:
\beq 0.04000 < \chi < 0.2493\ . \label{eq682} \eeq

The corresponding interval of variation of the relative Shafranov
shift is: \beq 0.01980 < \dt < 0.1177\ , \label{eq683} \eeq and that
of the equilibrium parameter $\lb$ is: \beq -0.2678 < \lb < \infty\
. \label{eq684} \eeq

The limits found for $\chi$, $\dt$ and $\lb$ are, of course, those
that stem from the approximation of order $N = 3$ to the solution to
the Grad-Shafranov boundary value problem. Although the figures
might change according to the order of approximation, we shall see
in the next Sections that it remains true that the allowed range of
variation of $\chi$ is defined to be: \beq \chi_P^{(N)} < \chi <
\chi_c^{(N)}\ , \label{eq685} \eeq where $\chi_P^{(N)}$ is the real
pole of the equilibrium function $F^{(N)}(\eps,\chi)$ of the least
positive value, and $\chi_c^{(N)}$ is the critical value of $\chi$
under the approximation of order $N$ to the flux function. Figure
6.9 represents graphically the dependence of the critical value of
the displacement variable on the inverse aspect ratio; also shown
are the curve for the maximal value of $\chi$ according to the
criterium that the magnetic axis must be internal to the toroidal
chamber, and the curve for the minimal value of $\chi$ according to
the criterium that the poloidal flux must be a negatively defined
quantity at the magnetic axis, $N$ assumed to be 3.

Figure 6.10 provides us with a view of the structural changes that
the bounding surface of a magnetic configuration of intermediate
inverse aspect ratio, as portrayed by a low order approximation to
the flux function, experiences when the value of the displacement
variable increases from subcritical to critical and then to
supercritical.

The two closed lines that appear on the two sides of Fig. 6.10(a),
one as the reflection of the other about the vertical centre line,
represent the (cartesian) plot  of the contour $Z  = Z (\rho)$ of
the plasma cross section as it is implied by the equation
$\widehat{\ovl\psi}^{(3)}(\rho, Z ) = 0$, for $\eps = 2/5$ and
$\chi$ a little smaller than the critical value $\chi_c^{(3)}$
\footnote{The lines that are seen on the top and at the bottom of
Fig. 6.10(a), which arise as branches of the level curve for
$\widehat{\ovl\psi}^{(3)}(\rho,Z) = 0$ in addition to the lines
representing the plasma contour, cannot be lent the meaning of
``fields lines'' even if we disregard the effect of the metallic
case in shielding the outside of the container from any magnetic
fields internally generated, for they do not close upon themselves
and extend to infinity, the magnitudes of the axial components of
the magnetic fields along their paths growing without limit as they
depart from the plasma neighborhood. Lacking physical sense as they
do, it is interesting however to keep them represented on the flux
maps to watch their advance towards the plasma vicinity as the value
of $\chi$ is increased bit to bit from $\chi < \chi_c$ to $\chi_c$,
then the emergence of singular points as $\chi$ reaches the critical
value $\chi_c$ and they meet the field lines running along the
plasma boundary (Fig. 6.10(b)), and finally the rupture of the
``magnetic wall'' around the plasma as $\chi$ is further increased
beyond $\chi_c$ and they connect to separate branches of the field
lines, which, once closed and encircling the plasma, appear now as
broken into disjoined pieces (Fig. 6.10(c)).}. Distorted as it is,
it remains nonetheless topologically equivalent to a circle and
concepts like error and deviation are still applicable.

\vv

\begin{center}

{\unitlength=1mm
\begin{picture}(80,65)
\put(0,0){\includegraphics[width=8cm]{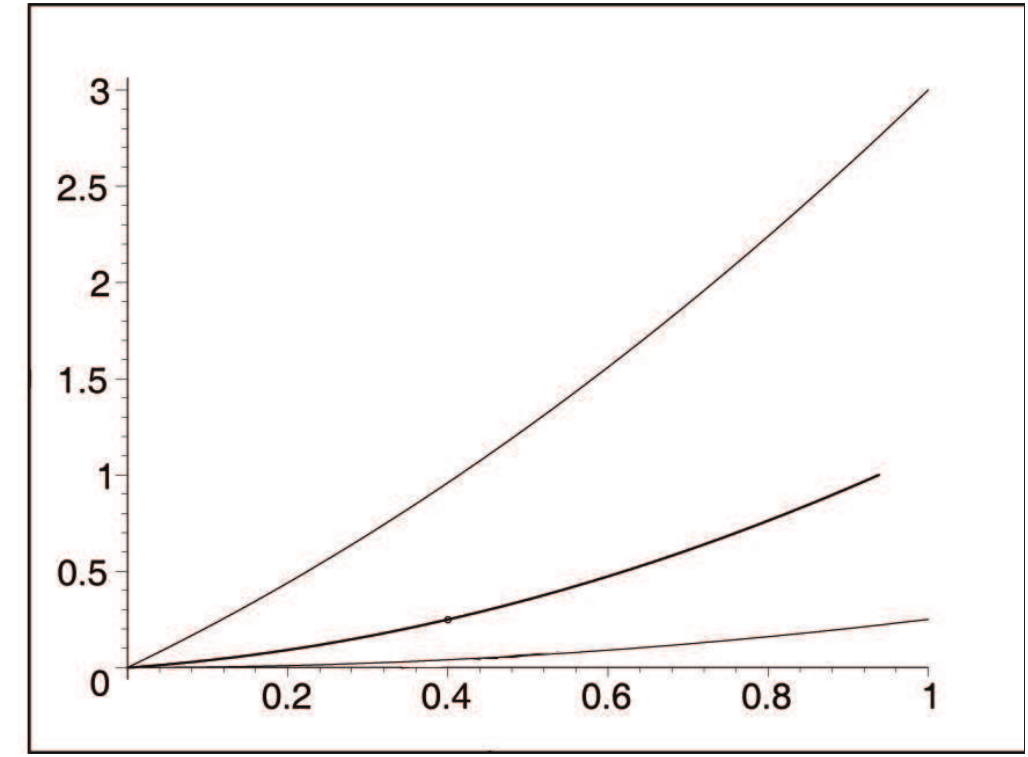}}
\put(70,14){{\footnotesize $\chi_{min}^{(3)}$}}
\put(65,25){{\footnotesize $\chi_{cr}^{(3)}$}}
\put(65,54){{\footnotesize
$\chi_{max}^{(3)}$}}\put(75,8){{\footnotesize $\eps$}}
\end{picture}

}

\vv

\begin{tabular}{lp{12cm}}
{\bf FIG. 6.9} & The curve in thick line represents the dependence
of the critical value of the displacement variable $\chi$ on the
inverse aspect ratio $\eps$ according to the approximation of order
$N=3$ to the flux function, as given implicitly by Eq.
(\ref{eq677}). The upper curve in thin line represents the variation
of the maximal allowed value of $\chi$, as defined in Eq.
(\ref{eq6.1}), and the lower curve in thin line, the variation of
the minimal value of $\chi$, as given by Eq. (\ref{eq624}), when the
inverse aspect ratio is increased from $\eps = 0$ to $\eps = 1$. The
point marked on the thick line defines the critical condition for
the configuration having $\eps = 2/5$.
\end{tabular}

\end{center}

\vv

\begin{figure}

{\unitlength=1mm

\nd (a) $\chi=0.24$

\begin{center}
\begin{picture}(80,70)
\put(0,80){\includegraphics[width=6cm,angle=-90]{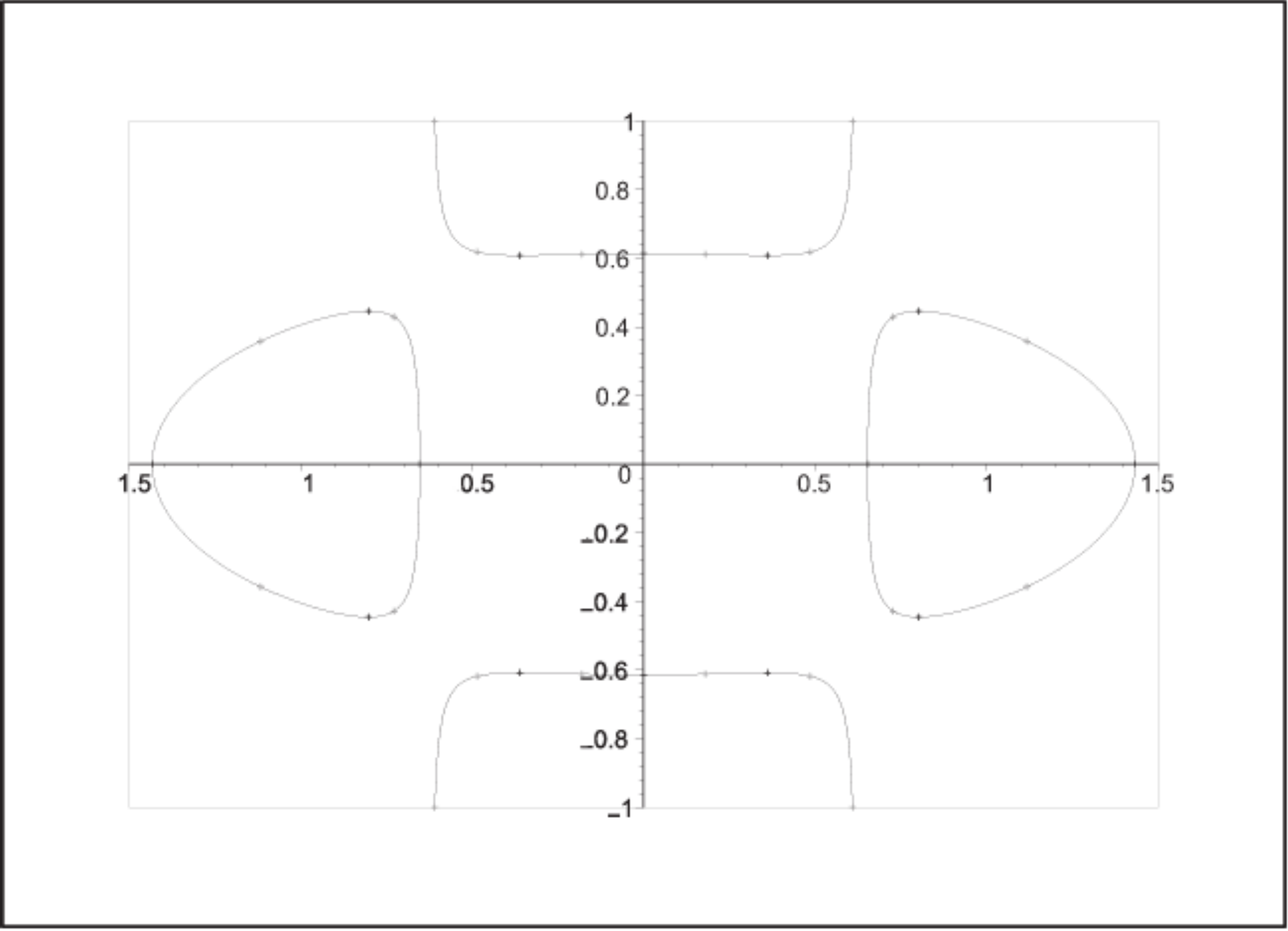}}
\put(32,58){{\footnotesize $Z$}} \put(64,54){{\footnotesize $\rho$}}
\put(58,67){{\footnotesize $\phi = 0$}} \put(12,67){{\footnotesize
$\phi = \pi$}} \put(20,54){{\footnotesize $\rho$}}
\end{picture}

\end{center}

\vspace{-17mm}

\nd (b) $\chi=0.2493$

\begin{center}

\begin{picture}(80,70)
\put(0,80){\includegraphics[width=6cm,angle=-90]{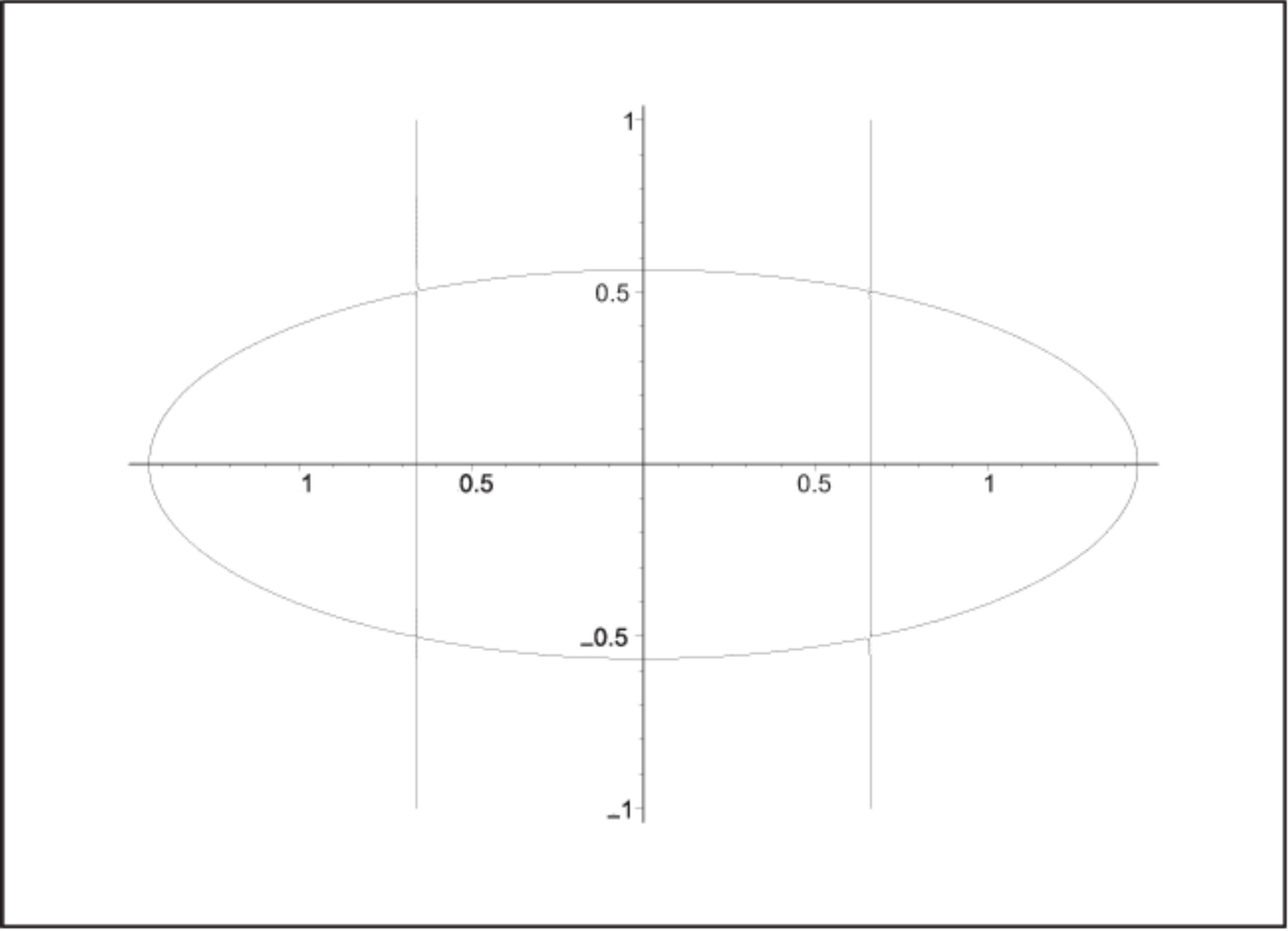}}
\put(30,55){{\footnotesize $Z$}} \put(64,53){{\footnotesize $\rho$}}
\put(62,75){{\footnotesize $\phi = 0$}} \put(5,75){{\footnotesize
$\phi = \pi$}} \put(20,53){{\footnotesize $\rho$}}
\end{picture}

\end{center}

\vspace{-17mm}

\nd (c) $\chi=0.25$

\begin{center}

\begin{picture}(80,70)
\put(0,80){\includegraphics[width=6cm,angle=-90]{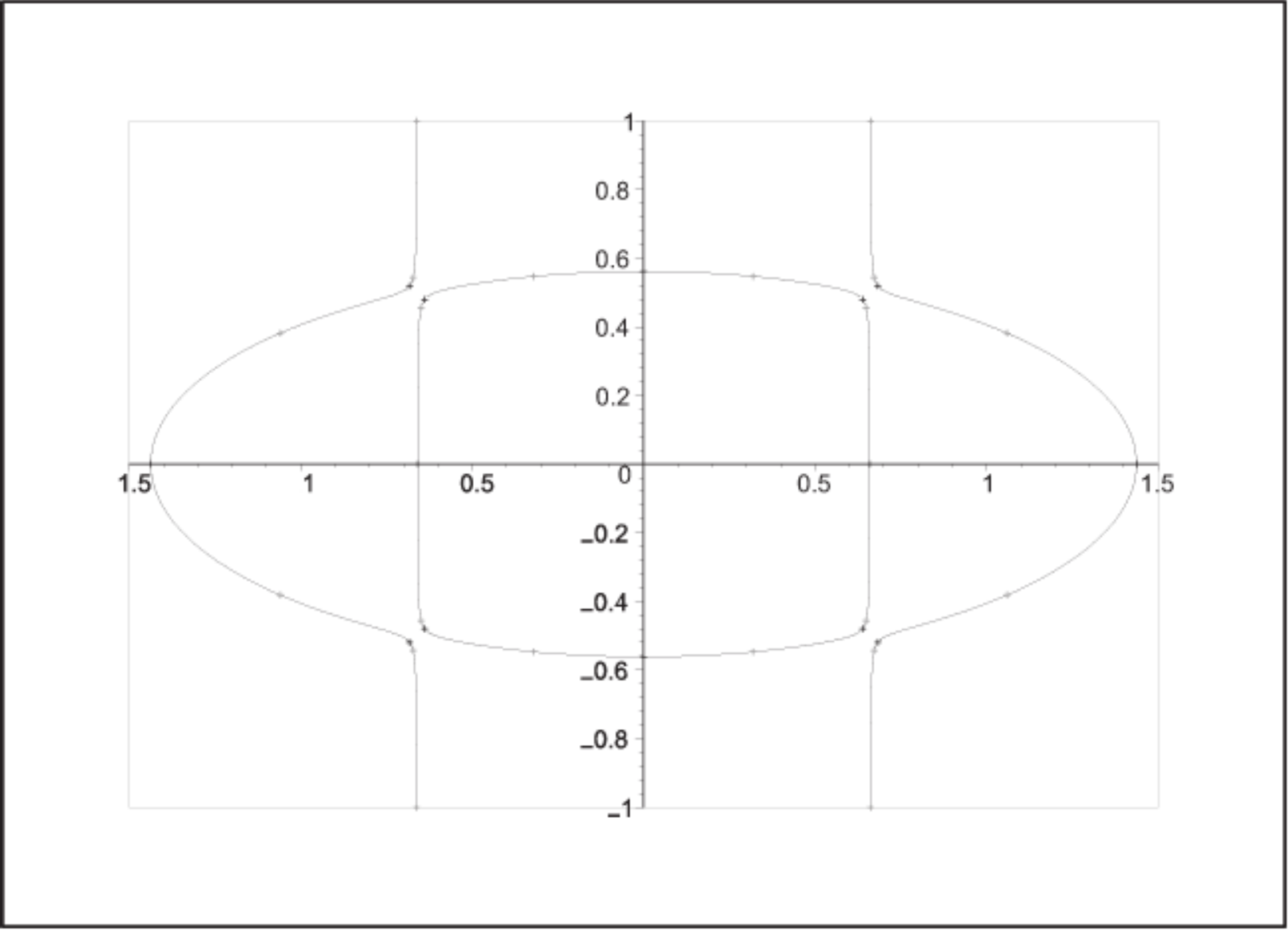}}
\put(30,57){{\footnotesize $Z$}} \put(63,54){{\footnotesize $\rho$}}
\put(62,67){{\footnotesize $\phi = 0$}} \put(11,67){{\footnotesize
$\phi = \pi$}} \put(19,54){{\footnotesize $\rho$}}
\end{picture}

\vspace{-17mm}

\begin{tabular}{lp{12cm}}
{\bf FIG. 6.10} &Level curves of $\widehat{\ovl\psi}^{(3)}(\rho,
Z;\chi,\eps=2/5)=0$ for (a) $\chi=0.24$;
 (b) $\chi=0.2493$ (critical value of the displacement variable);
(c) $\chi=0.25$.
\end{tabular}

\end{center}

}

\end{figure}

Figure 6.10(b) shows the configuration assumed by the field lines
when $\chi$ takes on the precise critical value $\chi_c$. The
singular points (which in Ref. \cite{um} are graphically referred to
as ``$X$-points'') appear as those, of coordinates $(\rho_c, \pm Z
_c)$, where the field lines coming from remote regions outside the
plasma region meet the field lines that run along the plasma
boundary, now converted into branches of the separatrix of the
configuration. At a singular point the field vanishes, providing the
plasma with an aperture where to escape from the trap contrived by
the magnetic architecture of the toroidal pinch; confinement is
thereby lost and we can no longer speak properly of equilibrium.

Finally Fig. 6.10(c) shows into which becomes the level curve
corresponding to $\widehat{\ovl\psi}^{(3)}(\rho, Z ) = 0$ for $\chi$
slightly greater than $\chi_c$. The branch that kept surrounding the
magnetic axis with progressive distortion in shape along the process
of gradual increase of $\chi$ since values close but smaller than
$\chi_c$ is now broken into two disconnected ones and any kinship
the boundary has once had with a toroid geometry is destroyed. With
the breakdown of the boundary condition the flux function ceases to
be the mathematical correspondent to the state of a physical system
and measures telling of accuracy like the relative error and the
relative deviation lose all meaning.

\vv\vv

\begin{center}
{\bf VII. SOLUTION TO THE GRAD-SHAFRANOV BOUNDARY VALUE PROBLEM
ACCORDING TO THE FOURTH ORDER OF APPROXIMATION TO THE FLUX FUNCTION}
\end{center}

\setcounter{section}{7} \setcounter{equation}{0}

\vv

In this Section and in the remaining ones in this paper we shall
adhere to the representation of the flux function that employs the
displacement variable ($\chi$)  as the quantity to characterize
globally an equilibrium configuration rather than to those that
employ instead the relative Shafranov shift ($\dt$) or the
equilibrium parameter $(\lb)$.

With the inclusion of a term proportional to the multipole solution
of order $n = 3$ in the expression for the partial flux, the
equilibrium function passes to be defined as: \beq
F^{(4)}(\eps,\chi) = - \frac{\left|\begin{array}{cccc} M_{11}(\eps)
& M_{12}(\eps) & M_{13}(\eps) & -P_1(\eps)\\
M_{21}(\eps) & M_{22}(\eps) & M_{23}(\eps) & -P_2(\eps) \\
M_{31}(\eps) & M_{32}(\eps) & M_{33}(\eps) & -P_3(\eps) \\
V_1'(\chi) & V_2'(\chi) & V_3'(\chi) & -P'(\chi)
\end{array}\right|}{D^{(4)}(\eps,\chi)} \label{eq8.1}
\eeq with \beq D^{(4)}(\eps,\chi) = \left|\begin{array}{cccc}
M_{11}(\eps)
& M_{12}(\eps) & M_{13}(\eps) & Q_1(\eps) \\
M_{21}(\eps) & M_{22}(\eps) & M_{23}(\eps) & Q_2(\eps) \\
M_{31}(\eps) & M_{32}(\eps) & M_{33}(\eps) & Q_3(\eps) \\
V_1'(\chi) & V_2'(\chi) & V_3'(\chi) & Q'(\chi)
\end{array}\right|\ . \label{eq8.2}
\eeq

\vv

Using the formulae for $P_i(x)$ ($i = 1, 2, 3$) and $P'(\chi)$, the
formulae for $Q_i(x)$ ($i = 1, 2, 3$) and $Q'(\chi)$ given in
Appendix A for the elements in the last column of one and the other
of the two above determinants respectively, the formulae for
$M_{ij}(x)$ ($i,j = 1, 2, 3$) given in Appendix B for the elements
$M_{ij}(\eps)$ in the first three rows and three columns of both
determinants, and those given in Appendix C for the elements
$V_i'(\chi)$ ($i = 1, 2, 3$) that belong to the first three columns
in the last rows \textit{idem}, we are able to arrive at the
following expression for the equilibrium function of the fourth
order: \beq F^{(4)}(\eps,\chi) = \frac{1}{3}\left(1 -
\frac{\eps^2}{2}\right)\frac{\chi^2 + \frac{{\dps \frac{16}{3} +
\frac{5}{4}\eps^2 + \frac{3}{8_{\,}}\eps^4}} {\dps {1 -
\frac{\eps^2}{2}}}\chi - \frac{{\dps \frac{20}{3}\eps^2 +
\frac{41}{48}\eps^4 + \frac{23}{96_{\,}}\eps^6}} {{\dps 1 -
\frac{\eps^2}{2_{\,}}}}}{\chi^2 - {\dps \left(\frac{16}{9} +
\frac{5}{12}\eps^2\right)\chi + \frac{4}{9}\eps^2 +
\frac{7}{48}\eps^4}}\ . \label{eq8.3} \eeq

\vpq

 This rational function of $\chi$ has two zeros, which are:
\vpq \beq \chi_{Z1, Z2}^{(4)} = \frac{8/3}{{\dps 1 -
\frac{\eps^2}{2}}} \!\left(\pm \sqrt{1 \!+\! \frac{45}{32}\eps^2
\!-\! \frac{627}{4096}\eps^4 \!+\! \frac{27}{4096}\eps^6 \!-\!
\frac{195}{16384}\eps^8} - 1 \!-\! \frac{15}{64}\eps^2 \!-\!
\frac{9}{128}\eps^4\!\right)\ , \label{eq8.4} \eeq the first one
corresponding to the positive sign in front of the square root on
the right hand side and the second to the negative sign. The zero
$\chi_{Z1}^{(4)}$ vanishes for $\eps = 0$, is positive for $0 < \eps
\le 1$ and for this range of values of $\eps$ falls in the interval
of variation of $\chi$ for which values this same variable is
capable to satisfy the demands of physical provenience that are put
on it. The zero $\chi_{Z2}^{(4)}$, in contradistinction, is large
and negative, and placed on the $\chi$-axis in a position far from
those that can be possibly associated with an equilibrium
configuration.

The equilibrium function $F^{(4)}(\eps,\chi)$ has also two poles,
which are: \beq \chi_{P1,P2}^{(4)} = \frac{8}{9}\left(1 \mp \sqrt{1
- \frac{3}{32}\eps^2 - \frac{531}{4096}\eps^4}\right) +
\frac{5}{24}\eps^2\ . \label{eq8.5} \eeq The pole $\chi_{P1}^{(4)}$,
the one for which the square root is to be taken with the minus
sign, is always positive, vanishing as $\eps$ goes to zero, and the
interval of variation it defines on the $\chi$-axis as $\eps$ ranges
from zero to unity covers the interval in which the displacement
variable assumes the values belonging to the set of those that can
be put in correspondence with a situation of physical significance.
The pole $\chi_{P2}^{(4)}$, also positive for all values of $\eps$,
lies on the $\chi$-axis beyond the point that defines the maximal
value the displacement variable can assume if the limits set on its
scale of variation by considerations of physical order are to be
respected.

Neither the zero $\chi_{Z2}^{(4)}$ nor the pole $\chi_{P2}^{(4)}$ of
the equilibrium function $F^{(4)}(\eps,\chi)$ have counterparts in
the structure of the equilibrium function $F^{(3)}(\eps,\chi)$, but
while the zero does not introduce a significant change in the aspect
of the graph of $F^{(4)}(\eps,\chi)$ with regard to that of
$F^{(3)}(\eps,\chi)$, the pole causes a discrepant behaviour between
the curves of the two functions, which, however, as we have seen, is
observed only in a domain of the displacement variable that lies
beyond that of physical interest.

The zero $\chi_{Z1}^{(4)}$ and the pole $\chi_{P1}^{(4)}$ of
$F^{(4)}(\eps,\chi)$ are essentially the same as the zero
$\chi_Z^{(3)}$ and the pole $\chi_P^{(3)}$ of $F^{(3)}(\eps,\chi)$,
only shifted in position on the $\chi$-axis by effect of the nonzero
value of the inverse aspect ratio. Indeed, using Eqs. (\ref{eq8.4})
and (\ref{eq616}) for the corresponding zeros of the two equilibrium
functions, and Eqs. (\ref{eq8.5}) and (\ref{eq617}) for the
corresponding poles, it is possible to show that: \beq
\chi_{Z1}^{(4)} = \chi_Z^{(3)}\left(1 - \frac{61}{320}\eps^2 +
O(\eps^4)\right) \label{eq8.6} \eeq and \beq \chi_{P1}^{(4)} =
\chi_P^{(3)}\left(1 + \frac{15}{64}\eps^2 + O(\eps^4)\right)\ .
\label{eq8.7} \eeq

In the interval of variation of $\chi$ of physical interest,
$F^{(4)}(\eps,\chi)$ does not pass through a maximum nor a minimum,
but exhibits an inflexion point whose position is given as the
solution to an equation of the third degree in $\chi$ with
$\eps$-dependent coefficients. For $\eps \ll 1$ it is possible to
find a representation in series of fractional powers of the inverse
aspect ratio for this solution, the first few terms of which are:
\beq \chi_I^{(4)} = \frac{4}{3} \eps^{2/3} - \eps^{4/3} + \eps^2 -
\frac{65}{144}\eps^{8/3} + \frac{5}{24}\eps^{10/3} + O(\eps^4)\ .
\label{eq8.6lin} \eeq

At the inflexion point the value assumed by the equilibrium
function, under the same assumption that the inverse aspect ratio is
small, can be expanded as: \beq F^{(4)}(\chi_I) = -1 - \eps^{2/3} +
\frac{3}{4}\eps^{4/3} + \frac{3}{4}\eps^2 + \frac{13}{96}\eps^{8/3}
- \frac{1}{256}\eps^{10/3} + O(\eps^{4})\ . \label{eq7.9} \eeq

Comparison of this expression with that of Eq. (\ref{eq618}) shows
that, for $\eps$ small, the value of $F^{(4)}(\chi_I)$ departs
little from that at which $F^{(3)}(\chi)$ reaches its asymptotic
limit as $\chi \to \infty$.

Figures 7.1(a), (b) and (c) show the graphical representations of
the equilibrium function $F^{(4)}(\eps,\chi)$ for $\eps = 1/50, \
 2/5$ and $4/5$ respectively. For small values of the inverse
aspect ratio an approximate representation of the equilibrium
function in the form of a partial fraction expansion proves to be
helpful to clarify the connection between this and the equilibrium
function of the third order. Keeping only terms of order $\eps^2$ in
the numerator and in the denominator we have: \beq F^{(4)}(\chi,
\eps) \simeq \frac{1}{3}\left(1 - \frac{\eps^2}{2}\right) +
\frac{\eps^2}{\dps\chi - \frac{\eps^2}{4}} + \frac{64}{27}
\frac{\dps 1 - \frac{5}{16}\eps^2}{\dps\chi - \frac{16}{9} -
\frac{\eps^2}{6}}\ .
 \label{eq8.7lin} \eeq

\begin{figure}

{\unitlength=1mm

\nd (a) $\eps=1/50$

\vspace{-17mm}

\begin{center}
\begin{picture}(80,65)
\put(0,0){\includegraphics[width=8cm]{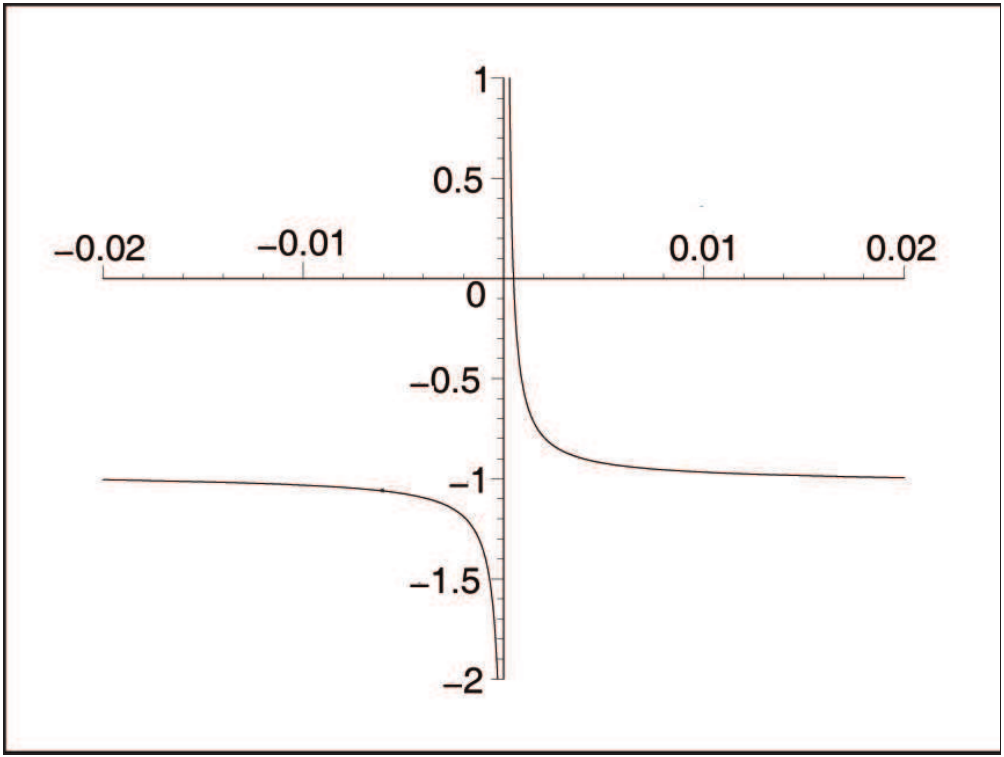}}
\put(73,35){{\footnotesize $\chi$}} \put(3,55){{\footnotesize
$F^{(4)}$}}
\end{picture}

\end{center}

\nd (b) $\eps=2/5$

\vspace{-17mm}

\begin{center}

\begin{picture}(80,65)
\put(0,0){\includegraphics[width=8cm]{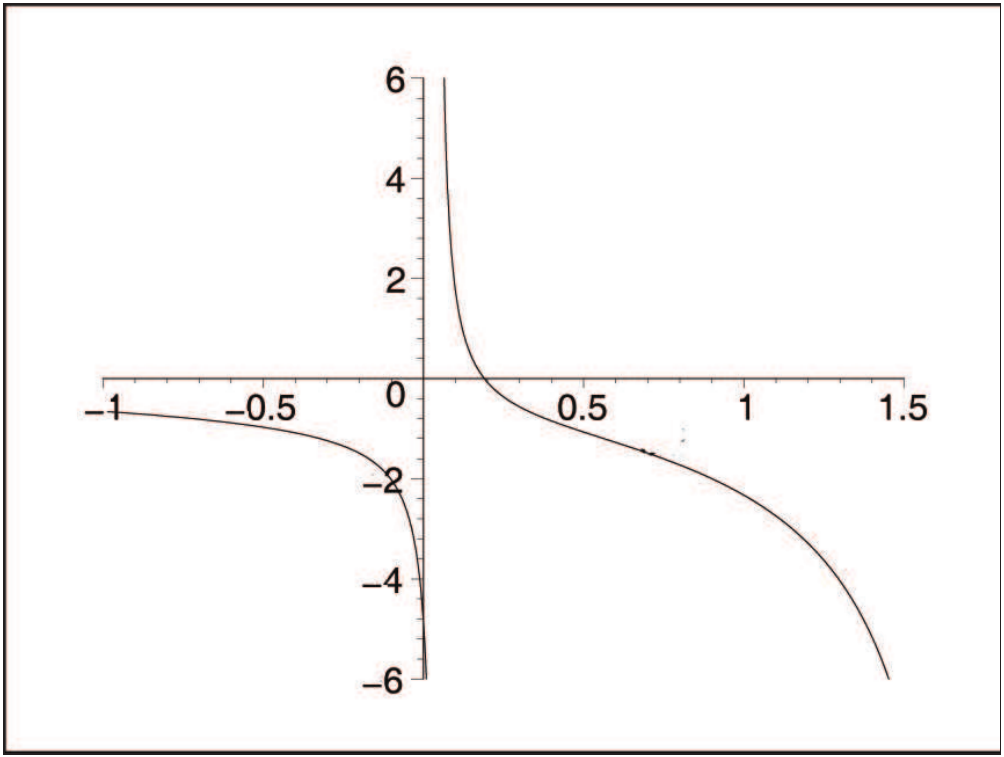}}
\put(3,55){{\footnotesize $F^{(4)}$}} \put(75,33){{\footnotesize
$\chi$}}
\end{picture}

\end{center}

\nd (c) $\eps=4/5$

\vspace{-17mm}

\begin{center}

\begin{picture}(80,65)
\put(0,0){\includegraphics[width=8cm]{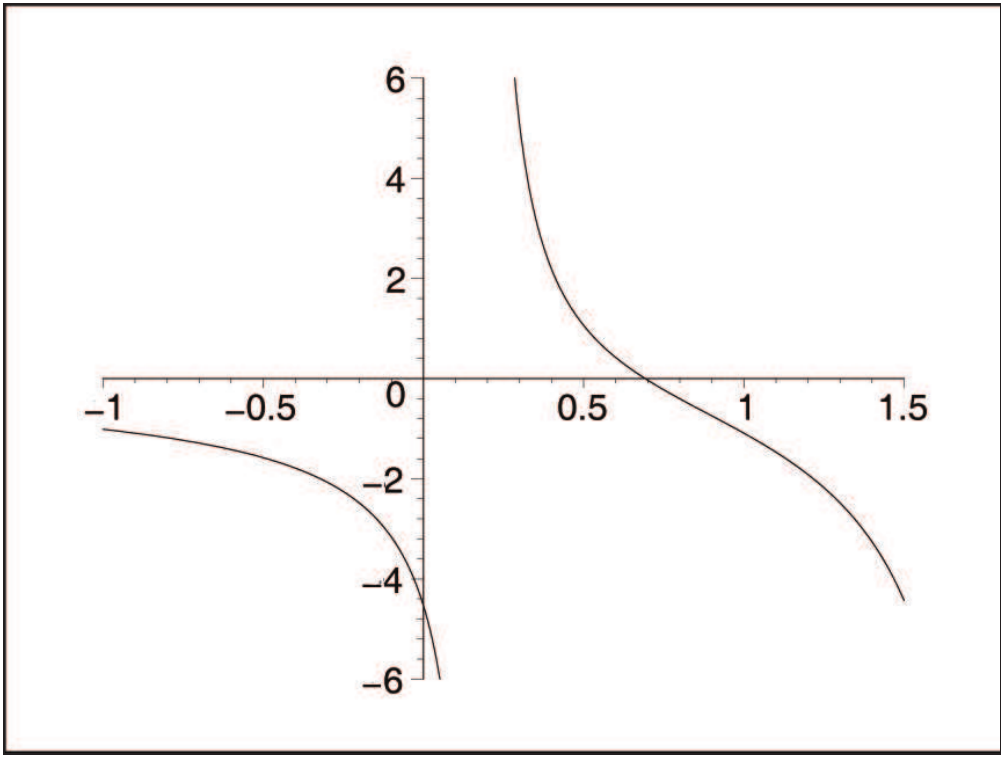}}
\put(3,55){{\footnotesize $F^{(4)}$}} \put(75,33){{\footnotesize
$\chi$}}
\end{picture}

\end{center}

}

\begin{center}

\begin{tabular}{lp{12cm}}
{\bf FIG. 7.1} & Plots of the equilibrium function of the fourth
order $F^{(4)}(\eps,\chi)$ for the inverse aspect ratio $\eps$ equal
to (a) 1/50; (b) 2/5; (c) 4/5.
\end{tabular}

\end{center}

\end{figure}

The second term on the right hand side of Eq. (\ref{eq8.7lin}),
which becomes the dominant one in the region of physical interest,
is essentially the same as the second term on the right hand side of
Eq. (\ref{eq618}). We note that, for negligible values of $\eps$ and
$\chi$, the first term on the right hand side of Eq.
(\ref{eq8.7lin}) adds to the third term to produce
$F^{(4)}(\eps,\chi) \sim -1$, which is the value taken by the first
term of $F^{(3)}(\eps,\chi)$ in the representation of Eq.
(\ref{eq618}) under equal assumptions. Thus it appears that the
first term in the equilibrium function in the passing from the third
to the fourth order of approximation unfolds into two terms whose
contributions to $F^{(4)}(\eps,\chi)$ in the region of physical
interest combine to reproduce approximately the value that would be
given by the unsplit term of the original equilibrium function. This
explains why the equilibrium equations as stated for the same values
of the equilibrium parameter on the right hand side and the two
equilibrium functions on the left hand side respectively do not lead
to widely discrepant values for the displacement variable, although
$F^{(3)}(\eps,\chi)$ and $F^{(4)}(\eps,\chi)$ are structurally
distinct functions.

According to Eq. (\ref{eqnova41}), the expression of the partial
flux for $N = 4$ comprises the particular solution and the multipole
solutions of the four lowest orders in combination. For the
evaluation of the constant $K_0^{(4)}(\eps, \chi)$ that multiplies
the multipole solution $\vf^{(0)}(x,\ta) = 1$ we refer to Eq.
(\ref{eqnova436}), and for the evaluation of the constants
$K_1^{(4)}(\eps, \chi)$, $K_2^{(4)}(\eps,  \chi)$ and
$K_3^{(4)}(\eps, \chi)$ that multiply the multipole solutions
$\vf^{(1)}(x,\ta)$, $\vf^{(2)}(x,\ta)$ and $\vf^{(3)}(x,\ta)$, to
Eq. (\ref{eqnova434}), applying the substitution scheme stated as
Eq. (\ref{eqnova442}) to write the elements in the last rows of the
determinants that appear as numerators and denominators in these two
equations as functions of $\chi$. By having recourse to the
collections of expressions that constitute the contents of
Appendices A, B and C we may then give definite forms to all the
elements of the determinants involved in the resulting formulae for
the four constants $K_i^{(4)}(\eps, \chi)$ ($i = 0, 1, 2, 3$). The
equilibrium parameter can be eliminated from the final expression
for the flux function in favour of $\chi$ if we substitute $\lb$ by
by $F^{(4)}(\eps, \chi)$, as given by Eq. (\ref{eq8.3}), in the
expression of the particular solution $\ovl\psi_p(x,\ta)$ (Eq.
(\ref{eq3.29nova})).

The flux function at the magnetic axis, $\ovl{\psi_{M}^{(4)}}(\eps,
\chi)$, can be determined either directly, by using the general
formula of Eq. (\ref{eqnova449}) appropriately specialized to the
case $N = 4$, or by evaluating the expression to which is reduced
that of the flux function, $\ovl{\psi^{(4)}}(x,\ta; \eps, \chi)$,
once it have been obtained, at the magnetic axis, where the
coordinate variables assume the values $\ta = 0$, $x = (1 +
\chi)^{1/2} - 1$. In a graphical representation,
$\ovl{\psi_{M}^{(4)}}(\eps, \chi)$ would show to participate of the
same general aspect as that of $\ovl{\psi_M^{(3)}}(\eps,\chi)$, how
it can be aprehended, for example, from the graph in Fig. 6.1, which
applies to $\eps = 2/5$. In particular, it would make apparent that
the flux at the magnetic axis is negative for $\chi
> \chi_{P_1}^{(4)}$ and positive for $\chi < \chi_{P_1}^{(4)}$.

The flux function normalized to minus the magnetic flux at the
magnetic axis is obtained as: \bey \widehat{\ovl\psi}^{(4)}(x, \ta;
\eps, \chi) &=& \frac{x^2 - \eps^2}{d^{(4)}}\Bigl[Y_0^{(4)}(x; \eps,
\chi) + xY_1^{(4)}(x; \eps,\chi)\cos\ta + x^2Y_2^{(4)}(x;
\eps,\chi)\cos 2\ta \nonumber \\
&&+ x^3Y_3^{(4)}(\eps,\chi)\cos 3\ta\Bigr] + \frac{1}{d^{(4)}}
\Bigl[x^4W_4^{(4)}(x;\eps,\chi)\cos 4\ta \nonumber \\
&&+ x^5W_5^{(4)}(\eps,\chi) \cos 5\ta + x^6W_6^{(4)}(\eps,\chi)\cos
6\ta\Bigr] \ , \label{eq8.11} \eey where \bey d^{(4)} &\equiv&
d^{(4)}(\eps,\chi) \nonumber \\
&=& \chi^4 \!-\! 4\eps^2\chi^3 \!-\! \left(8 \!-
\frac{55}{16}\eps^2\right)\eps^2\chi^2 \!- \frac{7}{8}\eps^6\chi
\!+\! 16\eps^4 \!- \frac{\eps^6}{4} \!+\! \frac{21}{128}\eps^8,
\label{eq8.12} \\
Y_0^{(4)}(x;\eps,\chi) &=& \frac{1}{8}(\chi - \eps^2)x^4 -
\left[\frac{15}{8}\chi^2 - \left(4 + \frac{17}{16}\eps^2\right)\chi
+ \frac{3}{2}\eps^2 + \frac{71}{128}\eps^4\right]x^2 \nonumber \\
&&- \left(12 - \frac{21}{8}\eps^2\right)\chi^2
-\frac{7}{8}\chi\eps^4 + 16\eps^2 - \frac{\eps^4}{4} +
\frac{21}{128}\eps^6 \ , \label{eq8.13} \\
Y_1^{(4)}(x;\eps, \chi) &=& -\frac{1}{2}(\chi - \eps^2)x^2 -
12\chi^2 + \left(16 + \frac{13}{4}\eps^2\right)\chi\ , \label{eq8.14} \\
Y_2^{(4)}(x; \eps,\chi) &=& -\frac{5}{16}(\chi - \eps^2)x^2 -
\frac{3}{2}\chi^2 + \left(4 - \frac{5}{16}\eps^2\right)\chi -
2\eps^2 + \frac{23}{32}\eps^4\ , \label{eq8.15} \\
Y_3^{(4)}(\eps,\chi) &=& \frac{15}{4}(\chi - \eps^2) \ , \label{eq8.16} \\
W_4^{(4)}(x; \eps,\chi) &=& \frac{7}{8}(\chi - \eps^2)x^2 +
\frac{3}{8}\chi^2 + \left(16 - \frac{15}{16}\eps^2\right)\chi -
\frac{33}{2}\eps^2 + \frac{107}{128}\eps^4 \ , \label{eq8.17} \\
W_5^{(4)}(\eps,\chi) &=& \frac{35}{4}(\chi - \eps^2)\ , \label{eq8.18} \\
W_6^{(4)}(\eps;\chi) &=& \frac{21}{16}(\chi - \eps^2)\ .
\label{eq8.19} \eey

Once we are in possession of the expressions for the equilibrium
function and for the flux function, the fourth order solution to the
equilibrium problem for any given value of the equilibrium parameter
proceeds in complete analogy with the third order solution as it was
developed in Section VI. Table 7.1 summarizes the results for the
equilibrium configurations having the inverse aspect ratio equal to
$\eps = 2/5$ and equilibrium parameters equal to $\lb = 1$, 0 and
$-1/5$ respectively.

\vv

\begin{center}

\begin{tabular}{|c||c|c|c|c|} \hline
&&&& \raisebox{-3mm}{$\cald_{|\max|}^{(4)} \times 10^2$} \\
$\lb$ & $\chi^{(4)}$ & $\dt^{(4)}$ & $\cale _{|\max|}^{(4)}\times 10^2$ & \\
&&&& (Angle (degrees))
\\ \hline\hline
&&&& \\
1 & 0.1214 & 0.05896 & $-5.500$ \ \ & 2.935 \ (180)\ \ \ \ \ \\
0 & 0.1905 & 0.09112 & 3.770 & 2.075 \ (132.92) \\
$\dps -1/5$\ \ \ & 0.2190 & 0.1041\,\,\, & 8.120 & 5.947 \ (133.83)
\\
&&&& \\ \hline
\end{tabular}

\vv

\begin{tabular}{lp{12cm}}
\textbf{Table 7.1} & Values assumed by the displacement variable
($\chi$) and by the relative Shafranov shift ($\dt$), extreme values
of the relative error $(\cale _{|\max|})$, relative deviations of
the normalized radial variable from $x = \eps$ at the surface
$\psi^{(4)} = 0$ whose absolute values are maximal
($\cald_{|\max|}$) and the poloidal angles where they are observed
for the configurations characterized by having the equilibrium
parameter $\lb$ equal to 1, 0 and $-1/5$ respectively and aspect
ratio fixed as $\eps = 2/5$, when the fourth order of approximation
to the flux function is employed. The errors $\cale _{|\max|}$ occur
at the poloidal angle $\ta = 0^{\rm o}$ for all the three values of
$\lb$ listed.
\end{tabular}

\end{center}

\vv

The comparison between the data exhibited in Table 7.1 with those
appearing in Tables 6.2 and 6.7 for the same values of $\lb$, and in
particular for $\lb = -1/5$, shows that the relative errors and the
relative deviations are much decreased when the order of
approximation to the flux function is increased from $N = 3$ to $N =
4$. Level curves of the flux function as obtained with the latter
order of approximation for $\eps = 2/5$ and the three reference
values of $\lb$ are given in Figs. 7.2(a), (b) and (c) respectively.

\begin{figure}
\begin{center}

\begin{tabular}{ll}
(a) $\lb = 1$ & (b) $\lb = 0$ \\
& \\
\includegraphics[width=7.08cm]{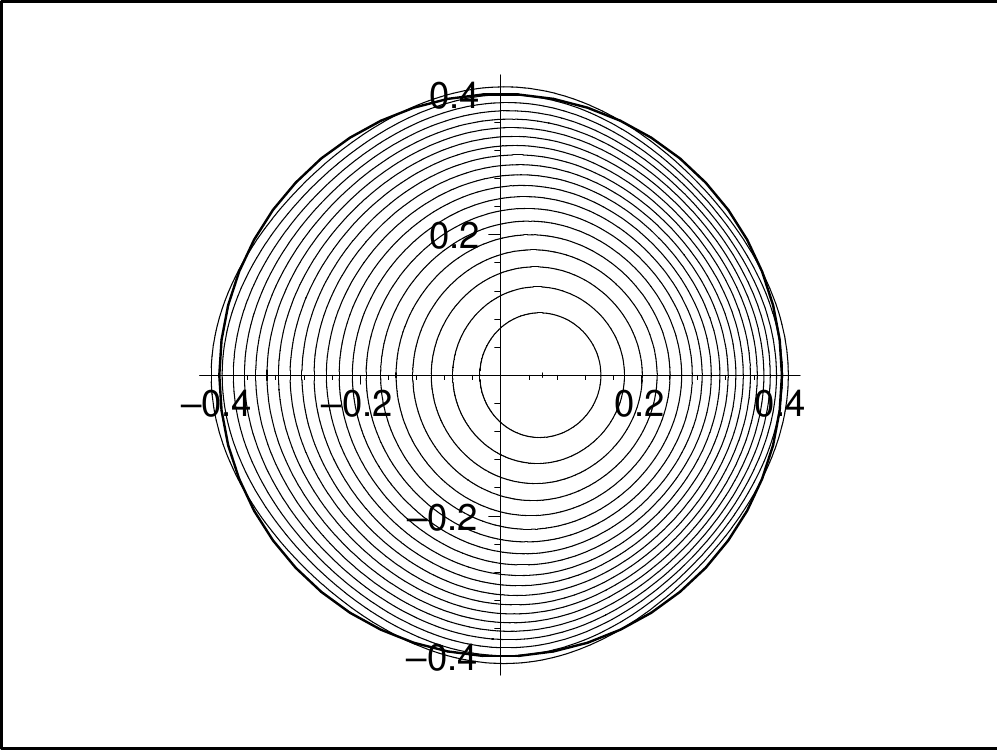} &
\includegraphics[width=7.08cm]{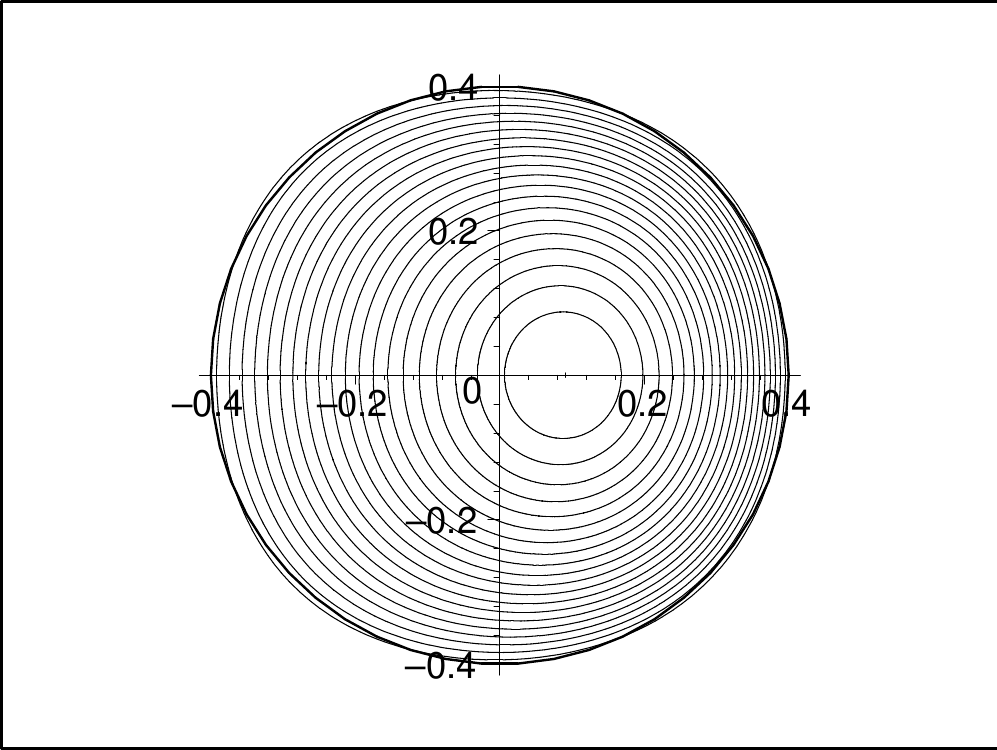} \\
& \\
& \\
\multicolumn{2}{c}{(c) $\lb=-1/5$} \\
& \\
\multicolumn{2}{c}{\includegraphics[width=7.08cm]{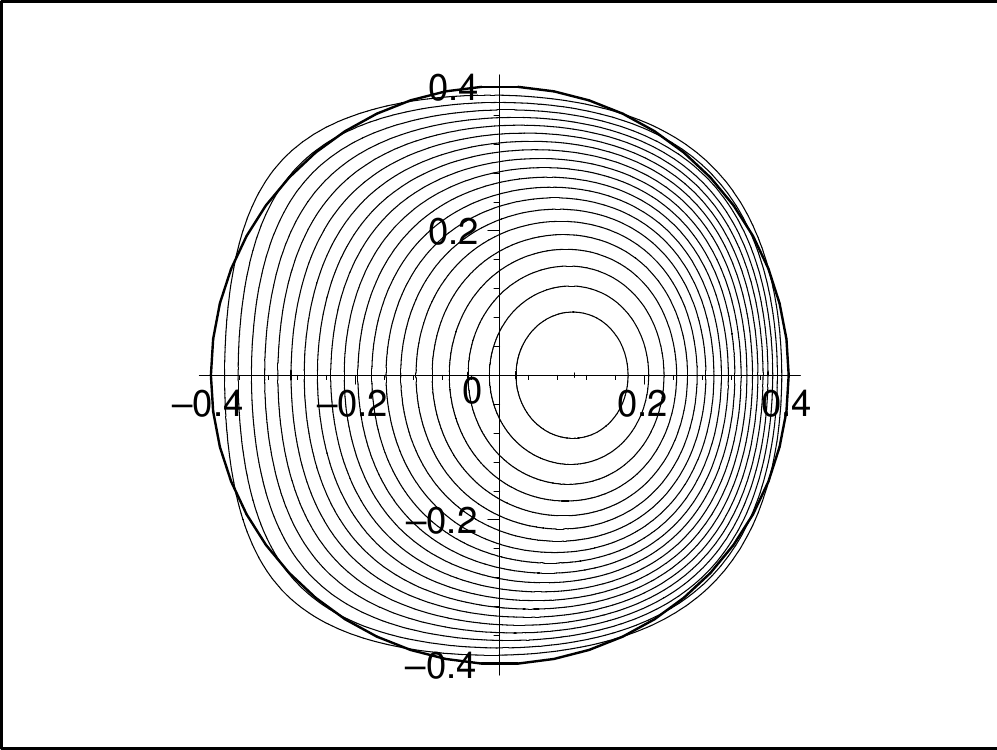}}
\end{tabular}

\vv

\begin{tabular}{lp{12cm}}
{\bf FIG. 7.2} & Flux surfaces according to the approximation of
order $N=4$ to the flux function for the equilibria in a torus of
inverse aspect ratio 2/5 and equilibrium parameter $\lb$ equal to
(a) 1; (b) 0; (c) $-1/5$.
\end{tabular}

\end{center}

\end{figure}

The expression for the flux function $\widehat{\ovl\psi}^{(4)}(x,
\ta; \eps,\chi)$, as given by Eq. (\ref{eq8.11}), shows that the
error at the boundary associated with the fourth harmonic can be
eliminated if the coefficient $W_4^{(4)}(x;\eps,\chi)$ is made to
vanish for $x = \eps$. From Eq. (\ref{eq8.17}) the condition by
which this can be made to happen translates as: \beq \chi^2 +
\left(\frac{128}{3} - \frac{\eps^2}{6}\right)\chi - 44\eps^2 -
\frac{5}{48}\eps^4 = 0 \ , \label{eq8.20} \eeq and thus $\chi$ must
assume the value: \beq \chi = \frac{64}{3}\left(\sqrt{1 +
\frac{91}{1024}\eps^2 + \frac{\eps^4}{4096}} - 1 \right) +
\frac{\eps^2}{12}\ . \label{eq8.21} \eeq

Similarly, the errors at the boundary coming from the terms of the
fifth and sixth harmonics can be simultaneously suppressed by
choosing for the displacement variable the value: \beq \chi =
\eps^2\ . \label{eq8.22} \eeq

Results obtained for the equilibrium configurations derived from
these choices for $\chi$ are summarized in Table 7.2.

To determine the equilibrium configuration for which branching
points appear on the boundary surface, as in the study undertaken
with the third order of approximation to the flux function, it is
preferable to use cylindrical rather than toroidal-polar
coordinates. The system to be solved is that which translates the
condition of simultaneous vanishing of
$\widehat{\ovl\psi}^{(4)}(\rho, Z; \chi, \eps)$ and of the two
components of its gradient. In this case of a partial flux of the
order $N = 4$ it is still possible to combine the three equations by
which it is constituted to establish a single relation connecting
the critical value of the displacement variable and the value of the
inverse aspect ratio of the configuration, the steps to be followed
to reach this end we pass to indicate briefly.

\vv

\begin{center}

\begin{tabular}{|p{4.2cm}|c|c|c|c|c|} \hline
Numbers of the har\-mon\-ics of the po\-loi\-dal angle that are
present in the function $\cale^{(4)}(\ta, \chi)$ &
\raisebox{-10mm}{$\lb$} & \raisebox{-10mm}{$\chi^{(4)}$} &
\raisebox{-10mm}{$\dt^{(4)}$} & \raisebox{-10mm}{$\cale
_{|\max|}^{(4)} \times 10^2$} &
\raisebox{-10mm}{$\cald_{|\max|}^{(4)}\times 10^2$} \\
\hline\hline &&&&& \\
\multicolumn{1}{|c|}{5, \ 6} & 0.2524 & 0.1645 & 0.07913 & 0.1152 &
0.08621
\\
&&&&& \\ &&&&& \\
\multicolumn{1}{|c|}{4} & 0.3067 & 0.1600
& 0.07703 & 0.4975 & 0.4167\,\,\, \\
&&&&& \\ \hline
\end{tabular}

\vv

\begin{tabular}{lp{12cm}}
\textbf{Table 7.2} & Data concerning the equilibrium configurations
of inverse aspect ratio $\eps = 2/5$ as portrayed by the fourth
order of approximation to the flux function when, by appropriate
choice of the values of the displacement variable, one or more of
the harmonics of the poloidal angle are made to disappear from the
function $\cale ^{(4)}(\ta; \eps,\chi)$ that describes the error at
the boundary $x = \eps$. The quantities displayed are the same as
those of Table 7.1. The maximum relative errors and the maximum
relative deviations occur at the poloidal angles $\ta = 0^{\rm o}$
and $\ta = 180^{\rm o}$ respectively.
\end{tabular}

\end{center}

\vv

We first substitute the coordinate variables in the three equations
of the original system by the variables $\Rho = \rho^2$ and $\zt =
Z^2$. Observing that the equation proceeding from the condition
$\ptl\psi^{(4)}(\rho,Z)/\ptl Z = 0$ is linear in $\zt$, we may solve
it to obtain a representation of $\zt$ in terms of $\varrho$ and
$\chi$. By replacing this expression for $\zt$ in the two other
equations, we are led to a system for the quantities $\Rho$ and
$\chi$ of the fourth degree in each of them. The appropriate
combination of these two equations to obtain a single equation for
$\chi$ can be achieved by a method due to Sylvester and known under
the name of dyalitic elimination \cite{onze}, which in this case
concerns specifically the variable $\Rho$. This task is facilitated
by the capability that has the program Maple 7 \cite{dez} to
construct, in obedience to the order of execution of the command
\textbf{SylvesterMatrix} contained in its package
\textbf{LinearAlgebra}, the proper Sylvester matrix, the determinant
of which furnishes the equation for $\chi$ that is satisfied by the
common root belonging to the two equations of the reduced system for
any given value of the inverse aspect ratio. After factoring out two
terms that would give rise to only spurious roots (of which the
distinct ones are in number of three), we are led to an equation of
the 18th degree in $\chi$, the coefficients of its terms being
polynomials in $\eps^2$ whose degrees vary from the third to the
20th. For the reader reference, the equation for $\eps = 2/5$ can be
found in Appendix E. Note that, of the fourteen roots that this
equation admits as real, only one is to meet the requirements that
make $\chi$ be critical according to our definitions.

A graph of the critical value of the displacement variable as a
function of the inverse aspect ratio, obtained by applying the
command \textbf{Implicitplot} of Maple 7 \cite{dez} to the equation
whose derivation we have just outlined, is given in Fig. 7.3. For
values of $\eps$ less than $1/4$, certainly because of the large
number of roots, many of them having small and close values, Maple
is not able to distinguish the path followed by a particular root
 as the value of $\eps$ is varied from the paths followed by the
 others, and all it appears on the plot is a blot.

 The high degree of the equation that gives the critical value of the
displacement variable for an arbitrary value of the inverse aspect
ratio within the framework of the approximation  of order $N=4$ to
the flux function warns us that equations of parallel purpose cannot
in practice be derived for partial flux functions  of order superior
to the fourth. The recourse that then remains to us is to fix the
numerical value of $\eps$ at the start and solve the system of three
equations that defines the value of $\chi$, $\rho$ and $Z$ at the
critical condition numerically. Results for $\eps = 2/5$, obtained
by using any of the methods that prove to be suitable for the order
$N = 4$, are displayed in Table 7.3. The level curve for the
critical boundary surface is presented in Fig. 7.4.

\vv

\begin{center}

{\unitlength=1mm
\begin{picture}(80,58)
\put(0,0){\includegraphics[width=8cm]{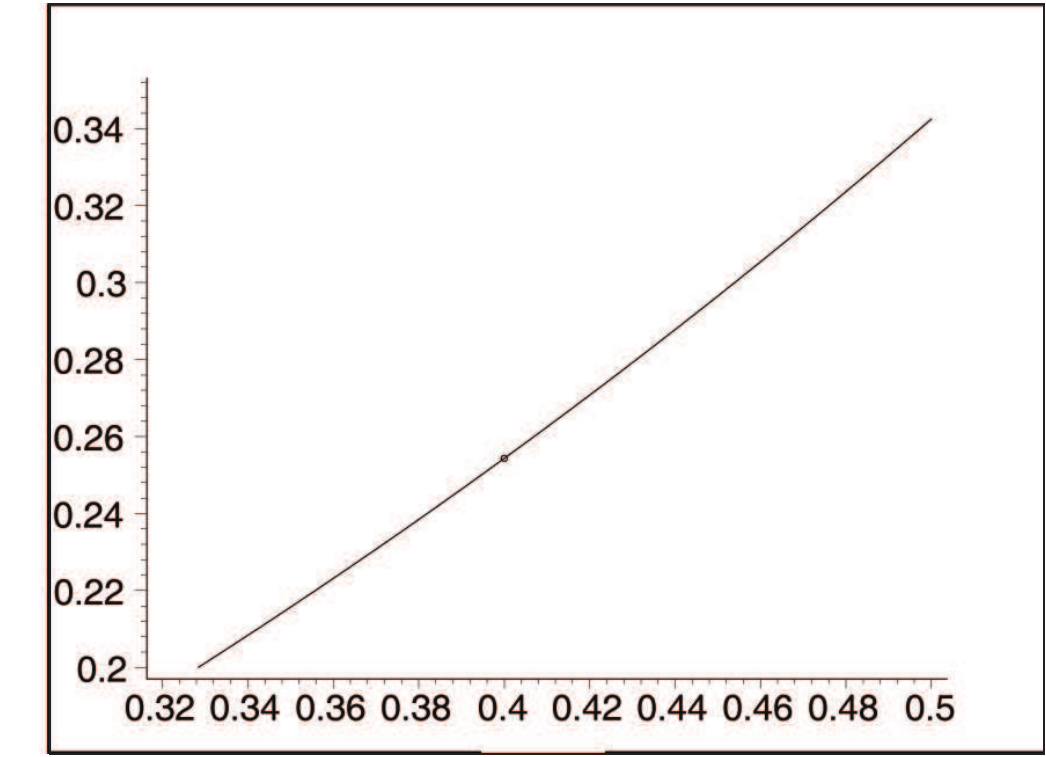}}
\put(6,53){{\small $\chi_{c}^{(4)}$}} \put(75,8){{\small $\eps$}}
\end{picture}
}

\vv\vv

\begin{tabular}{lp{12cm}}
{\bf FIG. 7.3} & Dependence of the critical value of the
displacement variable with the inverse aspect ratio within the
framework of the approximation  of order $N=4$ to the flux function.
The point marked on the curve defines the critical condition for the
configuration having $\eps = 2/5$.
\end{tabular}

\vspace{20mm}

\begin{tabular}{|c|c|c|c|c|} \hline
$\chi_c^{(4)}$ & $\dt_c^{(4)}$ & $\lb_c^{(4)}$ & $\rho_c^{(4)}$ & $Z_c^{(4)}$ \\
\hline\hline 0.2543 & 0.1200 & $-0.3840$ & 0.6241 & $\pm 0.3645$
\\ \hline
\end{tabular}

\vv\vv

\begin{tabular}{lp{12cm}}
\textbf{Table 7.3} & Critical values of the displacement variable
($\chi_c$), relative Shafranov shift ($\dt_c$) and equilibrium
parameter ($\lb_c$), and values of the radial and axial cylindrical
coordinates ($\rho_c$ and $Z_c$) of the branching points on the flux
surface $\psi^{(4)} = 0$ for $\eps = 2/5$, as obtained from the
fourth order of approximation to the flux function.
\end{tabular}

\newpage

{\unitlength=1mm

\begin{picture}(80,90)
\put(0,80){\includegraphics[width=6cm,angle=-90]{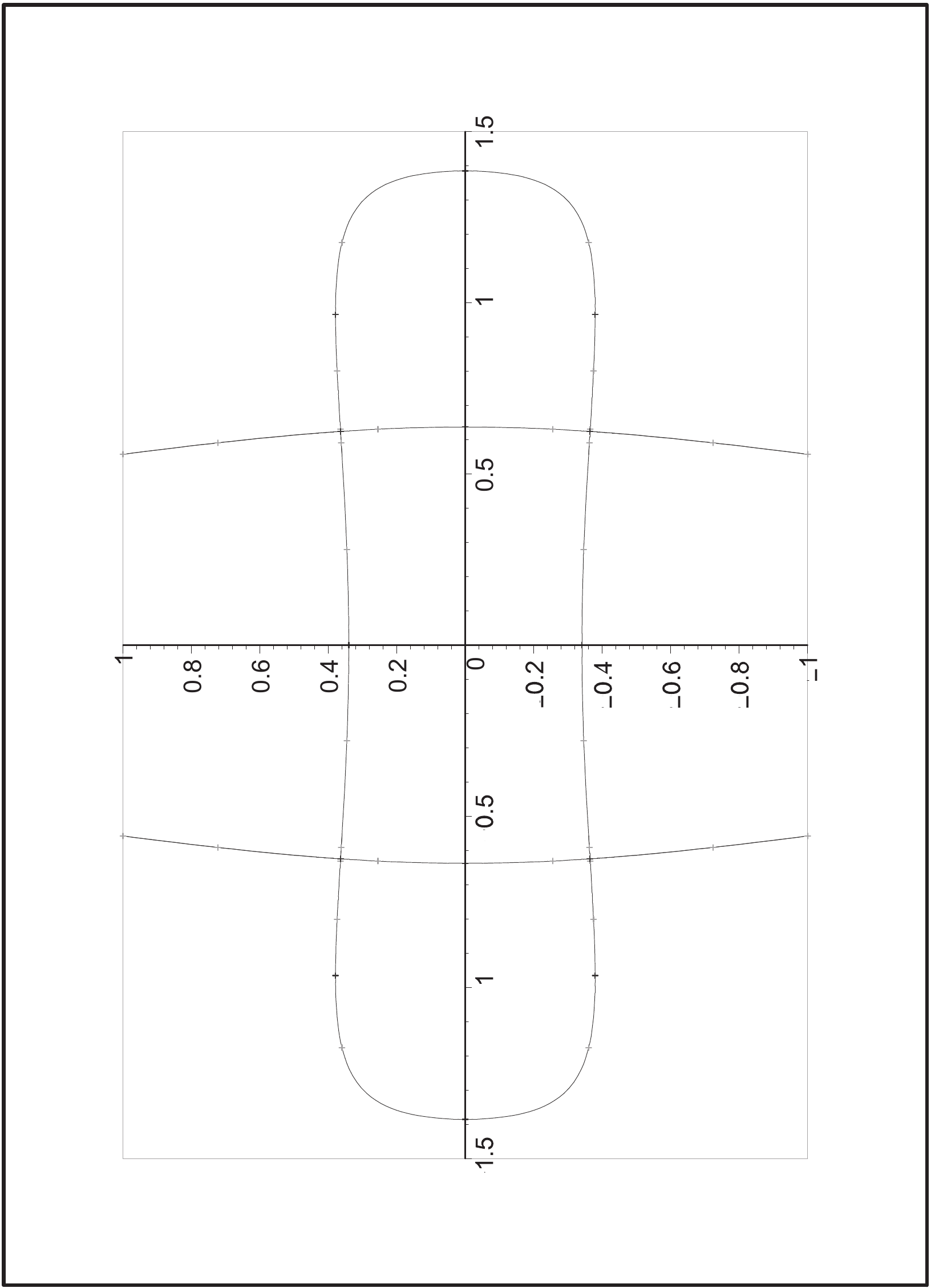}}
\put(33,61){{\footnotesize $Z$}} \put(63,53){{\footnotesize $\rho$}}
\put(20,53){{\footnotesize $\rho$}}
\end{picture}

}

\vspace{-16mm}

\begin{tabular}{lp{12cm}}
{\bf FIG. 7.4} & The level curve associated with the surface
$\widehat{\ovl\psi}^{(4)}(\rho, Z; \eps = 2/5, \chi) = 0$ at the
critical value of the displacement variable, $\chi = \chi_c^{(4)}$.
\end{tabular}
\end{center}

\vv\vv

\begin{center}
{\bf VIII. THE STRUCTURES OF THE EQUILIBRIUM FUNCTION AND OF THE
PARTIAL FLUX FUNCTION, THE DILATION OF THE LIMITS OF VALIDITY AND
THE EVOLUTION OF THE MEASURES OF ACCURACY OF THE APPROXIMATE
SOLUTIONS TO THE GRAD-SHAFRANOV BOUNDARY VALUE PROBLEM WITH THE
INCREASE OF THE ORDER OF APPROXIMATION TO THE PARTIAL FLUX FUNCTION}
\end{center}

\setcounter{section}{8} \setcounter{equation}{0}

\vv

By applying the procedure expounded in Section IV we have determined
the expressions for the equilibrium functions $F^{(N)}(\eps,\chi)$
of all orders ranging from $N = 3$ to $N = 10$. The expression for
$F^{(3)}(\eps, \chi)$ is given by Eq. (\ref{eq615}). For the
equilibrium functions of orders 5, 7, 9 and in general odd and
higher than the third, it was found that they can all be written
under the following common guise: \beq F^{(N)}(\eps, \chi)=
\cala^{(N)}(\eps^2)\frac{[\chi - \chi_1^{(N)}(\eps^2)][\chi -
\chi_Z^{(N)}(\eps^2)][\chi -
\chi_2^{(N)}(\eps^2)]f_{N-5}^{(N)}(\eps^2, \chi)}{[\chi -
\chi_P^{(N)}(\eps^2)]g_{N-3}^{(N)}(\eps^2, \chi)}\ ,
\label{eqnova8.1} \eeq where $f_{N-5}^{(N)}(\eps^2, \chi)$ and
$g_{N-3}^{(N)}(\eps^2, \chi)$ are polynomials of the form: \bey
f_{N-5}^{(N)}(\eps^2, \chi) &=& \chi^{N-5} + a_1(\eps^2)\chi^{N-6} +
\cdots +
a_{N-6}(\eps^2)\chi + a_{N-5}(\eps^2)\ , \label{eqnova8.2} \\
g_{N-3}^{(N)}(\eps^2, \chi) &=& \chi^{N-3} + b_1(\eps^2)\chi^{N-4} +
\cdots + b_{N-4}(\eps^2)\chi + b_{N-3}(\eps^2)\ , \label{eqnova8.3}
\eey with coefficients $a_1(\eps^2), \ldots, a_{N-5}(\eps^2)$ and
$b_1(\eps^2), \ldots, b_{N-3}(\eps^2)$ that depend on the square of
the inverse aspect ratio, and whose roots, for $\eps$ positive and
less than unity, are all complex. The function
$\cala^{(N)}(\eps^2)$, which we shall call the \textit{amplitude
function of order $N$}, equally depends on the square of the inverse
aspect ratio. The real zero $\chi_Z^{(N)}(\eps^2)$ and the real pole
$\chi_P^{(N)}(\eps^2)$, which are also functions of the square of
the inverse aspect ratio, are positive definite for $0 \le \eps \le
1$ and fall in the range of $\chi$ of physical interest, meaning
this that they are smaller than the critical value of the
displacement variable as evaluated in accordance with the order $N$
of approximation to the flux function. The real zero
$\chi_1^{(N)}(\eps^2)$ is large and negative, and, as $N$ is
increased to values larger than $N = 5$, it moves towards the
origin, remaining however always greater than unity in absolute
value. The real zero $\chi_2^{(N)}(\eps^2)$ is large and positive,
and also approaches the origin with the increase of $N$, but keeping
itself beyond the point of abscissa $\chi = 1$. Between the two real
and positive zeros, $\chi_Z^{(N)}(\eps^2)$ and
$\chi_2^{(N)}(\eps^2)$, the equilibrium function becomes negative,
and exhibits a point of minimum located by an abscissa on the
$\chi$-axis whose measure is larger than the critical value of the
displacement variable. This last fact precludes the possibility of
the arising of two magnetic axes for the same (negative) value of
the equilibrium parameter\footnote{The equilibrium function
$F^{(3)}(\eps, \chi)$ has only one zero, that is
$\chi_Z^{(3)}(\eps^2)$, and, therefore, no local minimum occurs for
$\chi_P^{(3)}(\eps^2) < \chi < \infty$. This makes its behaviour for
$\chi > \chi_Z^{(3)}$ be distinct from the behaviour for $\chi >
\chi_Z^{(N)}$ common to the equilibrium functions of odd orders
higher than the third, and also explains why the amplitude function
$\cala^{(3)}(\eps^2)$ is the only one to be negative among all the
amplitude functions of odd orders.}.

The equilibrium functions of even orders (for $N = 4, 6, 8, 10,
\ldots$) were found to obey in general a representation of the form:
\beq F^{(N)}(\eps, \chi) = \cala^{(N)}(\eps^2)\frac{[\chi -
\chi_1^{(N)}(\eps^2)][\chi -
\chi_Z^{(N)}(\eps^2)]h_{N-4}^{(N)}(\eps^2, \chi)}{[\chi -
\chi_P(\eps^2)][\chi - \chi_\pi^{(N)}(\eps^2)]j_{N-4}^{(N)}(\eps^2,
\chi)}\ , \label{eqnova8.4} \eeq where $h_{N-4}^{(N)}(\eps^2, \chi)$
and $j_{N-4}^{(N)}(\eps^2, \chi)$ are complete polynomials in $\chi$
of degrees $N-4$ with coefficients that are functions of $\eps^2$
save those for the highest power of $\chi$, which, as in
$f_{N-5}^{(N)}(\eps^2, \chi)$ and $g_{N-3}^{(N)}(\eps^2, \chi)$ in
Eqs. (\ref{eqnova8.2}) and (\ref{eqnova8.3}), are equal to unity.
The real zero $\chi_1^{(N)}(\eps^2)$, for a given even value of $N$,
locates itself on the $\chi$-axis between the two real zeros
$\chi_1^{(N-1)}(\eps^2)$ and $\chi_1^{(N+1)}$ of the equilibrium
functions of orders $N-1$ and $N+1$ respectively, as these are
defined for odd orders by Eq. (\ref{eqnova8.1}), and therefore, like
the two last-mentioned zeros, it approaches the origin from the left
as $N$ is increased to values above $N=4$. The pole
$\chi^{(N)}_\pi(\eps^2)$ of the equilibrium functions of even
orders, which is absent in the equilibrium functions of odd orders,
is real and positive, and approaches the origin as $N$ is increased
from $N=4$ to $N =10$, yet remaining larger than unity and therefore
outside the range of $\chi$ of physical interest. In the
representation of Eq. (\ref{eqnova8.4}) for the equilibrium
functions of even orders, the polynomials $h_{N-4}^{(N)}(\eps^2,
\chi)$ and $j_{N-4}^{(N)}(\eps^2, \chi)$ have only complex roots.

The amplitude function $\cala^{(N)}(\eps^2)$ is the limit, for a
given $\eps$, of the equilibrium function $F^{(N)}(\eps, \chi)$ as
$\chi \to \infty$. Except for $N=3$, it is positive definite and
less than unity, and, for a fixed $\eps$, decreases monotonically
with increasing $N$, with all appearances approaching a non-null
limiting value as $N$ tends to infinity; for $\eps = 2/5$ and $N =
10$, it is worth 0.1978.

The pole $\chi_P$ separates the values of the displacement variable
that are associated with the configurations for which $\ovl\psi$ is
positive at the magnetic axis (those that have $\chi < \chi_P$) from
the values of the displacement variable that are associated with the
configurations for which $\ovl\psi$ is negative at the magnetic axis
(those that have $\chi > \chi_P$). The zero $\chi_Z$ gives the value
of the displacement variable that makes the equilibrium
configuration be magnetically neutral, that is to say, with null
poloidal current density; equilibria for which $\chi < \chi_Z$ are
paramagnetic and those for which $\chi > \chi_Z$ are diamagnetic. As
$N$ is increased from $N = 3$ to $N = 10$, the pole and the zero of
the equilibrium function $F^{(N)}(\eps, x)$ that belong to the
interval of $\chi$ of physical interest show a clear tendency to
convergence as it can be seen in Table 8.1, where their values are
given for $\eps = 2/5$.

In a plot the equilibrium functions of even orders show to follow
the same pattern as that of $F^{(4)}(\eps,\chi)$, while those of odd
orders follow that of $F^{(5)}(\eps,\chi)$. In Fig. 8.1, graphs of
$F^{(9)}(\eps = 2/5,\chi)$ and $F^{(10)}(\eps = 2/5,\chi)$ for $\chi
> 0$ are shown superimposed one on the other. It can be seen that the point
where these two curves seem to depart from a common track to pursue
opposite directions is close to the point of inflexion of
$F^{(10)}(\eps = 2/5,\chi)$.

\vv

\begin{center}

\begin{tabular}{|c||c|c|} \hline
$N$ & $\chi_P^{(N)}(\eps = 2/5)$ & $\chi_Z^{(N)}(\eps = 2/5)$ \\
\hline\hline 3 & 0.04000000 & 0.19538462 \, \\
4 &  0.041512632 & 0.19053374 \\
5 &  0.041571493 & 0.19101171 \\
6 &  0.041578215 & 0.19094767 \\
7 &  0.041578812 & 0.19095680 \\
8 &  0.041578877 & 0.19095539 \\
9 &  0.041578884 & 0.19095561 \\
10 & 0.041578885 & 0.19095557 \\ \hline
\end{tabular}

\vv

\begin{tabular}{lp{12cm}}
\textbf{Table 8.1} & The pole and the zero of the equilibrium
function $F^{(N)}(\eps,\chi)$ within the range $0 < \chi <
\chi_c^{(N)}(\eps)$ for $\eps = 2/5$ and $N$ ranging from 3 to 10.
\end{tabular}

\end{center}

\vv

Also shown on the curves in Fig. 8.1 are the points $(\chi_c^{(9)},
\lb_c^{(9)})$ and $(\chi_c^{(10)}, \lb_c^{(10)})$ with abscissas and
ordinates equal to the critical values of the displacement variable
and of the equilibrium parameter respectively, at which branching
points break on the contours of $\widehat{\ovl\psi}^{(9)}(\rho, Z;
\eps = 2/5, \chi) = 0$ and $\widehat{\ovl\psi}^{(10)}(\rho, Z; \eps
= 2/5, \chi) = 0$; owing to their nearness on the common display,
they appear to be fused into a single one. To be noticed is that
these points of critical coordinates, which pinpoint the limits to
which the equilibrium variable can be pushed without causing the
boundary condition to collapse, are far moved back from the region
where the curves start to diverge one from the other. This means
that the useful parts of the equilibrium functions do not extend to
the domain of $\chi$ where that of even order and that of odd order
exhibit disparate behaviours. In general, in the ``useful'' domain
$\chi < \chi_c^{(N)}(\eps)$, the locations and the profiles of the
curves representing the equilibrium functions built from numbers of
multipole solutions moderately high and high, even and odd, are
sufficiently close and sufficiently similar to appear coincident on
a common display, as do the curves for $F^{(9)}(\eps = 2/5,\chi)$
and $F^{(10)}(\eps = 2/5,\chi)$ on a strip next to the vertical axis
in Fig. 8.1.

\begin{center}

{\unitlength=1mm

\begin{picture}(80,65)
\put(0,0){\includegraphics[width=8cm]{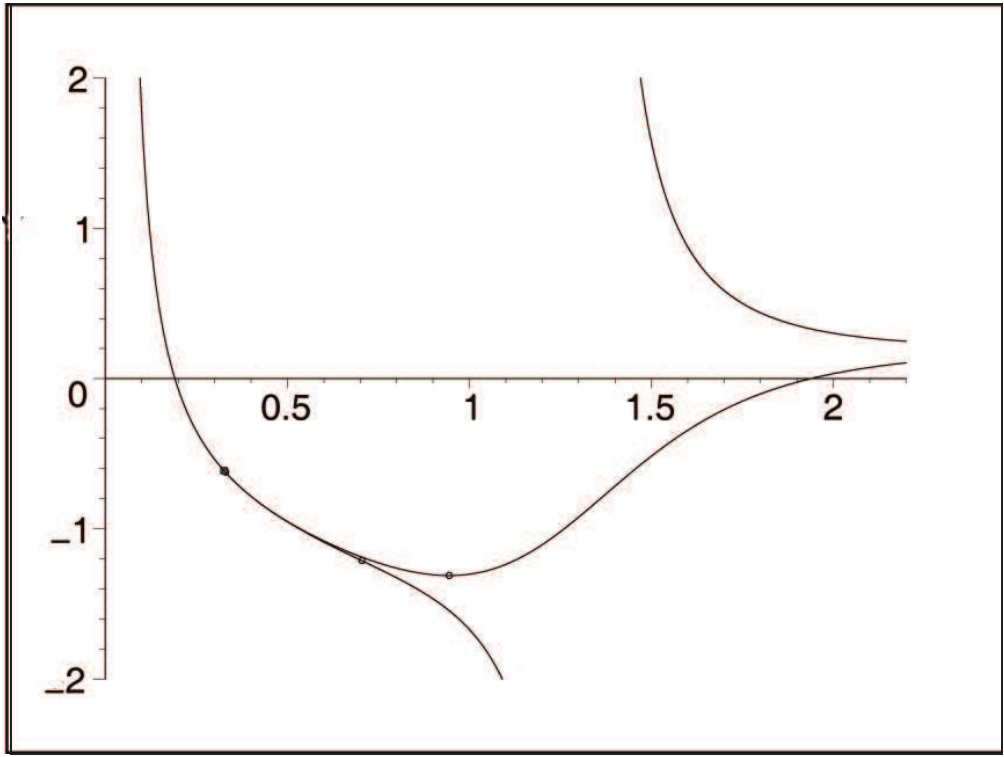}}
\put(10,56){{\footnotesize $F^{(9)}(\eps = 2/5,\chi)$ and
$F^{(10)}(\eps=2/5,\chi)$}} \put(72,26){{\footnotesize $\chi$}}
\put(19,8){{\footnotesize $N=10 \ \to$}} \put(48,18){{\footnotesize
$\leftarrow \ N=9$}}
\end{picture}

}

\vv\vv

\begin{tabular}{lp{12cm}}
{\bf FIG. 8.1} &The equilibrium functions
$F^{(9)}(\epsilon=2/5,\chi)$ and $F^{(10)}(\epsilon=2/5,\chi)$. In
order of increasing values of $\chi$ are seen on the curves: the
points defining the critical equilibria at which $X$-points occur on
the flux surfaces $\widehat{\ovl\psi}^{(9)}(\rho,
Z;\epsilon=2/5,\chi)=0$ and $\widehat{\ovl\psi}^{(10)}(\rho,
Z;\epsilon=2/5,\chi)=0$ (which appear as a single point on the scale
of the figure); the point of inflexion of
$F^{(10)}(\epsilon=2/5,\chi)$; the point of minimum of
$F^{(9)}(\epsilon=2/5,\chi)$.
\end{tabular}

\end{center}

\newpage

Table 8.2 summarizes the data concerning the equilibrium
configurations whose outermost magnetic surfaces exhibit singular
points as determined by use of the flux functions in the
approximations of order $N = 3$ to $N = 10$; the parameter $\eps$ is
taken to be equal to $2/5$. No pattern of convergence can be
recognized here: as the number $N$ increases the values of the
critical displacement and of the critical Shafranov shift are
monotonically increased, while those of the critical equilibrium
parameter become more and more negative.

\vv

\begin{center}

\begin{tabular}{|c||c|c|c|c|c|}\hline
$N$ & $\chi_c^{(N)}$ & $\dt_c^{(N)}$ & $\lb_c^{(N)}$ & $\rho_c^{(N)}$ & $Z_c^{(N)}$ \\
\hline\hline
3 & 0.2493 & 0.1177 & $-0.2678$ & 0.6584 & 0.5024 \\
4 & 0.2543 & 0.1200 & $-0.3840$ & 0.6241 & 0.3645 \\
5 & 0.2817 & 0.1321 & $-0.4717$ & 0.6099 & 0.2869 \\
6 & 0.2982 & 0.1394 & $-0.5323$ & 0.6042 & 0.2344 \\
7 & 0.3116 & 0.1453 & $-0.5717$ & 0.6024 & 0.1959 \\
8 & 0.3201 & 0.1490 & $-0.5966$ & 0.6022 & 0.1663 \\
9 & 0.3258 & 0.1514 & $-0.6122$ & 0.6025 & 0.1428 \\
10 & 0.3294 & 0.1530 & $-0.6219$ & 0.6029 & 0.1238
\\ \hline
\end{tabular}

\vv\vv

\begin{tabular}{lp{12cm}}
\textbf{Table 8.2} & Critical values of the displacement variable
($\chi_c$), relative Shafranov shift ($\dt_c$) and equilibrium
parameter ($\lb_c$) at which branching points, with normalized
cylindrical coordinates $(\rho_c, Z_c)$, appear on the surface
$\widehat{\ovl\psi}^{(N)}(\rho, Z; \eps = 2/5,\chi) = 0$, as
evaluated by use of the approximations to the flux function of
orders ranging from $N=3$ to $N = 10$.
\end{tabular}

\end{center}

\vv

The critical values of the displacement variable, as obtained from
the $N=3$ order of approximation to the flux function, grow with the
inverse aspect ratio faster than the critical values obtained from
the next higher order of approximation ($N=4$) and eventually
surpass the latter ones, but this trend of relative growths between
curves generated from representations to the flux function of two
successive orders of approximation is observed only when that
proceeding from the lowest of the two orders has it as $N =3$. For
higher orders of approximation to the flux function, for any given
value of $\eps$, the critical value of $\chi$ obtained by using the
order $N$ seems to be consistently larger than the critical value
obtained by use of the order $N-1$. Table 8.3 shows the evolution of
the critical values of the displacement variable with the order of
approximation, from $N=3$ to $N=5$, for a range of values of the
inverse aspect ratio.

\vv\vv

\begin{center}

{\unitlength=1mm
\begin{picture}(70,40)
\put(0,20){\begin{tabular}{|c||c|c|c|} \hline
$\eps$ \ \ \ \ \ \ \raisebox{3mm}{$N$} & \raisebox{1.2mm}{3} & \raisebox{1.2mm}{4} & \raisebox{1.2mm}{5} \\
\hline\hline 0.4 & 0.24927 & 0.25432 & 0.28171 \\ \hline 0.43323 &
\multicolumn{2}{|c|}{0.28194} & 0.30853 \\ \hline 0.5 & 0.35309 &
0.34250 & 0.36697 \\ \hline 0.6 & 0.47333 & 0.44658 & 0.46806 \\
\hline 0.7 & 0.60985 & 0.56771 & 0.58773 \\ \hline 0.8 & 0.76252 &
0.70672 & 0.72732 \\ \hline 0.9 & 0.93114 & 0.86429 & 0.88717 \\
\hline
\end{tabular}}
\put(0.5,42.3){\line(3,-1){18}}
\end{picture}}

\vv\vv

\begin{tabular}{lp{12cm}}
\textbf{Table 8.3} & Values of the critical displacement variable
for the configurations with inverse aspect ratios listed in the
first column on the left, as evaluated with the help of the
approximations to the flux function of the orders $N= 3, 4$ and 5.
For the value of the inverse aspect ratio in the second row, the
critical values of the displacement variable, according to the
orders of approximation $N= 3$ and $N=4$, are equal.
\end{tabular}
\end{center}

\vv

As a matter of interest, Fig. 8.2 provides us with a view of the
transformations undergone by the flux map, as drawn with the help of
a relatively high order representation to the flux function, for a
configuration having medium-sized inverse aspect ratio, when the
displacement variable is increased from a value slightly less than
the critical to the critical one, and then from this to a
supercritical value.

\newpage

\nd (a) $\chi = 0.32$

\begin{center}

{\unitlength=1mm

\begin{picture}(80,65)
\put(0,0){\includegraphics[width=10cm]{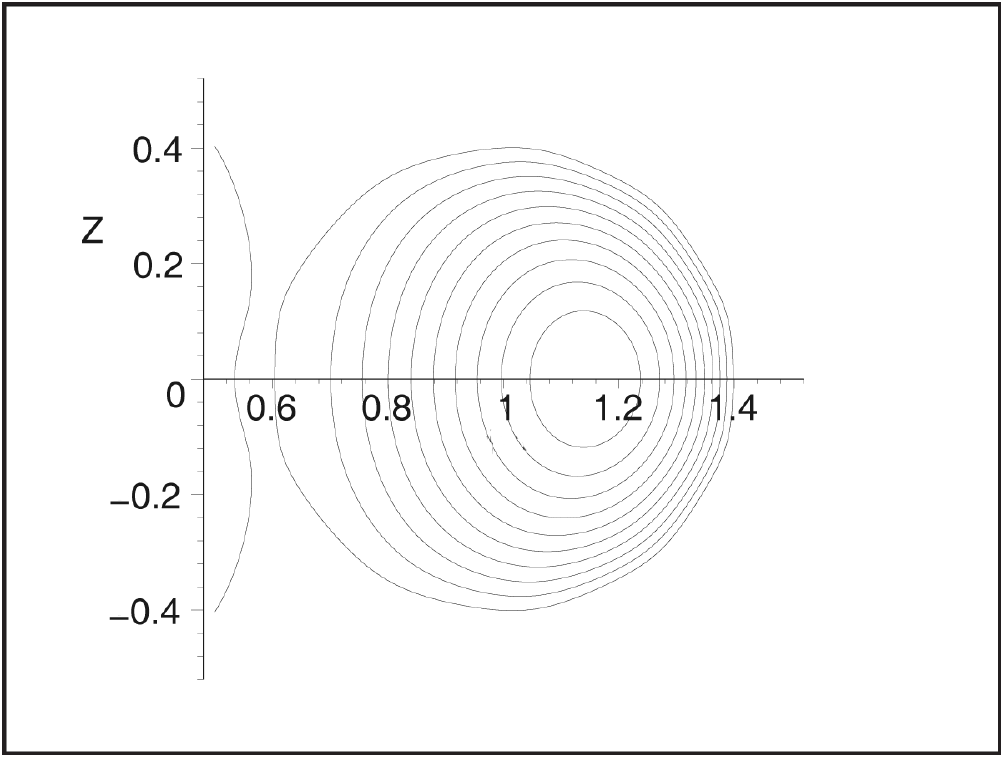}}
\put(72,54){$\psi=0$} \put(84,37){$\rho$} \put(71,50){$\swarrow$}
\end{picture}

}
\end{center}

\vv

\nd (b) $\chi \approx 0.3294$
\begin{center}

{\unitlength=1mm

\begin{picture}(80,65)
\put(0,0){\includegraphics[width=10cm]{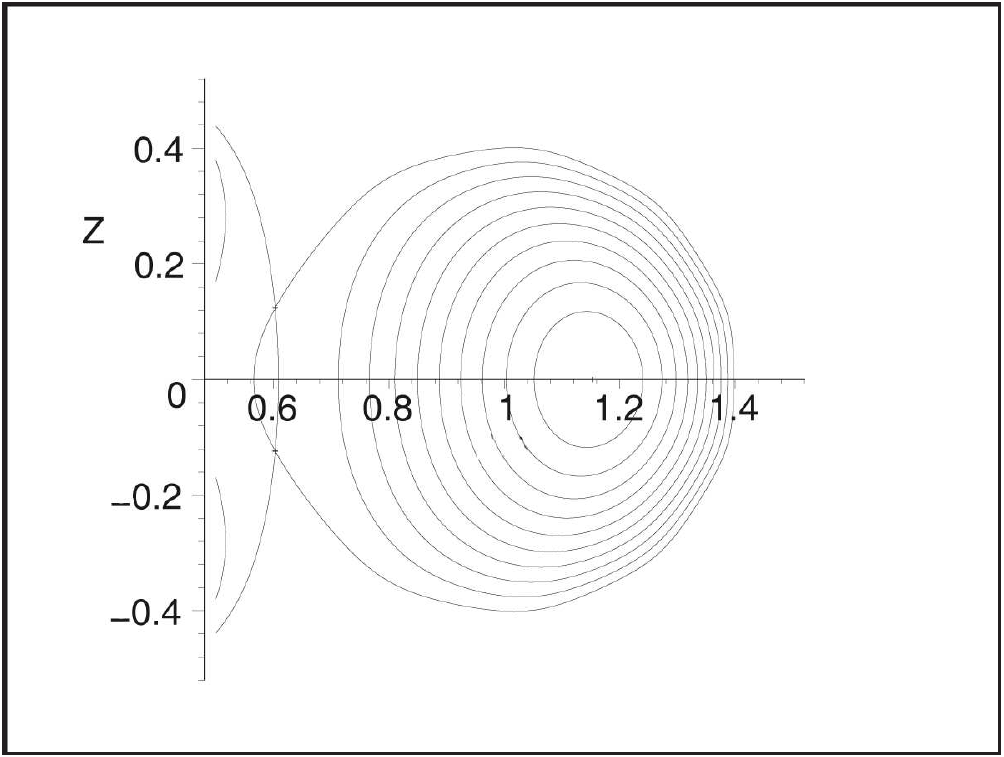}}
\put(72,54){$\psi=0$} \put(84,37){$\rho$}\put(71,50){$\swarrow$}
\end{picture}

}

\end{center}

\newpage

\nd (c) $\chi = 0.35$
\begin{center}

{\unitlength=1mm

\begin{picture}(80,65)
\put(0,0){\includegraphics[width=10cm]{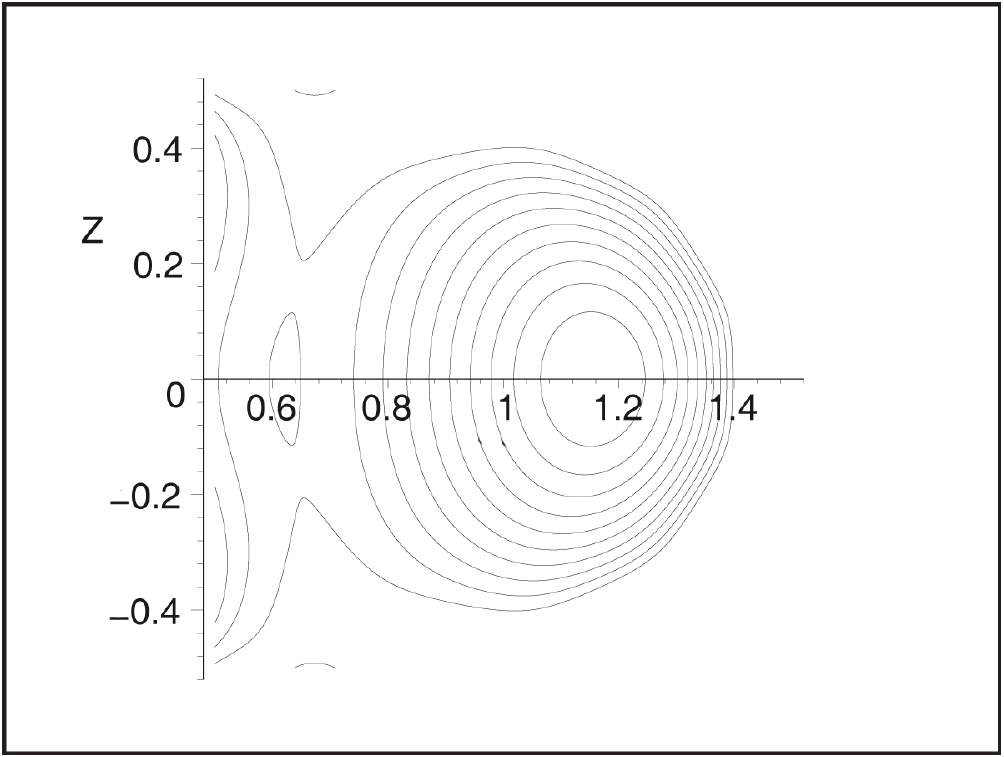}}
\put(72,54){$\psi=0$} \put(84,37){$\rho$} \put(71,50){$\swarrow$}
\end{picture}

}

\vv

\begin{tabular}{lp{12cm}}
{\bf FIG. 8.2} &Evolution of the flux map for the plasma equilibrium
in a torus of inverse aspect ratio $\eps = 2/5$ with the increase of
the value taken by the displacement variable, according to the
approximation of order $N=10$ to the flux function. In (a) $\chi$ is
less than and in (c) greater than the critical value $\chi_c$. The
flux map for the equilibrium at $\chi = \chi_c$ is shown in (b); the
point marked on the horizontal axis represents the trace of the
magnetic axis on the torus cross section, and the two points that
are marked on symmetrical positions above and below the horizontal
axis represent the branching points as they appear on the contour of
the intersection of the surface $\psi = 0$ with a meridian plane.
\end{tabular}

\end{center}

\vv\vv

We can now give form to a statement of convergence regarding the
sequence of values that the method generates for the displacement
variable. If the equilibrium equation (Eq. (\ref{eqnova444})) is
solved for a fixed $\lb$ and a growing order $N$ of approximation to
$F^{(N)}(\eps,\chi)$, up to the highest for which such function is
presently available, $\eps$ being kept constant along the process,
then the solutions found for $\chi$ show a clear tendency to
approach a definite limit, provided they remain always smaller than
the critical value $\chi_c^{(N)}$.

Before presenting the data in support to this claim we look at the
question of convergence brought about by the partial flux function
$\widehat{\ovl\psi}^{(N)}(x,\ta; \eps,\chi)$, which is built, as the
equilibrium function $F^{(N)}(\eps,\chi)$, from a number $N$ of
multipole solutions which, at least in principle, can be increased
indefinitely.

Evaluation of the partial fluxes from order $N = 3$ to $N = 10$
shows that they can in general be written in the form of a quotient,
\beq \ovl\psi^{(N)}(x,\ta;\eps,\chi) = \frac{\caln^{(N)}(x,\ta;
\eps,\chi)}{d^{(N)}(\eps,\chi)}\ , \label{eq9.2} \eeq the numerator
of which is given by the sum of two terms: \beq \caln^{(N)}(x,\ta;
\eps,\chi) = H^{(N)}(x,\ta; \eps,\chi) + R^{(N)}(x, \ta; \eps,\chi)\
, \label{eq9.3} \eeq whose forms are respectively: \beq
H^{(N)}(x,\ta; \eps,\chi) = (\eps^2 - x^2)\sum_{k=0}^N Y_k^{(N)}(x;
\eps,\chi)x^k\cos k\ta \label{eq9.4} \eeq and \beq R^{(N)}(x, \ta;
\eps,\chi) = \sum_{k=N+1}^{2N} W_k^{(N)}(x; \eps,\chi)x^k\cos k\ta\
, \label{eq9.5} \eeq such that only the first one is apt to satisfy
the vanishing condition at $x = \eps$ in any case. The functions
that appear under the symbols of summation in Eqs. (\ref{eq9.4}) and
(\ref{eq9.5}) are defined to be: \beq Y_k^{(N)}(x;\eps,\chi) =
\sum_{j=0}^l p_{k,2j}^{(N)} (\eps,\chi)x^{2j} \label{eq9.6} \eeq and
\beq W_k^{(N)}(x; \eps,\chi) = \sum_{j=0}^{l+1}q_{k,2j}^{(N)}
(\eps,\chi) x^{2j} \ , \label{eq9.8} \eeq with \beq l =
\left\{\begin{array}{ll} N - {\dps \frac{k}{2}}-1 &\hspace{2cm} (k \hbox{ even})  \\
\\
N - {\dps \frac{k+1}{2}} -1 &\hspace{2cm} (k \hbox{ odd}),
\end{array}\right. \label{eq9.9}
\eeq $p_{k,2j}^{(N)}(\eps,\chi)$ \ and \ $q_{k,2j}^{(N)}(\eps,\chi)$
being polynomials in $\chi$ with coefficients that are polynomials
in $\eps^2$. The denominator in Eq. (\ref{eq9.2}) is of the form:
\beq d^{(N)}(\eps,\chi) = [\chi
-\chi_P^{(N)}(\eps^2)]t^{(N)}(\eps,\chi)\ , \label{eq9.10} \eeq
$t^{(N)}(\eps,\chi)$ being a polynomial in $\chi$ whose coefficients
are polynomials in $\eps^2$.

The expression for the poloidal flux function normalized by minus
the poloidal flux at the magnetic axis writes in general as: \beq
\widehat{\ovl\psi}^{(N)}(x,\ta;\eps,\chi) =
-\frac{\caln^{(N)}(x,\ta;\eps,\chi)}{\caln^{(N)}(x=\dt(\chi),
\ta=0;\eps,\chi)}\ . \label{eqnew8.13} \eeq

All the results obtained from the application of the method to a
number of equilibrium configurations, as characterized by the
parameter $\lb$, are consistent with the supposition that, if the
values taken by the displacement variable in the expression for the
partial flux $\ovl\psi^{(N)}(x,\ta; \eps, \chi)$ are the same as
those provided by the solution of the equilibrium equation in which
the equilibrium function $F^{(N)}(\eps,\chi)$ is also taken to be of
order $N$, then the remainder $R^{(N)}(x,\ta; \eps,\chi)$ in Eq.
(\ref{eq9.3}) tends to zero as $N$ tends to infinity. In other
words, the sequence of partial fluxes constructed by increasing the
number of multipole solutions that are made to enter their
compositions, one by one since $N = 3$ until they include the
totality of those that are presently available ($N = 10$), seems to
obey a pattern of convergence in which the exact solution to the
Grad-Shafranov boundary value problem appears as the limit for $N$
tending to infinity.

The data resulting from the solution to the equilibrium equation and
those relative to the departure of the boundary flux surface
($\widehat{\ovl\psi} = 0$) from the circular shape as the order of
approximation to the flux function is varied from $N = 3$ to $N =
10$ are given for the equilibrium configurations having $\lb = 1$, \
0 and $-1/5$ with $\eps$ fixed at $2/5$ in Tables 8.4(a), 8.4(b) and
8.4(c) respectively. Note that the values of the displacement
variable (and thus of the relative Shafranov shift) in the sequence
that is generated when $N$ is increased by unity in succession
starting with $N = 3$ are alternately lower and higher than that of
the limit to which the sequence tends.

\newpage

\nd (a) \mbox{\boldmath${\lb = 1}$}

\begin{center}

\begin{tabular}{|c||c|c|r|c|c|} \hline
$N$ & $\chi^{(N)}$ & $\dt^{(N)}$ & $\vps_{|\max|}^{(N)} \times
10^2$ & $\cald_{|\max|}^{(N)} \times 10^2$ & $\ta_{\cald}^{(N)}$ (degrees) \\
\hline\hline 3 & 0.11921569 & 0.057929906 & -10.072\,\,\, &
8.522\,\,\, & 180.00 \\
4 & 0.12140418 & 0.058963729 & -5.500\,\,\, & 2.935\,\,\, & 180.00 \\
5 & 0.12123370 & 0.058883231 & 3.206\,\,\, & 1.3000 & \,\, 33.87 \\
6 & 0.12125217 & 0.058891954 & -2.051\,\,\, & 0.8257 & \,\,\,\,\, 0.00 \\
7 & 0.12125059 & 0.058891209 & 1.402\,\,\, & 0.5450 & \,\, 23.53 \\
8 & 0.12125078 & 0.058891297 & -1.005\,\,\, & 0.3983 & \,\,\,\,\, 0.00 \\
9 & 0.12125076 & 0.058891289 & 0.7474 & 0.2864 & \,\, 18.03 \\
10 & 0.12125076 & 0.058891290 & -0.5718 & 0.2246 & \,\,\,\,\, 0.00 \\
\hline
\end{tabular}

\end{center}

\vv

\nd (b) \mbox{\boldmath${\lb = 0}$}

\begin{center}

\begin{tabular}{|c||c|c|r|c|c|} \hline
$N$ & $\chi^{(N)}$ & $\dt^{(N)}$ & $\vps_{|\max|}^{(N)} \times
10^2$ & $\cald_{|\max|}^{(N)} \times 10^2$ & $\ta_{\cald}^{(N)}$ (degrees) \\
\hline\hline 3 & 0.19538462 & 0.093336460 & $-10.07$\,\,\,\,\,\, &
6.573\,\,\, & 119.60 \\
4 & 0.19053374 & 0.091115823 & 3.770\,\,\, & 2.075\,\,\, & 132.92
\\
5 & 0.19101171 & 0.091334831 & $-1.897$\,\,\, & 0.8221 & 141.15 \\
6 & 0.19094767 & 0.091305487 & 1.114\,\,\, & 0.3824 & \,\,27.79 \\
7 & 0.19095680 & 0.091309670 & $-0.7202$ & 0.2468 & 0 \\
8 & 0.19095539 & 0.091309025 & 0.4966 & 0.1664 & \,\,20.44 \\
9 & 0.19095561 & 0.091309127 & $-0.3587$ & 0.1222 & 0 \\
10 & 0.19095557 & 0.091309110 & 0.2684 & $-0.08959$\,\, & 0 \\
\hline
\end{tabular}

\end{center}

\vv

\nd (c) \mbox{\boldmath${\lb = -1/5}$}

\begin{center}

\begin{tabular}{|c||c|c|r|c|c|} \hline
$N$ & $\chi^{(N)}$ & $\dt^{(N)}$ & $\vps_{|\max|}^{(N)} \times
10^2$ & $\cald_{|\max|}^{(N)} \times 10^2$ & $\ta_{\cald}^{(N)}$ (degrees) \\
\hline\hline 3 & 0.23238095 & 0.11012655 & -19.786\,\,\, & 20.78\,\,\,\,\,\,\,\, & 121.87 \\
4 & 0.21903232 & 0.10409797 & 8.120\,\,\, & 5.947\,\,\, & 133.83 \\
5 & 0.22060622 & 0.10481049 & -4.290\,\,\, & 2.327\,\,\, & 141.64 \\
6 & 0.22035084 & 0.10469491 & 2.600\,\,\, & 1.030\,\,\, & 147.12 \\
7 & 0.22039421 & 0.10471454 & -1.716\,\,\, & 0.5712 & \,\,\,\,\,\,0.00 \\
8 & 0.22038637 & 0.10471099 & 1.201\,\,\, & 0.3908 & \,\,\,20.46 \\
9 & 0.22038783 & 0.10471165 & -0.8779 & 0.2886 & \,\,\,\,\,\,0.00 \\
10 & 0.22038755 & 0.10471152 & 0.6628 & 0.2125 & \,\,\,16.17 \\
\hline
\end{tabular}

\vv

\begin{tabular}{lp{9.5cm}}
\textbf{Tables 8.4 (a), (b), (c)} & The values taken by the
displacement variable ($\chi$), \end{tabular}
\begin{tabular}{p{1.3cm}p{12cm}}& the relative Shafranov shift
($\dt$), the relative error of maximum absolute value
($\vps_{|\max|}$), the relative deviation of maximum absolute value
($\cald_{|\max|}$) and by the poloidal angle where this deviation
occurs ($\ta_{\cald}$) as the order of approximation to the flux
function is increased from $N = 3$ to $N = 10$ for the equilibria
with $\eps = 2/5$ and (a) \mbox{\boldmath$\lb = 1$}, (b)
\mbox{\boldmath$\lb = 0$} and (c) \mbox{\boldmath$\lb = -1/5$}
respectively. In all cases the poloidal angle where $\vps_{|\max|}$
occurs is $\ta = 0^{\rm o}$.
\end{tabular}

\end{center}

\vv

As the order of approximation is increased while the value of the
equilibrium parameter is kept fixed as $\lb=\ovl\lb$, we observe
that the absolute values of the relative error and of the relative
deviation decrease monotonically and that the point with critical
coordinates $(\chi_c^{(N)}, \lb_c^{(N)})$ is removed away from the
point on the curve for $F^{(N)}(\eps,\chi)$ having ordinate equal to
$\ovl\lb$. For $N \ge 4$, the absolute value of the relative error
is always greater than that of the relative deviation. Of the three
reference equilibria, the largest values of one and the other
measures of inaccuracy of the approximate solutions are found for
$\lb = -1/5$. Note that, of the two configurations having $\lb = 1$
and $\lb = 0$ respectively, it is the former that has the smallest
relative Shafranov shift, but it is to the latter that belong the
smallest relative error and the smallest relative deviation whatever
be the value of $N$.

By judicious choice of the value of $\chi$ it is possible to
eliminate one of the harmonics of the poloidal angle in the
expression of the ``remainder'' $R^{(N)}(x, \ta; \eps,\chi)$ in Eq.
(\ref{eq9.5}) for $x = \eps$, in this way determining a
configuration that can be described with exceptionally high
accuracy. Of course the harmonic to be suppressed in order to obtain
the least possible maximum relative error varies according to the
order $N$ of approximation to the flux function. For $N = 10$ and
$\eps = 2/5$, the smallest relative error and the smallest relative
deviation we have been able to observe are those for the
configuration in which the eleventh harmonic is absent at the plasma
boundary; they are given together with the main parameters
characterizing such equilibrium in Table 8.5.

\begin{figure}

\begin{center}

{\tabcolsep=0.8mm
\begin{tabular}{|c|c|c|c|c|c|c|}\hline
&&&&&& \\
$\chi^{(10)}$ & $\dt^{(10)}$ & $\lb^{(10)}$ & $\vps_{|\max|}^{(10)}$
& $\ta_\vps^{(10)}$ & $\cald_{|\max|}^{(10)}$
& $\ta_{\cald}^{(10)}$ \\
&&&& {\footnotesize (degrees)} &&  {\footnotesize (degrees)} \\ \hline &&&&&& \\
0.16965874 & 0.081507621 & 0.19593095 & $-\!0.7225 \!\times\!
10^{-5}$ & 72.19 &
$0.3489 \!\times\! 10^{-5}$ & 106.46 \\
&&&&&& \\ \hline
\end{tabular}
}

\vv

\begin{tabular}{lp{12cm}}
\textbf{Table 8.5} & Data relative to the equilibrium configuration
for which the eleventh harmonic of the poloidal angle is missing in
the expression of order $N = 10$ for the partial flux function, for
$x = \eps=2/5$. The poloidal angles for which the relative error and
the relative deviation of maximum absolute values are observed are
$\ta_\vps$ and $\ta_{\cald}$ respectively.
\end{tabular}

\end{center}

\end{figure}

To conclude this Section, we note that, for any order of
approximation to the flux function, the relative departure of the
equilibrium parameter from the value corresponding to that of the
displacement variable as determined by the equilibrium equation, for
a small variation of the displacement variable, is in general much
larger than the relative variation of the displacement variable
itself.

If the value of the displacement variable is changed from $\chi$ by
$\Dt\chi$, and the corresponding change in the value of the
equilibrium parameter $\lb$ is $\Dt\lb$, we have, from Eq.
(\ref{eqnova444}), the following relation between the fractional
variations of both quantities: \beq \frac{\Dt\lb}{\lb} =
S^{(N)}(\eps,\chi)\frac{\Dt\chi}{\chi}\ , \label{eqnova8.13} \eeq
where $S^{(N)}(\eps, \chi)$, defined as: \beq S^{(N)}(\eps, \chi) =
\frac{\chi}{F^{(N)}(\eps,\chi)} \, \frac{\ptl F^{(N)}(\eps,
\chi)}{\ptl\chi}\ , \label{eqnova8.14} \eeq is a function with two
distinct poles in the interval of $\chi$ of physical interest, one
at the zero $\chi_Z^{(N)}$ and the other at the pole $\chi_P^{(N)}$
of $F^{(N)}(\eps,\chi)$.

Figure 8.3 is a plot of $S^{(N)}(\eps,\chi)$ for $\eps = 2/5$ and $N
= 10$, and Table 8.6 brings the numerical values taken by this
function for the three reference equilibria. Note that, according to
the graph of $S^{(N)}(\eps,\chi)$, as far as $\lb$ remains small,
any uncertainty in the value of $\chi$ will lead to enormous errors
in the value of $\lb$. This means that, for equilibria having $\lb
\sim 0$, it is virtually impossible to extract any reliable
information concerning the value of $\lb$ from the position taken up
by the magnetic axis.

\begin{center}

{\unitlength=1mm

\begin{picture}(80,65)
\put(0,0){\includegraphics[width=8cm]{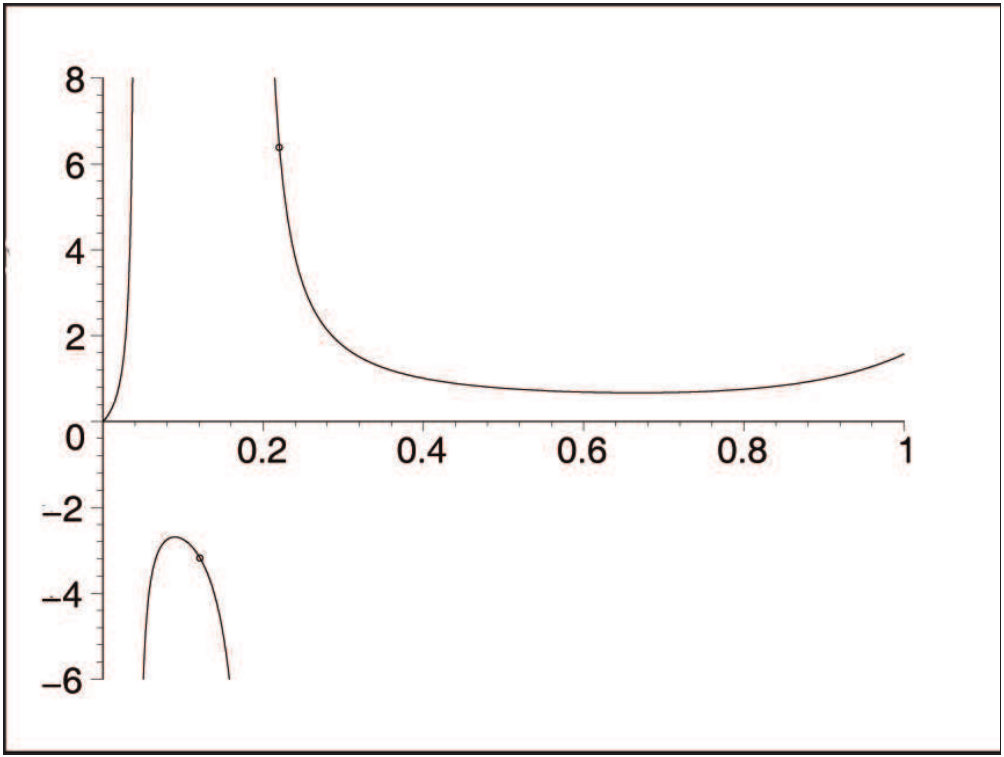}}
\put(7,56){{\footnotesize $S^{(10)}(\eps = 2/5, \chi)$}}
\put(73,29){{\footnotesize $\chi$}}
\end{picture}

}

\vv

\begin{tabular}{lp{12cm}}
{\bf FIG. 8.3} &Dependence of the sensitivity function
$S^{(N)}(\eps,\chi)$, as defined by Eq. (\ref{eqnova8.14}), on the
displacement variable $\chi$ for $\eps = 2/5$ and $N=10$. The point
with negative ordinate shown on the curve corresponds to the
equilibrium $\lb = 1$ and the point with positive ordinate to the
equilibrium $\lb = -1/5$.
\end{tabular}

\end{center}

\vv

\begin{center}
\begin{tabular}{|c||c|c|c|}\hline
$\lb$ & 1 & 0 & $-1/5$ \\ \hline $S^{(10)}(\eps = 2/5, \chi)$ &
$-3.18$ & $\pm\infty$ & 6.39 \\ \hline
\end{tabular}

\vv

\begin{tabular}{lp{12cm}}
{\bf Table 8.6} & Values of the sensitivity function
$S^{(N)}(\eps,\chi)$ according to the order $N=10$ of approximation
to the flux function for the three reference equilibria.
\end{tabular}
\end{center}

\vv\vv

\begin{center}
{\bf IX. DESCRIPTION OF THE EQUILIBRIUM CONFIGURATIONS OF REFERENCE
BY USE OF THE FLUX FUNCTION IN THE APPROXIMATION OF ORDER $N = 10$}
\end{center}

\setcounter{section}{9} \setcounter{equation}{0}

\vv

In this Section we shall use the approximate solution to the
Grad-Shafranov boundary value problem that comprises a superposition
of $N = 10$ multipole solutions to derive expressions for the fluid
pressure and for the magnetic fields and current densities, poloidal
and toroidal, in the equilibrium configurations we have been
referring to throughout for illustrative purposes, namely, those for
which the inverse aspect ratio is $\eps = 2/5$ and the equilibrium
parameter $\lb$ \ is equal to \ 1, 0 and $-1/5$ respectively; the
knowledge of the spatial distributions of those quantities of local
definition will then be used to determine the numerical values of
some parameters of global definition as poloidal beta and beta. Not
every quantity or parameter, local and global, that has a share in
the composition of a complete picture of the equilibrium or that is
currently employed in stability analysis has been evaluated as the
objective here merely amounts to illustrating the recourses that are
made available by the analytical representation of the flux
function.

The flux functions for $\lb = 1$, $\lb = 0$ and $\lb = -1/5$, with
the value of $\eps$ fixed at $2/5$, are given in the approximations
of order $N = 10$ in Appendix D. They are presented there in the
forms they take in the cylindrical coordinate system, since they
look simpler in this than in the toroidal-polar system. The
transformation from the former to the latter system can be achieved
with the help of the formulae also given in Appendix D. The use of
ordinary fractions for the coefficients of the powers of the axial
variable $Z$ was dictated by the intent of making to look less
cumbersome to the eye the representation of numbers that, in decimal
notation, should require a decimal point and ten places to have
fully expressed the degree of accuracy with which they were
generated.

\vv

\nd (1) \textit{Flux maps}

They are given in Fig. 9.1.

\vv

\begin{figure}

\begin{center}

\begin{tabular}{ll}
(a) $\lb = 1$ & (b) $\lb = 0$ \\
& \\
\includegraphics[width=7.08cm]{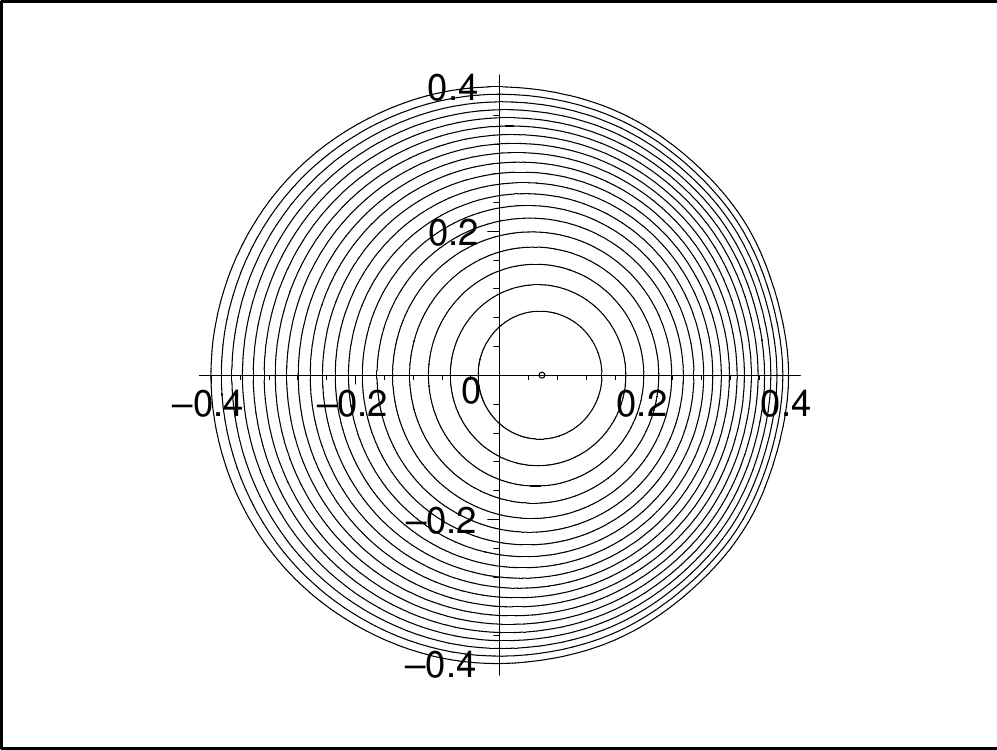}
& \includegraphics[width=7.08cm]{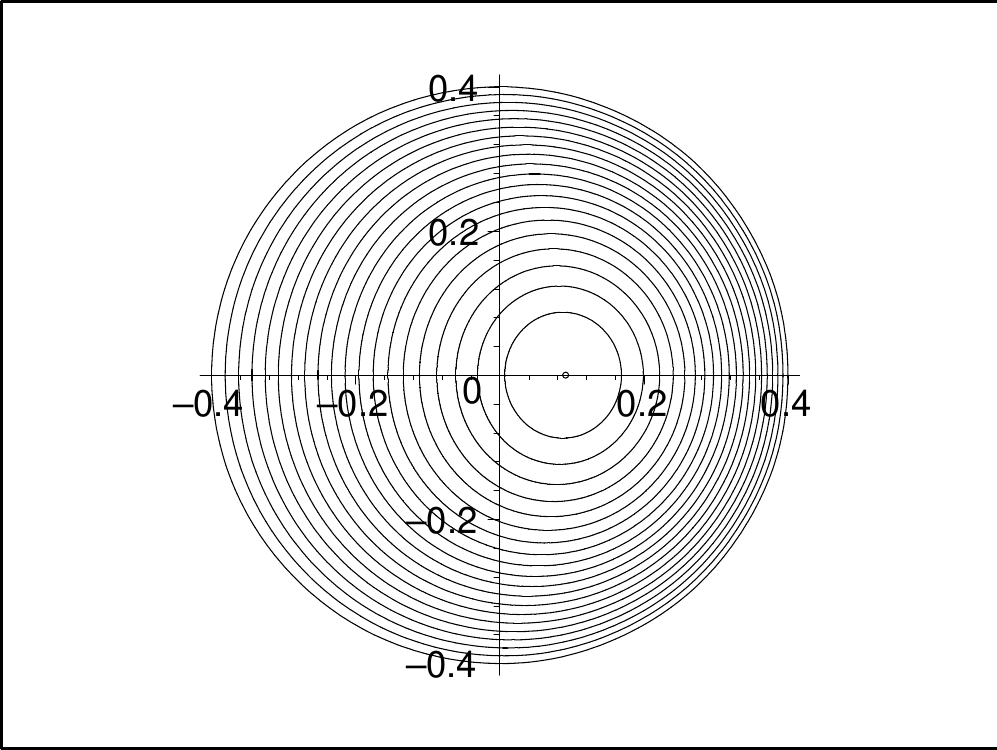} \\
& \\
& \\
\multicolumn{2}{c}{(c) $\lb=-1/5$} \\
& \\
\multicolumn{2}{c}{\includegraphics[width=7.08cm]{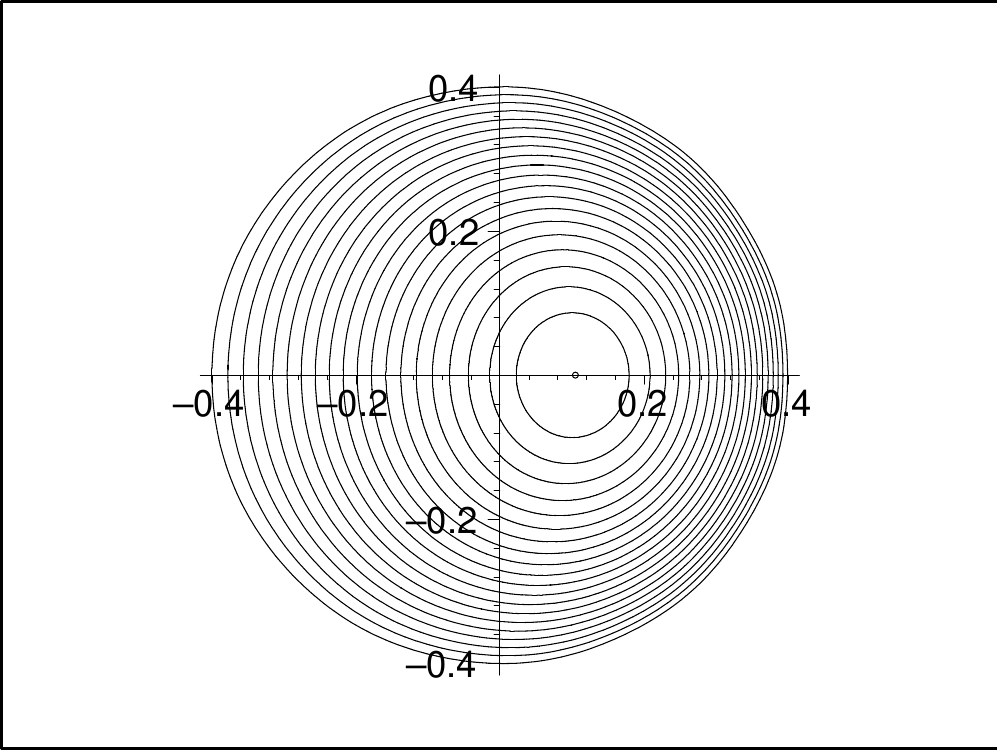}}
\end{tabular}

\vv

\begin{tabular}{lp{12cm}}
{\bf FIG. 9.1} &Flux surfaces as given by the $N=10$ order of
approximation to the flux function for the equilibrium
configurations characterized by having $\epsilon=2/5$ and (a)
$\lambda=1$; (b) $\lambda=0$; (c) $\lambda=-1/5$.
\end{tabular}

\end{center}

\end{figure}

\nd (2) \textit{The plasma pressure}

The plasma pressure profile in flux space can be obtained by
integrating Eq. (\ref{eq2.7}) under the assumption that $s_P$ is
constant, and since the plasma pressure must vanish at the plasma
boundary, where $\psi = 0$, the integration constant must be chosen
to be null, with the result for $\hat p$ that it is a quantity
proportional to $\psi$. From the normalizations for the flux
function introduced by Eq. (\ref{eq4.2nova}) and Eq. (\ref{eq6.4}),
we find that the connection between $\psi$ and $\widehat{\ovl\psi}$
is given by: \beq \psi(\rho,Z) =
s_P(-\ovl\psi_M)\widehat{\ovl\psi}(\rho, Z)\ , \label{eq10.1} \eeq
and we obtain a relation for the plasma pressure normalized by the
magnetic pressure at the magnetic axis, $\hat p$, as: \beq
\frac{\hat p}{s_P^2} = 2(-\ovl\psi_M)(-\widehat{\ovl\psi}(\rho,Z))\
. \label{eq10.2} \eeq

The ratio of the normalized plasma pressure $\hat p$ at any point
inside the cavity located by the coordinates $\rho$ and $Z$ to the
normalized plasma pressure at the torus centre line, $\hat p_C$, is:
\beq \frac{\hat p}{\hat p_C} = \frac{\widehat{\ovl\psi}(\rho,
Z)}{\widehat{\ovl\psi}_C}\ , \label{eqnova10.3} \eeq where \beq
\widehat{\ovl\psi}_C \equiv \widehat{\ovl\psi}(\rho=1, Z=0)
\label{eqnew94} \eeq is the normalized flux function at the torus
centre line.

Figure 9.2 shows the curves that represent the radial dependences of
the quantity defined by the right hand side of Eq.
(\ref{eqnova10.3}) on the equatorial plane of the torus for the
three reference configurations of equilibrium respectively, as
obtained by using, for $\widehat{\ovl\psi}_C$ the values given in
Table D.1 in Appendix D, and for $\widehat{\ovl\psi}(\rho,Z)$ the
expressions for the flux function to which those given in that same
Appendix are reduced by putting $Z$ equal to zero.

\begin{figure}
\begin{center}

{\unitlength=1mm

\begin{picture}(80,65)
\put(0,0){\includegraphics[width=8cm]{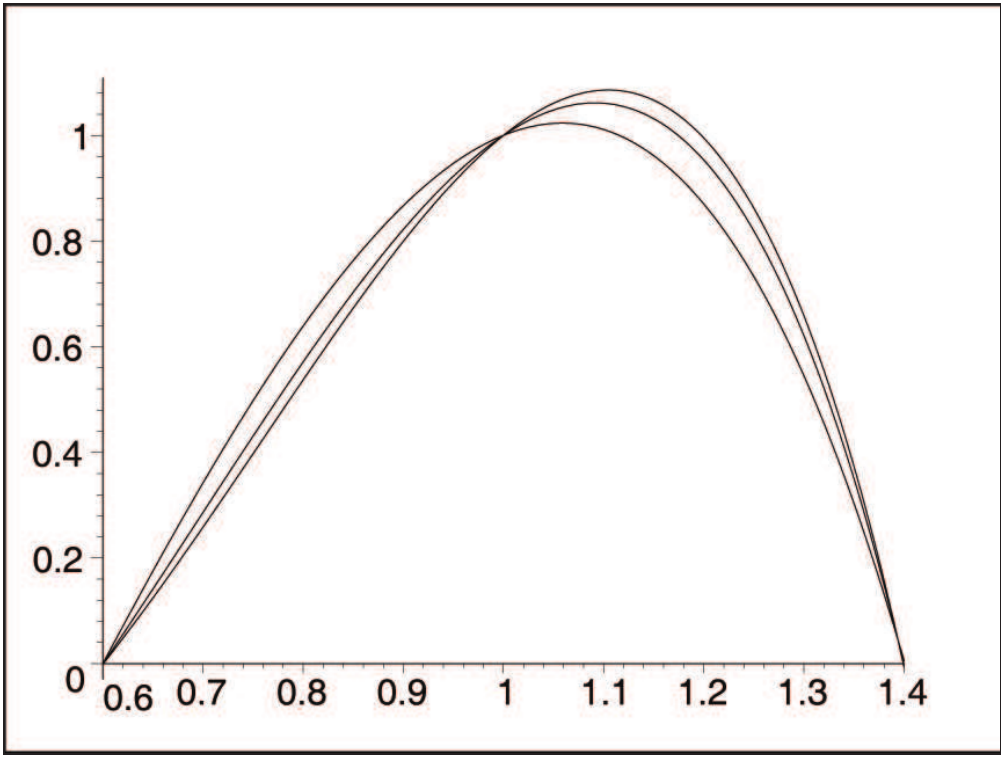}}
\put(10,51){{\footnotesize $\dps\frac{\widehat p}{\widehat
p_C}\Bigl|_{z=0}$}} \put(40,2){{\footnotesize $\rho$}}
\end{picture}

}

\vv

\begin{tabular}{lp{12cm}}
{\bf FIG. 9.2} &The ratios of the plasma pressures along the
intersection of the torus midplane with a meridian plane to the
values they take at the torus centre line, for the three reference
equilibria. In order of increasing peak values are seen the graphs
for $\lb = 1$, $\lb = 0$ and $\lb = -1/5$.
\end{tabular}

\end{center}

\end{figure}

\vv

\nd (3) \textit{The toroidal magnetic field}

The toroidal magnetic field can be obtained by integrating Eq.
(\ref{eq2.8}) in the variable $\psi$, considering that $s_I$ is
constant by assumption and recalling that the normalized toroidal
field function $\hat I(\psi)$ is defined by Eq. (\ref{eq2.5})
together with Eq. (\ref{eq1.2}). Replacing the flux $\psi$ by its
normalized version according to Eq. (\ref{eq10.1}), the integration
constant can be determined from the condition that at the centre of
the torus cross section, located by $R = R_C$, $Z = 0$, where
$\widehat{\ovl\psi} = \widehat{\ovl\psi}_C$, the toroidal field
$B_\phi$ equals $B_{\phi_C}$. We are in this way conducted to the
following expression for the toroidal field: \beq
\frac{B_\phi}{B_{\phi_C}} = \frac{1}{\rho} \sqrt{1 \mp
L^2(\widehat{\ovl\psi}(\rho, Z) - \widehat{\ovl\psi}_C)}\ ,
\label{eq10.4} \eeq where we have introduced the parameter: \beq L =
s_P\sqrt{2|\lb|(-\ovl\psi_M)}\ , \label{eq10.5} \eeq and where the
sign inside the symbol of square root depends on the sign of $\lb$
in accordance with a scheme of discrimination stated in the
discussion that follows. For $\lb = 0$, the parameter $L$ vanishes;
the toroidal field decays from the inner to the outer edge of the
chamber in conformity with the $1/R$ law, undisturbed by the plasma
interposed between them, the toroidal field function is constant,
and the poloidal current density is zero (see Eq. (\ref{eq10.13})
farther on) also as if there were no material medium filling the
cavity. For positive values of $\lb$, the sign to be taken in Eq.
(\ref{eq10.4}) is the negative one; the expression under the symbol
of square root then reaches its minimal value at the plasma
boundary, where $\widehat{\ovl\psi} = 0$, and its maximum value at
the magnetic axis, where $\widehat{\ovl\psi} = -1$. This means an
enhancement of the toroidal field intensity at all points in the
plasma bulk with respect to that which would exist within a
magnetically inert configuration generated by a toroidal field
function of equal strength at the plasma bounding surface, and the
plasma as a whole thus behaves as a paramagnetic body: the poloidal
current that is now induced flows in such a sense that the magnetic
field it originates adds constructively to the applied toroidal
field. Note that, as the parameter $L$ is increased, the toroidal
field at the plasma boundary decreases and vanishes for $L =
L_{\max,1}$, where \beq L_{\max,1} \equiv
\frac{1}{\sqrt{|\widehat{\ovl\psi}_C|}}\ . \label{eqnova10.5} \eeq
For example, for $\lb=1$, using for $\widehat{\psi}_C$ the numerical
value given in Table D.1 in Appendix D, we obtain that the maximal
allowed value for $L$ is $L_{\max,1} = 1.0115$, to which, by Eq.
(\ref{eq10.5}) and the numerical value of $\ovl\psi_M$ given in that
same Table, there corresponds a maximal allowed value for the
pressure gradient parameter equal to $s_{P_{\max,1}} = 2.5195$.  For
negative values of $\lb$, the sign to be taken in Eq. (\ref{eq10.4})
is the positive one, and the equilibrium configuration reproduces
the characteristics of a diamagnetic body, one that tends to cancel
the applied magnetic field in its interior. In this case the
expression under the sign of square root reaches its minimal value
at the magnetic axis as $\rho$ runs between its two edge values, and
remains positive at this and all locations inside the plasma as far
as the value of the parameter $L$ is smaller than $L_{\max,2}$,
defined to be: \beq L_{\max,2} \equiv \frac{1}{\sqrt{1 -
|\widehat{\ovl\psi}_C|}}\ . \label{eqnova10.6} \eeq For example, for
$\lb = -1/5$, using the data for the fluxes displayed in Table D.1
in Appendix D, we obtain that $L_{\max, 2} = 3.5582$, and thus the
value of the pressure parameter cannot exceed $s_{P_{\max,2}} =
30.4134$.

Figures 9.3(a) and 9.3(b) illustrate the behaviour of the profiles
of the toroidal field intensities along the intersection of a
meridian plane with the equatorial plane of the torus for a
paramagnetic ($\lb > 0$) and for a diamagnetic ($\lb < 0$)
configurations respectively, together with those for the
magnetically inert configurations ($\lb = 0$) having the same
respective field strengths at the plasma-wall interface.

\vv

\nd (4) \textit{The poloidal field}

The total magnetic field in the plasma can be written as: \beq \vec
B = \vec B_\phi + \vec B_p\ , \label{eq10.5lin} \eeq where $\vec
B_p$ \ is the poloidal field and $\vec B_\phi = B_\phi\vec e_\phi$ \
is the toroidal field. From the fact that the divergence of the unit
vector pointing in the azimuthal direction $\vec e_\phi$ is null,
and from the constancy of the toroidal field strength $B_\phi$ with
the azimuthal angle, we can establish that the divergence of the
toroidal field is zero. We thus arrive at that which can be called
the fundamental property of the fields in axisymmetric
configurations: the divergences of the toroidal and poloidal fields
vanish separately. This means that Eqs. (1.4) and (1.5) in Ref.
\cite{quatro} for the radial and the axial components of a multipole
field still hold for the poloidal field generated by a
current-carrying plasma subjected to a toroidal field. Using vector
notation we may thus express the poloidal field of the equilibrium
toroidal pinch in generality as:

\newpage

{\unitlength=1mm

\nd (a) $\lb = 1$

\vspace{-7mm}

\begin{center}
\hspace{25mm}\begin{picture}(80,60)
\put(0,0){\includegraphics[width=8cm]{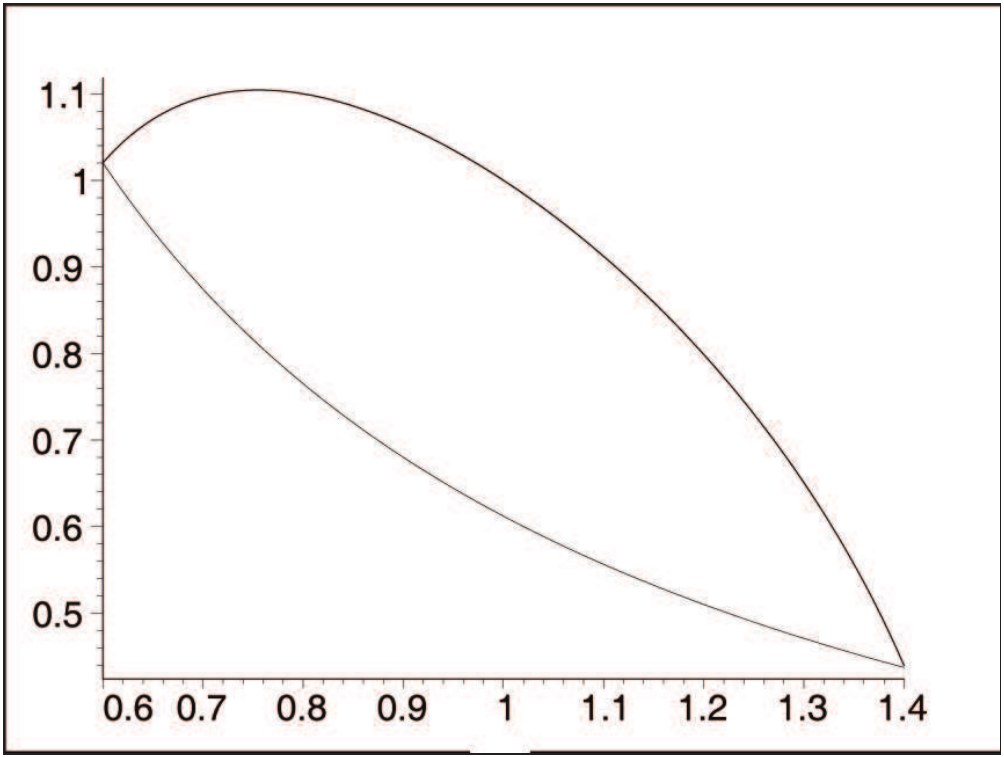}}
\put(-15,55){{\footnotesize
$\dps\frac{B_\phi}{B_{\phi_C}}\Bigl|_{z=0}$}}
\put(75,9){{\footnotesize $\rho$}}
\end{picture}
\end{center}

\nd (b) $\lb = -1/5$

\vspace{-9mm}

\begin{center}
\hspace{25mm}\begin{picture}(80,60)
\put(0,0){\includegraphics[width=8cm]{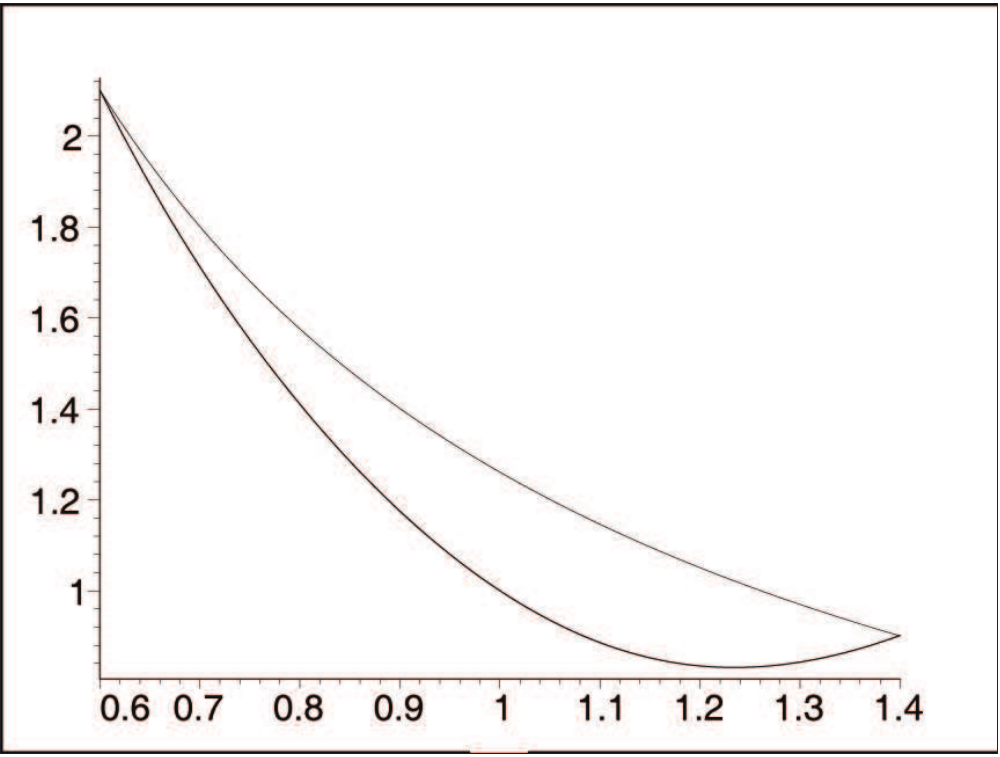}}
\put(-15,55){{\footnotesize
$\dps\frac{B_\phi}{B_{\phi_C}}\Bigl|_{z=0}$}}
\put(75,9){{\footnotesize $\rho$}}
\end{picture}
\end{center}

}

\begin{center}
\begin{tabular}{lp{12.4cm}}
{\bf FIG. 9.3} &Profiles of the normalized toroidal field
intensities along the intersection of the midplane with the torus
cross section for $L=4/5$, $L$ being the parameter defined by Eq.
(\ref{eq10.5}), and (a) $\lambda=1$; (b) $\lambda=-1/5$. The thin
lines represent the $1/R -$ decays of the toroidal field intensities
in case of constant toroidal field functions, for which $\lambda=0$.
\end{tabular}

\end{center}

\beq \vec B_p = \frac{1}{R}\vec e_\phi \times
\mbox{\boldmath${\nabla}$}\Psi \ . \label{eq10.6lin} \eeq

In terms of normalized quantities, the polar component of the
poloidal field in the toroidal-polar coordinate system with the pole
coinciding with the centre of the torus cross section is given by:
\beq (\vec B_p)_\ta =
s_PB_{\phi_C}(-\ovl\psi_M)b_\ta(x,\ta)\label{eq10.7} \eeq with \beq
b_\ta(x,\ta) =
\frac{1}{1+x\cos\ta}\frac{\ptl\widehat{\ovl\psi}}{\ptl x}\ .
\label{eq10.8} \eeq

In units of the applied vertical field (which is identical with the
polar component of the poloidal field at the centre line of the
torus), the polar component of the poloidal field at any point in
the plasma interior is expressed by: \beq \frac{(\vec B_p)_\ta}{B_v}
= \frac{b_\ta(x,\ta)}{b_\ta(x = 0, \ta = 0)}\ . \label{eq10x} \eeq
Figure 9.4 displays the dependences of the right-hand side of Eq.
(\ref{eq10x}) on the poloidal angle at the plasma boundary for the
three reference equilibria. Note that, although the Shafranov shift
is a positive quantity for all of them, for the paramagnetic
configuration ($\lb = 1$) the poloidal field has the intensity
increased as the inner edge of the torus is approached from the
outer edge by following the path described by a field line close to
the plasma boundary, while for the magnetically inert ($\lb = 0$)
and the diamagnetic ($\lb = -1/5$) configurations the tendency it
manifests to observation along the same path is the opposite one.

Since the flux function is constructed from the combination of a
number of multipole solutions that is necessarily finite, the
poloidal magnetic field it generates according to the formulae in
Eqs. (\ref{eq10.7}) and (\ref{eq10.8}) will not be rigorously
tangent to the boundary, but will admit a small radial component at
the points lying on it. The ratio of this radial component  to the
tangential component of the poloidal magnetic field at the plasma
toroidal surface can be calculated as: \beq \frac{(\vec
B_p)_r}{(\vec B_p)_\ta}\left|\raisebox{-10mm}{\rm\tiny Plasma \
surface} = -\frac{1}{x}\frac{{\dps
\frac{\ptl\widehat{\ovl\psi}(x,\ta)}{\ptl\ta}}}{{\dps
\frac{\ptl\widehat{\ovl\psi}(x,\ta)}{\ptl x}}}\right|_{x=\eps}\ .
\label{eq10.9} \eeq

\vv

\begin{center}

{\unitlength=1mm

\begin{picture}(80,65)
\put(0,0){\includegraphics[width=8cm]{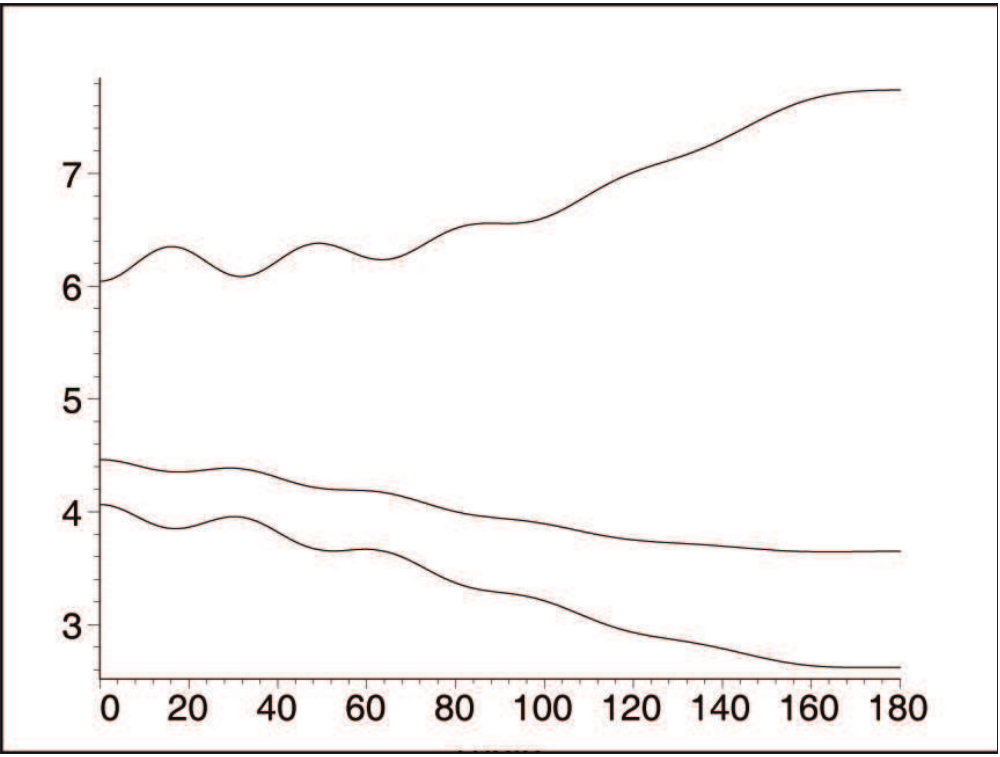}}
\put(-18,55){{\footnotesize
$\left.\dps\frac{(B_p)_\ta}{B_v}\right|_{x=\eps}$}}
\put(62,10.5){{\footnotesize $\ta$ (degrees)}}
\put(39,50){\footnotesize $\lb = 1$} \put(43,47){$\searrow$}
\put(25,27){\footnotesize $\lb = 0$} \put(29,23){$\searrow$}
\put(12,10){\footnotesize $\lb = -1/5$} \put(16,14){$\nearrow$}
\end{picture}

}

\vv

\begin{tabular}{lp{12cm}}
{\bf FIG. 9.4} &Magnitudes of the normalized poloidal field at the
plasma boundary as functions of the poloidal angle  for the three
reference equilibria. The upper curve corresponds to the
paramagnetic configuration ($\lambda=1$), the lowest one to the
diamagnetic configuration ($\lambda=-1/5$), and the curve between
both to the magnetically inert configuration ($\lambda=0$).
\end{tabular}

\end{center}

\vv

Figure 9.5 is the plot of the ratio on the left hand side of Eq.
(\ref{eq10.9}) as obtained by using the expression of the $N = 10$
order of approximation to the normalized flux function
$\widehat{\ovl\psi}(x,\ta)$, with $\eps = 2/5$ and $\lb$ taken to be
zero, on the right hand side. Since the poloidal field is a quantity
physically more relevant than the magnetic flux is, Eq.
(\ref{eq10.9}) embodies a measure of error which is perhaps more
suggestive of the merit of the partial flux function of order $N$ in
approximating the solution for a given equilibrium than are the
other measures we have introduced previously. Table 9.1 presents the
errors attached to the description of the three reference equilibria
obtained from the approximation of order $N = 10$ to the flux
function according to this view.

Because of the up-down symmetry of the configuration, the
longitudinal component $B_\rho$ of the poloidal field vanishes on
the equatorial plane of the torus. Figure 9.6 \ shows the variations
of the normalized axial components of the poloidal fields on the
equatorial plane with the normalized radial coordinate $\rho$ from
the inner to the outer edge of the torus for the three reference
equilibria. It is worth noting the remarkably linear pattern of
decaying followed by the magnitude of the poloidal field as the
magnetic axis is neared from any of the two extreme positions of the
equator line on a meridian section of the torus in the equilibrium
having $\lb = 1$.

\vv \vv

\begin{center}

{\unitlength=1mm

\begin{picture}(80,56)
\put(0,0){\includegraphics[width=8cm]{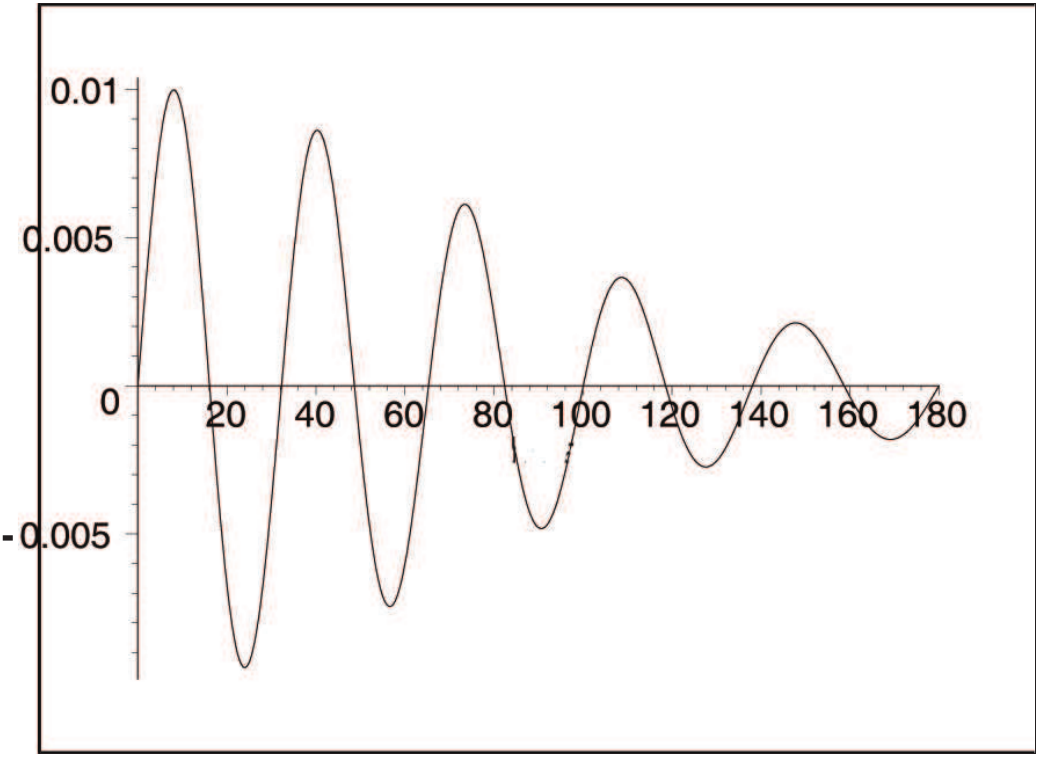}}
\put(-14,50){{\footnotesize
$\left.\dps\frac{(B_p)_r}{(B_p)_\ta}\right|_{x=\eps}$}}
\put(62,35){{\footnotesize $\ta$ (degrees)}}
\end{picture}

}

\vv

\begin{tabular}{lp{12cm}}
{\bf FIG. 9.5} &The ratio of the normal to the tangential component
of the poloidal field at the plasma boundary ($x=\epsilon$) for the
equilibrium defined by   $\epsilon=2/5$ and $\lambda=0$, according
to the approximation of order  $N=10$ to the flux function.
\end{tabular}

\end{center}

\vv\vv\vv

\begin{center}

{\tabcolsep=4mm
\begin{tabular}{|c||c|c|} \hline
$\lb$ & Error $\times 10^{2}$ & $\ta$ (degrees) \\ \hline\hline
1 & $-2.42$ & 7.88 \\
0 & $0.998$ & 8.06 \\
$-1/5$ & 2.34 & 8.13 \\
\hline
\end{tabular}
}

\vv\vv

\begin{tabular}{lp{12cm}}
\textbf{Table 9.1} & Values of the errors introduced by the
approximation of order $N = 10$ to the flux function for the
equilibria defined by $\lb = 1$, 0 and $-1/5$ respectively, as
measured by the maxima of the ratios of the normal to the polar
components of the poloidal field at the boundary $x = \eps = 2/5$,
and the angular positions where they occur.
\end{tabular}

\end{center}

\newpage

\begin{center}

{\unitlength=1mm

\begin{picture}(80,65)
\put(0,0){\includegraphics[width=8cm]{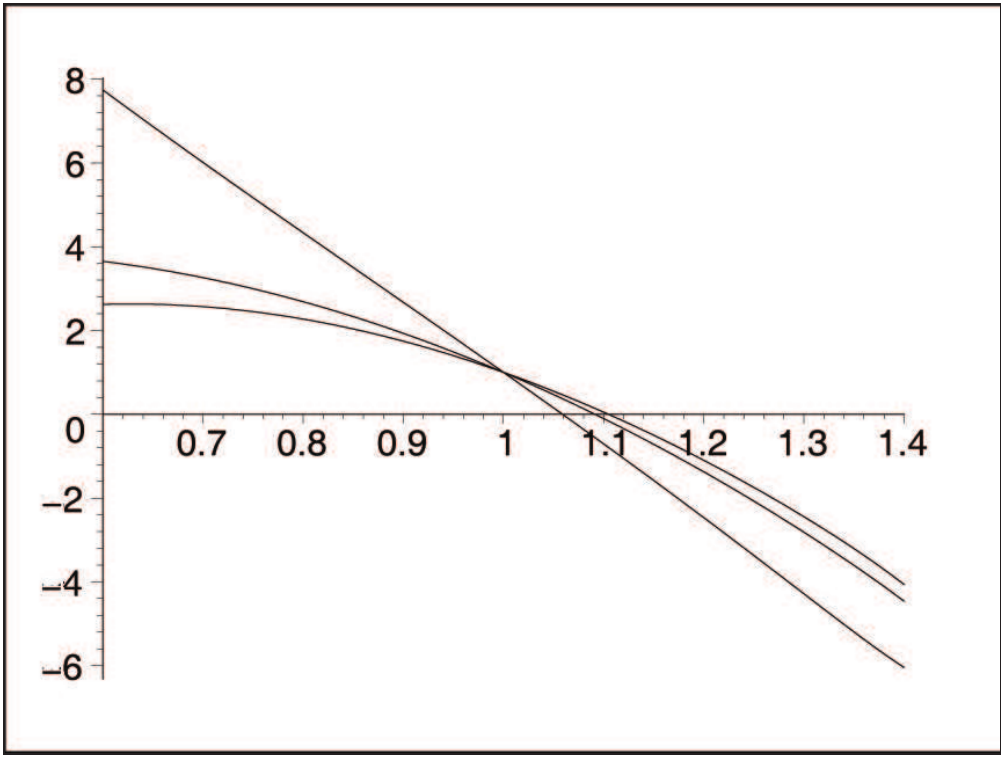}}
\put(-14,52){{\footnotesize
$\left.\dps\frac{B_z}{B_{z_C}}\right|_{z=0}$}}
\put(72,31){{\footnotesize $\rho$}} \put(14,51.5){\footnotesize $\lb
= 1$} \put(17.5,48.5){\footnotesize $\downarrow$}
\put(10,43){\footnotesize $\lb = 0$} \put(13.5,40.5){\footnotesize
$\downarrow$} \put(14,29){\footnotesize $\lb = -1/5$}
\put(17.5,32){\footnotesize $\uparrow$}
\end{picture}

}

\vv

\begin{tabular}{lp{12cm}}
{\bf FIG. 9.6} &Variations of the axial components of the normalized
poloidal fields with the normalized longitudinal coordinate $\rho$
along the intersection of the torus midplane with a meridian plane,
for the three reference equilibria. The normalization quantity
$B_{zC}$ is the value taken by the axial component of the poloidal
field $B_z$ at the torus centre line. In order of increasing values
of the abscissas of the magnetic axes are seen the curves for the
configurations having $\lb = 1$, $\lb = 0$ and $\lb = -1/5$
respectively.
\end{tabular}

\end{center}

\vv\vv

\nd (5) \textit{The poloidal current density}

To be applicable to the toroidal pinch, Eq. (1.2) in  Ref.
\cite{quatro} has to be replaced by Amp\`ere's law in its general
form for static fields, which is: \beq
\mbox{\boldmath${\nabla}$}\times\vec B = \mu_0\vec J\ ,
\label{eq10.10} \eeq where $\mu_0$ is the permeability of vacuum and
$\vec J$ is the total current density: \beq \vec J = \vec J_\phi +
\vec J_p \ , \label{eq10.11} \eeq $\vec J_\phi = J_\phi\vec e_\phi$
\ being the toroidal current density and $\vec J_p$ the poloidal
current density.

By writing down the longitudinal and the axial components of Eq.
(\ref{eq10.10}) in the cylindrical coordinate system $(R, \phi, z)$
(see Fig. 1 in Ref. \cite{quatro}), we see immediately that the
poloidal current density vector can be expressed as: \beq \mu_0\vec
J_p = -\mbox{\boldmath${\nabla}$}\phi \times
\mbox{\boldmath${\nabla}$}I \ , \label{eq10.12} \eeq where we have
used the fact that \mbox{\boldmath${\nabla}$}$\phi = \vec e_\phi/R$.
Since $I \equiv RB_\phi$ \ is a function of the flux function only,
we may rewrite Eq. (\ref{eq10.12}) as: \bey \mu_0\vec J_p &=&
-\frac{\dif I}{\dif
\Psi}\mbox{\boldmath${\nabla}$}\phi\times \mbox{\boldmath${\nabla}$}\Psi \nonumber \\
&=& -\frac{\dif I}{\dif \Psi}\vec B_p\ , \label{eq10.13} \eey the
second equality in Eq. (\ref{eq10.13}) following from the expression
for $\vec B_p$ stated in Eq. (\ref{eq10.6lin}). The poloidal current
density and the poloidal field are therefore parallel, the lines of
force of one vector field coinciding with those of the other field,
and of both with the level curves in the flux map. This is of course
a statement that applies to any and all axisymmetric equilibria
(except to the magnetically inert ones, which have constant toroidal
field function and carry no poloidal current). The profiles of the
longitudinal and axial components of the poloidal current density
and the profiles of the longitudinal and axial components of the
poloidal field along any direction in space, however, are not
respectively parallel in general, since the ratio between
corresponding components depends on the shape of the function
$I(\Psi)$. Along the intersection of a meridian plane with the
equatorial plane of the torus the longitudinal component of the
poloidal current density is zero and the axial component is
expressed in terms of normalized quantities as: \beq \mu_0J_{pz} =
s_P^2\frac{B_{\phi_C}}{R_C} \frac{\lb(-\ovl\psi_M)}{\sqrt{1 \mp
L^2(\widehat{\ovl\psi} - \widehat{\ovl\psi}_C)}}.
\frac{1}{\rho}\frac{\ptl\widehat{\ovl\psi}}{\ptl\rho}\Biggl|_{Z =
0}\ , \label{eq10.14} \eeq where the minus and the plus signs inside
the square root symbol, as in Eq. (\ref{eq10.4}), apply to the
paramagnetic and the diamagnetic configurations respectively. The
same as the $z$-component of the poloidal field, $J_{pz}$ vanishes
at the magnetic axis. The graphs in Fig. 9.7 show the variations of
the ratios $J_{pz}/J_{pzC}$, where $J_{pzC}$ is the axial component
of the poloidal current density at the torus centre line, along the
equator line for the two reference equilibria having non-null values
of $\lb$, assuming that the parameter $L$ takes the value $4/5$ for
both.

\begin{figure}

\begin{center}

{\unitlength=1mm

\begin{picture}(80,65)
\put(0,0){\includegraphics[width=8cm]{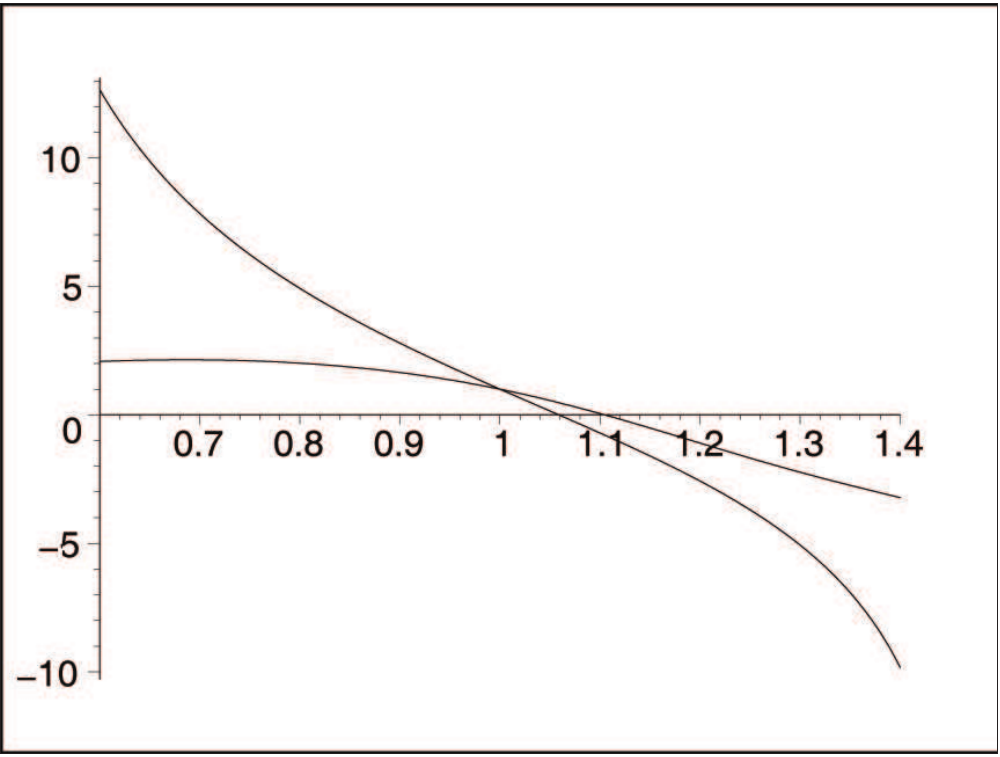}}
\put(-16,52){{\footnotesize
$\left.\dps\frac{J_{pz}}{J_{pz_C}}\right|_{z=0}$}}
\put(72,31){{\footnotesize $\rho$}} \put(14,47){\footnotesize $\lb =
1$} \put(17.5,44.5){\footnotesize $\downarrow$}
\put(9.5,36){\footnotesize $\lb = -\frac{1}{5}$}
\put(13,33.5){\footnotesize $\downarrow$}
\end{picture}
}

\vv

\begin{tabular}{lp{12cm}}
{\bf FIG. 9.7} &Dependences of the axial components  of the poloidal
current densities  normalized to the values they respectively take
at the torus centre line, $J_{pz}/J_{pzC}$, on the normalized
longitudinal coordinate $\rho$ along the intersection of the torus
midplane with a meridian plane, for the reference equilibria having
$\lb = 1$ (curve showing the smallest value of the abscissa for
which $J_{pz}$ is null) and $\lb = -1/5$ respectively, and a common
value of the parameter $L$ equal to $4/5$.
\end{tabular}

\end{center}

\end{figure}

Writing the toroidal field as: \beq \vec B_\phi =
I(\Psi)\mbox{\boldmath${\nabla}$}\phi \label{eq10.15} \eeq and then
taking the curl, by subsequent use of Eq. (\ref{eq10.13}) for the
poloidal current density, we can establish that: \beq
\mbox{\boldmath${\nabla}$} \times \vec B_\phi = \mu_0\vec J_p \ .
\label{eq10.16} \eeq

The poloidal current density thus acts as the source to the toroidal
field, as we indeed should expect intuitively from the rotational
invariance property of the toroidal  pinch in equilibrium. Equation
(\ref{eq10.16}) also implies that Amp\`ere's law for the toroidal
pinch breaks into two separate laws of similar structures in which
the toroidal and the poloidal symmetries belong to either one of the
two respective vector fields they put in relation and which
alternate between them the fields to which one and other of both
symmetries are ascribed. The partial version of Amp\`ere's law that
makes a pair with that expressed by Eq. (\ref{eq10.16}) states that
the toroidal current density is the sole source to the poloidal
field, and reads: \beq \mbox{\boldmath${\nabla}$}\times \vec B_p =
\mu_0\vec J_\phi\ . \label{eq10.17} \eeq

\vv

\nd (6) \textit{The toroidal current density}

The expression for the toroidal current density that is best suited
to the purpose of numerical evaluation is the one that follows
directly from the pressure balance condition: \beq
\mbox{\boldmath${\nabla}$}p = \vec J \times \vec B\ .
\label{eq10.18} \eeq

Considering that the pressure is a function of $\Psi$ and that the
poloidal current density is parallel to the poloidal field, Eq.
(\ref{eq10.18}) can be written as: \beq \frac{\dif
p}{\dif\Psi}\mbox{\boldmath${\nabla}$}\Psi = \vec J_\phi \times \vec
B_p + \vec J_p \times \vec B_\phi \ . \label{eq10.19} \eeq By taking
the cross product of the above relation with
$\mbox{\boldmath${\nabla}$}\phi$ on the left, and then developing
the two terms that constitute the right hand side of the ensuing
equation in accordance with the rule of the vector triple product,
we obtain: \beq \frac{\dif p}{\dif \Psi}\vec B_p =
-\frac{1}{R}J_\phi \vec B_p + \frac{1}{R}B_\phi \vec J_p\ ,
\label{eq10.20} \eeq where in the left hand side we have made use of
Eq. (\ref{eq10.6lin}) for the poloidal field. We next substitute
$\vec J_p$ on the right hand side by the expression for it given by
Eq. (\ref{eq10.13}). Cancelling the factor $\vec B_p$ that then
appears in all terms of the resulting equation, and isolating the
term of the toroidal current density, we have: \beq \mu_0J_\phi =
-\mu_0 R\frac{\dif p}{\dif\Psi} - \frac{I}{R}\frac{\dif
I}{\dif\Psi}\ . \label{eq10.21} \eeq Note that, if the sources to
the equilibrium are constant, as we are considering them to be,
$J_\phi$ is independent of the axial coordinate.

In terms of normalized quantities the above expression can be made
to become: \beq \frac{J_\phi}{J_{\phi_C}} = \frac{1}{1+\lb}
\left(\rho + \frac{\lb}{\rho}\right) \label{eq10.22} \eeq where \beq
J_{\phi_C} = (1+\lb)\frac{s_PB_{\phi_C}}{\mu_0R_C} \label{eq10.23}
\eeq is the toroidal current density at the torus centre line.

For $\lb = 0$ \ the toroidal current density grows linearly from the
inner to the outer side of the torus. For $\lb > 0$ \ it is positive
at all points in the plasma and passes through a minimum at the
coordinate: \beq \rho = \rho_m \equiv \sqrt\lb \label{eq10.24} \eeq
where it assumes the value: \beq (J_\phi)_{\min} =
\frac{2\sqrt\lb}{1 + \lb} J_{\phi_C}\ . \label{eq10.25} \eeq For
$\lb = 1$ \ as the value of the equilibrium parameter we have been
taking to consider a concrete example of a paramagnetic
configuration, the minimum value of the toroidal current density is
reached at the torus centre line. For $\lb < 0$ \ no local minimum
of $J_\phi$ can occur; there can be current reversion, however, at
the point of normalized longitudinal coordinate given by: \beq \rho
= \rho_R \equiv \sqrt{-\lb}\ , \label{eq10.26} \eeq provided, of
course, that this point falls inside the chamber for the equilibrium
considered. Figure 9.8 shows the profiles of the toroidal current
densities along the equatorial line on the torus cross section for
the three reference equilibria.

\vv

\begin{center}

{\unitlength=1mm

\begin{picture}(80,65)
\put(0,0){\includegraphics[width=8cm]{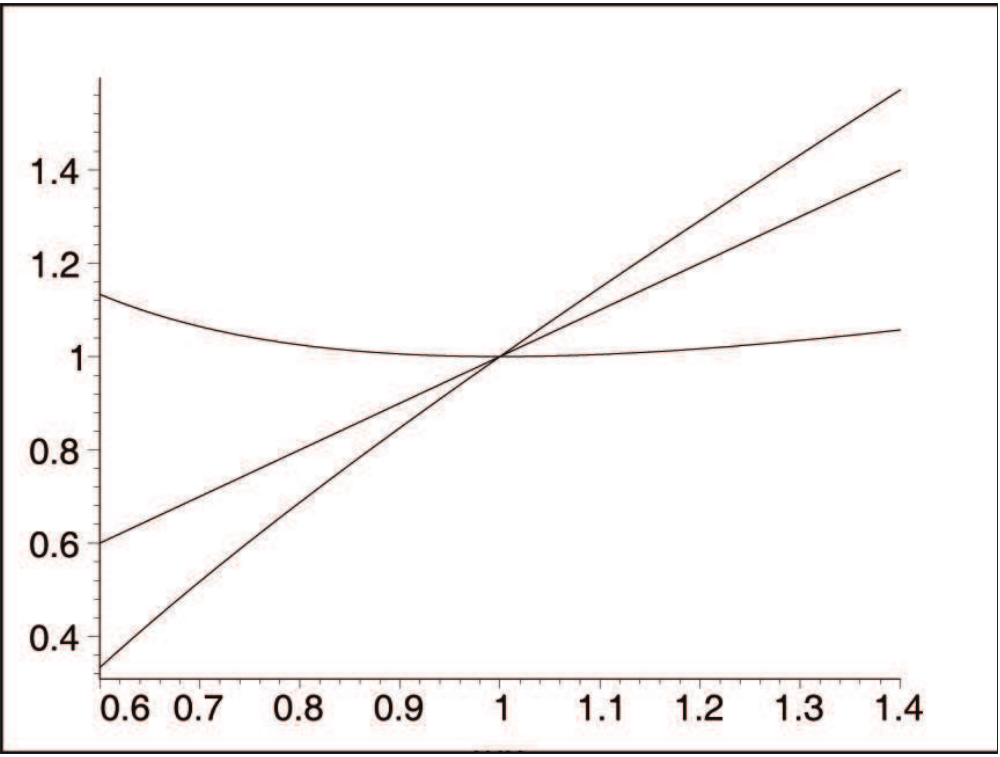}}
\put(12,52){{\footnotesize $\dps\frac{J_\phi}{J_{\phi_C}}$}}
\put(72,11){{\footnotesize $\rho$}} \put(25,38){\footnotesize $\lb =
1$} \put(28,35){$\downarrow$} \put(60,35){\footnotesize $\lb = 0$}
\put(61,38){$\nwarrow$} \put(26,15){\footnotesize $\lb = -1/5$}
\put(27,18){$\nwarrow$}
\end{picture}

}

\vv

\begin{tabular}{lp{12cm}}
{\bf FIG. 9.8} &Profiles of the magnitudes of the normalized
toroidal current densities, $J_\phi/J_{\phi_C}$,  on a meridian
plane of the torus, for the three reference equilibria. The
normalization quantity $J_{\phi_C}$ is the toroidal current density
on the cylindrical surface having the axis coincident with the axis
of rotational symmetry of the torus and radius equal to $R_C$. In
order of increasing values of the ordinates at the abscissa
associated with the location of the inner edge of the torus ($\rho =
0.6$) are seen the curves relative to the equilibria having $\lb =
-1/5$, $\lb = 0$ and $\lb = 1$ respectively.
\end{tabular}

\end{center}

\vv

Before we pass to the consideration of examples of calculation of
macroscopic parameters associated with a given state of equilibrium,
let us observe that, alternative to the formula stated in Eq.
(\ref{eq10.21}), a relation for the toroidal current density which
relies on operations on the flux function only can be established by
exploiting the information contained in Amp\`ere's law for the
poloidal field. Indeed, by taking the dot product of Eq.
(\ref{eq10.17}) with $\vec e_\phi$, we have: \bey \mu_0J_\phi &=&
R\mbox{\boldmath${\nabla}$}\phi.\mbox{\boldmath${\nabla}$} \times \vec B_p \nonumber \\
&=& R\mbox{\boldmath${\nabla}$}.(\vec B_p \times
\mbox{\boldmath${\nabla}$}\phi) \ , \label{eq10.27} \eey where, to
write the second equality, we have recalled a well known identity
for the divergence of the cross product of two vectors. With $\vec
B_p$ written as in Eq. (\ref{eq10.6lin}), application of the rule of
the triple vector product to the expression between parentheses in
Eq. (\ref{eq10.27}) conducts us at once to the result: \beq
\mu_0J_\phi =
R\mbox{\boldmath${\nabla}$}.\left(\frac{1}{R^2}\mbox{\boldmath${\nabla}$}\Psi\right)\
. \label{eq10.28} \eeq

The equation that comes up from the comparison of Eq.
(\ref{eq10.28}) with Eq. (\ref{eq10.21}) is the Grad-Shafranov's.

\vv

\nd (7) \textit{Poloidal beta}

From a physical standpoint poloidal beta is probably the most useful
of the macroscopic parameters that are costumarily associated with
an equilibrium configuration, since there is an univocal
correspondence between its numerical value and that of the
equilibrium parameter $\lb$, while embodying a physical connotation
of more prompt appeal than the latter. Loosely speaking, it is a
measure of the ratio of the plasma thermal energy to the magnetic
energy content of the poloidal field. A convenient mathematical
definition is the one that adopts for itself the form of a term that
arises naturally in the derivation of the average integral version
of Grad-Shafranov's equation. Following Ref. \cite{doze} we then
define the poloidal beta as: \beq \bt_p \equiv \frac{{\dps
\frac{1}{V_S}\int_{V_S}p\dif V}}{\lge B_p^2\rge/2\mu_0}\ ,
\label{eq10.29} \eeq where the integral of the pressure is taken
over the plasma volume, $V_S$ is the plasma volume, which for a
torus of major radius $R_C$ and inverse aspect ratio $\eps$ is given
by: \beq V_S = 2\pi^2\,R_C^3\,\,\eps^2\ , \label{eq10.30} \eeq and
$\lge B_p^2\rge$ means the average of the squared poloidal field
magnitude over the plasma contour in the following sense: \beq \lge
B_p^2\,\rge \equiv \frac{{\dps \oint B_p\dif l}}{{\dps \oint
\frac{\dif l}{B_p}}}\ , \label{eq10.31} \eeq the integrations being
performed along the poloidal contour of the plasma surface (which in
our case coincides with the circular trace of the inner wall on the
cross section of the containing chamber).

In terms of the normalized quantities we have been using throughout
the expression for poloidal beta introduced by Eq. (\ref{eq10.29})
translates as: \beq \bt_p = \frac{4h}{\eps^2(-\ovl\psi_M) \lge
b_\ta^2(\eps)\rge} \ , \label{eq10.32} \eeq where \beq h =
-\frac{1}{2\pi}\int_0^\eps \dif x\int_0^{2\pi}\dif\ta
\,\,\widehat{\ovl\psi}(x,\ta)\,\,x\,\,(1 + x\cos\ta) \label{eq10.33}
\eeq and \beq \lge\, b_\ta^2(\eps)\,\rge = \frac{{\dps
\int_0^{2\pi}b_\ta(x=\eps, \ta)\dif\ta}}{{\dps
\int_0^{2\pi}\frac{\dif\ta}{b_\ta(x=\eps, \ta)}}}\ , \label{eq10.34}
\eeq

\vspace{2mm}

\nd $b_\ta(x,\ta)$ \ being given by Eq. (\ref{eq10.8}). The
integrals appearing in the definitions of $h$ and $\lge
\,b_\ta^2(\eps)\,\rge$ are easily computed with the help of Maple V.
Table 9.2 displays the results obtained for poloidal beta for our
three reference equilibria. It should be noted that, based on the
balancing the terms in the expression of the integral average of
Grad-Shafranov's equation derived in Ref. \cite{doze} must keep
among themselves, the figures for $\bt_p$ that we find in Table 9.2
for the three kinds of magnetically distinct configurations are all
of the orders of magnitude we expect them to be, to know, $\bt_p <
1$ for $\lb > 0$, $\bt_p \sim 1$ for $\lb = 0$ and $\bt_p > 1$ for
$\lb < 0$.

\vv

\begin{center}

{\tabcolsep=4mm
\begin{tabular}{|c||c|} \hline
$\lb$ & $\bt_p$ \\ \hline\hline  1 & 0.487 \\ 0 & 1.035 \\
 $-1/5$ & 1.358 \\ \hline
\end{tabular}
}

\vv

\begin{tabular}{lp{12cm}}
\textbf{Table 9.2} & Values of poloidal beta ($\bt_p$) corresponding
to the three configurations of reference as defined by the values of
the equilibrium parameter $\lb$.
\end{tabular}

\end{center}

\vv

\nd (8) \textit{Beta}

The global parameter known as beta is probably the stick of
commonest use in assessing the merit of a thermonuclear plasma
confining device as a prospective reactor, and in ruling the
discussions on engineering and economic aspects of fusion research.
In general terms it is a measure of the proportion of the plasma
thermal energy to the total magnetic energy spent in confining the
plasma. Since it is not put forward by any mathematical statement
meaning a balance between physical quantities, as the poloidal beta
is, its definition remains somewhat arbitrary. In this paper we
shall adopt the same as that of Refs. \cite{um} and \cite{seis},
which is: \beq \bt \equiv \frac{{\dps \frac{1}{V_S}\int_{V_S}
p\,\dif V}}{B_{\phi_C}^2/2\mu_0}\ , \label{eq10.35} \eeq
where$B_{\phi_C}$ is the toroidal magnetic field at the centre of
the plasma cross section. In terms of normalized quantities Eq.
(\ref{eq10.35}) writes as: \beq \bt =
\frac{4s_P^2(-\ovl\psi_M)h}{\eps^2}\ , \label{eq10.36} \eeq where \
$h$ is defined as in Eq. (\ref{eq10.33}). Note that the value of
beta, distinctly from that of poloidal beta, is not uniquely defined
by the value of the equilibrium parameter $\lb$, and thus cannot be
linked in an unequivocal way to any particular equilibrium
configuration and to the geometrical characteristics typefying the
level curves in a given flux map. In other words, beta is
\textit{not} a state function and the value it assumes depends on
the way the plasma is driven from its early formative stages to the
final equilibrium state. Besides the implicit references it contains
to $\lb$ through the dependences that $\ovl\psi_M$ and $h$ keep on
that parameter, a formula for beta in general will call for the
presence of another equilibrium quantity (in addition to that of the
normalization constants, $R_C$ and $B_{\phi_C}$), which can be
chosen among a multitude of possibilities. In the formula stated as
Eq. (\ref{eq10.36}) this is $s_P$, but of course equivalent
expressions could be written as well in which, in place of $s_P$, we
should have $s_I$ or the pressure at the centre line $\hat p_C$ or
the toroidal current, etc. The convenient one to use in each case is
that which incorporates as second parameter the quantity that is
kept constant during the beta-raising process. Table 9.3 shows the
values of beta for the three equilibrium configurations of reference
as evaluated by means of Eq. (\ref{eq10.36}) assuming that they have
$s_P$ equal to unity.

\vv

\begin{center}

{\tabcolsep=4mm
\begin{tabular}{|c||c|} \hline
$\lb$ & $\bt \times 10^2$ \\ \hline\hline  1 & 8.08 \\ 0 & 4.12
\\ $-1/5$ & 3.33 \\ \hline
\end{tabular}
}

\vspace{4mm}

\begin{tabular}{lp{12cm}}
\textbf{Table 9.3} & Values of beta in unities of $s_P^2$ for the
reference equilibrium configurations.
\end{tabular}

\end{center}

\vv\vv

\begin{center}
{\bf X. SUMMARY}
\end{center}

\setcounter{section}{11} \setcounter{equation}{0}

\vv

The problem considered in this paper is an instance of that of the
plasma equilibrium in toroidal confinement systems as described by
the Grad-Shafranov equation, the constraint under which it is to be
solved being that the outermost of the flux surfaces in the
\textit{continuum} that structures the magnetic configuration
represented by its solution must conform to the toroidal shape of
the surface of the plasma bulk, with which that surface identifies
itself. In terms of normalized quantities, as they are defined in
Eqs. (\ref{eq2.2}) -- (\ref{eq2.5}), the Grad-Shafranov equation
writes in the toroidal-polar coordinate system as in Eq.
(\ref{eq2.6}), and the boundary condition on its solution as in Eq.
(\ref{eqnova49}). The sources to the equilibrium -- the pressure
gradient and the gradient of the squared toroidal field function --
are formally defined by Eqs. (\ref{eq2.7}) and (\ref{eq2.8}). Under
the assumption that these two quantities are constant in flux space,
by introducing a flux function normalized by the pressure gradient
parameter as in Eq. (\ref{eq4.2nova}), the problem can be made to
admit of a single datum of input, the equilibrium parameter, defined
in Eq. (\ref{eq3.30}).

For the application of the method that is the theme of the present
paper, the solution to the Grad-Shafranov equation is written as the
sum of the particular solution and a linear combination of $N$
multipole solutions of the orders $n = 0, 1, 2, \ldots, N-1$ in the
polar coordinate system having the pole located at the intersection
of the centre line of the torus with a meridian plane, the azimuthal
coordinate being ignorable (Eq. (\ref{eqnova41})). The particular
solution is derived in Section III by the method of expansion of the
flux function in series of Chebyshev polynomials, and is given under
normalization by the pressure gradient constant in Eq.
(\ref{eq3.29nova}) (in toroidal-polar coordinates) and in Eq.
(\ref{eq3.31}) (in cylindrical coordinates). Imposition that the
flux be null at the boundary generates a system of $2N-1$ linear
equations for the $N$ constants $K_0^{(N)}, K_1^{(N)}, \ldots,
K_{N-1}^{(N)}$ multiplying the multipole solutions that enter the
combination designated to represent the flux function. This is an
overdetermined system, and, to extract a meaningful solution from
it, the strict number of $N$ equations is retained, the $N-1$
abandoned ones being chosen as those that, in remaining unsatisfied,
the error effect they will have on the value taken by the assumed
flux function at the frontier of the domain of integration of the
partial differential equation will be the least. The solution of the
truncated system of $N$ equations for the $N$ constants $K_j^{(N)}$
is given by Eqs. (\ref{eqnova416}) and (\ref{eqnova415}) in
conjunction with Eq. (\ref{eqnova414}) in terms of the equilibrium
parameter $\lb$. It may be found advantageous to replace $\lb$ by
the relative Shafranov shift $\dt$, this being defined by Eq.
(\ref{eqnova418}), as the input datum to the problem. In this case
the constants $K_j^{(N)}$ pass to be given by Eqs. (\ref{eqnova436})
and (\ref{eqnova434}) in conjunction with Eq. (\ref{eqnova431}). The
connection between $\lb$ and $\dt$ is given by Eq. (\ref{eqnova432})
together with Eq. (\ref{eqnova433}).

A significant simplification to the mathematical expressions
involved in the development of the solution can be gained by further
substituting $\dt$ by the displacement variable $\chi$ as the input
datum to the problem, this latter quantity being related to the
former through Eq. (\ref{eqnova438}). The value of $\chi$ can be
evaluated from the value of $\lb$ (or vice versa) by means of the
equilibrium equation (Eq. (\ref{eqnova444})), which equates the
equilibrium function $F^{(N)}(\eps, \chi)$ to $\lb$. The equilibrium
function itself is calculated from the formula given in Eq.
(\ref{eqnova433}) in conjunction with that of Eq. (\ref{eqnova431}),
replacing quantities defined in terms of $\dt$ by quantities defined
in terms of $\chi$ according to the rules of transformation
expounded in the paragraph containing Eq. (\ref{eqnova442}).
Similarly, the constants $K_j^{(N)}$ are still given by the formulae
that apply when the input parameter is $\dt$, provided that the
quantities appearing in them be subjected to this same scheme of
substitution. Use of Eq. (\ref{eqnova444}) in Eq. (\ref{eq3.29nova})
permits us also to express the particular solution in terms of
$\chi$.

The allowed interval of variation of $\chi$ is that specified by Eq.
(\ref{eq622}). The lower bound of it finds a reasonably good
estimate in Eq. (\ref{eq624}), provided that the torus is not too
thick. There seems to be no physical constraint to set an upper
bound to $\chi$, so that the extreme value it can take is that by
which the magnetic axis is placed at the outer edge of the torus. In
practice, however, as a feature of the method, branching points
erupt on the boundary surface as $\chi$ reaches a critical value
$\chi_c^{(N)}$, and this fixes the maximal value that can be
attributed to $\chi$ in describing an equilibrium configuration
within the framework of the order $N$ of approximation to the flux
function. It is found that the value of $\chi_c^{(N)}$, which,
together with those of the coordinates of the branching points, is
obtained by simultaneously solving Eq. (\ref{eq670}), Eq.
(\ref{eq671}) and Eq. (\ref{eq672}) (with the superscript 3 replaced
by $N$), increases with the increase of $N$, apparently without
tending to a limiting value. A last normalization to the poloidal
flux function (by the absolute value of the flux at the magnetic
axis) is introduced by Eq. (\ref{eq6.4}). Section V also introduces
the quantities that will be used in the subsequent ones to evaluate
the accuracy of a solution obtained through the use of the method,
to know, the relative error (defined by Eq. (\ref{eq6.5})) and the
relative deviation (defined by Eq. (\ref{eq6.6})).

Three configurations, for which the common inverse aspect ratio is
$2/5$ and the equilibrium parameters are 1, 0 and $-1/5$
respectively, are elected to serve the purpose of testing the use of
the method of series of multipole solutions. Solutions to the
boundary value problem for these reference configurations are
derived in Section VI by using the $N = 3$ order of approximation,
and again in Section VII by using the $N = 4$ order of approximation
to the flux function. Also obtained are the measures of accuracy of
the solutions and the allowed intervals of variation of the
displacement variables in each of these two orders of approximation.

In Section VIII the findings of the two previous Sections are
generalized to characterize the structures of the equilibrium
function and of the poloidal flux function that stem from a
representation of the solution to the boundary value problem
comprising any finite number $N$ of multipole solutions. The general
forms of $F^{(N)}(\eps, \chi)$ for $N$ odd and $N$ even are given by
Eqs. (\ref{eqnova8.1}) and (\ref{eqnova8.4}) (with Eqs.
(\ref{eqnova8.2}) and (\ref{eqnova8.3})), and that of
$\ovl\psi^{(N)}(x, \ta; \eps, \chi)$ by Eq. (\ref{eq9.2}) (with Eqs.
(\ref{eq9.3}) -- (\ref{eq9.10})). A systematic application of the
method of solution by series of multipole solutions to the
equilibria of reference with the order of approximation to the flux
function ranging from $N = 3$ to $N = 10$ allows us to infer that
the series is convergent.

Finally in Section IX the $N = 10$ order of approximation to the
poloidal flux function is employed to derive expressions for a
number of quantities of interest in characterizing an equilibrium
configuration (the plasma pressure, the toroidal and the poloidal
fields, the toroidal and the poloidal current densities, beta and
poloidal beta). These expressions are then applied to obtain a
graphical and a numerical description of the equilibria chosen to
illustrate the use of the method as it was presented here.

\vv\vv

\begin{center}
{\bf APPENDIX A: EXPRESSIONS OF THE FUNCTIONS $P_i(x)$ AND $Q_i(x)$
($i = 0, 1, 2, 3, 4$) CONSTITUENTS OF THE RADIAL DEPENDENT
COEFFICIENTS $S_i(x)$ IN THE REPRESENTATION OF THE NORMALIZED
PARTICULAR SOLUTION $\ovl\psi_p(x,\ta)$ AS A COMBINATION OF
HARMONICS OF THE POLOIDAL ANGLE $\ta$. EXPRESSIONS OF THE FUNCTIONS
$P'(\dt)$ AND $Q'(\dt)$ AND OF THE FUNCTIONS $P'(\chi)$ AND
$Q'(\chi)$ REQUIRED FOR THE EVALUATION OF THE EQUILIBRIUM FUNCTIONS,
$F^{(N)}(\dt)$ AND $F^{(N)}(\chi)$, AND OF THE CONSTANTS MULTIPLYING
THE MULTIPOLE SOLUTIONS, $K_j^{(N)}(\dt)$ AND $K_j^{(N)}(\chi)$,
$(j= 0, 1, 2, \ldots, N-1)$, THAT ENTER THE REPRESENTATIONS OF THE
PARTIAL FLUX OF ORDER $N$ IN TERMS OF THE RELATIVE SHAFRANOV SHIFT
$\dt$ AND IN TERMS OF THE DISPLACEMENT VARIABLE $\chi$,
RESPECTIVELY.}
\end{center}

\setcounter{equation}{0}
\def\theequation{A.\arabic{equation}}

\vv

For the normalized particular solution $\ovl\psi_p(x,\ta)$ written
as in Eq. (\ref{eq3.32nova}), and its $x$-dependent average value
$S_0(x)$ and coefficient functions $S_i(x)$ of the harmonics of the
poloidal angle, $\cos i\ta$ ($i = 1, 2, 3, 4$), expressed under the
form of a two-term combination linear in $\lb$, as in Eq.
(\ref{eq3.33nova}), the radial functions that constitute the terms
in each of the combinations are respectively given by:

\begin{picture}(120,10)
\put(113,-10){$\Biggl\{$}
\end{picture}
\vspace{-12mm}
\bey P_0(x) &=& \frac{1}{4}x^2 + \frac{7}{128}x^4 \ , \label{a1} \\
Q_0(x) &=& \frac{1}{4}x^2 - \frac{1}{128}x^4\ , \label{a2}  \eey

\newpage

\begin{picture}(120,10)
\put(113,-30){$\Biggl\{$}
\end{picture}
\vspace{-5mm} \bey
\hspace{-14mm} P_1(x) &=& \frac{5}{16}x^3 \ , \label{a3} \\
\hspace{-14mm} Q_1(x) &=& \frac{1}{16}x^3 \ , \label{a4}  \eey

\begin{picture}(120,10)
\put(113,-30){$\Biggl\{$}
\end{picture}
\vspace{-5mm} \bey
\hspace{-14mm} P_2(x) &=& \frac{1}{32}x^4 \ , \label{a5} \\
\hspace{-14mm} Q_2(x) &=& \frac{1}{32}x^4 \ , \label{a6}  \eey

\begin{picture}(120,10)
\put(113,-30){$\Biggl\{$}
\end{picture}
\vspace{-5mm} \bey
\hspace{-12mm} P_3(x) &=& -\frac{1}{16}x^3 \ , \label{a7} \\
\hspace{-12mm} Q_3(x) &=& \frac{3}{16}x^3 \ , \label{a8}  \eey

\begin{picture}(120,10)
\put(113,-30){$\Biggl\{$}
\end{picture}
\vspace{-5mm} \bey
\hspace{-11mm} P_4(x) &=& -\frac{3}{128}x^4 \ , \label{a9} \\
\hspace{-11mm} Q_4(x) &=& \frac{5}{128}x^4 \ . \label{a10} \eey

The functions $P(x)$ and $Q(x)$, according to the definitions of Eq.
(\ref{eqnova423}) and Eq. (\ref{eqnova424}), assume then the
following specific forms: \bey P(x) &=& P_0(x) +
P_1(x) + P_2(x) + P_3(x) + P_4(x) \nonumber \\
&=& \frac{1}{4}x^2 + \frac{1}{4}x^3 + \frac{1}{16}x^4 \nonumber \\
&=& \frac{1}{16}x^2(x+2)^2 \ , \label{a11}
\\
Q(x) &=& Q_0(x) + Q_1(x) + Q_2(x) + Q_3(x) + Q_4(x) \nonumber \\
&=& \frac{1}{16}x^2(x+2)^2 = P(x) \ . \label{a12} \eey

For the evaluation of the equilibrium function $F^{(N)}(\eps,\dt)$
and of the constants $K_j^{(N)}(\eps,\dt)$ ($j = 1, 2, \ldots, N-1$)
and $K_0(\eps,\dt)$ that enter the making up of the partial flux of
order $N$, the formulae stated in Eqs. (\ref{eqnova431}),
(\ref{eqnova433}), (\ref{eqnova434}) and (\ref{eqnova436}) require
the knowledge of the derivatives of the two above functions at the
coordinate $x = \dt$. We have: \bey P'(\dt) &=& Q'(\dt) \nonumber \\
&=& \frac{1}{4}\dt(\dt + 1)(\dt + 2) \ , \label{a13} \eey where
primes denote derivatives with respect to the argument.

If instead of the relative Shafranov shift it is the displacement
variable the quantity favoured for playing the role of working
parameter along the analytical development of the solution, the
aforementioned formulae can still be used, provided that the
functions $V_j'(\dt)$ ($j = 1, 2, \ldots, N-1$), $P'(\dt)$ and
$Q'(\dt)$ that constitute the elements in the bottom rows of the
determinants in those formulae be first substituted by the functions
$V_j'(\chi)$, $P'(\chi)$ and $Q'(\chi)$, as explained in the
following of Eq. (\ref{eqnova442}). By having recourse to Eq.
(\ref{eqnova438}) to achieve the conversion of the representation in
terms of $\dt$ to a representation in terms of $\chi$, we obtain for
the functions $P$ and $Q$ proceeding from the particular solution:
\bey P(\chi) &\equiv& P(x = \dt(\chi)) \nonumber \\
&=& \frac{1}{16}\chi^2 \ , \label{a14} \\
Q(\chi) &\equiv& Q(x = \dt(\chi)) \nonumber \\
&=& \frac{1}{16}\chi^2 \ , \label{a15} \eey and thus \beq P'(\chi) =
Q'(\chi) = \frac{1}{8}\chi\ . \label{a16} \eeq

\newpage

\begin{center}
{\bf APPENDIX B: THE FUNCTIONS $M_{ij}(x)$ REQUIRED FOR THE
EVALUATION OF THE EQUILIBRIUM FUNCTION AND OF THE CONSTANTS THAT
MULTIPLY THE MULTIPOLE SOLUTIONS IN THE EXPRESSIONS OF THE PARTIAL
FLUXES OF ORDER $N=3$ TO ORDER $N=10$}
\end{center}

\setcounter{equation}{0}
\def\theequation{B.\arabic{equation}}

\vv

In this Appendix we shall denote the matrix element $M_{ij}(x)$ of
the main text by $M(i,j)$, the indices there appearing as arguments
here, and the argument $x$ there being suppressed here. \bey
M(0, 0) &=& 1\ . \label{eqb1} \\
M(0, 1) &=& {\displaystyle \frac {x^{2}}{8}} \ .
\label{eqb2} \\
M(1, 1) &=& {\displaystyle \frac {x}{2}} \ . \label{eqb3}
\\
M(2, 1) &=& {\displaystyle \frac {x^{2}}{8}} \ .
\label{eqb4} \\
 M(3,1) &=& M(4,1) = M(5,1) = M(6,1) = M(7,1) \nonumber \\
 &=& M(8,1) = M(9,1) = M(10,1) = M(11,1) \nonumber \\
 &=& M(12,1) = M(13,1) = M(14,1) = M(15,1) \nonumber \\
 &=& M(16,1) = M(17,1) = M(18,1) = 0\ .
\label{eqb5} \\
M(0, 2) &=& {\displaystyle \frac {x^{4}}{32}} \ .
\label{eqb6} \\
M(1, 2) &=&  - {\displaystyle \frac {x^{3}}{4}} \ .
\label{eqb7} \\
M(2, 2) &=&  - {\displaystyle \frac {1}{8}} \,x^{4} - x
^{2} \ . \label{eqb8} \\
M(3, 2) &=&  - {\displaystyle \frac {3\,x^{3}}{4}} \ .
\label{eqb9} \\
M(4, 2) &=&  - {\displaystyle \frac {5\,x^{4}}{32}} \ .
\label{eqb10} \\
M(5,2) &=& M(6,2) = \cdots = M(18,2) = 0 \ . \label{eqb11} \\
M(0, 3) &=& {\displaystyle \frac {x^{6}}{128}} \ .
\label{eqb12} \\
M(1, 3) &=&  - {\displaystyle \frac {x^{5}}{32}} \ .
\label{eqb13} \\
M(2, 3) &=&  - {\displaystyle \frac {5}{256}} \,x^{6}
 + {\displaystyle \frac {1}{4}} \,x^{4}
\ . \label{eqb14} \\
M(3, 3) &=& {\displaystyle \frac {15}{64}} \,x^{5} + x ^{3}
\ . \label{eqb15} \\
M(4, 3) &=& {\displaystyle \frac {7}{128}} \,x^{6} +
{\displaystyle \frac {5}{4}} \,x^{4} \ . \label{eqb16} \\
M(5, 3) &=& {\displaystyle \frac {35\,x^{5}}{64}} \ .
\label{eqb17} \\
M(6, 3) &=& {\displaystyle \frac {21\,x^{6}}{256}} \ .
\label{eqb18} \\
M(7,3) &=& M(8,3) = \cdots = M(18,3) = 0\ . \label{eqb19} \\
M(0, 4) &=& {\displaystyle \frac {x^{8}}{512}} \ .
\label{eqb20} \\
M(1, 4) &=&  - {\displaystyle \frac {x^{7}}{160}} \ .
\label{eqb21} \\
M(2, 4) &=&  - {\displaystyle \frac {7}{1600}} \,x^{8}
 + {\displaystyle \frac {1}{40}} \,x^{6}
\ . \label{eqb22} \\
M(3, 4) &=& {\displaystyle \frac {21}{800}} \,x^{7} -
{\displaystyle \frac {1}{5}} \,x^{5} \ . \label{eqb23} \\
M(4, 4) &=& {\displaystyle \frac {21}{3200}} \,x^{8} -
{\displaystyle \frac {7}{25}} \,x^{6} - {\displaystyle \frac {4}{
5}} \,x^{4} \ . \label{eqb24} \\
M(5, 4) &=&  - {\displaystyle \frac {21}{160}} \,x^{7}
 - {\displaystyle \frac {7}{5}} \,x^{5}
\ . \label{eqb25} \\
M(6, 4) &=&  - {\displaystyle \frac {33}{1600}} \,x^{8}
 - {\displaystyle \frac {189}{200}} \,x^{6}
\ . \label{eqb26} \\
M(7, 4) &=&  - {\displaystyle \frac {231\,x^{7}}{800}} \ .
\label{eqb27} \\
M(8, 4) &=&  - {\displaystyle \frac {429\,x^{8}}{12800} } \
. \label{eqb28} \\
M(9,4) &=& M(10,4) = \cdots =M(18,4) = 0 \ . \label{eqb29} \\
M(0, 5) &=& {\displaystyle \frac {x^{10}}{2048}} \ .
\label{eqb30} \\
M(1, 5) &=&  - {\displaystyle \frac {5\,x^{9}}{3584}} \ .
\label{eqb31} \\
M(2, 5) &=&  - {\displaystyle \frac {15}{14336}} \,x^{ 10}
+ {\displaystyle \frac {1}{224}} \,x^{8} \ . \label{eqb32} \\
M(3, 5) &=& {\displaystyle \frac {9}{1792}} \,x^{9} -
{\displaystyle \frac {1}{56}} \,x^{7} \ . \label{eqb33} \\
M(4, 5) &=& {\displaystyle \frac {33}{25088}} \,x^{10}
 - {\displaystyle \frac {3}{112}} \,x^{8} + {\displaystyle
\frac {1}{7}} \,x^{6} \ . \label{eqb34} \\
M(5, 5) &=&  - {\displaystyle \frac {165}{12544}} \,x^{ 9} +
{\displaystyle \frac {15}{56}} \,x^{7} + {\displaystyle \frac
{4}{7}} \,x^{5} \ . \label{eqb35} \\
M(6, 5) &=&  - {\displaystyle \frac {429}{200704}} \,x ^{10} +
{\displaystyle \frac {297}{1568}} \,x^{8} + {\displaystyle
\frac {9}{7}} \,x^{6} \ . \label{eqb36} \\
M(7, 5) &=& {\displaystyle \frac {429}{7168}} \,x^{9}
 + {\displaystyle \frac {33}{28}} \,x^{7}
\ . \label{eqb37} \\
M(8, 5) &=& {\displaystyle \frac {715}{100352}} \,x^{10 } +
{\displaystyle \frac {429}{784}} \,x^{8} \ . \label{eqb38} \\
M(9, 5) &=& {\displaystyle \frac {6435\,x^{9}}{50176}} \ .
\label{eqb39} \\
M(10, 5) &=& {\displaystyle \frac {2431\,x^{10}}{200704 }}
\ . \label{eqb40} \\
M(11,5) &=& M(12,5) = \cdots = M(18,5) = 0\ . \label{eqb41} \\
M(0, 6) &=& {\displaystyle \frac {x^{12}}{8192}} \ .
\label{eqb42} \\
M(1, 6) &=&  - {\displaystyle \frac {x^{11}}{3072}} \ .
\label{eqb43} \\
M(2, 6) &=&  - {\displaystyle \frac {11}{43008}} \,x^{ 12}
+ {\displaystyle \frac {5}{5376}} \,x^{10} \ . \label{eqb44} \\
M(3, 6) &=& {\displaystyle \frac {55}{50176}} \,x^{11}
 - {\displaystyle \frac {1}{336}} \,x^{9}
\ . \label{eqb45} \\
M(4, 6) &=& {\displaystyle \frac {715}{2408448}} \,x^{ 12} -
{\displaystyle \frac {11}{2352}} \,x^{10} + {\displaystyle \frac
{1}{84}} \,x^{8} \ . \label{eqb46} \\
M(5, 6) &=&  - {\displaystyle \frac {715}{301056}} \,x ^{11} +
{\displaystyle \frac {55}{2352}} \,x^{9} - {\displaystyle
\frac {2}{21}} \,x^{7} \ . \label{eqb47} \\
M(6, 6) &=&  - {\displaystyle \frac {715}{1806336}} \,x ^{12} +
{\displaystyle \frac {429}{25088}} \,x^{10} - {\displaystyle \frac
{11}{49}} \,x^{8} - {\displaystyle \frac {8 }{21}} \,x^{6} \ .
\label{eqb48} \\
M(7, 6) &=& {\displaystyle \frac {715}{129024}} \,x^{11 } -
{\displaystyle \frac {143}{672}} \,x^{9} - {\displaystyle \frac
{22}{21}} \,x^{7} \ . \label{eqb49} \\
M(8, 6) &=& {\displaystyle \frac {2431}{3612672}} \,x^{ 12} -
{\displaystyle \frac {715}{7056}} \,x^{10} - {\displaystyle \frac
{715}{588}} \,x^{8} \ . \label{eqb50} \\
M(9, 6) &=&  - {\displaystyle \frac {2431}{100352}} \,x ^{11} -
{\displaystyle \frac {3575}{4704}} \,x^{9} \ . \label{eqb51}
\\
M(10, 6) &=&  - {\displaystyle \frac {4199}{1806336}} \,x^{12} -
{\displaystyle \frac {60775}{225792}} \,x^{10} \ .
\label{eqb52} \\
M(11, 6) &=&  - {\displaystyle \frac {46189\,x^{11}}{
903168}} \ . \label{eqb53} \\
M(12, 6) &=&  - {\displaystyle \frac {4199\,x^{12}}{
1032192}} \ . \label{eqb54} \\
M (13,6) &=& M(14,6) = M(15,6) = M(16,6) = M(17,6) \nonumber \\
&=& M(18,6) = 0 \ . \label{eqb55} \\
M(0, 7) &=& {\displaystyle \frac {x^{14}}{32768}} \ .
\label{eqb56} \\
M(1, 7) &=&  - {\displaystyle \frac {7\,x^{13}}{90112} } \
. \label{eqb57} \\
M(2, 7) &=&  - {\displaystyle \frac {91}{1441792}} \,x ^{14} +
{\displaystyle \frac {7}{33792}} \,x^{12} \ . \label{eqb58}
\\
M(3, 7) &=& {\displaystyle \frac {91}{360448}} \,x^{13}
 - {\displaystyle \frac {5}{8448}} \,x^{11}
\ . \label{eqb59} \\
M(4, 7) &=& {\displaystyle \frac {455}{6488064}} \,x^{ 14} -
{\displaystyle \frac {65}{67584}} \,x^{12} + {\displaystyle \frac
{1}{528}} \,x^{10} \ . \label{eqb60} \\
M(5, 7) &=&  - {\displaystyle \frac {1625}{3244032}} \, x^{13} +
{\displaystyle \frac {65}{16896}} \,x^{11} - {\displaystyle
\frac {1}{132}} \,x^{9} \ . \label{eqb61} \\
M(6, 7) &=&  - {\displaystyle \frac {1105}{12976128}} \,x^{14} +
{\displaystyle \frac {65}{22528}} \,x^{12} - {\displaystyle \frac
{13}{704}} \,x^{10} + {\displaystyle \frac {
2}{33}} \,x^{8} \ . \label{eqb62} \\
M(7, 7) &=& {\displaystyle \frac {1547}{1622016}} \,x^{ 13} -
{\displaystyle \frac {455}{25344}} \,x^{11} + {\displaystyle \frac
{91}{528}} \,x^{9} + {\displaystyle \frac {8 }{33}} \,x^{7} \ .
\label{eqb63} \\
M(8, 7) &=& {\displaystyle \frac {4199}{35684352}} \,x ^{14} -
{\displaystyle \frac {221}{25344}} \,x^{12} + {\displaystyle \frac
{325}{1584}} \,x^{10} + {\displaystyle \frac {26}{33}} \,x^{8}
\ . \label{eqb64} \\
M(9, 7) &=&  - {\displaystyle \frac {4199}{1982464}} \, x^{13} +
{\displaystyle \frac {1105}{8448}} \,x^{11} +
{\displaystyle \frac {195}{176}} \,x^{9} \ . \label{eqb65} \\
M(10, 7) &=&  - {\displaystyle \frac {29393}{142737408} } \,x^{14} +
{\displaystyle \frac {104975}{2230272}} \,x^{12} +
{\displaystyle \frac {5525}{6336}} \,x^{10} \ . \label{eqb66} \\
M(11, 7) &=& {\displaystyle \frac {29393}{3244032}} \,x ^{13} +
{\displaystyle \frac {20995}{50688}} \,x^{11} \ .
\label{eqb67} \\
M(12, 7) &=& {\displaystyle \frac {52003}{71368704}} \, x^{14} +
{\displaystyle \frac {29393}{247808}} \,x^{12} \ .
\label{eqb68} \\
M(13, 7) &=& {\displaystyle \frac {676039\,x^{13}}{
35684352}} \ . \label{eqb69} \\
M(14, 7) &=& {\displaystyle \frac {185725\,x^{14}}{
142737408}} \ . \label{eqb70} \\
M(15,7) &=& M(16,7)  = M(17,7)  = M(18,7)  = 0\ . \label{eqb71} \\
M(0, 8) &=& {\displaystyle \frac {x^{16}}{131072}} \ .
\label{eqb72} \\
M(1, 8) &=&  - {\displaystyle \frac {x^{15}}{53248}} \ .
\label{eqb73} \\
M(2, 8) &=&  - {\displaystyle \frac {5}{319488}} \,x^{ 16}
+ {\displaystyle \frac {7}{146432}} \,x^{14} \ . \label{eqb74} \\
M(3, 8) &=& {\displaystyle \frac {35}{585728}} \,x^{15}
 - {\displaystyle \frac {7}{54912}} \,x^{13}
\ . \label{eqb75} \\
M(4, 8) &=& {\displaystyle \frac {119}{7028736}} \,x^{ 16} -
{\displaystyle \frac {35}{164736}} \,x^{14} + {\displaystyle \frac
{5}{13728}} \,x^{12} \ . \label{eqb76} \\
M(5, 8) &=&  - {\displaystyle \frac {595}{5271552}} \,x ^{15} +
{\displaystyle \frac {125}{164736}} \,x^{13} -
{\displaystyle \frac {1}{858}} \,x^{11} \ . \label{eqb77}\\
M(6, 8) &=&  - {\displaystyle \frac {2261}{115974144}} \,x^{16} +
{\displaystyle \frac {85}{146432}} \,x^{14} - {\displaystyle \frac
{5}{1716}} \,x^{12} + {\displaystyle \frac {
2}{429}} \,x^{10} \ . \label{eqb78} \\
M(7, 8) &=& {\displaystyle \frac {11305}{57987072}} \,x ^{15} -
{\displaystyle \frac {119}{41184}} \,x^{13} + {\displaystyle \frac
{35}{2574}} \,x^{11} - {\displaystyle \frac {16}{429}} \,x^{9}
\ . \label{eqb79} \\
M(8, 8) &=& {\displaystyle \frac {11305}{463896576}}   x^{16} -
{\displaystyle \frac {323}{226512}}  x^{14} + {\displaystyle \frac
{85}{5148}}  x^{12} - {\displaystyle \frac {160}{1287}}  x^{10} -
{\displaystyle \frac {64}{429}}  x ^{8} \ .
\label{eqb80} \\
M(9, 8) &=&  - {\displaystyle \frac {2261}{6443008}}   x^{15} +
{\displaystyle \frac {1615}{151008}}  x^{13} - {\displaystyle \frac
{51}{286}}  x^{11} - {\displaystyle \frac {
80}{143}}  x^{9} \ . \label{eqb81} \\
M(10, 8) &=&  - {\displaystyle \frac {52003}{1507663872 }}  x^{16} +
{\displaystyle \frac {56525}{14496768}}  x^{14} - {\displaystyle
\frac {8075}{56628}}  x^{12} - {\displaystyle \frac
{1190}{1287}}  x^{10} \ . \label{eqb82} \\
M(11, 8) &=& {\displaystyle \frac {52003}{68530176}}   x^{15} -
{\displaystyle \frac {11305}{164736}}  x^{13} -
{\displaystyle \frac {2261}{2574}}  x^{11} \ . \label{eqb83} \\
M(12, 8) &=& {\displaystyle \frac {185725}{3015327744} }  x^{16} -
{\displaystyle \frac {52003}{2617472}}  x^{14} -
{\displaystyle \frac {79135}{151008}}  x^{12} \ . \label{eqb84} \\
M(13, 8) &=&  - {\displaystyle \frac {185725}{57987072} }  x^{15} -
{\displaystyle \frac {364021}{1812096}}  x^{13} \ .
\label{eqb85} \\
M(14, 8) &=&  - {\displaystyle \frac {37145}{167518208} }  x^{16} -
{\displaystyle \frac {9100525}{188457984}}  x^{14} \ .
\label{eqb86} \\
M(15, 8) &=&  - {\displaystyle \frac {557175 x^{15}}{
83759104}} \ . \label{eqb87} \\
M(16, 8) &=&  - {\displaystyle \frac {1077205 x^{16}}{
2680291328}} \ . \label{eqb88} \\
M(17,8) &=& M(18,8) = 0 \ . \label{eqb89} \\
M(0, 9) &=& {\displaystyle \frac {x^{18}}{524288}} \ .
\label{eqb90} \\
M(1, 9) &=&  - {\displaystyle \frac {3 x^{17}}{655360} } \
. \label{eqb91} \\
M(2, 9) &=&  - {\displaystyle \frac {51}{13107200}}  x ^{18} +
{\displaystyle \frac {3}{266240}}  x^{16} \ . \label{eqb92}
\\
M(3, 9) &=& {\displaystyle \frac {153}{10649600}}  x^{ 17}
- {\displaystyle \frac {21}{732160}}  x^{15} \ . \label{eqb93} \\
M(4, 9) &=& {\displaystyle \frac {969}{234291200}}  x ^{18} -
{\displaystyle \frac {357}{7321600}}  x^{16} +
{\displaystyle \frac {7}{91520}}  x^{14} \ . \label{eqb94} \\
M(5, 9) &=&  - {\displaystyle \frac {6783}{257720320}}  x^{17} +
{\displaystyle \frac {119}{732160}}  x^{15} -
{\displaystyle \frac {1}{4576}}  x^{13} \ . \label{eqb95} \\
M(6, 9) &=&  - {\displaystyle \frac {47481}{10308812800 }}  x^{18} +
{\displaystyle \frac {20349}{161075200}}  x^{16}
 - {\displaystyle \frac {51}{91520}}  x^{14} \nonumber \\
 &&+ {\displaystyle
\frac {1}{1430}}  x^{12} \ . \label{eqb96} \\
M(7, 9) &=& {\displaystyle \frac {110789}{2577203200}}  x^{17} -
{\displaystyle \frac {2261}{4026880}}  x^{15} + {\displaystyle \frac
{119}{57200}}  x^{13} - {\displaystyle \frac
{2}{715}}  x^{11} \ . \label{eqb97} \\
M(8, 9) &=& {\displaystyle \frac {364021}{67007283200} }  x^{18} -
{\displaystyle \frac {2261}{8053760}}  x^{16} + {\displaystyle \frac
{323}{125840}}  x^{14} - {\displaystyle \frac
{34}{3575}}  x^{12} \nonumber \\
&&+ {\displaystyle \frac {16}{715}}  x ^{10} \ .
\label{eqb98} \\
M(9, 9) &=&  - {\displaystyle \frac {468027}{6700728320 }}  x^{17} +
{\displaystyle \frac {6783}{4026880}}  x^{15} - {\displaystyle \frac
{8721}{629200}}  x^{13} + {\displaystyle \frac
{306}{3575}}  x^{11} \nonumber \\
&&+ {\displaystyle \frac {64}{715}}  x ^{9} \ .
\label{eqb99} \\
M(10, 9) &=&  - {\displaystyle \frac {37145}{5360582656 }}  x^{18} +
{\displaystyle \frac {52003}{83759104}}  x^{16} - {\displaystyle
\frac {2261}{201344}}  x^{14} \nonumber \\
&&+ {\displaystyle \frac {2261}{15730}}  x^{12}+ {\displaystyle
\frac {272}{715}} x^{10}
\ . \label{eqb100} \\
M(11, 9) &=& {\displaystyle \frac {7429}{60915712}}  x ^{17} -
{\displaystyle \frac {52003}{9518080}}  x^{15} + {\displaystyle
\frac {15827}{114400}}  x^{13} + {\displaystyle
\frac {2584}{3575}}  x^{11} \ . \label{eqb101} \\
M(12, 9) &=& {\displaystyle \frac {66861}{6700728320}}  x^{18} -
{\displaystyle \frac {66861}{41879552}}  x^{16} +
{\displaystyle \frac {1092063}{13087360}}  x^{14} \nonumber \\
&&+ {\displaystyle
\frac {31654}{39325}}  x^{12} \ . \label{eqb102} \\
M(13, 9) &=&  - {\displaystyle \frac {66861}{257720320} }  x^{17} +
{\displaystyle \frac {52003}{1610752}}  x^{15} + {\displaystyle
\frac {364021}{629200}}  x^{13} \ . \label{eqb103}
\\
M(14, 9) &=&  - {\displaystyle \frac {1938969}{ 107211653120}}
 x^{18} + {\displaystyle \frac {3276189}{ 418795520}}  x^{16} +
{\displaystyle \frac {364021}{1308736}}
x^{14} \ . \label{eqb104} \\
M(15, 9) &=& {\displaystyle \frac {5816907}{5360582656} }  x^{17} +
{\displaystyle \frac {468027}{5234944}}  x^{15} \ .
\label{eqb105} \\
M(16, 9) &=& {\displaystyle \frac {3535767}{53605826560 }}  x^{18} +
{\displaystyle \frac {1938969}{104698880}}  x^{16} \ .
\label{eqb106} \\
M(17, 9) &=& {\displaystyle \frac {60108039 x^{17}}{
26802913280}} \ . \label{eqb107} \\
M(18, 9) &=& {\displaystyle \frac {1178589 x^{18}}{ 9746513920}} \ .
\label{eqb108} \eey

\newpage

\begin{center}
{\bf APPENDIX C: THE FUNCTIONS $V_j'(\dt)$ AND $V_j'(\chi)$ THAT
APPEAR AS ELEMENTS OF THE DETERMINANTS IN THE NUMERATORS AND
DENOMINATORS OF THE EXPRESSIONS FOR THE EQUILIBRIUM FUNCTIONS
$F^{(N)}(\eps, \dt)$ AND $F^{(N)}(\eps, \chi)$ RESPECTIVELY, AND IN
THE NUMERATOR AND DENOMINATOR OF THE EXPRESSION FOR THE CONSTANTS
$K_i^{(N)}$ THAT MULTIPLY THE MULTIPOLE SOLUTIONS $\vf^{(i)}(x,\ta)$
IN THE REPRESENTATIONS OF THE FLUX FUNCTIONS OF THE ORDERS $N = 3$
TO $N = 10$.}
\end{center}

\setcounter{equation}{0}
\def\theequation{C.\arabic{equation}}

\vv

The functions $V_j^{(N)}(x)$ are formally defined by Eq.
(\ref{eqnova427}), which, here reproduced, writes as:
$$V_j^{(N)}(x) = \sum_{n=0}^{2N-2} M_{nj}(x) \ \ \  (j = 1, 2, \ldots,
N-1). \eqno (4.28)$$

Considering that, according to Eq. (\ref{eqnova48}) the matrix
elements $M_{nj}$ in general vanish for $n$ greater than $2j$, the
definition of $V_j^{(N)}(x)$ can be restated as: \beq V_j^{(N)}(x) =
\sum_{n=0}^{2j}M_{nj}(x) \ \ \  (j = 1, 2, \ldots, N-1).
\label{eqc1} \eeq

The replacement of $2N-2$ by $2j$ as the upper limit of the
summation is permissible because, even for the least value of $N$,
which is $3$, all non-null matrix elements $M_{nj}(x)$ that are
summoned to contribute to the sum according to the original
definition of $V_j^{(N)}(x)$ are also taken into account by this
newest one. Thus, except for the specification of the range of
variation of $j$ and of the number $N-1$ of functions that are
implied by a given order $N$ of approximation, the definition of
$V_j^{(N)}(x)$ is actually independent of $N$, and, the same as for
the functions $P(x)$ and $Q(x)$ of Appendix A, we can omit the
superscript $N$ without incurring the risk that any ambiguity be
introduced into the meaning of the symbol that denotes the quantity
defined by Eq. (\ref{eqc1}). Another consequence of this lack of
dependence of $V_j^{(N)}(x)$ on $N$ is that, in passing from the
order $N$ to the order $N+1$ of approximation to the flux function,
there is need to carry out the calculation as new only of the
function $V_N^{(N+1)}(x)$, all the functions of lower subscripts,
$V_1^{(N+1)}(x)$, $V_2^{(N+1)}(x), \ldots$, $V_{N-1}^{(N+1)}(x)$,
remaining the same as those previously calculated for the order $N$,
to know, $V_1^{(N)}(x)$, $V_2^{(N)}(x), \ldots$, $V_{N-1}^{(N)}(x)$.
The functions listed in the present Appendix are sufficient in
number to compute the flux functions $\psi^{(N)}(x,\ta)$ up to the
order $N = 10$.

It may be helpful to observe that, adopting the pattern of notation
used in Appendix B to denote the matrix elements, the evaluation of
the sums to obtain the functions $V_j(x)$ according to the rule of
Eq. (\ref{eqc1}) can be most conveniently accomplished by use of
MAPLE's command \textbf{add}.

\vv

\nd \textit{(a) The functions $V_j(\dt)$ and their derivatives
$V_j'(\dt)$ for $1 \le j \le 9$}

\vv

The functions $V_j(\dt)$ can be obtained from the functions
$V_j(\chi)$ given in \textit{(b)} ahead by the replacement $\chi \to
\dt(\dt + 2)$. \bey V_1'(\dt) &=& \frac{1}{2}(\dt + 1)\ ,
\label{eqc11}
\\
V_2'(\dt) &=& -\dt(\dt+1)(\dt + 2)\ , \label{eqc12} \\
V_3'(\dt) &=& \frac{3}{4}\dt^2(\dt + 1)(\dt + 2)^2\ , \label{eqc13}
\\
V_4'(\dt) &=& -\frac{2}{5}\dt^3(\dt + 1)(\dt + 2)^3\ , \label{eqc14}
\\
V_5'(\dt) &=& \frac{5}{28}\dt^4(\dt + 1)(\dt + 2)^4\ , \label{eqc15}
\\
V_6'(\dt) &=& -\frac{1}{14}\dt^5(\dt + 1)(\dt + 2)^5\ ,
\label{eqc16}
\\
V_7'(\dt) &=& \frac{7}{264}\dt^6(\dt + 1)(\dt + 2)^6\ ,
\label{eqc17}
\\
V_8'(\dt) &=& -\frac{4}{429}\dt^7(\dt + 1)(\dt + 2)^7\ ,
\label{eqc18}
\\
V_9'(\dt) &=& \frac{9}{2860}\dt^8(\dt + 1)(\dt + 2)^8\ .
\label{eqc19} \eey

\vv

\nd \textit{(b) The functions $V_j(\chi)$ and their derivatives
$V_j'(\chi)$ for $1 \le j \le 9$}
\bey V_1(\chi) &=& \frac{1}{4}\chi\ , \label{eqc20} \\
V_2(\chi) &=& -\frac{1}{4}\chi^2\ , \label{eqc21} \\
V_3(\chi) &=& \frac{1}{8}\chi^3\ , \label{eqc22} \\
V_4(\chi) &=& -\frac{1}{20}\chi^4\ , \label{eqc23} \\
V_5(\chi) &=& \frac{1}{56}\chi^5\ , \label{eqc24} \\
V_6(\chi) &=& -\frac{1}{168}\chi^6\ , \label{eqc25} \\
V_7(\chi) &=& \frac{1}{528}\chi^7\ , \label{eqc26} \\
V_8(\chi) &=& -\frac{1}{1716}\chi^8\ , \label{eqc27} \\
V_9(\chi) &=& \frac{1}{5720}\chi^9\ , \label{eqc28} \eey \dotfill
\bey
V_1'(\chi) &=& \frac{1}{4}\ , \label{eqc29} \\
V_2'(\chi) &=& -\frac{1}{2}\chi\ , \label{eqc30} \\
V_3'(\chi) &=& \frac{3}{8}\chi^2\ , \label{eqc31} \\
V_4'(\chi) &=& -\frac{1}{5}\chi^3\ , \label{eqc32} \\
V_5'(\chi) &=& \frac{5}{56}\chi^4\ , \label{eqc33} \\
V_6'(\chi) &=& -\frac{1}{28}\chi^5\ , \label{eqc34} \\
V_7'(\chi) &=& \frac{7}{528}\chi^6\ , \label{eqc35} \\
V_8'(\chi) &=& -\frac{2}{429}\chi^7\ , \label{eqc36} \\
V_9'(\chi) &=& \frac{9}{5720}\chi^8\ . \label{eqc37} \eey

\newpage

\begin{center}
{\bf APPENDIX D: VALUES OF THE NORMALIZED MAGNETIC POLOIDAL FLUX AT
THE MAGNETIC AXIS AND VALUES OF THE NORMALIZED MAGNETIC POLOIDAL
FLUX AT THE CENTRE LINE OF THE TORUS FOR THE EQUILIBRIUM
CONFIGURATIONS WITH INVERSE ASPECT RATIO $\eps = 2/5$ AND VALUES OF
THE EQUILIBRIUM PARAMETER $\lb$ EQUAL TO 1,0 AND $-1/5$
RESPECTIVELY. EXPRESSIONS FOR THE NORMALIZED PARTIAL FLUX FUNCTIONS
OF THE ORDER $N=10$ IN NORMALIZED CYLINDRICAL COORDINATES $(\rho, Z,
\phi)$ FOR THE THREE REFERENCE EQUILIBRIA.}
\end{center}

\setcounter{equation}{0}
\def\theequation{D.\arabic{equation}}

\vv

\nd (a) The dimensionless normalized magnetic poloidal flux at the
magnetic axis is defined as: \beq \ovl\psi_M \equiv \frac{\psi(x =
\dt, \ta = 0)}{s_P}\ , \label{eqxxx} \eeq where $\psi(x,\ta)$ is
related to the dimensional magnetic poloidal flux $\Psi(x,\ta)$
through Eq. (\ref{eq2.3}). The dimensionless normalized magnetic
poloidal flux at the centre line of the torus is defined as: \beq
\hat{\ovl\psi}_C \equiv \frac{\ovl\psi(x = 0,\hbox{(independent of)
} \ta)}{(-\ovl\psi_M)} \ \hbox{ or } \ \frac{\ovl\psi(\rho=1, Z =
0)}{(-\ovl\psi_M)}\ , \label{eqyyy} \eeq where $\ovl\psi(x,\ta)$ is
related to the dimensionless magnetic poloidal flux $\psi(x,\ta)$
through Eq. (\ref{eq4.2nova}) and $\ovl\psi(\rho, Z)$ is its
representation in terms of the normalized cylindrical coordinates
$(\rho, Z)$. These definitions hold for any order $N$ of
approximation to the poloidal flux function. Table D.1 to follow
displays the numerical values for the two normalized fluxes
$\ovl\psi_M$ and $\hat{\ovl\psi}_C$ for the three reference
equilibria as evaluated by use of the approximation of order $N =
10$ to the poloidal flux function.

\vv

\begin{center}
\begin{tabular}{|c||c|c|} \hline
$\lb$ & $\ovl\psi_M$ & $\hat{\ovl\psi}_C$ \\ \hline \hline 1 &
$-0.08058778256$ & $-0.9773960062$ \\ \hline 0 & $-0.04180788277$ &
$-0.9422073888$ \\ \hline $-1/5$ & $-0.03421957455$ & $-0.9210167079$ \\
\hline
\end{tabular}

\vv

\begin{tabular}{lp{12cm}}
{\bf Table D.1} & Values of the normalized dimensionless flux at the
magnetic axis $\ovl\psi_M$ and values of the normalized
dimensionless flux at the centre line of the torus
$\hat{\ovl\psi}_C$ for the three reference equilibria according to
the approximation of order $N = 10$ to the flux function.
\end{tabular}
\end{center}

\vv

\nd (b) The normalized cylindrical coordinates $\rho$ and $Z$ are
defined in the following of Eq. (\ref{eq3.31}). In these coordinates
the expression for the normalized flux in the approximation of order
$N = 10$ writes in general as: \beq \hat{\ovl\psi}^{(10)}(\rho, Z,
\phi) = \calc_0 + \calc_2\rho^2 + \calc_4\rho^4 + \cdots +
\calc_{18}\rho^{18}\ . \label{eqd1} \eeq The coefficients $\calc_l$
($l = 0, 2, \ldots, 18$) are functions of $Z^2$ and depend on the
values of the displacement variable $\chi$ and of the squared
inverse aspect ratio $\eps^2$. The formulae for the $\calc_l$'s
given here apply all to $\eps = 2/5$.
\bey &&\hbox{\textit{(b.1)} $\lb = 1$} \nonumber \\
&& \nonumber \\
\calc_0 &=& {\displaystyle \frac {31202}{5029}Z^{2}} +
{\displaystyle \frac {6669}{4402}} \ , \label{eqd2}
\\
&& \nonumber \\
\calc_2 &=& - {\displaystyle \frac {245677}{260681}} Z^{16}
 - {\displaystyle \frac {62049}{7630}}  Z^{14} - {\displaystyle
\frac {299573}{9654}}  Z^{12} - {\displaystyle \frac {444275}{
6471}}  Z^{10} - {\displaystyle \frac {400731}{4133}}  Z^{8}
\nonumber \\
&& -{\displaystyle \frac {409266}{4519}}  Z^{6} - {\displaystyle
\frac {38252}{683}}  Z^{4} - {\displaystyle \frac {180008}{8747}}
Z^{2} - {\displaystyle \frac {83601}{14006}}
\ , \label{eqd3} \\
&& \nonumber \\
\calc_4 &=& {\displaystyle \frac {2326383}{82282}} Z^{14} +
{\displaystyle \frac {130801}{707}}
 Z^{12} + {\displaystyle \frac {649742}{1269}}  Z^{10} +
{\displaystyle \frac {487374}{ 631}}  Z^{8} + {\displaystyle \frac
{612877}{903}}  Z^{6} \nonumber \\
&& + {\displaystyle \frac {429281}{1264}}  Z^{4} + {\displaystyle
\frac {1042381}{12408}}  Z^{2} + {\displaystyle
\frac {173077}{25848}}  \ , \label{eqd4} \\
&& \nonumber \\
\calc_6 &=&  - {\displaystyle \frac {2818793}{13147}} Z^{12}
 - {\displaystyle \frac {2125655}{2089}}  Z^{10} -
{\displaystyle \frac {1067543}{556}}  Z^{8} - {\displaystyle \frac
{702869}{390}}  Z^{6} - {\displaystyle \frac {941713}{1110 }}  Z^{4}
\nonumber \\
&&- {\displaystyle \frac {509941}{3003}}  Z^{2} - {\displaystyle
\frac {9563}{1366}} \ , \label{eqd5} \\
&& \nonumber \\
\calc_8 &=& {\displaystyle \frac {3180978}{5395}} Z^{10} +
{\displaystyle \frac {8282194}{4341}}  Z^{8} + {\displaystyle \frac
{835538}{373}}  Z^{6} + {\displaystyle \frac {886471}{787} } Z^{4} +
{\displaystyle \frac {539364}{2543}}  Z^{2} \nonumber \\
&&+ {\displaystyle \frac{72219}{10207}} \ , \label{eqd6} \\
&& \nonumber \\
\calc_{10} &=&  - {\displaystyle \frac {847057}{1277}} Z^{8}
 - {\displaystyle \frac {1312826}{983}}  Z^{6} - {\displaystyle
\frac {4409255}{5249}}  Z^{4} - {\displaystyle \frac {263069}{
1557}}  Z^{2} - {\displaystyle \frac {53719}{10131}} \ ,
\label{eqd7} \\
&& \nonumber \\
\calc_{12} &=& {\displaystyle \frac {531185}{1716}} Z^{6} +
{\displaystyle \frac {912167}{2732}}  Z^{4} + {\displaystyle \frac
{92822}{1105}}  Z^{2} + {\displaystyle \frac {50147}{17808 }}\ ,
\label{eqd8} \\
&& \nonumber \\
\calc_{14} &=& - {\displaystyle \frac {81367}{1472}} Z^{4} -
{\displaystyle \frac {92557}{3881}}  Z^{2} - {\displaystyle \frac
{46411}{46410}} \ , \label{eqd9} \\
&& \nonumber \\
\calc_{16} &=& {\displaystyle \frac {50119}{16925}Z^{2}} +
{\displaystyle \frac {3987}{18724}}\ , \label{eqd10} \\
&& \nonumber \\
\calc_{18} &=& {\displaystyle \frac {-2795}{135916}}\ .
\label{eqd11} \\
&& \nonumber \\
&&\hbox{\textit{(b.2)} $\lb = 0$} \nonumber \\
&& \nonumber \\
\calc_0 &=&  - {\displaystyle \frac
{Z^{2}}{250156325}} + {\displaystyle \frac {32177}{39153}} \ ,
\label{eqd12} \\
&& \nonumber \\
\calc_2 &=& {\displaystyle \frac {19849}{44309}}
 Z^{16} + {\displaystyle \frac {37471}{9662}}  Z^{14} +
{\displaystyle \frac {1084410}{72947}}  Z^{12} + {\displaystyle
\frac {62779}{ 1896}}  Z^{10} + {\displaystyle \frac {93829}{1985}}
 Z^{8} +
{\displaystyle \frac {216053}{4795}}  Z^{6} \nonumber \\
&& + {\displaystyle \frac {138829}{4735}}  Z^{4} + {\displaystyle
\frac {185449}{10569}}  Z^{2} - {\displaystyle \frac {59431}{27950}}
\ , \label{eqd13} \\
&& \nonumber \\
\calc_4 &=&  - {\displaystyle \frac {54885}{4084}} Z^{14} -
{\displaystyle \frac {519755}{5891}}
 Z^{12} - {\displaystyle \frac {222473}{907}}  Z^{10} -
{\displaystyle \frac {470470}{ 1263}}  Z^{8} - {\displaystyle \frac
{4223723}{12765}}  Z^{6} \nonumber \\
&& - {\displaystyle \frac {368180}{2179}}  Z^{4} - {\displaystyle
\frac {157535}{3582}}  Z^{2} - {\displaystyle
\frac {38186}{27339}}  \ , \label{eqd14} \\
&& \nonumber \\
\calc_6 &=& {\displaystyle \frac {190169}{1866}}
 Z^{12} + {\displaystyle \frac {1037966}{2139}}  Z^{10} +
{\displaystyle \frac {2037394}{2215}}  Z^{8} + {\displaystyle \frac
{345061}{ 397}}  Z^{6} + {\displaystyle \frac {656803}{1588}}
 Z^{4} \nonumber \\
 &&+ {\displaystyle \frac {549651}{6506}}  Z^{2}
+ {\displaystyle \frac {20861}{5692}}  \ , \label{eqd15} \\
&& \nonumber \\
\calc_8 &=&  - {\displaystyle \frac {279419}{997}} Z^{10} -
{\displaystyle \frac {2915185}{3204}}  Z^{8} - {\displaystyle \frac
{1607533}{1498}}  Z^{6} - {\displaystyle \frac {620371}{ 1142}}
 Z^{4} - {\displaystyle \frac {105986}{1025}}  Z^{2} \nonumber \\
 && - {\displaystyle \frac {79295}{22526}} \ , \label{eqd16} \\
&& \nonumber \\
\calc_{10} &=& {\displaystyle \frac {826381}{2621}} Z^{8} +
{\displaystyle \frac {647091}{1016}}  Z^{6} + {\displaystyle \frac
{459161}{1141}}  Z^{4} + {\displaystyle \frac {120679}{ 1481}}
 Z^{2} + {\displaystyle \frac {36560}{14143}}
\ , \label{eqd17} \\
&& \nonumber \\
\calc_{12} &=&  - {\displaystyle \frac {334441}{2273}} Z^{6}
 - {\displaystyle \frac {350773}{2203}}  Z^{4} - {\displaystyle
\frac {590873}{14683}}  Z^{2} - {\displaystyle \frac {34775}{
25606}} \ , \label{eqd18} \\
&& \nonumber \\
\calc_{14} &=& {\displaystyle \frac {142407}{5420}} Z^{4} +
{\displaystyle \frac {148159}{13027}}  Z^{2} + {\displaystyle \frac
{50073}{104521}} \ , \label{eqd19} \\
&& \nonumber \\
\calc_{16} &=&  - {\displaystyle \frac {99455}{70658}Z^{2}}
 - {\displaystyle \frac {13315}{131122}}
\ , \label{eqd20} \\
&& \nonumber \\
\calc_{18} &=& {\displaystyle \frac {6100}{624061}} \ .\label{eqd21}
\\
&& \nonumber \\
&&\hbox{\textit{(b.3)} $\lb = -1/5$} \nonumber \\
&& \nonumber \\
\calc_0 &=& - {\displaystyle \frac {25012}{8559}Z^{2}} +
{\displaystyle \frac {57683}{117405}} \ ,
\label{eqd22} \\
&& \nonumber \\
\calc_2 &=& {\displaystyle \frac {46241}{42012}}
 Z^{16} + {\displaystyle \frac {107618}{11309}}  Z^{14} +
{\displaystyle \frac {30075}{826}}  Z^{12} + {\displaystyle \frac
{303631}{3754 }}  Z^{10} + {\displaystyle \frac {355945}{3096}}
 Z^{8} \nonumber \\
&& + {\displaystyle \frac {337021}{3100}}  Z^{6} + {\displaystyle
\frac {450663}{6497}}  Z^{4} + {\displaystyle \frac {172946}{4883}}
Z^{2} - {\displaystyle \frac
{12927}{42241}}  \ , \label{eqd23} \\
&& \nonumber \\
\calc_4 &=& - {\displaystyle \frac {523959}{15868}} Z^{14}
 - {\displaystyle \frac {206317}{953}}  Z^{12} - {\displaystyle
\frac {613388}{1021}}  Z^{10} - {\displaystyle \frac {856237}{ 941}}
 Z^{8} - {\displaystyle \frac {228559}{284}}  Z^{6} \nonumber \\
&& - {\displaystyle \frac {2245946}{5509}}  Z^{4} - {\displaystyle
\frac {156591}{1505}}  Z^{2} -
{\displaystyle \frac {72781}{13992}}  \ , \label{eqd24} \\
&& \nonumber \\
\calc_6 &=& {\displaystyle \frac {278195}{1111}}
 Z^{12} + {\displaystyle \frac {2111123}{1773}}  Z^{10} +
{\displaystyle \frac {3706011}{1645}}  Z^{8} + {\displaystyle \frac
{1895975}{ 893}}  Z^{6} + {\displaystyle \frac {544236}{541}}
 Z^{4} \nonumber \\
&& + {\displaystyle \frac {268054}{1315}}  Z^{2}
+ {\displaystyle \frac {62489}{7207}}  \ , \label{eqd25} \\
&& \nonumber \\
\calc_8 &=& - {\displaystyle \frac {647974}{941}} Z^{10} -
{\displaystyle \frac {1308289}{586}}  Z^{8} - {\displaystyle \frac
{1235337}{470}}  Z^{6} - {\displaystyle \frac {578559}{436 }}
 Z^{4} - {\displaystyle \frac {462500}{1839}}  Z^{2} \nonumber \\
&&- {\displaystyle \frac {108003}{12716}} \ , \label{eqd26} \\
&& \nonumber \\
\calc_{10} &=& {\displaystyle \frac {232403}{300}} Z^{8} +
{\displaystyle \frac {854853}{547}}  Z^{6} + {\displaystyle \frac
{2027464}{2057}}  Z^{4} + {\displaystyle \frac {371618}{ 1867}}
 Z^{2} + {\displaystyle \frac {79592}{12659}}
\ , \label{eqd27} \\
&& \nonumber \\
\calc_{12} &=&  - {\displaystyle \frac {1031043}{2852}} Z^{6}
 - {\displaystyle \frac {1100213}{2816}}  Z^{4} -
{\displaystyle \frac {254394}{2581}}  Z^{2} - {\displaystyle \frac
{67984}{20493}} \ , \label{eqd28} \\
&& \nonumber \\
\calc_{14} &=& {\displaystyle \frac {198059}{3068}} Z^{4} +
{\displaystyle \frac {113359}{4062}}  Z^{2} + {\displaystyle \frac
{43360}{36953}} \ , \label{eqd29} \\
&& \nonumber \\
\calc_{16} &=& - {\displaystyle \frac {245199}{70900}Z^{2}}
 - {\displaystyle \frac {17064}{68483}}
\ , \label{eqd30} \\
&& \nonumber \\
\calc_{18} &=& {\displaystyle \frac {605}{25191}} \ . \label{eqd31}
\eey

Transformation from the normalized cylindrical coordinates $(\rho,
Z, \phi)$ to the normalized toroidal-polar coordinates $(x, \ta,
\phi)$ can be accomplished by means of the relations: \beq
\left.\begin{array}{l}
\rho = 1+x\cos\ta\ , \\
\\
Z = x\sin\ta\ , \\
\\
\phi = \phi\ . \end{array} \right\} \label{eqd32} \eeq

\vv\vv\vv\vv

\begin{center}
{\bf APPENDIX E: EQUATION FOR THE CRITICAL VALUE OF THE DISPLACEMENT
VARIABLE FOR AXISYMMETRIC EQUILIBRIA OF INVERSE ASPECT RATIO 2/5
UNDER THE APPROXIMATION OF ORDER $N=4$ TO THE FLUX FUNCTION}
\end{center}

\setcounter{equation}{0}
\def\theequation{E.\arabic{equation}}

\vv

The equation is: \bey \chi ^{18} &&\hspace{-5mm}- 119.6972128\chi
^{17} + 64.65882500\chi
^{16} + 6630.186141\chi ^{15} + 1633.052456\chi ^{14} \nonumber \\
&&\hspace{-5mm} - 20702.02206\chi ^{13} + 5737.175591\chi ^{12} +
15141.43088\chi ^{11} - 7192.561804\chi ^{10} \nonumber  \\
&&\hspace{-5mm} - 3778.002560\chi ^{9} + 2659.627640\chi ^{8} +
166.2328382\chi ^{7} - 378.2158119\chi ^{6} \nonumber  \\
&&\hspace{-5mm} + 48.19578700\chi ^{5} + 17.57526666\chi ^{4} -
5.141304959\chi ^{3} + 0.3449459767\chi ^{2} \nonumber  \\
&&\hspace{-5mm} + 0.01733079721\chi  - 0.002048736713=0 \ .
\label{eqe1}\eey

The root with physical significance is: \beq \chi \equiv
\chi_c^{(4)} = 0.2543144486\ . \label{eqe2}
\eeq


\newpage

\begin{center}
{\bf ACKNOWLEDGEMENTS}
\end{center}

\vv

The author is grateful to Professor S. W. Song and to Professor I.
Simon (deceased) for permission to use the computational facilities
and the working station of the Instituto de Matem\'atica e Estat\'\i
stica of Universidade de S\~ao Paulo where this research started.

The author benefited from the tutorials by A. Lymberopoulos on the
use of the program Maple and from the assistance of C. E. Telles de
Menezes on multiple questions relative to the handling of the
computer during the development of this work.

\vv\vv

\begin{center}
{\bf DEDICATION}
\end{center}

\vv

This paper is dedicated to the memory of Professor Gumercindo Lima
who, as professor of the author in high school, taught him the
principles of linear algebra that are applied in the theory here
presented.

\vv\vv

\end{document}